\pdfoutput=1

\documentclass[12pt,a4paper]{article}

\usepackage{ifthen} 
\newboolean{pdflatex}
\setboolean{pdflatex}{true} 

\newboolean{articletitles}
\setboolean{articletitles}{true} 

\newboolean{uprightparticles}
\setboolean{uprightparticles}{false} 

\def\paperauthors{LHCb collaboration} 
\def\paperasciititle{Angular analysis of B0->K*0e+e+ decays} 
\def\papertitle{Angular analysis of $B^0\rightarrow K^{*0}e^{+}e^{-}$ decays} 
\def\paperkeywords{{High Energy Physics}, {LHCb}} 
\def\papercopyright{\the\year\ CERN for the benefit of the LHCb collaboration} 
\def\paperlicence{CC BY 4.0 licence}
\def\paperlicenceurl{https://creativecommons.org/licenses/by/4.0/}

\newif\ifEnableSectionTOCLinks
\EnableSectionTOCLinksfalse 

\usepackage[top=1in, bottom=1.25in, left=1in, right=1in]{geometry}

\usepackage{microtype}
\usepackage{lineno}  
\usepackage{xspace} 
                   
\usepackage{caption}

\usepackage{graphicx}  
\usepackage{color}
\usepackage{colortbl}
\graphicspath{{./figs/}} 

\usepackage{amsmath} 
\usepackage{amssymb}
\usepackage{amsfonts}
\usepackage{upgreek}

\newcommand*\patchAmsMathEnvironmentForLineno[1]{
\expandafter\let\csname old#1\expandafter\endcsname\csname #1\endcsname
\expandafter\let\csname oldend#1\expandafter\endcsname\csname
end#1\endcsname
 \renewenvironment{#1}
   {\linenomath\csname old#1\endcsname}
   {\csname oldend#1\endcsname\endlinenomath}
}
\newcommand*\patchBothAmsMathEnvironmentsForLineno[1]{
  \patchAmsMathEnvironmentForLineno{#1}
  \patchAmsMathEnvironmentForLineno{#1*}
}
\AtBeginDocument{
\patchBothAmsMathEnvironmentsForLineno{equation}
\patchBothAmsMathEnvironmentsForLineno{align}
\patchBothAmsMathEnvironmentsForLineno{flalign}
\patchBothAmsMathEnvironmentsForLineno{alignat}
\patchBothAmsMathEnvironmentsForLineno{gather}
\patchBothAmsMathEnvironmentsForLineno{multline}
\patchBothAmsMathEnvironmentsForLineno{eqnarray}
}

\usepackage{hyperxmp}

\usepackage[pdftex,
            pdfauthor={\paperauthors},
            pdftitle={\paperasciititle},
            pdfkeywords={\paperkeywords},
            pdfcopyright={Copyright (C) \papercopyright},
            pdflicenseurl={\paperlicenceurl}]{hyperref}

\usepackage[colorinlistoftodos,textsize=scriptsize]{todonotes}

\usepackage[bottom,flushmargin,hang,multiple]{footmisc}

\usepackage[all]{hypcap} 

\usepackage{xspace} 
\usepackage{upgreek}

\def\lhcb   {\mbox{LHCb}\xspace}

\def\MagUp {\mbox{\em Mag\kern -0.05em Up}\xspace}

\ifthenelse{\boolean{uprightparticles}}%
{

 \def\Pmu         {\ensuremath{\upmu}\xspace}

 \def\Ppi         {\ensuremath{\uppi}\xspace}

 \def\Ppsi        {\ensuremath{\uppsi}\xspace}

 \def\PDelta      {\ensuremath{\Delta}\xspace}                 
 \def\PXi         {\ensuremath{\Xi}\xspace}                 
 \def\PLambda     {\ensuremath{\Lambda}\xspace}                 
 \def\PSigma      {\ensuremath{\Sigma}\xspace}                 
 \def\POmega      {\ensuremath{\Omega}\xspace}                 
 \def\PUpsilon    {\ensuremath{\Upsilon}\xspace}
 \let\oldPi\Pi
 \def\PPi         {\ensuremath{\oldPi}\xspace}

 \def\PB      {\ensuremath{\mathrm{B}}\xspace}                 
 \def\PD      {\ensuremath{\mathrm{D}}\xspace}                 
 \def\PJ      {\ensuremath{\mathrm{J}}\xspace}                 
 \def\PK      {\ensuremath{\mathrm{K}}\xspace}                 
 \def\Pb      {\ensuremath{\mathrm{b}}\xspace}

 \def\Pe      {\ensuremath{\mathrm{e}}\xspace}                 
 \def\Ps      {\ensuremath{\mathrm{s}}\xspace}

 \def\thebaroffset{0.0em}
}
{

 \def\Pmu         {\ensuremath{\mu}\xspace}

 \def\Ppi         {\ensuremath{\pi}\xspace}

 \def\Ppsi        {\ensuremath{\psi}\xspace}                 
                  
 \mathchardef\PDelta="7101
 \mathchardef\PXi="7104
 \mathchardef\PLambda="7103
 \mathchardef\PSigma="7106
 \mathchardef\POmega="710A
 \mathchardef\PUpsilon="7107
 \mathchardef\PPi="7105
 \def\PB      {\ensuremath{B}\xspace}                 
 \def\PD      {\ensuremath{D}\xspace}                 
 \def\PJ      {\ensuremath{J}\xspace}                 
 \def\PK      {\ensuremath{K}\xspace}                 
 \def\Pb      {\ensuremath{b}\xspace}

 \def\Pe      {\ensuremath{e}\xspace}                 
 \def\Ps      {\ensuremath{s}\xspace}

 \def\thebaroffset{0.18em}
}
\newcommand{\offsetoverline}[2][\thebaroffset]{\kern #1\overline{\kern -#1 #2}}%

\makeatletter
\ifcase \@ptsize \relax
  \newcommand{\miniscule}{\@setfontsize\miniscule{4}{5}}
\or
  \newcommand{\miniscule}{\@setfontsize\miniscule{5}{6}}
\or
  \newcommand{\miniscule}{\@setfontsize\miniscule{5}{6}}
\fi
\makeatother

\DeclareRobustCommand{\optbar}[1]{\shortstack{{\miniscule (\rule[.5ex]{1.25em}{.18mm})}
  \\ [-.7ex] $#1$}}

\def\ep         {{\ensuremath{\Pe^+}}\xspace}

\def\epem       {{\ensuremath{\Pe^+\Pe^-}}\xspace}

\def\mup        {{\ensuremath{\Pmu^+}}\xspace}
\def\mun        {{\ensuremath{\Pmu^-}}\xspace} 

\def\squark    {{\ensuremath{\Ps}}\xspace}

\def\bquark    {{\ensuremath{\Pb}}\xspace}

\def\pion   {{\ensuremath{\Ppi}}\xspace}

\def\pip    {{\ensuremath{\pion^+}}\xspace}
\def\pim    {{\ensuremath{\pion^-}}\xspace}

\def\kaon    {{\ensuremath{\PK}}\xspace}
\def\Kbar    {{\ensuremath{\offsetoverline{\PK}}}\xspace}

\def\KorKbar {\kern \thebaroffset\optbar{\kern -\thebaroffset \PK}{}\xspace}

\def\Kp      {{\ensuremath{\kaon^+}}\xspace}
\def\Km      {{\ensuremath{\kaon^-}}\xspace}

\def\Kstarz  {{\ensuremath{\kaon^{*0}}}\xspace}
\def\Kstarzb {{\ensuremath{\Kbar{}^{*0}}}\xspace}
\def\Kstar   {{\ensuremath{\kaon^*}}\xspace}

\def\Dbar    {{\ensuremath{\offsetoverline{\PD}}}\xspace}
\def\D       {{\ensuremath{\PD}}\xspace}

\def\DorDbar {\kern \thebaroffset\optbar{\kern -\thebaroffset \PD}\xspace}

\def\Dzb     {{\ensuremath{\Dbar{}^0}}\xspace}
\def\Dp      {{\ensuremath{\D^+}}\xspace}
\def\Dm      {{\ensuremath{\D^-}}\xspace}

\def\DpDm    {\ensuremath{\Dp {\kern -0.16em \Dm}}\xspace}

\def\B       {{\ensuremath{\PB}}\xspace}
\def\Bbar    {{\ensuremath{\offsetoverline{\PB}}}\xspace}

\def\BorBbar {\kern \thebaroffset\optbar{\kern -\thebaroffset \PB}\xspace}
\def\Bz      {{\ensuremath{\B^0}}\xspace}
\def\Bzb     {{\ensuremath{\Bbar{}^0}}\xspace}
\def\Bd      {{\ensuremath{\B^0}}\xspace}

\def\BdorBdbar {\kern \thebaroffset\optbar{\kern -\thebaroffset \Bd}\xspace}
\def\Bu      {{\ensuremath{\B^+}}\xspace}

\def\Bs      {{\ensuremath{\B^0_\squark}}\xspace}

\def\BsorBsbar {\kern \thebaroffset\optbar{\kern -\thebaroffset \Bs}\xspace}

\def\jpsi     {{\ensuremath{{\PJ\mskip -3mu/\mskip -2mu\Ppsi}}}\xspace}

\def\Y#1S{\ensuremath{\PUpsilon{(#1S)}}\xspace}

\def\Lz          {{\ensuremath{\PLambda}}\xspace}

\def\LorLbar     {\kern \thebaroffset\optbar{\kern -\thebaroffset \PLambda}\xspace}

\def\Lb           {{\ensuremath{\Lz^0_\bquark}}\xspace}

\newcommand{\decay}[2]{\mbox{\ensuremath{#1\!\to #2}}\xspace} 

\def\to                 {\ensuremath{\rightarrow}\xspace}

\def\CP                {{\ensuremath{C\!P}}\xspace}

\def\BdKstee  {\decay{\Bd}{\Kstarz\epem}}

\def\bsll     {\decay{\bquark}{\squark \ell^+ \ell^-}}
\def\AFB      {\ensuremath{A_{\mathrm{FB}}}\xspace}
\def\FL       {\ensuremath{F_{\mathrm{L}}}\xspace}
\def\AT#1     {\ensuremath{A_{\mathrm{T}}^{#1}}\xspace}           

\def\ctl       {\ensuremath{\cos{\theta_\ell}}\xspace}
\def\ctk       {\ensuremath{\cos{\theta_K}}\xspace}

\def\C#1      {\ensuremath{\mathcal{C}_{#1}}\xspace}                       
\def\Cp#1     {\ensuremath{\mathcal{C}_{#1}^{'}}\xspace}                    
\def\Ceff#1   {\ensuremath{\mathcal{C}_{#1}^{\mathrm{(eff)}}}\xspace}        
\def\Cpeff#1  {\ensuremath{\mathcal{C}_{#1}^{'\mathrm{(eff)}}}\xspace}       
\def\Ope#1    {\ensuremath{\mathcal{O}_{#1}}\xspace}                       
\def\Opep#1   {\ensuremath{\mathcal{O}_{#1}^{'}}\xspace}                    

\newcommand{\nospaceunit}[1]{\ensuremath{\text{#1}}}       
\newcommand{\aunit}[1]{\ensuremath{\text{\,#1}}}       

\newcommand{\tev}{\aunit{Te\kern -0.1em V}\xspace}
\newcommand{\gev}{\aunit{Ge\kern -0.1em V}\xspace}
\newcommand{\mev}{\aunit{Me\kern -0.1em V}\xspace}
\newcommand{\kev}{\aunit{ke\kern -0.1em V}\xspace}
\newcommand{\ev}{\aunit{e\kern -0.1em V}\xspace}
 
\newcommand{\mevc}{\ensuremath{\aunit{Me\kern -0.1em V\!/}c}\xspace}
\newcommand{\gevc}{\ensuremath{\aunit{Ge\kern -0.1em V\!/}c}\xspace}
\newcommand{\mevcc}{\ensuremath{\aunit{Me\kern -0.1em V\!/}c^2}\xspace}
\newcommand{\gevcc}{\ensuremath{\aunit{Ge\kern -0.1em V\!/}c^2}\xspace}
\newcommand{\gevgevcccc}{\ensuremath{\gev^2\!/c^4}\xspace} 

\def\mum  {\ensuremath{\,\upmu\nospaceunit{m}}\xspace}

\def\fb   {\ensuremath{\aunit{fb}}\xspace}
\def\invfb   {\ensuremath{\fb^{-1}}\xspace}

\def\gsim{{~\raise.15em\hbox{$>$}\kern-.85em
          \lower.35em\hbox{$\sim$}~}\xspace}
\def\lsim{{~\raise.15em\hbox{$<$}\kern-.85em
          \lower.35em\hbox{$\sim$}~}\xspace}

\def\pt         {\ensuremath{p_{\mathrm{T}}}\xspace}

\def\tell1  {TELL1\xspace}
\def\ukl1   {UKL1\xspace}

\newcommand{\lhcborcid}[1]{\href{https://orcid.org/#1}{\hspace*{0.1em}\raisebox{-0.45ex}{\includegraphics[width=1em]{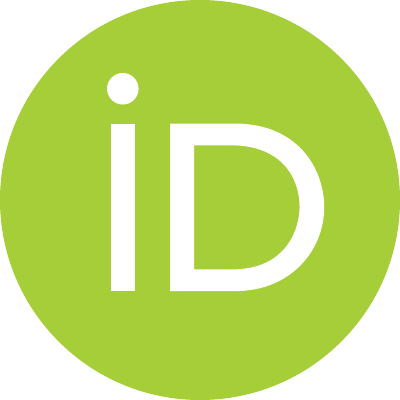}}}}

\hypersetup{
  colorlinks   = true, 
  urlcolor     = blue,
  linkcolor    = blue, 
  citecolor    = red   
}

\ifEnableSectionTOCLinks
    \usepackage[explicit]{titlesec} 
    
    \let\oldcontentsline\contentsline
    \renewcommand

    \titleformat{\section}{\normalfont\Large\bf}{\hyperlink{tocsection.\thesection}{{\thesection} \parbox[t]{\dimexpr\textwidth-1pc}{#1}}}{1pc}{}

    \titleformat{\subsection}{\normalfont\bf}{\hyperlink{tocsubsection.\thesubsection}{{\thesubsection} \parbox[t]{\dimexpr\textwidth-1pc}{#1}}}{1pc}{}

    \titleformat{name=\section,numberless}[display]{}{}{0pt}{\normalfont\Huge\bfseries #1}
\fi

\usepackage{cite} 
\usepackage{mciteplus}

\usepackage{longtable} 
\usepackage{siunitx}

\begin{document}

\renewcommand{\thefootnote}{\fnsymbol{footnote}}
\setcounter{footnote}{1}

\begin{titlepage}
\pagenumbering{roman}

\vspace*{-1.5cm}
\centerline{\large EUROPEAN ORGANIZATION FOR NUCLEAR RESEARCH (CERN)}
\vspace*{1.5cm}
\noindent
\begin{tabular*}{\linewidth}{lc@{\extracolsep{\fill}}r@{\extracolsep{0pt}}}
\ifthenelse{\boolean{pdflatex}}
{\vspace*{-1.5cm}\mbox{\!\!\!\includegraphics[width=.14\textwidth]{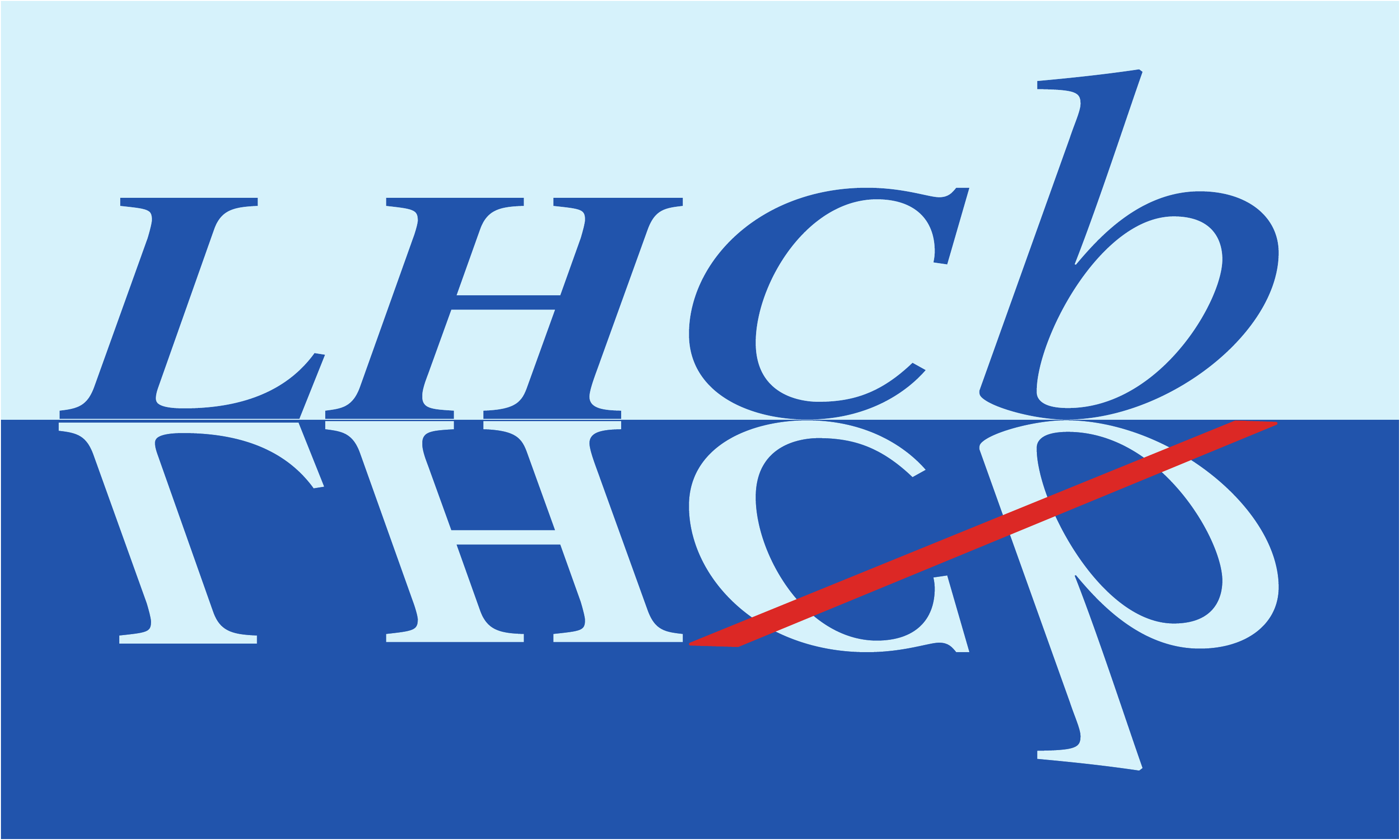}} & &}%
{\vspace*{-1.2cm}\mbox{\!\!\!\includegraphics[width=.12\textwidth]{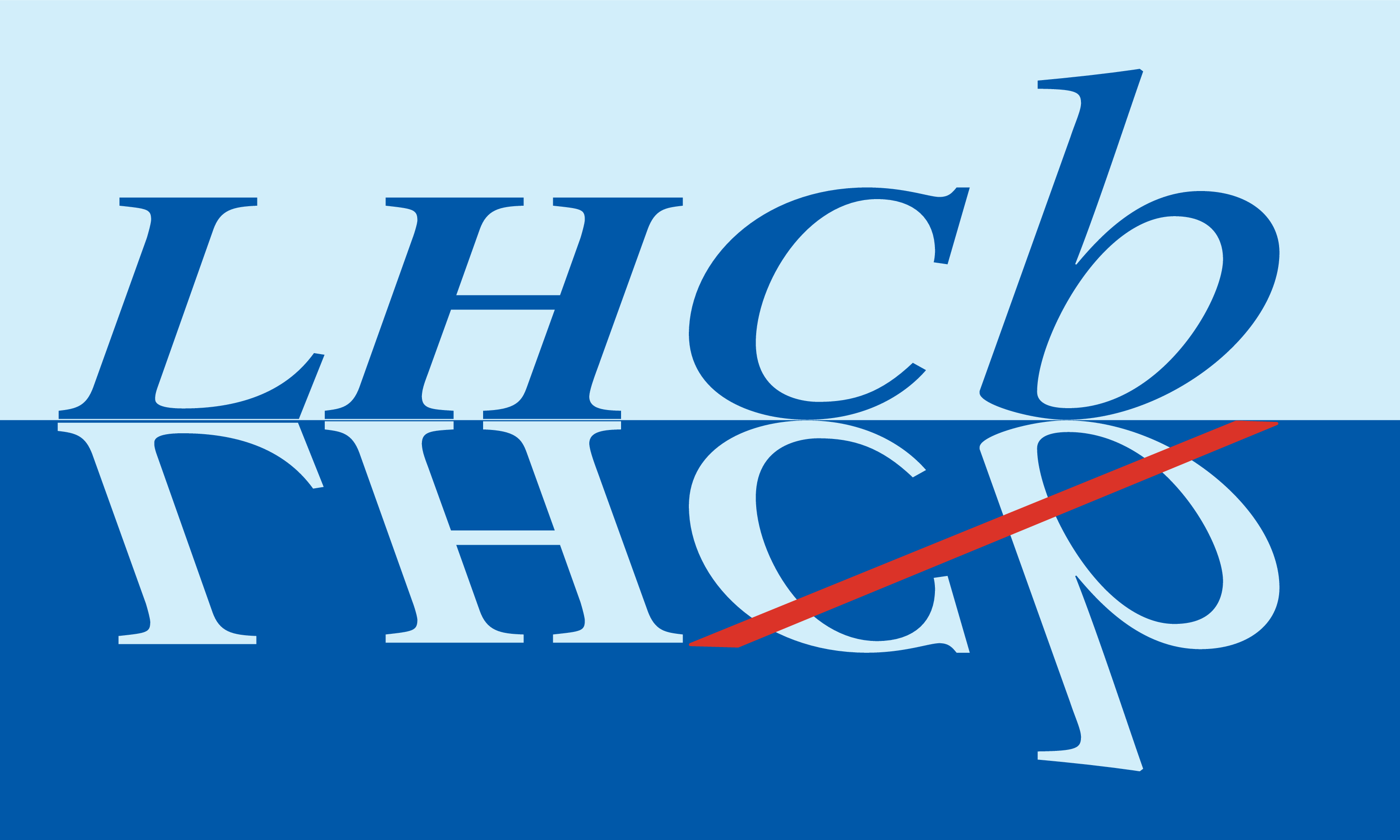}} & &}%
\\
 & & CERN-EP-2025-001 \\ 
 & & LHCb-PAPER-2024-022 \\ 
 & & September 08, 2025 \\
 & & \\
\end{tabular*}

\vspace*{4.0cm}

{\normalfont\bfseries\boldmath\huge
\begin{center}
  \papertitle 
\end{center}
}

\vspace*{2.0cm}

\begin{center}
\paperauthors\footnote{Authors are listed at the end of this paper.}
\end{center}

\vspace{\fill}

\begin{abstract}
  \noindent
  An angular analysis of \BdKstee decays is presented using proton-proton collision data collected by the $\lhcb$ experiment at centre-of-mass energies of 7, 8 and 13\tev, corresponding to an integrated luminosity of $9\,\invfb$.
  The analysis is performed in the region of the dilepton invariant mass squared of 1.1--6.0\gevgevcccc. 
  In addition, a test of lepton flavour universality is performed by comparing the obtained angular observables with those measured in $\Bz \to \Kstarz \mu^+\mu^-$ decays.
  In general, the angular observables are found to be consistent with the Standard Model expectations as well as with global analyses of other $ b \to s \ell^+ \ell^-$ processes, where $\ell$ is either a muon or an electron.    
  No sign of lepton-flavour-violating effects is observed. 
  
\end{abstract}

\vspace*{2.0cm}

\begin{center}
  Published in JHEP 06 (2025) 140
\end{center}

\vspace{\fill}

{\footnotesize 
\centerline{\copyright~\papercopyright. \href{\paperlicenceurl}{\paperlicence}.}}
\vspace*{2mm}

\end{titlepage}

\newpage
\setcounter{page}{2}
\mbox{~}

\renewcommand{\thefootnote}{\arabic{footnote}}
\setcounter{footnote}{0}

\cleardoublepage

\pagestyle{plain} 
\setcounter{page}{1}
\pagenumbering{arabic}

\section{Introduction}

Decays mediated by the $\bsll$ quark transition are suppressed in the Standard Model (SM) due to the absence of flavour changing neutral currents (FCNCs) at tree level. 
Contributions from physics beyond the SM (BSM) can cause deviations of various observables from their SM expectation values.
Previous studies of $b\to s\mup \mun$ transitions have revealed tensions with SM predictions in 
branching fractions~\cite{LHCb-PAPER-2014-006,LHCb-PAPER-2015-023,LHCb-PAPER-2015-009,LHCb-PAPER-2016-012} and angular observables~\cite{CMS:2017rzx, ATLAS:2018gqc,LHCb-PAPER-2020-041,LHCb-PAPER-2020-002, LHCb-PAPER-2021-022,CMS:2024atz}.
In particular, a long-standing anomaly has been found in measurements of the differential branching fraction~\cite{LHCb-PAPER-2016-012} and angular observables of the $\Bz\to\Kstarz\mup\mun$ decay mode~\cite{LHCb-PAPER-2020-002}.\footnote{The inclusion of charge-conjugate processes is implied throughout the paper.}
This set of anomalies is typically interpreted, within an effective field theory approach, as a modification of the effective coupling corresponding to the vector leptonic current operator, known as the Wilson coefficient $C_9^{(\mu)}$~\cite{ Bobeth:2012vn, Gubernari:2022hxn, Ciuchini:2015qxb, Alguero:2023jeh}.
In addition, recent tests of lepton flavour universality (LFU) in $B^{+,\,0}\to K^{+,\,*0}\ell^+\ell^-$ decays, where $\ell=e,\,\mu$, 
indicate that LFU is respected by the measured ratios of branching fractions~\cite{LHCb-PAPER-2022-045,LHCb-PAPER-2022-046}.
Taken together, these results could be explained by a lepton-flavour-universal BSM contribution or by unexpectedly large SM hadronic effects~\cite{Jager:2014rwa, Ciuchini:2022wbq, Khodjamirian:2010vf, LHCb-PAPER-2016-045, Lyon:2014hpa, Ciuchini:2017mik}.
In particular, contributions from four-quark operators which produce a pair of leptons via the coupling to the electromagnetic current can result in a shift of the $C_{9}$ Wilson coefficient.
Significant effort has been made to quantify the impact of these contributions on the measured observables~\cite{Gubernari:2020eft,Gubernari:2022hxn, Bobeth:2017vxj, Chrzaszcz:2018yza, Ciuchini:2015qxb, Lyon:2014hpa, Ciuchini:2017mik}. 
Recently, $\Bz\to\Kstarz\mu^+\mu^-$ decays have been reanalysed using a data-driven approach to explicitly model hadronic effects~\cite{LHCb-PAPER-2023-032,LHCb-PAPER-2023-033,LHCb-PAPER-2024-011}. The results of these analyses indicate that the tension seen in the existing measurements cannot be fully explained by nonlocal hadronic contributions.
This motivates the angular analysis of $\Bz\to\Kstarz e^+e^-$ decays, which would allow for a test of LFU using angular observables. The results of this test could provide additional information to distinguish LFU BSM contributions from SM hadronic effects~\cite{Capdevila:2016ivx,Serra:2016ivr,Alguero:2022wkd}.

This paper presents an angular analysis of the $\Bz\to\Kstarz e^+ e^-$ decay in the central region of the dilepton invariant mass squared ($q^2$) of $1.1$--$6.0\gevgevcccc$, using proton-proton ($pp$) data collected by the LHCb experiment corresponding to an integrated luminosity of $9\,\invfb$. The symbol $\Kstarz$ refers to the vector meson $\Kstar(892)^0$, which is reconstructed in the $\Kp \pim$ final state. The data were collected at centre-of-mass energies of $7$ and $8\,\tev$ in 2011 and 2012, respectively (Run1), and at $13\,\tev$ in the years from 2015 to 2018 (Run2), which are split, for the purpose of this analysis, into the periods of 2015--2016 (Run2p1) and 2017--2018 (Run2p2). 

The $\Bz\to\Kstarz e^+e^-$ decay rate can be described using $q^2$ and the angles $\theta_{\ell}$, $\theta_{K}$ and $\phi$. Here, $\theta_{\ell}$ is defined as the angle between the $e^+$ ($e^-$) direction in the dielectron rest frame and the direction of the dielectron in the $\Bz$ ($\Bzb$) rest frame, $\theta_{K}$ is the angle between the kaon direction in the $\Kstarz$ ($\Kstarzb$) rest frame and the direction of the $\Kstarz$ ($\Kstarzb$) meson in the $\Bz$ ($\Bzb$) rest frame, and $\phi$ is the angle between the decay planes of the $\Kstarz$ ($\Kstarzb$) and the dielectron system in the $\Bz$ ($\Bzb$) rest frame~\cite{LHCb-PAPER-2013-019}.
Averaging over the differential decay rates $\Gamma$ and $\overline{\Gamma}$ of $\Bz$ and $\Bzb$ mesons, the angular distribution of the final-state particles for a given $q^2$ region can be expressed as
\begin{equation}
\begin{split}
  \frac{1}{ \mathrm{d} (\Gamma + \overline{\Gamma}) /  \mathrm{d} q^2 }   \frac{\mathrm{d^4} (\Gamma + \overline{\Gamma}) }{ \mathrm{d} q^2 \,\mathrm{d} \vec{\Omega} }    =  \frac{9}{32\pi} & 
\Big[ { \tfrac{3}{4}} (1 - \FL)  \sin^{2} \theta_{K} +  \FL  \cos^{2}\theta_{K} \\
&   \phantom{\Big[} + \tfrac{1}{4}   (1-
\FL)  \sin^2 \theta_{K} \cos 2\theta_{\ell}  \\ 
&   \phantom{\Big[} - \FL  \cos^{2} \theta_{K} \cos 2\theta_{\ell}  + S_3 \sin^{2}\theta_{K} \sin^2 \theta_{\ell} \cos 2\phi  \\ 
&   \phantom{\Big[} + S_4 \sin 2\theta_{K} \sin 2\theta_{\ell} \cos\phi  +  S_5  \sin 2\theta_{K} \sin\theta_{\ell}\cos\phi  \\ 
&   \phantom{\Big[} + \tfrac{4}{3} \AFB  \sin^2\theta_{K}  \ctl  +  S_7   \sin 2\theta_{K} \sin\theta_{\ell} \sin\phi   \\ 
&   \phantom{\Big[}   + S_8 \sin 2\theta_{K} \sin 2\theta_{\ell}\sin\phi  + S_9 \sin^2 \theta_{K} \sin^2 \theta_{\ell} \sin 2\phi    \Big]     \,  .
\end{split}
\label{eq:fullangular}
\end{equation}
Here $\FL$ is the fraction of longitudinally polarised $\Kstarz$ mesons, $\AFB$ is the forward-backward asymmetry of the dielectron system, and $S_{i}$, with $i=3, 4, 5, 7, 8$ and 9, are \CP-averaged $\it{S}$-basis observables as discussed in Refs.~\cite{Altmannshofer:2008dz, Gratrex:2015hna}. In addition to the resonant $\Kstarz$ (P-wave), the reconstructed $\Kp\pim$ system can also originate from a nonresonant decay, or from the decays of scalar resonances. These S-wave contributions modify the angular distribution, and can be described by introducing six additional terms to Eq.~\ref{eq:fullangular}~\cite{LHCb-PAPER-2015-051}. Nevertheless, given the limited signal yield in the studied data sample, the S-wave contributions are neglected and treated as a source of systematic uncertainty (Sec.~\ref{sec:systematic_uncertainties}). 
The $\it{S}$-basis observables can be used to construct a set of optimised $\it{P}$-basis observables~\cite{Descotes-Genon:2012isb}, for which the $\Bz\to\Kstarz$ form-factor uncertainties cancel at leading order~\cite{Descotes-Genon:2013vna}. These are given by
\begin{equation}
\begin{split}
P_1 &  =    \frac{2 S_3}{(1 - \FL)}~,\\
P_2  &  =    \frac{2}{3} \frac{ \AFB}{(1 - \FL)}~,\\
P_3  &  =     \frac{ - S_9 }{(1 - \FL)}~,\\
P^\prime_{4,5,6,8}  &  =   \frac{ S_{4,5,7,8} }{  \sqrt{ \FL (1 - \FL)} }~.
\end{split}
\end{equation}
Finally, the differences of the angular observables between the muon and electron channels $Q_{i}=P_{i}^{(\mu)}-P_{i}^{(e)}$ can be determined to directly test LFU in the angular distributions of the decays, as they are expected to be close to zero in the SM~\cite{Capdevila:2016ivx}. Information from the muon channel is extracted from data samples analysed in Ref.~\cite{LHCb-PAPER-2020-002}, which is comprised of data recorded by the LHCb detector in Run1 ($3\,\invfb$), and 2016 ($1.7\,\invfb$).

The angular observables of $\Bz\to\Kstarz e^+e^-$ decays have been measured by the $\lhcb$ collaboration in the low-$q^2$ region of 0.0008--0.257\gevgevcccc. They strongly constrain the Wilson coefficient $C^{\prime}_7$ to SM expectations~\cite{LHCb-PAPER-2020-020}. In addition, the observables $P_4^{\prime}$ and $P_5^{\prime}$ have been measured by the Belle collaboration for the decays $B^{+,0}\to K^{*+,*0} \ell^+\ell^-$ (where $\ell=e,\,\mu$) in different bins of $q^2$ in the 0.1--19.0\gevgevcccc~\cite{Belle:2016fev} range, including the central-$q^2$ bin of 1.0--6.0\gevgevcccc. 
The same analysis also determined $Q_4$ and $Q_5$, and obtained results consistent with SM predictions.

Electron reconstruction at LHCb is challenging due to substantial energy loss caused by photon emission.
While part of the lost energy can be recovered using a dedicated algorithm~\cite{LHCB-PAPER-2017-013}, some degradation of the momentum resolution, and therefore of the invariant masses of the dielectron and $\Bz$ candidates, cannot be avoided. This leads to large background contamination, which complicates the signal extraction.
Consequently, the angular observables are only determined in a large $q^2$ region with sufficient signal yield. 
Moreover, there are significant differences between the measured $q^2$ and the true $q^2$, for which SM predictions are calculated.
To minimise these differences, the constrained $q^2$ is used to define the measurement region. This quantity is determined from a fit of the decay chain in which the $\Bz$ candidate is constrained to originate from its associated primary vertex (PV) and its invariant mass is constrained to the known mass of the $\Bz$ meson~\cite{PDG2024}.
In addition, the signal simulation is used to parametrise the relationship between the reconstructed angles and $q^2$ and their true values (Sec.~\ref{sec:effective_acceptance}).
The use of the constrained $q^2$ also reduces contamination from the radiative tail of the $J/\psi$ resonance. This allows the same analysis strategy to be applied directly to a larger $q^2$ region of $1.1$--$7.0\gevgevcccc$. The results of the additional measurement in this larger $q^2$ region are reported in Appendix~\ref{sec:large_q2}.
The decay of $\Bz\to\Kstarz J/\psi(\to e^+ e^-)$, selected within the $q^2$ window of $7.0$--$11.0\gevgevcccc$, is used as a control mode to reduce differences between simulation and data at various stages of the analysis.

The structure of the paper is as follows. Section~\ref{sec:the_lhcb_detector} describes the experimental apparatus and the production of simulation samples. Section~\ref{sec:reconstruction_and_selection} details the reconstruction and selection of $\BdKstee$ decays. The method used to correct the measured angular and $q^2$ distributions of the signal is explained in Sec.~\ref{sec:effective_acceptance}. Section~\ref{sec:mass_and_angular_models} introduces the main background components present in the data sample and describes the methods used to parametrise them. Section~\ref{sec:angular_fit} discusses the angular fit.
Section~\ref{sec:systematic_uncertainties} describes and quantifies the various sources of systematic uncertainties.
The results of this analysis are shown and discussed in Sec.~\ref{sec:results}. The conclusions are presented in Sec.~\ref{sec:conclusions}.

\section{The LHCb detector}
\label{sec:the_lhcb_detector}
The LHCb detector~\cite{LHCb-DP-2008-001,LHCb-DP-2014-002}  is a single-arm forward spectrometer covering the pseudorapidity range $2 < \eta < 5$, designed for the study of particles containing $b$ or $c$ quarks. The detector includes a high-precision tracking system consisting of a silicon strip vertex detector surrounding the $pp$ interaction region~\cite{LHCb-DP-2014-001}, a large-area silicon-strip detector located upstream of a dipole magnet with a bending power of about $4\,\mathrm{T\,m}$, and three stations of silicon-strip detectors and straw drift tubes~\cite{LHCb-DP-2013-003, LHCb-DP-2017-001} placed downstream of the magnet. The tracking system provides a measurement of the momentum, $p$, of charged particles with a relative uncertainty that varies from $0.5\%$ at low momentum to $1.0\%$ at $200\,\gevc$. The minimum distance of a track to a PV, the impact parameter (IP), is measured with a resolution of $(15 + 29/\pt)\mum$, where \pt is the component of the momentum transverse to the beam, in $\gevc$. Different types of charged hadrons are distinguished using information from two ring-imaging Cherenkov detectors (RICH)~\cite{LHCb-DP-2012-003}. Photons, electrons and hadrons are identified by a calorimeter system consisting of scintillating-pad and preshower detectors, an electromagnetic calorimeter (ECAL) and a hadronic calorimeter\cite{LHCb-DP-2020-001}. Muons are identified by a system composed of alternating layers of iron and multiwire proportional chambers~\cite{LHCb-DP-2012-002}. The online event selection is performed by a trigger~\cite{LHCb-DP-2012-004}, which consists of a hardware stage, based on information from the calorimeter and muon systems, followed by a software stage, which applies a full event reconstruction.

The hardware electron trigger requires the presence of an ECAL cluster corresponding to a transverse energy deposition that exceeds a threshold of around 2--3\gev, which varies depending on the data-taking period.
In addition, the presence of hit(s) in the preshower detector and at least one hit in the scintillating-pad detector in front of the ECAL cluster are required.
Events are retained if at least one electron fulfils the electron hardware trigger requirements. 
Alternatively, they are retained if one or more particles from the rest of the event, identified as muons, electrons or hadrons, satisfy their respective trigger requirements. 
These particles may originate from the decay of the other $b$-hadron from the $b\bar{b}$ pair produced in the $pp$ collision that leads to the signal decay.
At the software trigger stage, events are retained based on the presence of a two-, three- or four-track secondary vertex that is significantly displaced from any PV, and at least one high-momentum track that has a large IP with respect to all PVs in the event. Multivariate algorithms~\cite{BBDT, LHCb-PROC-2015-018} are used for the identification of secondary vertices that are consistent with $b$-hadron decays.

Simulation is required to model the effects of the detector acceptance, selection requirements and resolution, as well as the signal and background distributions. In the simulation, $pp$ collisions are generated using Pythia~\cite{Sjostrand:2007gs} with a specific $\lhcb$ configuration~\cite{LHCb-PROC-2010-056}. Decays of unstable particles are described by EvtGen~\cite{Lange:2001uf}, in which final-state radiation is generated using PHOTOS~\cite{Golonka:2005pn}. The interaction of the generated particles with the detector, and its response, are implemented using the Geant4 toolkit~\cite{Allison:2006ve} as described in Ref.~\cite{LHCb-PROC-2011-006}. 
Data-driven corrections are applied to the simulation to improve the modelling of particle identification (PID) variables, trigger efficiency, track multiplicity per event and the kinematic properties of the $\Bz$ meson. The PID variables are corrected using calibration samples consisting of decay modes that can be selected without the use of PID information~\cite{LHCb-PUB-2016-021, Poluektov:2014rxa}. 
Corrections to trigger efficiencies are subsequently applied using per-event weights obtained via a tag-and-probe approach\cite{LHCb-PUB-2014-039}. 
Finally, a multivariate boosted decision tree reweighter~\cite{jonas_eschle_2018} is trained using simulated control-mode decays and background-subtracted $\Bz\to\Kstarz J/\psi(\to e^+e^-)$ data to reduce simulation-data differences for quantities related to the track multiplicity of the event, the kinematics of the $\Bz$ meson and the PV and $\Bz$ decay-vertex (DV) fits.
An alternative simulation produced without running PHOTOS and the subsequent simulation of the detector (generator-level simulation), which only contains the effect of the physics model, is also used in the parametrisation of the functions that correct for distortions of the signal distributions (Sec.~\ref{sec:effective_acceptance}).

\section{Reconstruction and selection}
\label{sec:reconstruction_and_selection}
Signal candidates are reconstructed by combining pairs of oppositely charged tracks identified as electrons, with $\Kstarz$ candidates. The electron tracks are required to form a good-quality vertex and to have $\pt>500\mevc$ and $p>3000\mevc$. The $\Kstarz$ candidates are reconstructed from oppositely charged tracks, identified as pions and kaons, with $\pt > 250\mevc$. 
Their invariant mass must lie within $\pm100\mevcc$ of the known $\Kstarz$ mass~\cite{PDG2024}.
All tracks are required to be of good quality, and inconsistent with originating from any PV. 
For events with multiple PVs, the one with the smallest $\chi^2_{\mathrm{IP}}$, defined as the difference in $\chi^2$ between the PV fit with and without the \Bz candidate, is associated with the \Bz candidate.
The four tracks must form a good-quality vertex that is significantly displaced from any PV. 
The cosine of the direction angle, which is defined as the angle between the momentum vector of the \Bz candidate and the displacement vector from its PV to its DV, is required to be close to one.
The kaon, pion and electron candidates are required to be within the acceptance of the RICH detectors. The electron candidates are also required to be within the acceptance of the ECAL, and therefore have associated energy clusters. To reduce instances of partly or fully duplicated tracks, minimum values are imposed on the angles between pairs of final-state particle tracks associated with the \Bz decay.
To exclude phase-space regions where the acceptance is not well modelled due to the nonuniform generator-level distribution or very low selection efficiency, candidates with $|\ctl|<0.9$ and $\ctk<0.9$ are removed (Sec.~\ref{sec:effective_acceptance}).

Two types of PID variables are used for background suppression, as was done in Refs.~\cite{LHCb-PAPER-2022-045,LHCb-PAPER-2022-046}. 
The first type of variables is the difference in log-likelihoods between particle hypotheses $x$ and the pion hypothesis ($\mathrm{DLL}_{x\pi}$), which uses information from the RICH, calorimeter and muon systems to quantify the likelihood of a track being associated with particle $x$, such as a proton, kaon or electron, relative to its likelihood of being associated with a pion.
The second is the output of artificial neural networks trained using information from all subdetectors, which can be interpreted as the probability of a track to be associated with particle $x$ ($\mathrm{ProbNN}x$). 
The misidentification of pions as kaons is suppressed by a minimum requirement on $\mathrm{DLL}_{K\pi}$. Additionally, combinations of $\mathrm{ProbNN}K$, $\mathrm{ProbNN}p$ and $\mathrm{ProbNN}\pi$ are used to suppress proton-to-kaon misidentification and the misidentification of kaons and protons as pions. Electron misidentification is suppressed via requirements on $\mathrm{DLL}_{e\pi}$ and $\mathrm{ProbNN}e$. 

Specific sources of backgrounds are suppressed using dedicated selection requirements (vetoes). Vetoes detailed in Refs.~\cite{LHCb-PAPER-2022-045,LHCb-PAPER-2022-046} are used to suppress \mbox{$\Bs\to \phi (\to\Kp\Km) e^+e^-$} decays with kaon-to-pion misidentification, semileptonic decays such as \mbox{$\Bz\to \Dzb(\to \Kp \pim)\pim e^{+} \nu_{e}$} and $\Bz\to \Dm(\to\Kstarz (\to \Kp\pim) \pim)e^{+} \nu_{e}$ with pion-to-electron misidentification, $B^{+}\to \Kp e^{+}e^{-}$ decays reconstructed with the addition of a random pion track, and $\Bz\to\Kstarz J/\psi(\to e^+e^-)$ and $\Bz\to\Kstarz\psi(2S)(\to e^+e^-)$ decays with the misidentification of a hadron as an electron and vice versa. In addition, semileptonic $\Bs$ decays featuring $D_s^-$ mesons with kaon-to-electron misidentification are vetoed in analogy to the other semileptonic decays listed above, and double misidentification of the kaon and pion tracks of a signal decay is reduced by requiring the $\mathrm{DLL}_{K\pi}$ of the kaon to be larger than the $\mathrm{DLL}_{K\pi}$ of the pion.
The contamination from $\Lb \to p\Km e^+e^-$ decays is found to be negligible using simulation. 
For the control mode, to suppress the partially reconstructed background which is distributed in a lower mass range than the signal, an invariant mass is calculated by constraining the \Bz candidate to originate from its associated PV and the dielectron invariant mass to the known \jpsi mass, and is required to be larger than $5150\,\mevcc$.

After applying the aforementioned requirements, the dominant source of background is comprised of candidates reconstructed from random tracks.
This combinatorial background is suppressed using Boosted Decision Tree (BDT) classifiers~\cite{DBLP:journals/corr/ChenG16}.
One BDT is trained for each run period.
Simulated signal decays are used as a proxy for the signal, where the simulation is corrected to better describe the data,
and data candidates with an invariant mass above $5600\mevcc$ are used as the background proxy.
To increase the background sample size available for training, the baseline PID requirements are loosened and the $\Kstarz$ mass window is enlarged to $\pm200\mevcc$. The $k$-folding technique~\cite{kFold} is used with ten folds to make use of the full simulation and data samples available.
Fourteen variables are selected as inputs to the classifier through an optimisation procedure that systematically examines the performance of the BDT classifiers trained with different sets of variables. This final set includes variables related to the $\Bz$ candidate, namely its $\pt$ and $\chi^2_{\mathrm{IP}}$, the $\chi^{2}$ of the distance between its PV and DV ($\chi^2_{\mathrm{FD}}$),
the fit quality of its DV ($\chi^2_{\mathrm{DV}}$), the cosine of its direction angle, and the quality of the constrained kinematic fit of the decay chain ($\chi^{2}_{\mathrm{DTF}}$). The $\chi^2_{\mathrm{DV}}$ of the $\Kstarz$ candidate and that of the dielectron are also used. The other variables are the minimum $\pt$ and $\chi^2_{\mathrm{IP}}$ of the final-state hadrons, and the minimum and maximum $\pt$ and $\chi^2_{\mathrm{IP}}$ of the two electrons. The threshold on the classifier output is set in order to minimise the statistical uncertainty of $P_5^{\prime}$.
One threshold is used for all run periods and the same classifier is used for both the signal and control modes.

\begin{figure}[!tb]
\centering
    \includegraphics[width=.7\textwidth, trim={0 0 0 0},clip]{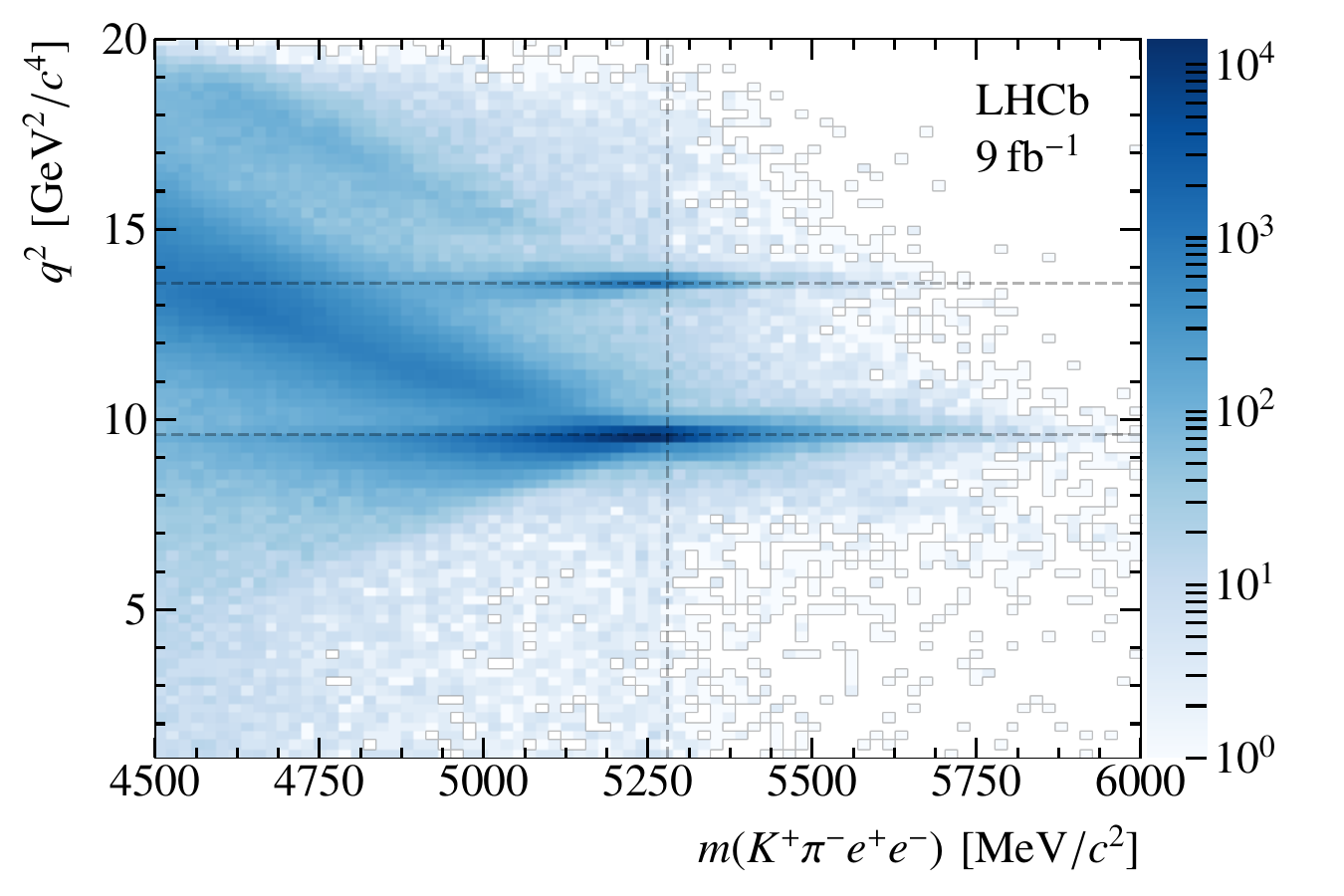} 
  \caption{Distribution of $q^2$ and \Bz invariant mass of signal candidates in data.
 Signal decays lie in a vertical band close to the known $\Bz$ mass (vertical dashed line)~\cite{PDG2024}. 
The decays $\Bz\to\Kstarz J/\psi(\to e^+e^-)$ and $\Bz\to\Kstarz\psi(2S)(\to e^+e^-)$ have an invariant mass close to that of the $\Bz$ meson, and $q^2$ values close to the square of the known $J/\psi$ and $\psi(2S)$ masses, respectively (horizontal dashed lines). 
They are visible as horizontal bands. 
The two diagonal bands contain combinatorial background comprised of genuine $J/\psi$ or $\psi(2S)$ mesons combined with random kaon and pion tracks.}
\label{fig:q2c_vs_mass}
\end{figure}

The invariant mass of the $\Bz$ candidate is calculated from a fit to the decay chain where the \Bz meson is constrained to originate from its associated PV, and used to separate signal from backgrounds.
The distribution of the invariant mass versus $q^2$ for the \Bz candidates is shown in Fig.~\ref{fig:q2c_vs_mass} (the analogous distribution for the unconstrained $q^2$ is shown in Fig.~\ref{fig:q2_vs_mass} of Appendix~\ref{sec:q2_choice}). 
The number of control-mode candidates that leak into the signal $q^2$ region is further reduced by requiring the invariant mass of the \Bz candidates to lie within the restricted range of $4900$--$5700\mevcc$. 
A mass window of $4500$--$6200\mevcc$ is used for the control mode.

After applying all the aforementioned requirements, less than one percent of the signal and control-mode events in simulation and data have multiple candidates. In these cases, one randomly selected candidate is retained.

\section{Effective acceptance}
\label{sec:effective_acceptance}
The measured distributions of the decay angles and $q^2$ are distorted by final state radiation, bremsstrahlung and detector resolution, as well as triggering, reconstruction, and selection processes. 
These effects are included in the signal simulation, which is used to obtain an effective acceptance function that corrects for all distortions simultaneously. 
First, a function that describes the generator-level sample is obtained, and is used to assign weights to each selected candidate in the simulation. The weighted distribution is then parametrised by a function which represents the effective acceptance.
The inverse of the value of this function, evaluated for each signal candidate, is used as a weight (effective acceptance weight) to correct the signal distribution.
The effective acceptance function obtained in this way depends on the physics model used in the simulation, which is a source of systematic uncertainty that is discussed in Sec.~\ref{sec:systematic_uncertainties}.

The effective acceptance is parametrised using a sum of Legendre polynomials and trigonometric functions in terms of the measured angles and $q^2$, with the latter variable scaled to the range of --1 to 1:
\begin{equation}
\epsilon(\ctl,\ctk,\phi,q^2) = \sum_{klmn} c_{klmn} L_k(\ctl) L_l(\ctk) F_m(\phi) L_n(q^{2})~,
\end{equation}
where $L_{a}(x)$ is the Legendre polynomial of order $a$ for the variable $x$, and the $F_{m}(\phi)$ term is given by 
\begin{equation}\label{equ:fourier_terms}
F_{m}(\phi)=
\begin{cases} 
\cos\dfrac{m\phi}{2}    &   \quad \mathrm{if}  \quad m~\mathrm{even} ~, \\
\noalign{\vskip8pt}
\sin\dfrac{(m+1)\phi}{2}   &    \quad \mathrm{if}  \quad  m~\mathrm{odd}~,\\
\end{cases}
\end{equation}
where $m$ is zero or a positive integer.
The use of sine and cosine terms for $\phi$ is motivated by the Fourier expansion in this angle.
The $c_{klmn}$ coefficients are calculated through a principal moments analysis of simulated signal decays, exploiting the orthogonality property of the Legendre polynomials and trigonometric functions. 

These functions are first used to parametrise the generator-level simulation, in terms of the true angles and $q^2$. 
In this case, orders of five, four, four, and nine are used for $\cos\theta_{K}$, $\cos\theta_{\ell}$, $\phi$ and $q^2$, respectively.
A $q^2$ range of $0.5$--$8.0\gevgevcccc$ is used to avoid pathologies that can occur near the edges of the range used in the measurement.
For the effective acceptance function, parametrised in terms of the measured angles and $q^2$, orders of five, four, four and three are used to describe the weighted $\ctk$, $\ctl$, $\phi$ and $q^2$ distributions, respectively.
One function is parametrised for each of the three run periods using the corresponding simulation samples. 
The one-dimensional projections of the functions used to describe the generator-level distributions and the effective acceptance are shown in Appendix~\ref{sec:effective_acceptance_functions}.

The quality of the parametrisation is checked by comparing the values of the angular observables obtained by a maximum-likelihood fit of the generator-level simulation using Eq.~\ref{eq:fullangular}, and those from a fit performed to the selected candidates of the signal simulation, where effective acceptance weights are included.
This check is performed separately for the three run periods, using a $k$-folding strategy.
The results show good retrieval of the generator-level observable values (Appendix~\ref{sec:effective_acceptance_functions}), with residual differences originating mainly from the imperfect description of some selection requirements, such as the requirements used to suppress semileptonic $\Bz$ decays and $B^+\to \Kp e^+e^-$ decays (Sec.~\ref{sec:reconstruction_and_selection}). The systematic uncertainty due to imperfect parametrisation is quantified in Sec.~\ref{sec:systematic_uncertainties}. 

\section{Invariant-mass and angular models}
\label{sec:mass_and_angular_models}
The angular observables are determined by fitting the distributions of the invariant mass of the \Bz candidate, \ctk, \ctl and $\phi$ weighted with effective acceptance weights. 
In addition to the signal component, the fit model includes four sources of background: combinatorial background; double-semileptonic decays of $b$-hadrons with two electrons in the final state; partially reconstructed decays mediated by $\bsll$ transitions with more than two hadrons in the final state,
dominated by $B^{+}\to K^{+}\pi^{+}\pi^{-}e^{+}e^{-}$ decays; and a mixture of $b$-hadrons decays with the misidentification of one or more hadrons as electrons, with or without missing energy.
 
The signal invariant-mass model is parametrised using simulation, and its peak position and width are subsequently corrected to match those in data. These corrections are determined by fitting the invariant-mass distribution of the control mode.
Four components are included in that case: \mbox{$\Bz\to\Kstarz\jpsi(\to e^+e^-)$} decays, \mbox{$\Bs\to\Kstarzb\jpsi(\to e^+e^-)$} decays, combinatorial background, and $\Lb\to p\Km\jpsi(\to e^+e^-)$ decays with proton-to-pion misidentification. 

Effective acceptance weights are included in the parametrisation of the invariant-mass distribution of the signal component, as well as the invariant-mass and angular distributions of the backgrounds. 
Different signal invariant-mass models, and background invariant-mass and angular models are used for each run period with the exception of the misidentified hadronic background, for which the same model is used for the periods of Run2p1 and Run2p2 due to the limited size of the data samples.
The models used to describe the background components of the signal and control modes are discussed below and illustrated in Appendix~\ref{sec:component_models}.

\subsection{Signal and control modes}
\label{sec:signal_and_control_mode_mass_distributions}
The signal angular distribution is described by Eq.~\ref{eq:fullangular}, and its invariant-mass distribution is described by the sum of two Crystal Ball functions~\cite{Skwarnicki:1986xj} (DCB) that share common parameters describing the mean and width of the peak, but have independent power-law tails on opposite sides of the peak. The impact of photon energy recovery on the signal mass distribution is significant, therefore separate models are used depending on the numbers of electrons that receive corrections (zero, one or two).
The full mass model is given by the sum of the three separate contributions.

For the control mode, due to a larger \Bz mass window, additional Gaussian components are used in the modelling of the contributions where one or both electrons receive corrections to improve the description of the tails of the distributions. 
Due to residual differences between simulation and data, the mean and width parameters determined from simulation are modified via a shift and a scaling parameter, respectively, which are determined 
using the control mode and are subsequently fixed in the signal-mode fits.

\subsection{Combinatorial and double-semileptonic backgrounds}
\label{sec:combinatorial_and_DSL_backgrounds}
Candidates reconstructed using unrelated electron or hadron tracks, hereafter referred to as the combinatorial background, have smooth, factorisable invariant-mass and angular distributions, which can be described by the product of an exponential function and three one-dimensional polynomials. Factorisation holds due to the random nature of this background, and is only broken in a small phase-space region by the requirements to veto $B^+\to K^+e^+e^-$ decays, for which a dedicated systematic uncertainty is assigned (Sec.~\ref{sec:systematic_uncertainties}).

Double-semileptonic (DSL) decays refer to semileptonic decays of a beauty hadron to a charm hadron that subsequently decays semileptonically, most often, into a strange hadron. When the two semileptonic decays produce two electrons, a kaon and a pion in the final state, the corresponding signal candidate can satisfy all selection requirements.
The most important contribution is expected to originate from the \mbox{$\Bz\to D^{-}(\to\Kstarz(\to K^{+}\pi^{-})e^{-}\bar{\nu}_{e})e^{+}\nu_{e}$} mode, which has a relatively large branching fraction of $\mathcal{O}(10^{-3})$~\cite{PDG2024}. 
Due to the missing neutrinos, the corresponding invariant-mass distribution can be described by an exponential function.
This makes the separation of combinatorial and DSL backgrounds using mass information alone challenging. 
However, the DSL background is characterised by an asymmetric $\ctl$ distribution peaking close to one.
Contributions from decays such as $B_{s}^{0}\to D_{s}^{-}(\to \Kstarzb e^{-}\bar{\nu}_{e})e^{+}\nu_{e}$, on the other hand, are negligbly small.
The DSL decays cannot be well described by a fully factorised approach due to non-negligible correlation between $\ctk$ and $\phi$. Therefore this component is described by Eq.~\ref{eq:fullangular}, integrated over $\ctl$.

Additional background contributions include DSL decays with the misidentification of one or more final-state particles, or reconstructed with one or more random tracks.
Given the complexity of these backgrounds, a two-step data-driven approach is used, where effective DSL and combinatorial models are obtained using background $\Bz$ candidates reconstructed from the $\Kp \pim \ep \mun$ final state. 
In the first step, the DSL angular model is obtained using a sample enriched in DSL decays selected with a stringent BDT requirement and falling within a lower mass window of 4500--$5200\mevcc$. The $\ctl$ distribution is modelled using kernel density estimation (KDE)~\cite{CRANMER2001198} while $\ctk$ and $\phi$ are described by integrating Eq.\,\ref{eq:fullangular} over $\ctl$. In the second step, candidates passing the baseline BDT requirement and lying in the signal mass window are fitted to determine the angular parameters of the combinatorial background, and the slopes of the exponential mass distributions of the DSL and combinatorial components. In this fit, the DSL angular parameters are fixed and polynomials up to second order are used for the three angles. Candidates used in the first step are excluded from the second. In this way, contributions that show characteristics intermediate between the combinatorial and DSL backgrounds are split between the effective DSL and combinatorial background models. The assumption that the ratio and the shapes of these two contributions are the same for the $\Kp\pim e^+e^-$ and $\Kp\pim e^+ \mu^-$ final states is a source of systematic uncertainty, which is quantified in Sec.~\ref{sec:systematic_uncertainties}.

\subsection{Misidentified hadronic decays}
\label{sec:misidentified_hadronic_decays}
Several hadronic decays with the misidentification of kaons or pions as electrons are known to satisfy the selection criteria~\cite{LHCb-PAPER-2022-045,LHCb-PAPER-2022-046}. Decays of the type $\Bz\to\Kstarz\pi^-(\pi^0,\gamma)X$, where $X$ represents any possible final-state particles, can contribute. These predominantly populate the lower mass region, although some contributions, such as the fully reconstructed, misidentified decays of $\Bz \to \Kstarz \Kp\Km$ and $\Bz \to \Kstarz \pip\pim$, peak close to the known mass of the \Bd meson. As a result, they cannot be described by the combinatorial model.

The data-driven approach developed in Refs.~\cite{LHCb-PAPER-2022-045,LHCb-PAPER-2022-046} is used to estimate the yields and determine the model for this type of background. Data samples enriched in misidentified hadrons are obtained by inverting the baseline electron PID criteria. The region in the electron PID space defined by these inverted criteria is hereafter referred to as the control region.
These samples contain a mixture of fully reconstructed misidentified decays, partially reconstructed decays with or without misidentification, combinatorial background, and residual signal and control-mode decays. 
Yield ratios are calculated in regions of transverse momentum and pseudorapidity using the number of kaon and pion candidates from PID calibration samples that are within the control and baseline regions.
The resulting maps, or transfer functions, are used to extrapolate the yields and distributions of this type of background in the baseline region.

Residual signal and control-mode decays are present in the control-region sample.
The contribution from signal decays is subtracted using weights based on the yield obtained from an initial data fit where the misidentified hadronic background component is ignored.
Similarly, control-mode candidates are subtracted based on the yield of the control-mode fit.

Adaptive KDEs from Refs.~\cite{LHCb-PAPER-2022-045,LHCb-PAPER-2022-046} are used to model the invariant-mass distributions of the control-region candidates.
Due to similarities between the angular distributions of these candidates and that of the effective DSL background, the same functions are also used in this case: 
a KDE is used for $\ctl$ and Eq.~\ref{eq:fullangular}, integrated over $\ctl$, is used for $\ctk$ and the $\phi$ angle.

\subsection{Partially reconstructed background}
The main sources of the partially reconstructed background include decays to heavier kaon resonances, such as $B^{+}\to K_{1}(1270)^{+}e^+e^-$ or \mbox{$B^{+}\to K^{*}_{2}(1430)^{+}e^+e^-$}, where the $K_{1}(1270)^{+}$ and $K^{*}_{2}(1430)^{+}$ mesons decay to a kaon, a pion and one or more additional pions.
Due to the missing particle(s), the reconstructed $\Bz$ invariant-mass distribution shows a broad peak centred in the lower mass region. 

The relatively large number of contributing kaon resonances motivates the use of a data-driven approach, where the simulation of $B^{+}\to K^+\pi^+\pi^- e^+e^-$ phase-space decays is corrected using weights obtained from efficiency-corrected and background-subtracted \mbox{$B^{+}\to K^{+}\pi^{+}\pi^{-}J/\psi(\to \mup\mun)$} data~\cite{LHCb-PAPER-2024-046,LHCb-PAPER-2022-045,LHCb-PAPER-2022-046}. These weights are assigned to the simulated (background) candidates based on their $K^{+}\pi^{+}\pi^{-}$, $K^{+}\pi^{-}$ and $\pi^{+}\pi^{-}$ invariant masses. The weighted angular and invariant-mass distributions are then used to model the corresponding background. A KDE is used to model the invariant-mass distribution, and factorised second-order polynomials are used for the three angles.

\subsection{Control-mode backgrounds}
\label{sec:control_mode_backgrounds}
Backgrounds of the control mode can be divided into those with an invariant-mass distribution that can be described by an exponential function, which includes combinatorial, DSL and partially reconstructed decays, and those that peak close to the known \Bz mass, dominated by 
$\Bs\to\Kstarzb \jpsi(\to e^{+}e^{-})$ and misidentified $\Lb\to p\Km\jpsi(\to e^+e^-)$ decays. 
The \Bs decays are suppressed with respect to the \Bz decays by the ratio of hadronisation fractions, $f_s/f_d$~\cite{PDG2024}, and branching fractions.
They form a peak centred at the known \Bs mass. The same model used for $\Bz\to\Kstarz \jpsi(\to e^{+}e^{-})$ decays is used to describe this component, with a shift in the mean value of $87\mevcc$~\cite{PDG2024}, which corresponds to the mass difference between the \Bz and \Bs mesons. 
Background from misidentified $\Lb$ decays, which is also subdominant, is modelled using simulated $\Lb\to p\Km\jpsi(\to e^{+}e^{-})$ phase-space decays with data-driven corrections obtained from background-subtracted $\Lambda_{b}^{0}\to p\Km J/\psi(\to \mup\mun)$ data~\cite{LHCb-PAPER-2015-029,LHCb-PAPER-2019-040,LHCb-PAPER-2022-045,LHCb-PAPER-2022-046}. These weights are assigned to simulated candidates based on their $\Km p$ and $J/\psi p$ invariant masses. Their weighted invariant-mass distribution is modelled using a KDE.

\section{Invariant-mass and angular fit} 
\label{sec:angular_fit}
A weighted maximum-likelihood fit is performed simultaneously to Run1, Run2p1 and Run2p2 data to determine the angular observables.
The parameters that describe the signal invariant-mass distribution and the angular distributions of background components are obtained as discussed in Sec.~\ref{sec:mass_and_angular_models}. Different shift and scaling parameter values are obtained for each run period from control-mode fits, in which the fractions of \mbox{$\Bs\to\Kstarzb \jpsi(\to e^{+}e^{-})$} and misidentified $\Lb\to p\Km\jpsi(\to e^+e^-)$ decays are fixed to expected values, and the slope of the combinatorial background is allowed to vary. The result of these fits are shown in Fig.~\ref{fig:data_fit_control_mass} of Appendix~\ref{sec:component_models}.

\begin{figure}[!tb]
\centering
    \includegraphics[width=.45\textwidth, trim={0 0 0 0},clip]{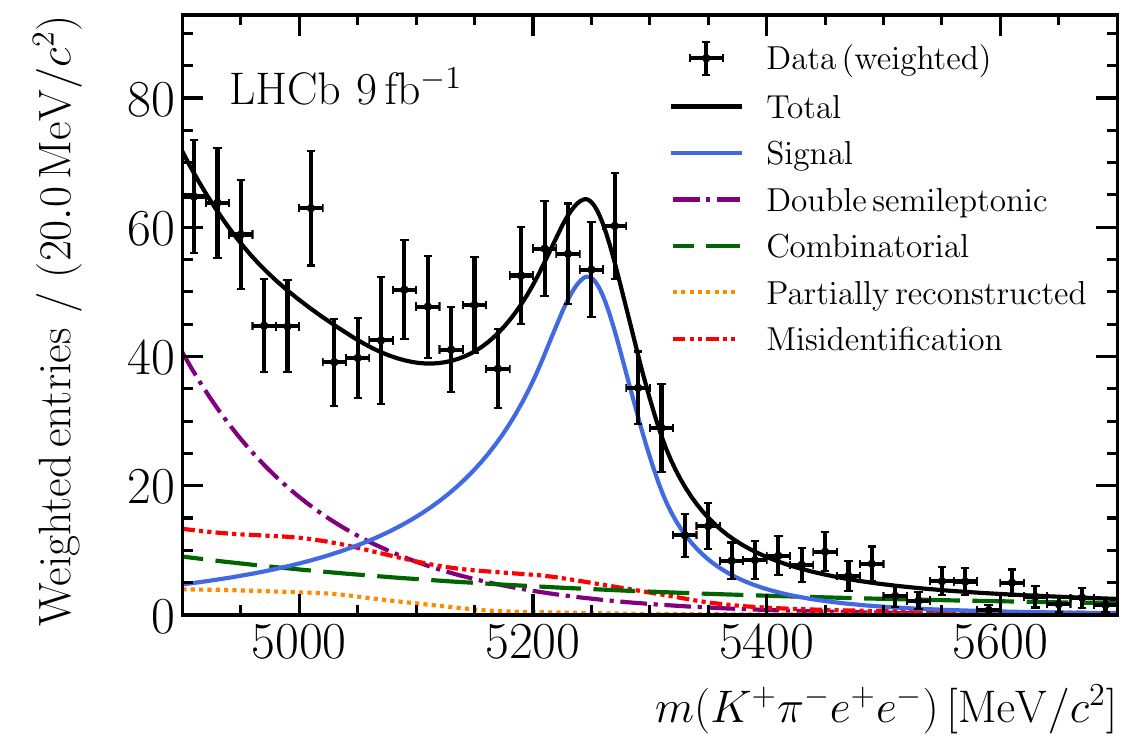} 
    \includegraphics[width=.45\textwidth, trim={0 0 0 0},clip]{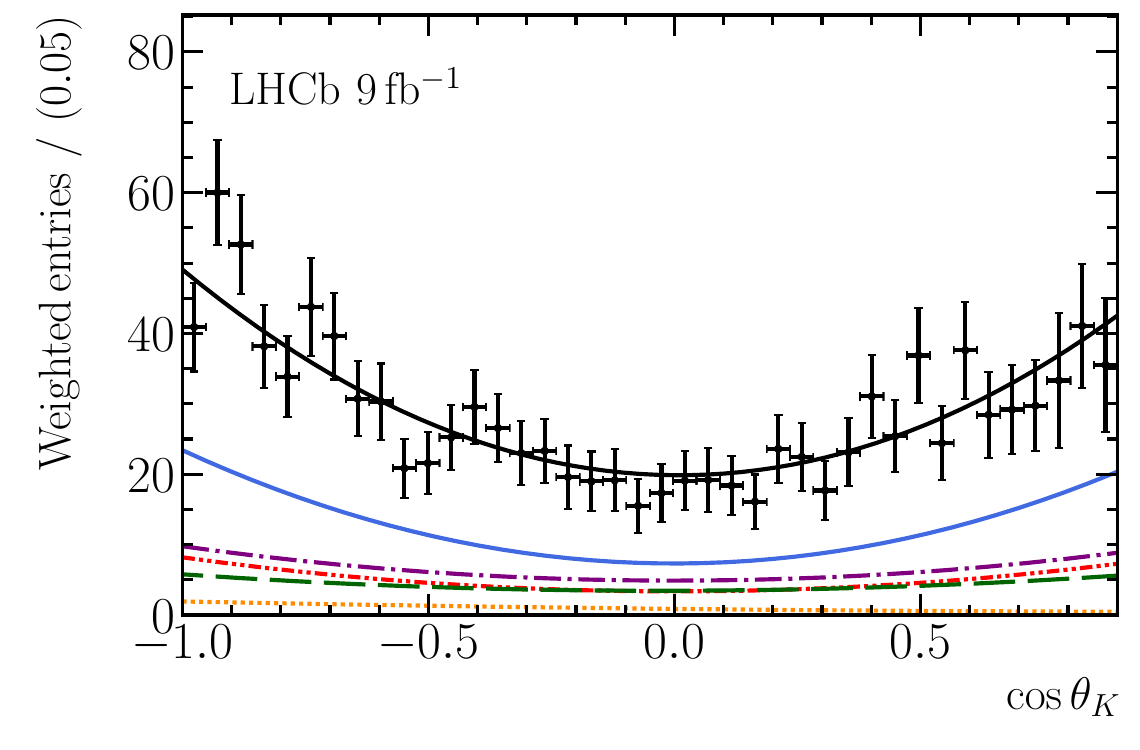} 
    \includegraphics[width=.45\textwidth, trim={0 0 0 0},clip]{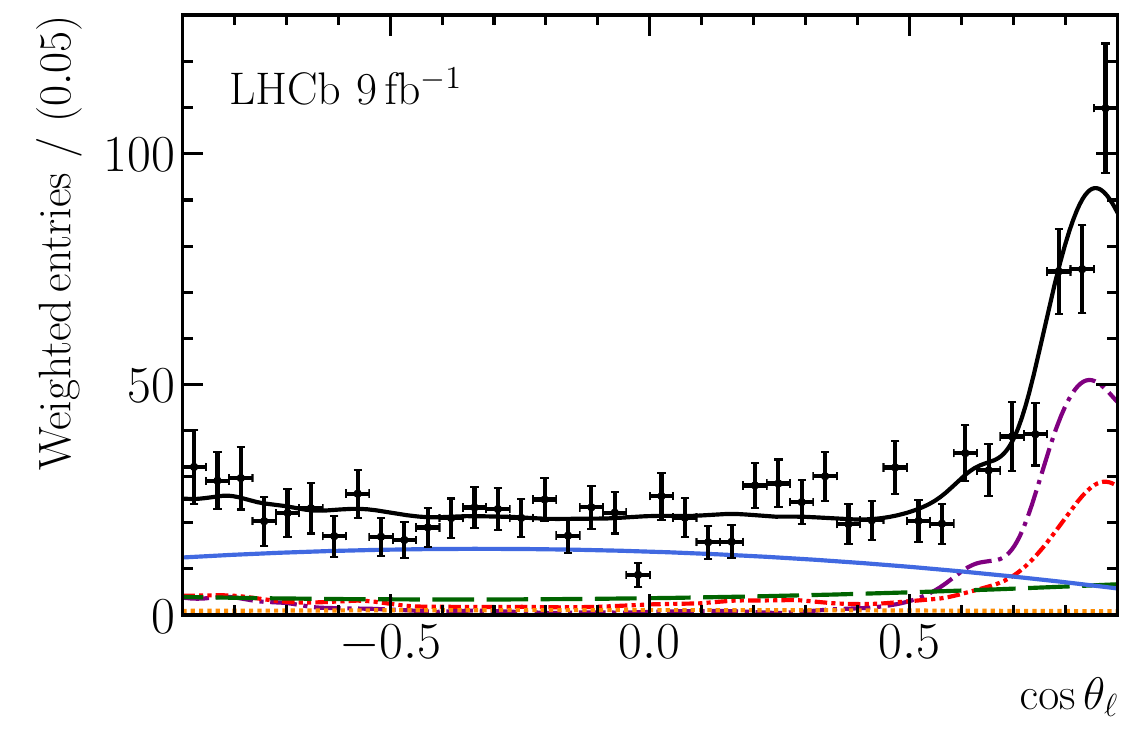} 
    \includegraphics[width=.45\textwidth, trim={0 0 0 0},clip]{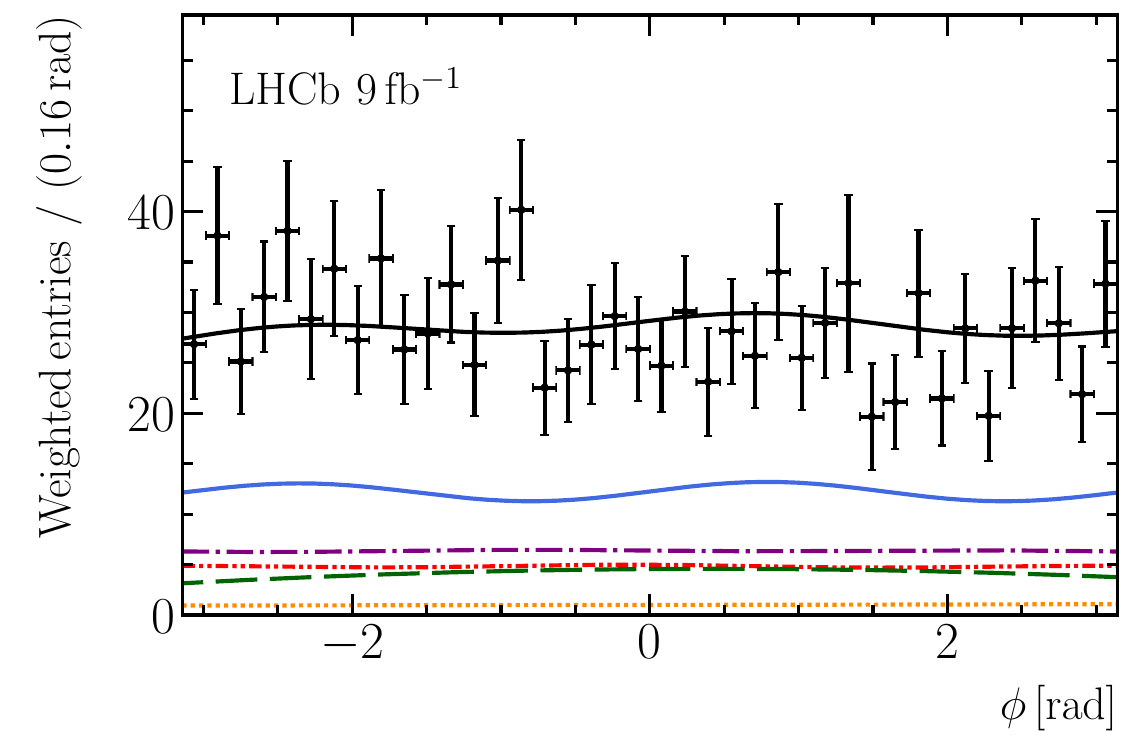} 
  \caption{Invariant-mass and angular distributions of selected candidates after the application of effective acceptance weights. 
  The signal distribution is shown with a solid blue line, and the background components are shown with dashed, dotted and dash-dotted lines. The solid black line corresponds to the full fit function.
  }
\label{fig:data_fit_q2c_small_S}
\end{figure}

The fractions of signal and DSL decays are allowed to vary freely for each run period, while those of the misidentified hadronic decays are allowed to vary, but are constrained based on their expected yields.
To improve fit stability, the fraction of partially reconstructed decays is expressed by the signal fraction multiplied by a factor that is shared among all run periods and allowed to vary. 
The sum of all the fractions, including the combinatorial background, is constrained to be one.
The slopes of the combinatorial mass distributions are allowed to vary separately for each run period.
As the fit is performed using weighted events, the asymptotically correct approach described in Ref.~\cite{Langenbruch:2019nwe} 
is used to calculate uncertainties on the fitted parameters.
The fit projections are shown in Fig.~\ref{fig:data_fit_q2c_small_S}.

The fit strategy is validated by means of pseudoexperiments generated with the baseline model, using the observables and other parameter values determined from the data fit.
The results show no sizable biases in the angular observables, or significant overestimation or underestimation of uncertainties. Nevertheless, all biases found are taken into account as systematic uncertainties, and the widths of the pull distributions are used to correct the uncertainties of the data fit. 

\section{Systematic uncertainties}
\label{sec:systematic_uncertainties}
Sources of systematic uncertainties include the choices made in the modelling of backgrounds, the determination of the effective acceptance, the description of the signal invariant-mass distribution, the impact of neglecting backgrounds that contain $J/\psi$ mesons and the S-wave contribution, the impact of the requirements to suppress $\Bu\to K^+ e^+e^-$ decays, and the fit biases. The effect of all sources is quantified using pseudoexperiments. In most cases, an alternative model is defined, and pseudoexperiments are generated with this model. Then, they are fitted with both this alternative and the baseline model, which results in two sets of observable values. The differences between these values are calculated for each observable and pseudoexperiment. The systematic uncertainty for a given observable is then obtained from the sum in quadrature of the mean and Gaussian width of the distribution of the resulting differences.
When fitting with an alternative model proves infeasible, pseudoexperiments generated with the alternative configuration are fitted using the baseline model only, and the resulting biases are taken as systematic uncertainties.

Sources of systematic uncertainties for the $S$- and $P$-basis observables are summarised in Tables~\ref{tab:rare_mode_syst_S_sq2c} and~\ref{tab:rare_mode_syst_P_sq2c}, respectively, where the values correspond to the systematic uncertainties divided by the statistical uncertainties. They are discussed in detail below.

All sources of systematic uncertainties discussed below are expected to be uncorrelated, and the total uncertainty is the sum in quadrature of all sources. 
Correlations may arise between different observables. The correlation matrix of total systematic uncertainties on the observables is given in Appendix~\ref{sec:app_correlation_matrices}.

\begin{table}[!tb]
\caption{Summary of the systematic uncertainties on the $\it{S}$-basis angular observables. All values are given as fractions of the statistical uncertainties.}
\label{tab:rare_mode_syst_S_sq2c}
\vspace{0.2cm}
\sisetup{separate-uncertainty}
\centering
\begin{tabular}{l|cccccccc}
\multicolumn{1}{l}{Source} & \multicolumn{1}{c}{$F_{L}$} & \multicolumn{1}{c}{$S_{3}$} & \multicolumn{1}{c}{$S_{4}$} & \multicolumn{1}{c}{$S_{5}$} & \multicolumn{1}{c}{$\AFB$} & \multicolumn{1}{c}{$S_{7}$} & \multicolumn{1}{c}{$S_{8}$} & \multicolumn{1}{c}{$S_{9}$} \\\hline

\multicolumn{1}{l}{Comb. and DSL backgrounds} & \multicolumn{1}{c}{0.69} & \multicolumn{1}{c}{0.37} & \multicolumn{1}{c}{0.30} & \multicolumn{1}{c}{0.32} & \multicolumn{1}{c}{0.45} & \multicolumn{1}{c}{0.18} & \multicolumn{1}{c}{0.67} & \multicolumn{1}{c}{0.29} \\

\multicolumn{1}{l}{Part. reco. background} & \multicolumn{1}{c}{0.21} & \multicolumn{1}{c}{0.05} & \multicolumn{1}{c}{0.07} & \multicolumn{1}{c}{0.11} & \multicolumn{1}{c}{0.09} & \multicolumn{1}{c}{0.04} & \multicolumn{1}{c}{0.04} & \multicolumn{1}{c}{0.05} \\

\multicolumn{1}{l}{Misid. had. background} & \multicolumn{1}{c}{0.38} & \multicolumn{1}{c}{0.25} & \multicolumn{1}{c}{0.11} & \multicolumn{1}{c}{0.18} & \multicolumn{1}{c}{0.15} & \multicolumn{1}{c}{0.34} & \multicolumn{1}{c}{0.13} & \multicolumn{1}{c}{0.14} \\

\multicolumn{1}{l}{Effective acceptance} & \multicolumn{1}{c}{0.39} & \multicolumn{1}{c}{0.27} & \multicolumn{1}{c}{0.44} & \multicolumn{1}{c}{0.43} & \multicolumn{1}{c}{0.34} & \multicolumn{1}{c}{0.53} & \multicolumn{1}{c}{0.43} & \multicolumn{1}{c}{0.23} \\

\multicolumn{1}{l}{Signal mass modelling} & \multicolumn{1}{c}{0.26} & \multicolumn{1}{c}{0.06} & \multicolumn{1}{c}{0.07} & \multicolumn{1}{c}{0.11} & \multicolumn{1}{c}{0.12} & \multicolumn{1}{c}{0.05} & \multicolumn{1}{c}{0.05} & \multicolumn{1}{c}{0.06} \\

\multicolumn{1}{l}{$J/\psi$ backgrounds} & \multicolumn{1}{c}{0.17} & \multicolumn{1}{c}{0.04} & \multicolumn{1}{c}{0.04} & \multicolumn{1}{c}{0.06} & \multicolumn{1}{c}{0.14} & \multicolumn{1}{c}{0.03} & \multicolumn{1}{c}{0.03} & \multicolumn{1}{c}{0.05} \\

\multicolumn{1}{l}{S-wave component} & \multicolumn{1}{c}{0.35} & \multicolumn{1}{c}{0.05} & \multicolumn{1}{c}{0.13} & \multicolumn{1}{c}{0.08} & \multicolumn{1}{c}{0.10} & \multicolumn{1}{c}{0.16} & \multicolumn{1}{c}{0.01} & \multicolumn{1}{c}{0.10} \\

\multicolumn{1}{l}{$B^+$ veto} & \multicolumn{1}{c}{0.50} & \multicolumn{1}{c}{0.13} & \multicolumn{1}{c}{0.15} & \multicolumn{1}{c}{0.18} & \multicolumn{1}{c}{0.24} & \multicolumn{1}{c}{0.16} & \multicolumn{1}{c}{0.15} & \multicolumn{1}{c}{0.12} \\

\multicolumn{1}{l}{Fit bias} & \multicolumn{1}{c}{0.01} & \multicolumn{1}{c}{0.01} & \multicolumn{1}{c}{0.03} & \multicolumn{1}{c}{0.03} & \multicolumn{1}{c}{0.05} & \multicolumn{1}{c}{0.01} & \multicolumn{1}{c}{0.01} & \multicolumn{1}{c}{0.03} \\\hline

\multicolumn{1}{l}{Total} & \multicolumn{1}{c}{1.13} & \multicolumn{1}{c}{0.55} & \multicolumn{1}{c}{0.59} & \multicolumn{1}{c}{0.62} & \multicolumn{1}{c}{0.68} & \multicolumn{1}{c}{0.70} & \multicolumn{1}{c}{0.82} & \multicolumn{1}{c}{0.44} \\\hline

\end{tabular}
\end{table}
\begin{table}[!tb]
\caption{Summary of the systematic uncertainties on the $\it{P}$-basis angular observables. All values are given as fractions of the statistical uncertainties.}
\label{tab:rare_mode_syst_P_sq2c}
\vspace{0.2cm}
\sisetup{separate-uncertainty}
\centering
\begin{tabular}{l|cccccccc}
\multicolumn{1}{l}{Source} & \multicolumn{1}{c}{$F_{L}$} & \multicolumn{1}{c}{$P_{1}$} & \multicolumn{1}{c}{$P_{4}^{\prime}$} & \multicolumn{1}{c}{$P_{5}^{\prime}$} & \multicolumn{1}{c}{$P_{2}$} & \multicolumn{1}{c}{$P_{6}^{\prime}$} & \multicolumn{1}{c}{$P_{8}^{\prime}$} & \multicolumn{1}{c}{$P_{3}$} \\\hline

\multicolumn{1}{l}{Comb and DSL backgrounds} & \multicolumn{1}{c}{0.69} & \multicolumn{1}{c}{0.87} & \multicolumn{1}{c}{0.49} & \multicolumn{1}{c}{0.61} & \multicolumn{1}{c}{0.95} & \multicolumn{1}{c}{0.24} & \multicolumn{1}{c}{0.81} & \multicolumn{1}{c}{0.71} \\

\multicolumn{1}{l}{Part. reco. background} & \multicolumn{1}{c}{0.21} & \multicolumn{1}{c}{0.17} & \multicolumn{1}{c}{0.14} & \multicolumn{1}{c}{0.22} & \multicolumn{1}{c}{0.20} & \multicolumn{1}{c}{0.06} & \multicolumn{1}{c}{0.07} & \multicolumn{1}{c}{0.16} \\

\multicolumn{1}{l}{Misid. had. background} & \multicolumn{1}{c}{0.38} & \multicolumn{1}{c}{0.57} & \multicolumn{1}{c}{0.18} & \multicolumn{1}{c}{0.26} & \multicolumn{1}{c}{0.34} & \multicolumn{1}{c}{0.41} & \multicolumn{1}{c}{0.17} & \multicolumn{1}{c}{0.36} \\

\multicolumn{1}{l}{Effective acceptance} & \multicolumn{1}{c}{0.39} & \multicolumn{1}{c}{0.49} & \multicolumn{1}{c}{0.52} & \multicolumn{1}{c}{0.51} & \multicolumn{1}{c}{0.55} & \multicolumn{1}{c}{0.62} & \multicolumn{1}{c}{0.50} & \multicolumn{1}{c}{0.40} \\

\multicolumn{1}{l}{Signal mass modelling} & \multicolumn{1}{c}{0.26} & \multicolumn{1}{c}{0.16} & \multicolumn{1}{c}{0.14} & \multicolumn{1}{c}{0.18} & \multicolumn{1}{c}{0.31} & \multicolumn{1}{c}{0.06} & \multicolumn{1}{c}{0.06} & \multicolumn{1}{c}{0.15} \\

\multicolumn{1}{l}{$J/\psi$ backgrounds} & \multicolumn{1}{c}{0.18} & \multicolumn{1}{c}{0.13} & \multicolumn{1}{c}{0.06} & \multicolumn{1}{c}{0.11} & \multicolumn{1}{c}{0.29} & \multicolumn{1}{c}{0.04} & \multicolumn{1}{c}{0.04} & \multicolumn{1}{c}{0.12} \\

\multicolumn{1}{l}{S-wave component} & \multicolumn{1}{c}{0.35} & \multicolumn{1}{c}{0.10} & \multicolumn{1}{c}{0.18} & \multicolumn{1}{c}{0.11} & \multicolumn{1}{c}{0.29} & \multicolumn{1}{c}{0.21} & \multicolumn{1}{c}{0.01} & \multicolumn{1}{c}{0.20} \\

\multicolumn{1}{l}{$B^+$ veto} & \multicolumn{1}{c}{0.50} & \multicolumn{1}{c}{0.41} & \multicolumn{1}{c}{0.28} & \multicolumn{1}{c}{0.37} & \multicolumn{1}{c}{0.52} & \multicolumn{1}{c}{0.22} & \multicolumn{1}{c}{0.21} & \multicolumn{1}{c}{0.37} \\

\multicolumn{1}{l}{Fit bias} & \multicolumn{1}{c}{0.01} & \multicolumn{1}{c}{0.00} & \multicolumn{1}{c}{0.04} & \multicolumn{1}{c}{0.03} & \multicolumn{1}{c}{0.08} & \multicolumn{1}{c}{0.02} & \multicolumn{1}{c}{0.02} & \multicolumn{1}{c}{0.02} \\\hline

\multicolumn{1}{l}{Total} & \multicolumn{1}{c}{1.14} & \multicolumn{1}{c}{1.25} & \multicolumn{1}{c}{0.84} & \multicolumn{1}{c}{0.97} & \multicolumn{1}{c}{1.38} & \multicolumn{1}{c}{0.84} & \multicolumn{1}{c}{0.99} & \multicolumn{1}{c}{1.02} \\\hline
\end{tabular}
\end{table}

\paragraph{Combinatorial and DSL background modelling}
Systematic uncertainties can arise from different sources, namely the limited size of the $\Kp \pim \ep \mun$ data sample, the choice of the models, and the assumption of factorisation.
The impact of the limited sample size is studied using bootstrapping techniques~\cite{efron:1979}. Systematic uncertainties related to model choice are quantified using alternative models. 
To examine the impact of the parametrisation strategy, alternative functions are used to describe both components. For the combinatorial background, \ctk and \ctl are described by polynomials up to third order, $\phi$ is described using trigonometric terms up to fourth order, and a Gaussian function is used to describe its invariant-mass distribution. 
For the DSL background, $\ctk$ and $\phi$ are described by an unfactorised model consisting of polynomials and trigonometric terms up to second and third order, respectively, \ctl is described by a parametric model composed of four Gaussian functions, and a Gaussian function is used to describe its invariant-mass distribution. 
To quantify systematic uncertainties related to the use of the $\Kp \pim \ep \mun$ sample to model backgrounds in the $\Kp \pim e^+ e^-$ final state, for the DSL background, an alternative model is obtained using simulated $\Bz\to D^{-}(\to\Kstarz e^-\bar{\nu}_{e})e^+\nu_{e}$ decays.
For the combinatorial background, a different model is obtained via an alternative data fit for $\ctl$ and $\phi$, and from the same-sign $\Kp\pim e^{\pm}e^{\pm}$ data for $\ctk$.
The impact of the factorisation assumption is studied using a fully unfactorised model for both backgrounds determined from the $\Kp \pim \ep \mun$ data sample.

\paragraph{Partially reconstructed background modelling}
Systematic uncertainties associated with the partially reconstructed background are relatively small as its contribution is limited by the narrow signal mass window.
Furthermore, given the large sample size available for parametrisation, the only relevant sources of uncertainties are the parametrisation strategy, the choice of the physics model, and the assumption of factorisation.
The impact of the parametrisation strategy is quantified using an alternative model, where the maximum polynomial orders are increased to three for $\ctk$ and $\ctl$, trigonometric functions up to fourth order are used for $\phi$, and a parametric model composed of two Gaussian functions and one exponential function is used to describe the invariant-mass distribution.
The systematic uncertainty associated with the choice of the physics model is quantified using an alternative model obtained from simulated $B^{+}\to K_{1}(1270)^{+}e^{+}e^{-}$ and $B^{+}\to K^*_{2}(1430)^{+} e^{+}e^{-}$ decays.
The impact of the factorisation assumption is assessed by bootstrapping the $B^{+}\to K^{+}\pi^{+}\pi^{-}e^+e^-$ simulation.

\paragraph{Misidentified hadronic background modelling}
Sources of systematic uncertainties related to the modelling of misidentified hadronic decays include the parametrisation method, the definition of the control region, the impact of potential dependencies on the event occupancy, and the size of the control-region sample.
The systematic uncertainties of the parametrisation method is quantified using an alternative parametric model composed of two Gaussian functions and an exponential function, a model which was used in Refs.~\cite{LHCb-PAPER-2022-045,LHCb-PAPER-2022-046}. Polynomials of up to sixth order are used to describe the $\ctl$ distribution. For $\ctk$ and $\phi$, an alternative unfactorised model consisting of a product of polynomials of up to second order is used. 
To quantify systematic uncertainties associated with the choice of the baseline control region, a different electron PID requirement is used.
Alternative transfer functions that depend on the transverse momentum and the number of hits in the scintillating-pad detector are made to quantify systematic uncertainties associated with potential dependencies on the event occupancy.
The systematic uncertainty related to the size of the control-region sample is assessed by bootstrapping.

\paragraph{Effective acceptance functions}
Sources of systematic uncertainties associated with the effective acceptance functions include the choice of the polynomial order used in their parametrisation, the size of the simulation samples used to calculate the coefficients, the model dependence of the resolution correction, and the strategy used to correct for differences between simulation and data. These are evaluated using signal-only pseudoexperiments.
The systematic uncertainties associated with the choice of the polynomial order are quantified by increasing all polynomial orders by three. The impact of the limited size of the simulation samples is assessed by bootstrapping to produce alternative effective acceptance functions. To quantify the systematic uncertainties due to the resolution correction, alternative effective acceptance functions are parametrised after modifying the simulated distributions using weights that remove the effects of the baseline physics model, and introduce those of the alternative physics model.
Three plausible BSM scenarios are considered: a) $\delta C_{9}=-1$, b) $\delta C_{9}=-\delta C_{10}=-0.7$ and c) $\delta C_{9}=-1.4$. Systematic uncertainties are quantified for each case separately. The largest uncertainty found in scenarios a) to c) for each observable is taken as the systematic uncertainty.
The simulation correction strategy affects the analysis primarily through the impact of the resulting per-event weights on the shape of the effective acceptance functions. 
The approach in Refs.~\cite{LHCb-PAPER-2022-045,LHCb-PAPER-2022-046} is used to quantify systematic uncertainties associated with the baseline simulation correction strategy.
The full set of per-event weights from this alternative correction strategy is used in the parametrisation of alternative effective acceptance functions.
The limited size of the PID calibration sample for the electron mode introduces another source of systematic uncertainty. This is quantified using alternative acceptance functions parametrised with the baseline approach, with the exception that all PID requirements are made on alternative PID variables obtained by bootstrapping.

\paragraph{Signal invariant-mass model}
The invariant-mass distribution of signal candidates is modelled using DCB functions with shift and scaling parameters determined from the control-mode fit.
This choice, as well as aspects of the control-mode fit and the assumption of factorisation between invariant mass and angles, are sources of systematic uncertainties. 
An alternative model obtained by applying KDE to all signal candidates
of each run period
is used to quantify the systematic uncertainties associated with the parametrisation strategy.
Two alternative sets of shift and scaling parameters are used to quantify the systematic uncertainties of the baseline choice.
The first is obtained by making a control-mode fit where the requirement used to suppress partially reconstructed background is removed and additional backgrounds are included.
The second is obtained by making a fit in the same mass range as the signal.
The systematic uncertainty due to the imperfect modelling of invariant-mass distributions in data due to residual simulation-data differences is assessed using alternative mass models, where these differences are removed using weights. The assumption of factorisation 
is broken by electron energy loss, which introduces correlations between the $\Bz$ invariant mass and $\ctl$. The signal simulation is bootstrapped to quantify the impact of this assumption.

\paragraph{$\boldsymbol{J/\psi}$ backgrounds}
Two types of backgrounds with distinct invariant-mass and angular distributions that are neglected in the fit are $\Bz\to\Kstarz J/\psi(\to e^+e^-)$ decays that leak into the signal $q^2$ region due to significant energy loss, and $J/\psi$ mesons combined with random kaon and pion tracks. In each case, the impact of neglecting the component is assessed by generating pseudoexperiments with its inclusion, and fitting them with the baseline strategy.

\paragraph{S-wave component}
The angular function used in the fit describes decays where the $\Kp\pim$ system originates from the $\Kstar(892)^0$ vector meson, and does not include S-wave related terms.
The impact of neglecting these contributions is assessed using signal-only pseudoexperiments generated with six additional angular terms, and fitted with and without their inclusion. Fits are performed to the $\Bz\to\Kstarz\mup\mun$ candidates from the data samples analysed in Ref.~\cite{LHCb-PAPER-2020-002}, to obtain realistic values for the additional angular observables.

\paragraph{$\boldsymbol{B^{+}}$ veto}
The requirements to veto $\Bu \to\Kp e^+e^-$ decays distort the background distribution above the known $\Bz$ mass, and introduce correlations between invariant mass and $\ctk$.
Furthermore, due to the removal of events in a region of the phase space, the distortions caused by this veto cannot be corrected properly using the effective acceptance functions.
To quantify the associated systematic uncertainties, an alternative data fit is made without this veto, using effective acceptance functions and signal and background models obtained likewise without it. The result of this fit is used to generate pseudoexperiments, which are fitted once using the no-veto models and weights. Then, candidates are removed using binned efficiency values from a histogram model that describes the effect of this veto in three dimensions ($q^2$, $\ctk$ and $\Bz$ invariant mass)~\cite{LHCb-PAPER-2023-032}. The pseudoexperiments are fitted again using the baseline setup.

\paragraph{Fit bias}
The biases found in the validation of the fit strategy using pseudoexperiments generated with data-fit observable values are all limited in size, with the largest found for $P_{2}$  at around $8\%$ of the statistical uncertainty.
These values are taken as systematic uncertainties.

\paragraph{Summary of systematic uncertainties}
The systematic uncertainties, summarised in Tables~\ref{tab:rare_mode_syst_S_sq2c} and~\ref{tab:rare_mode_syst_P_sq2c}, are not negligible compared to the statistical uncertainties.
For all $\it{S}$-basis observables, they amount to more than 40\% of their statistical uncertainties. The most affected observable is $\FL$, which is easily biased by differences in \ctk when changing between the baseline and alternative background models, neglecting correlation between $\ctk$ and \Bd invariant mass, and differences between the baseline and alternative effective acceptance corrections for $\ctk$. 
The observable $\AFB$ is sensitive to effects that are not symmetric in $\ctl$; in particular, it is biased by the difference between the width of the $\ctl$ peak of the baseline and alternative DSL model.
Other observables, such as $S_4$ and $S_5$, have more complex dependencies on the decay angles, and so the impact of one-dimensional effects is reduced. However, multidimensional effects due to the correlated effective acceptance functions as well as angular correlations and correlations between angles and \Bz invariant mass present for some backgrounds can lead to non-negligible systematic uncertainties. 
The $\it{P}$-basis observables, which incorporate $\FL$ in their definitions, have much larger systematic uncertainties of over 80\% of their respective statistical uncertainties in all cases. 

\section{Results}\label{sec:results}

\begin{figure}[!b]
\centering
    \includegraphics[width=.45\textwidth, trim={0 0 0 0},clip]{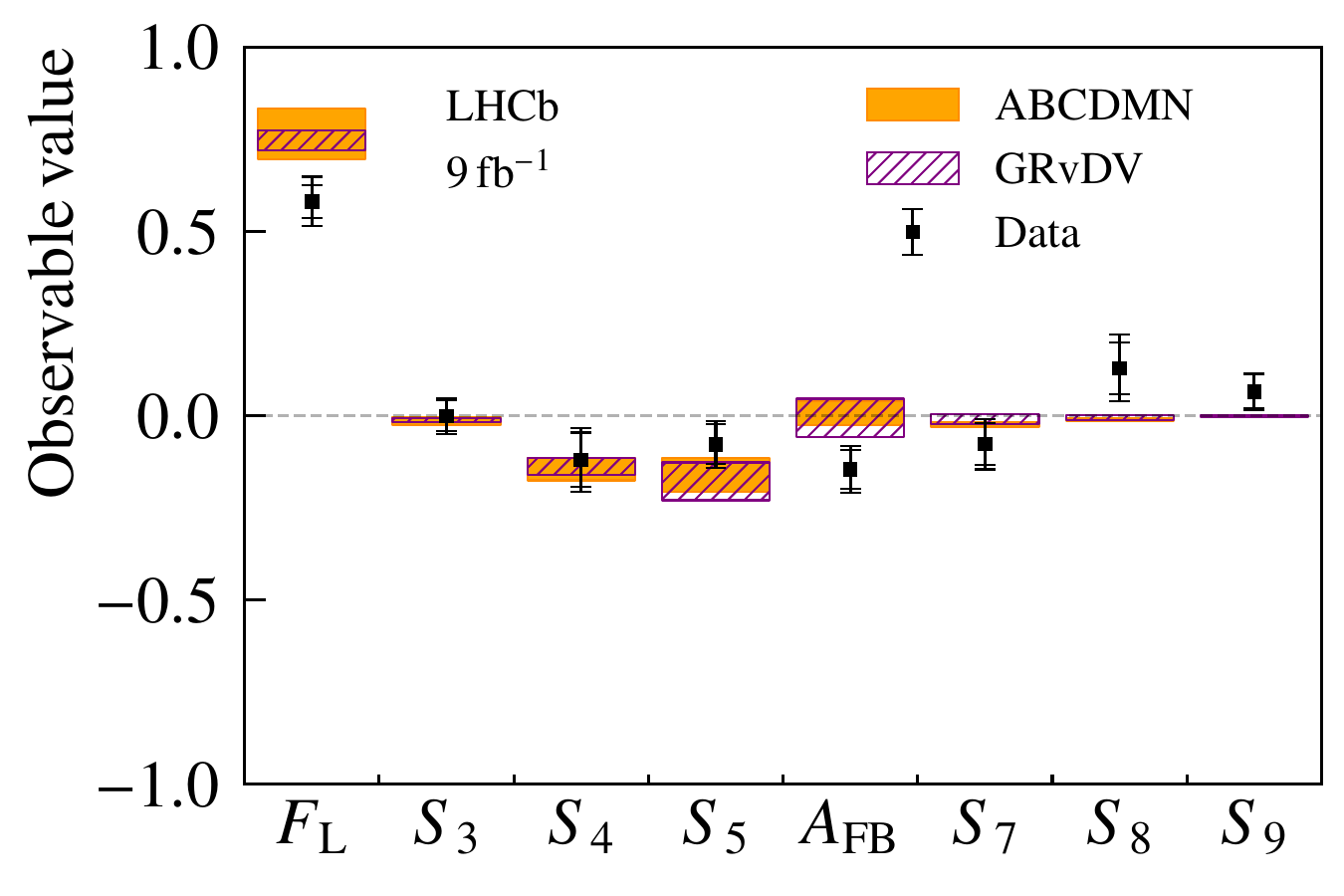} 
    \includegraphics[width=.45\textwidth, trim={0 0 0 0},clip]{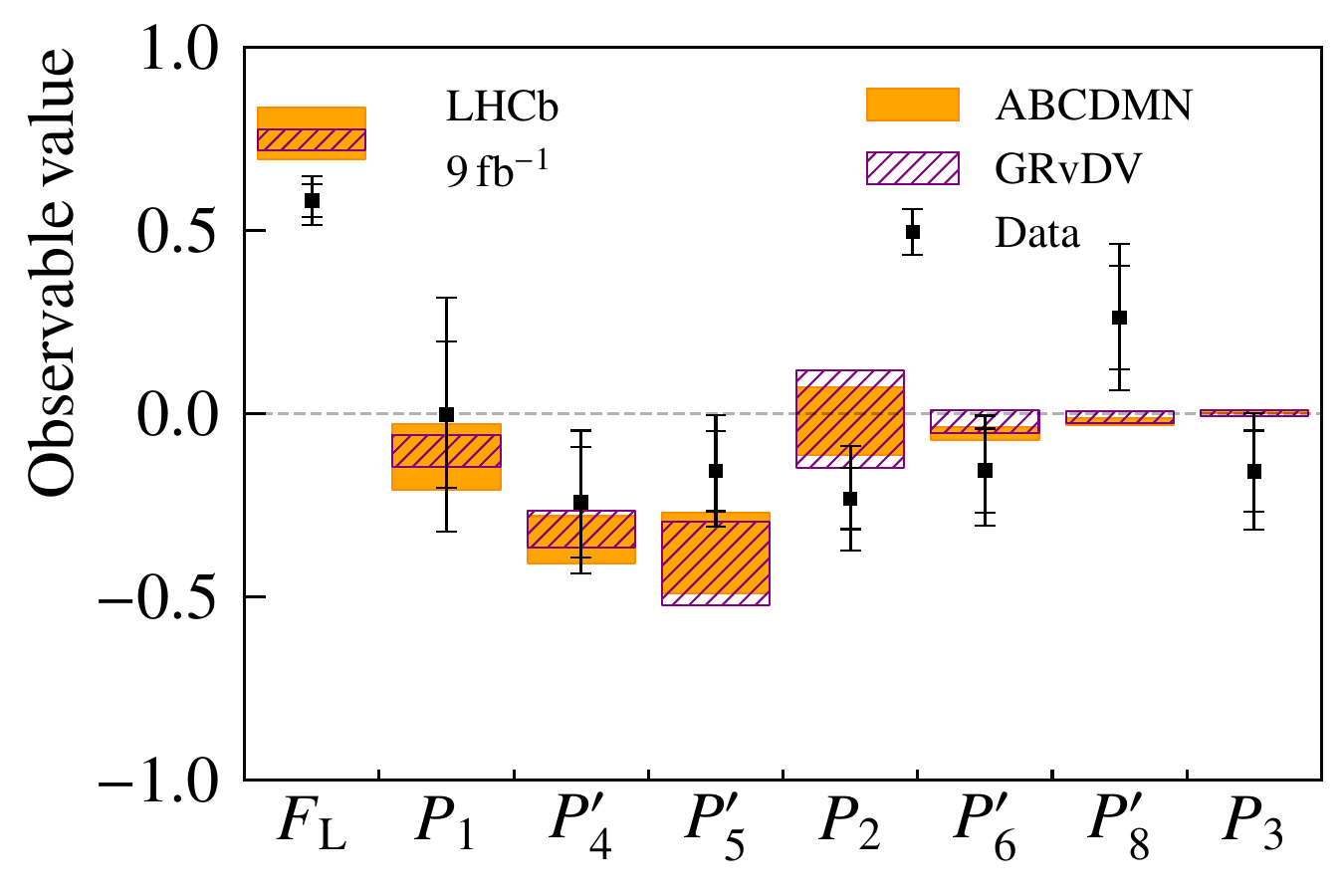} 
  \caption{The (left) $\it{S}$- and (right) $\it{P}$-basis angular observables. The overlapping error bars show statistical and total uncertainties. The orange and hatched purple boxes correspond to SM predictions based on Ref.~\cite{Alguero:2023jeh} and Refs.~\cite{EOSAuthors:2021xpv,Gubernari:2022hxn}, respectively. 
  }
\label{fig:summary_plots}
\end{figure}

\begin{table}[!tb]
\sisetup{separate-uncertainty}
\centering
\caption{Values for the (left) $\it{S}$- and (right) $\it{P}$-basis angular observables. The first uncertainty is statistical and the second is systematic.}
\begin{tabular}{l r l r}
  \multicolumn{4}{c}{}  \\ \hline
 \multicolumn{4}{c}{Angular observables}  \\ \hline
  $\FL$ &$ 0.58 \pm 0.04  \pm 0.05 $ & $\FL$ & $ 0.58 \pm 0.04  \pm 0.05 $ \\ 
 $S_{3}$ &$ -0.00 \pm 0.04  \pm 0.02 $ &$P_{1}$ &$ -0.00 \pm 0.20  \pm 0.25 $ \\
 $S_{4}$ &$ -0.12 \pm 0.07  \pm 0.04 $ & $P_{4}^{\prime}$ &$ -0.24 \pm 0.15  \pm 0.13 $ \\
 $S_{5}$ &$ -0.08 \pm 0.05  \pm 0.03 $ & $P_{5}^{\prime}$ &$ -0.16 \pm 0.11  \pm 0.11 $ \\
 $\AFB$ &$ -0.15 \pm 0.05  \pm 0.04 $ &$P_{2}$ &$ -0.23 \pm 0.08  \pm 0.12 $ \\
 $S_{7}$ &$ -0.08 \pm 0.06  \pm 0.04 $ & $P_{6}^{\prime}$ &$ -0.16 \pm 0.11  \pm 0.10 $ \\
 $S_{8}$ &$ 0.13 \pm 0.07  \pm 0.06 $ & $P_{8}^{\prime}$ &$ 0.26 \pm 0.14  \pm 0.14 $ \\
 $S_{9}$ &$ 0.07 \pm 0.05  \pm 0.02 $ &$P_{3}$ &$ -0.16 \pm 0.11  \pm 0.11 $ \\ \hline
\end{tabular}
\label{tab:observables_results}
\end{table}

The angular observables of $\Bz\to\Kstarz e^+e^-$ decays, measured in the $q^2$ region of \mbox{$1.1$--$6.0\gevgevcccc$}, are shown in Fig.~\ref{fig:summary_plots} together with the SM predictions based on Refs.~\cite{Alguero:2023jeh,EOSAuthors:2021xpv,Gubernari:2022hxn}.
This is the most precise measurement of the $\Bz\to\Kstarz e^+e^-$ angular observables in the central-$q^2$ region to date, improving the precision on $P_4^{\prime}$ and $P_5^{\prime}$, previously measured by Belle~\cite{Belle:2016fev}, by more than a factor of two. 
The measured observables are broadly in agreement with the SM predictions, with the largest differences of around $2\sigma$ found for $\FL$ and $\AFB$, followed by differences of around $1.5\sigma$ for $S_8$ and $P_8^{\prime}$.
The numerical results of the fit are given in Table~\ref{tab:observables_results}, with statistical and systematic uncertainties shown separately, and fit correlation matrices are reported in Tables~\ref{tab:corr_sq2_S} and~\ref{tab:corr_sq2_P} of Appendix~\ref{sec:app_correlation_matrices}. 
Statistical correlations between the observables are generally small.
The largest correlation of around 14\% is found between $S_{7}$ and $S_9$ for the $\it{S}$-basis observables, and 20\% between $F_{\rm L}$ and $P_2$ for the $\it{P}$-basis observables.
Correlations between systematic uncertainties are discussed in Appendix~\ref{sec:app_correlation_matrices}.

\begin{figure}[!tb]
\centering
    \includegraphics[width=.45\textwidth, trim={0 0 0 0},clip]{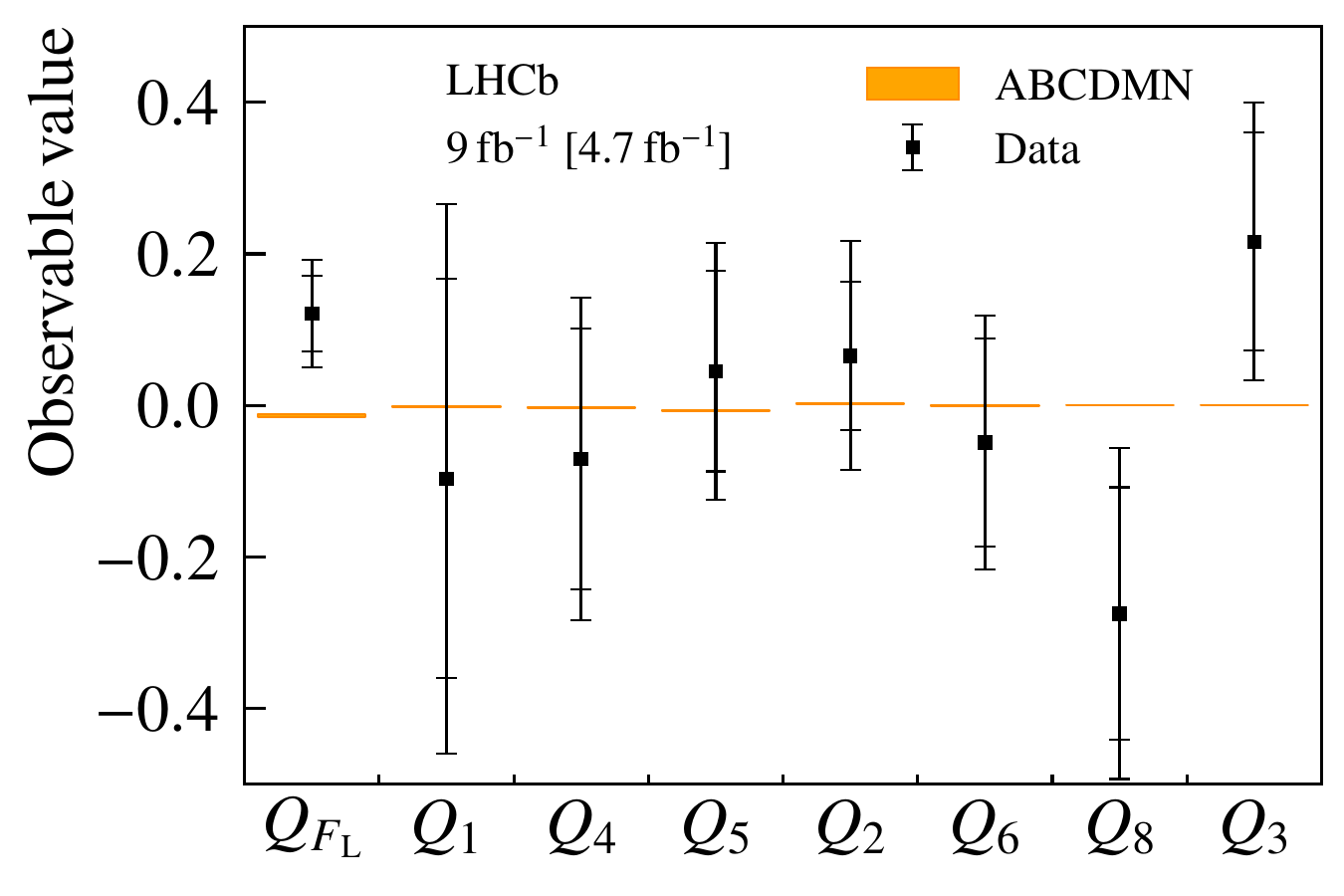} 
  \caption{LFU observables $Q_{i}$ calculated using the $\it{P}$-basis angular observables of the muon and electron modes.
  The overlapping error bars show statistical and total uncertainties. The SM predictions (orange boxes) are based on Ref.~\cite{Alguero:2023jeh}. 
  }
\label{fig:summary_plots_Qi}
\end{figure}

\begin{figure}[!tb]
\centering
    \includegraphics[width=.45\textwidth, trim={0 0 0 0},clip]{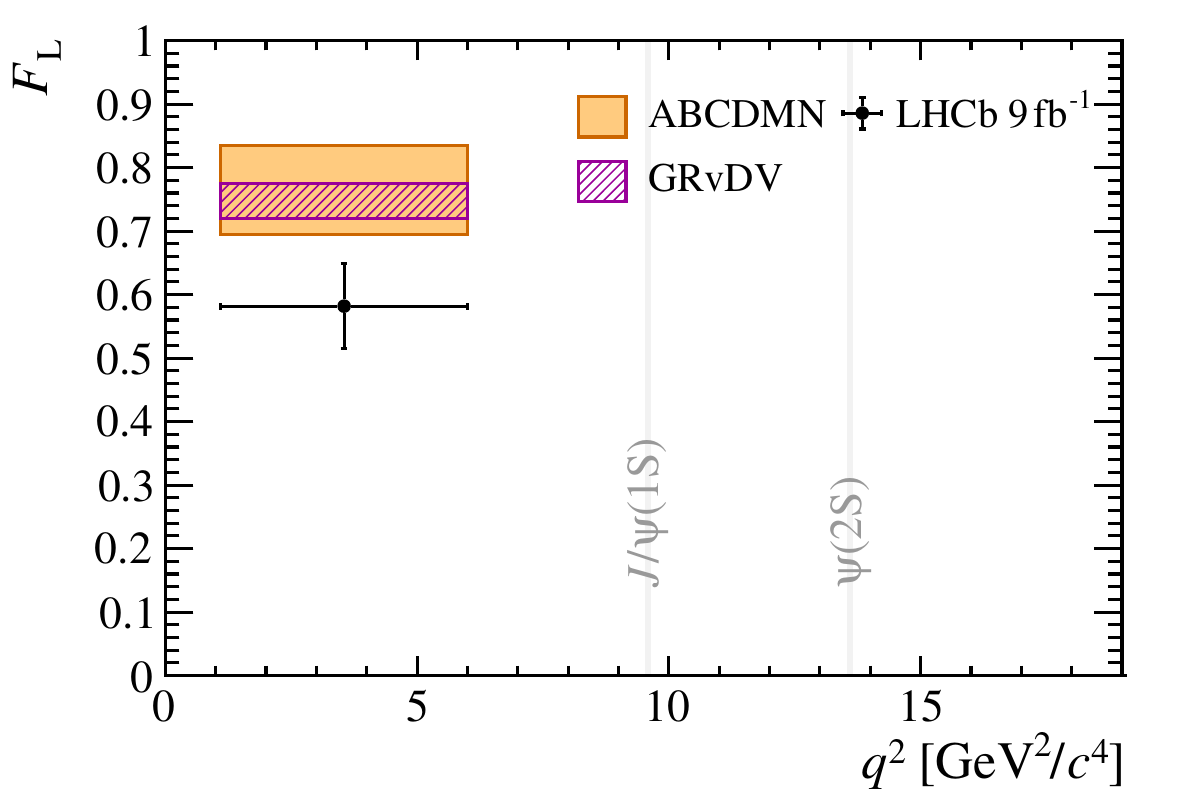} 
    \includegraphics[width=.45\textwidth, trim={0 0 0 0},clip]{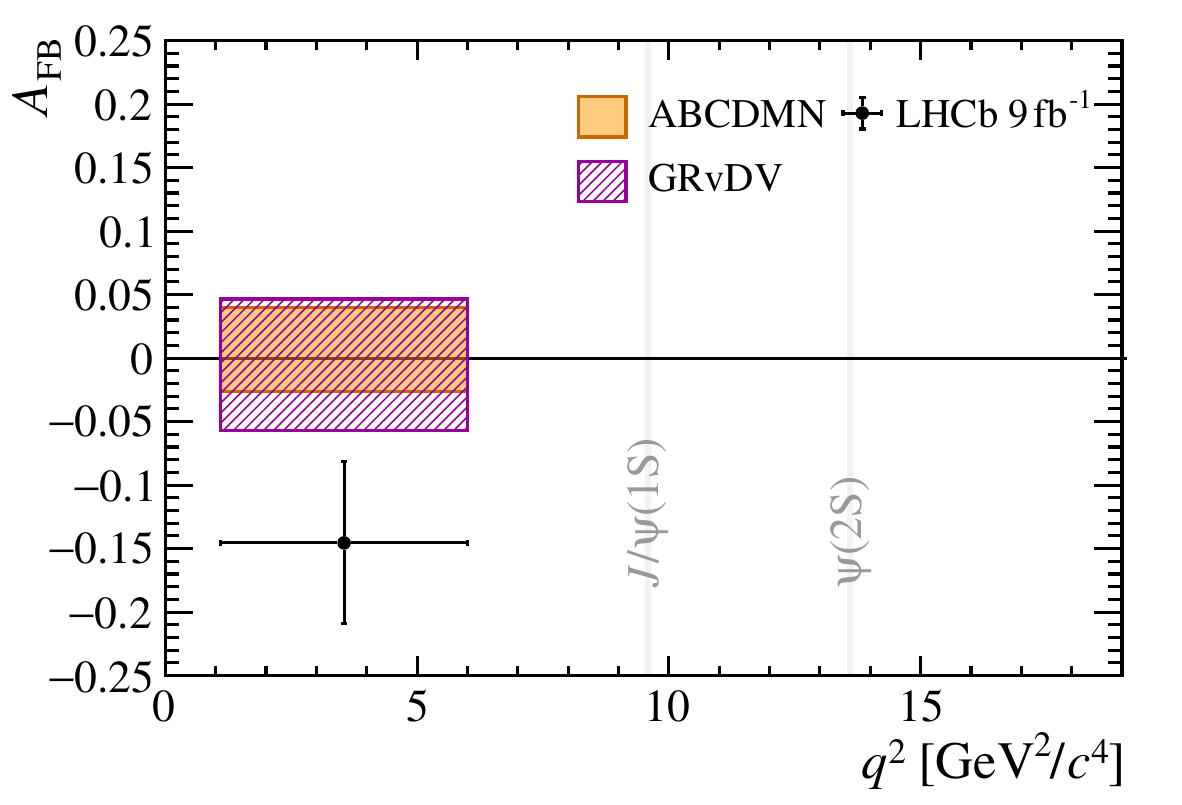} 
    \includegraphics[width=.45\textwidth, trim={0 0 0 0},clip]{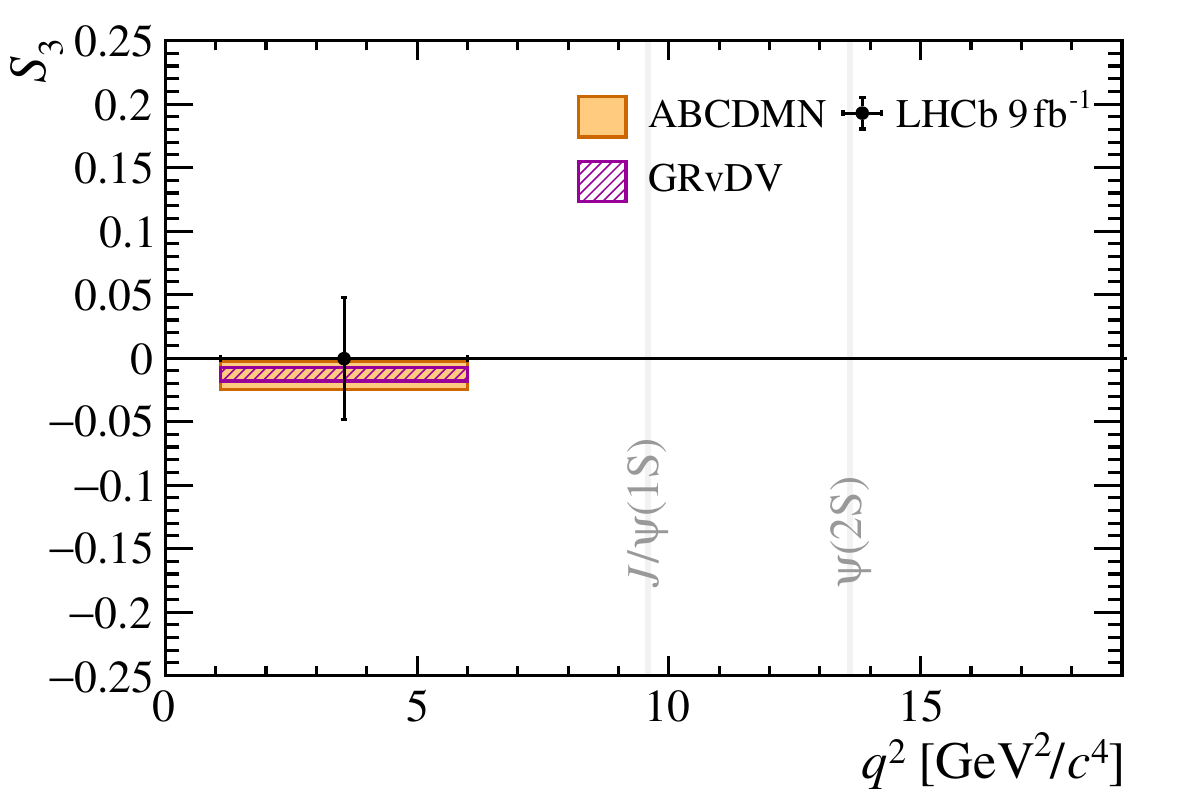} 
    \includegraphics[width=.45\textwidth, trim={0 0 0 0},clip]{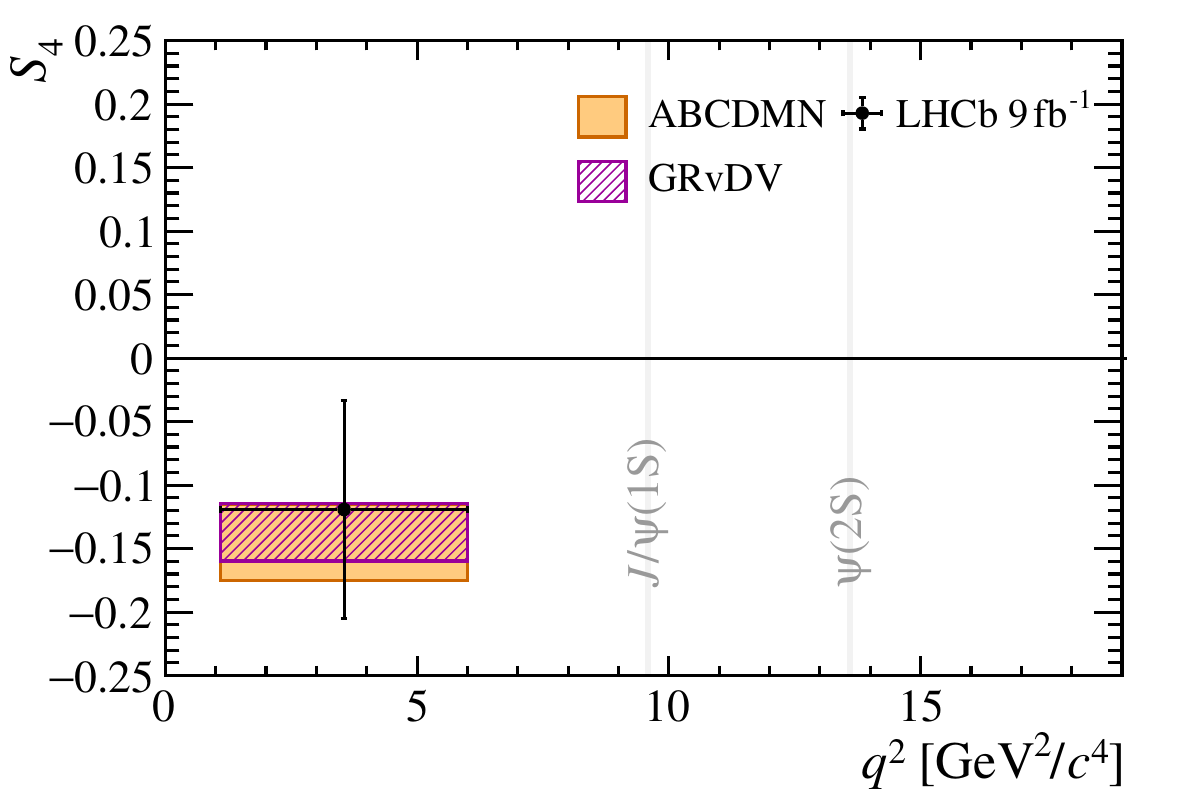} 
    \includegraphics[width=.45\textwidth, trim={0 0 0 0},clip]{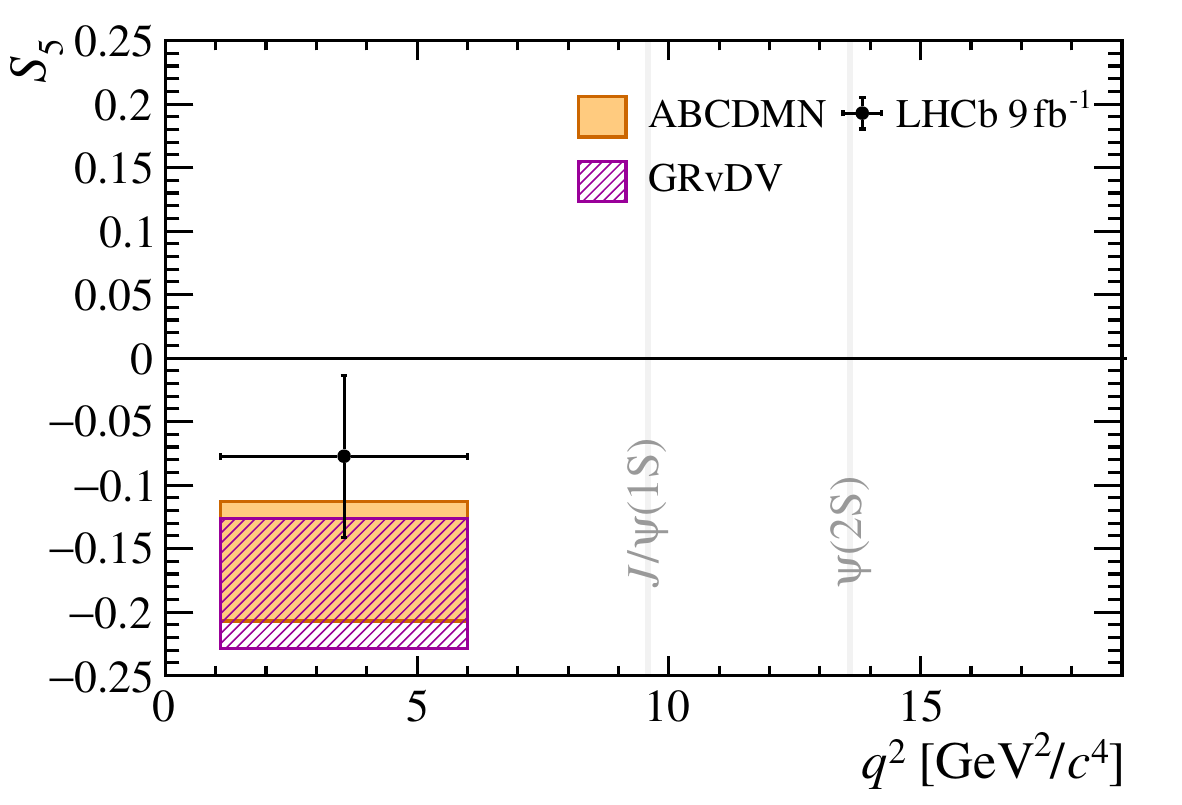} 
    \includegraphics[width=.45\textwidth, trim={0 0 0 0},clip]{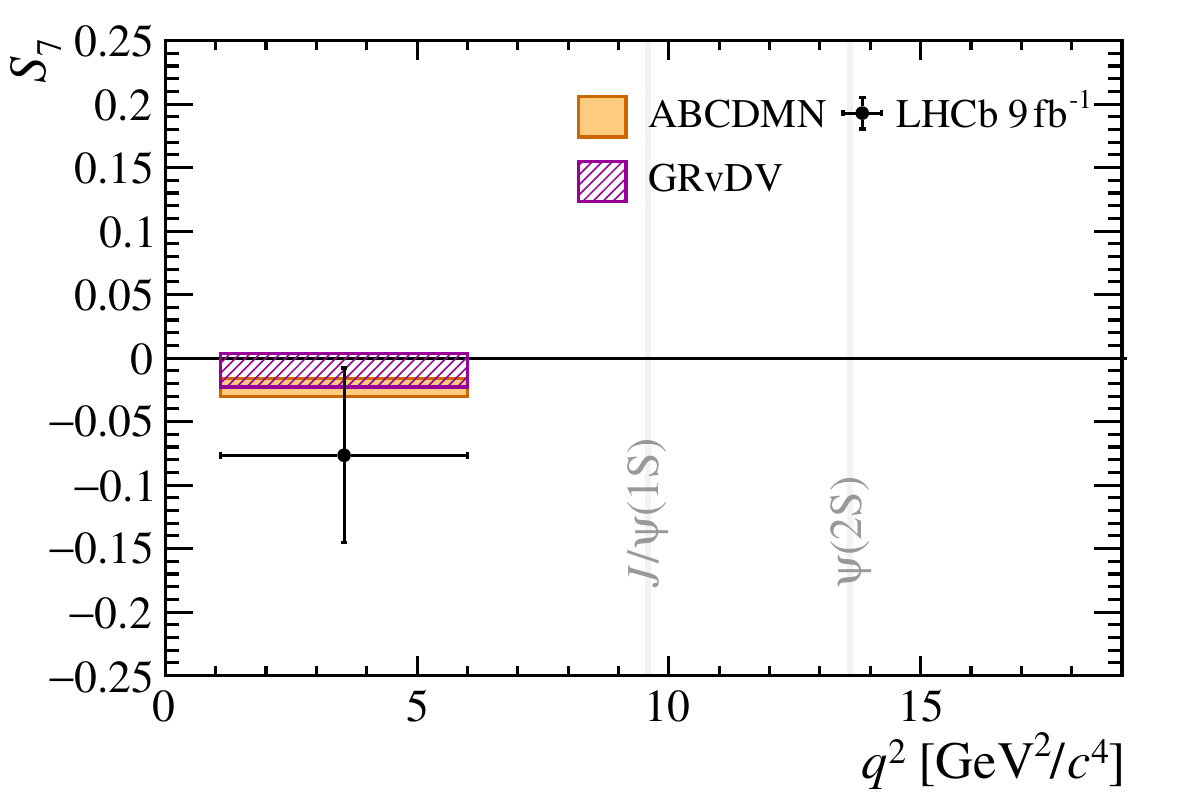} 
    \includegraphics[width=.45\textwidth, trim={0 0 0 0},clip]{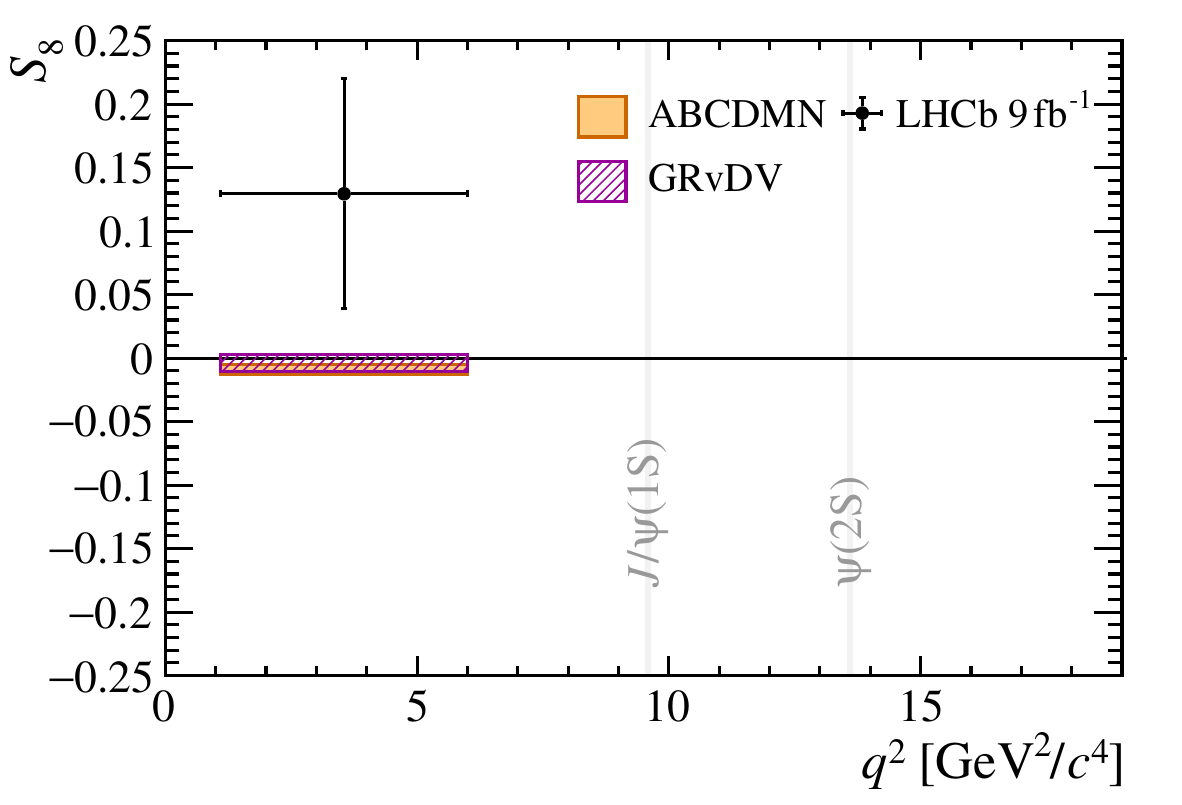} 
    \includegraphics[width=.45\textwidth, trim={0 0 0 0},clip]{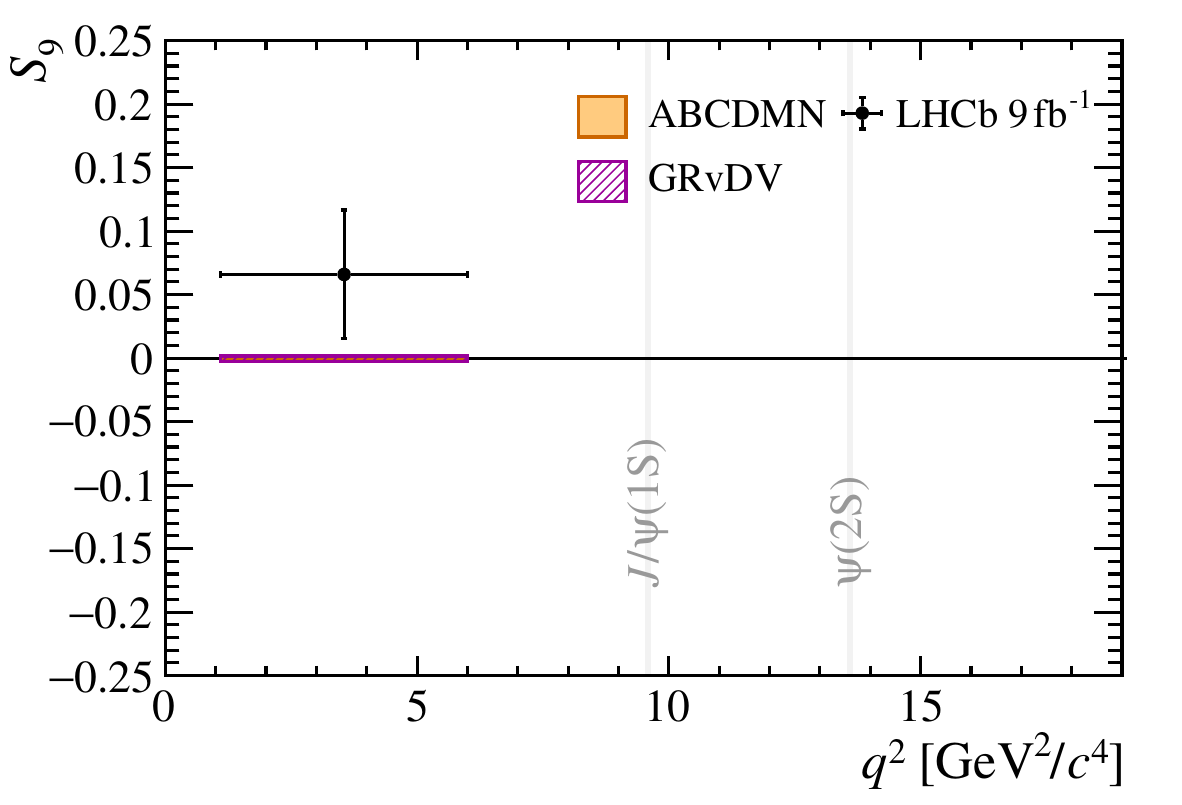} 
  \caption{Measured $\it{S}$-basis angular observables.
  The orange and hatched purple boxes correspond to SM predictions based on Ref.~\cite{Alguero:2023jeh} and Refs.~\cite{EOSAuthors:2021xpv,Gubernari:2022hxn}, respectively.
  }
\label{fig:Si_plots}
\end{figure}

\begin{figure}[!tb]
\centering
    \includegraphics[width=.45\textwidth, trim={0 0 0 0},clip]{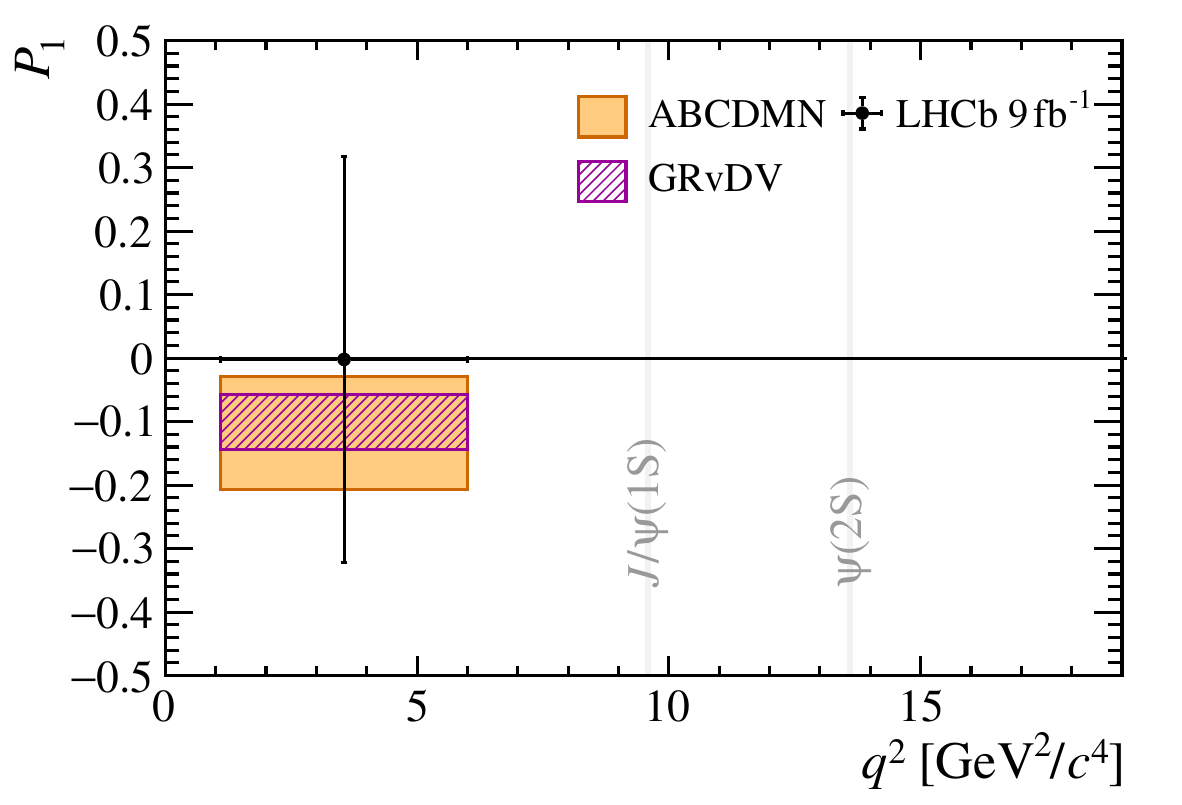} 
    \includegraphics[width=.45\textwidth, trim={0 0 0 0},clip]{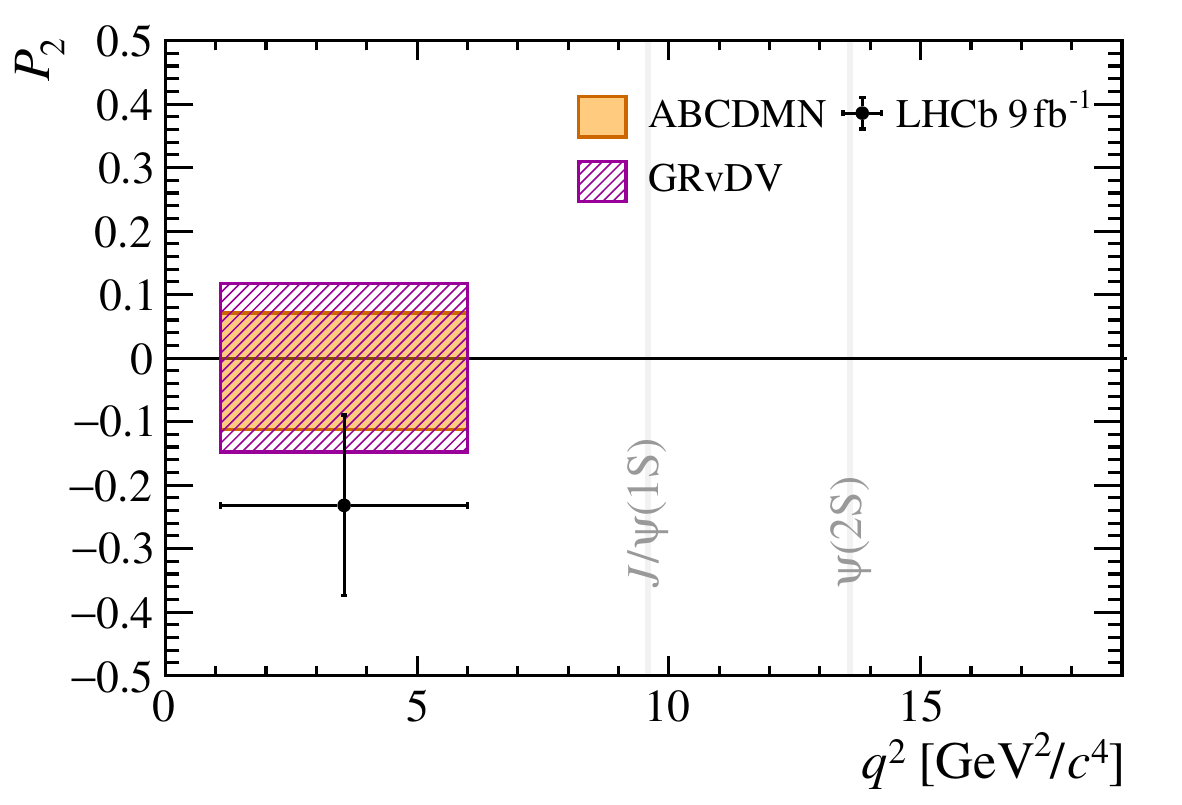} 
    \includegraphics[width=.45\textwidth, trim={0 0 0 0},clip]{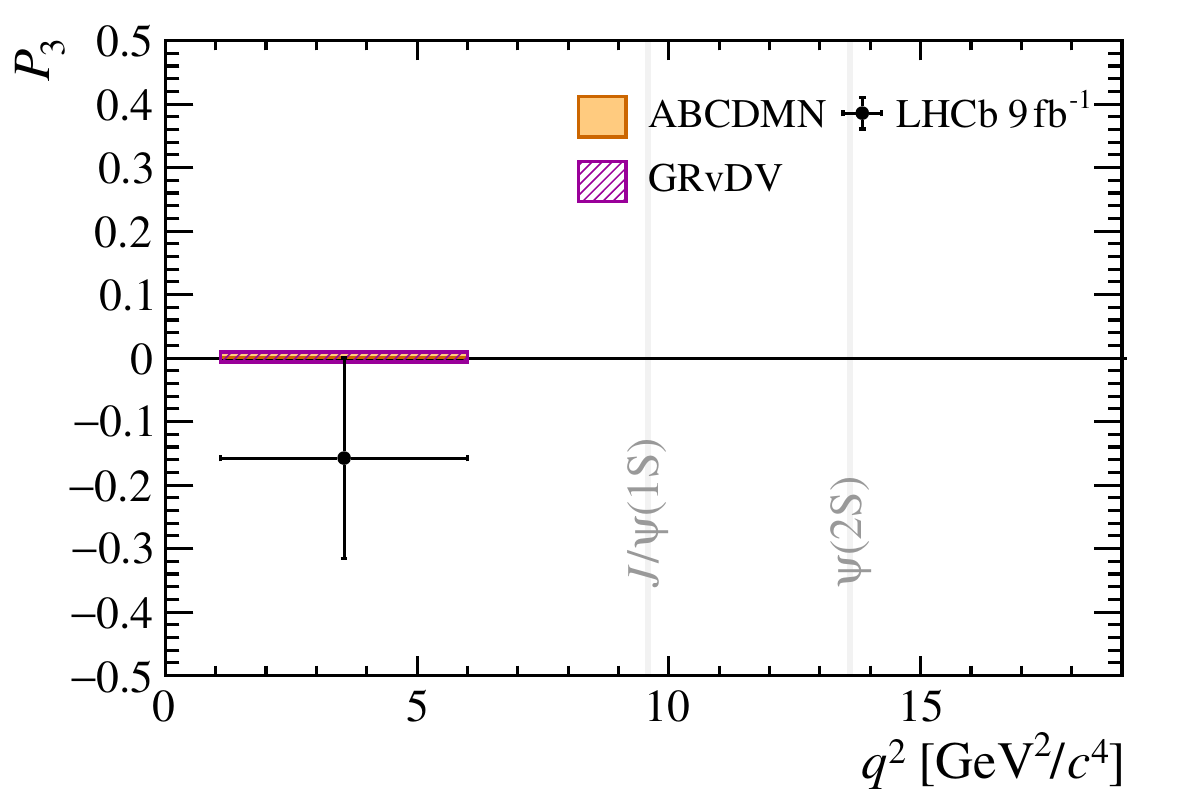} 
    \includegraphics[width=.45\textwidth, trim={0 0 0 0},clip]{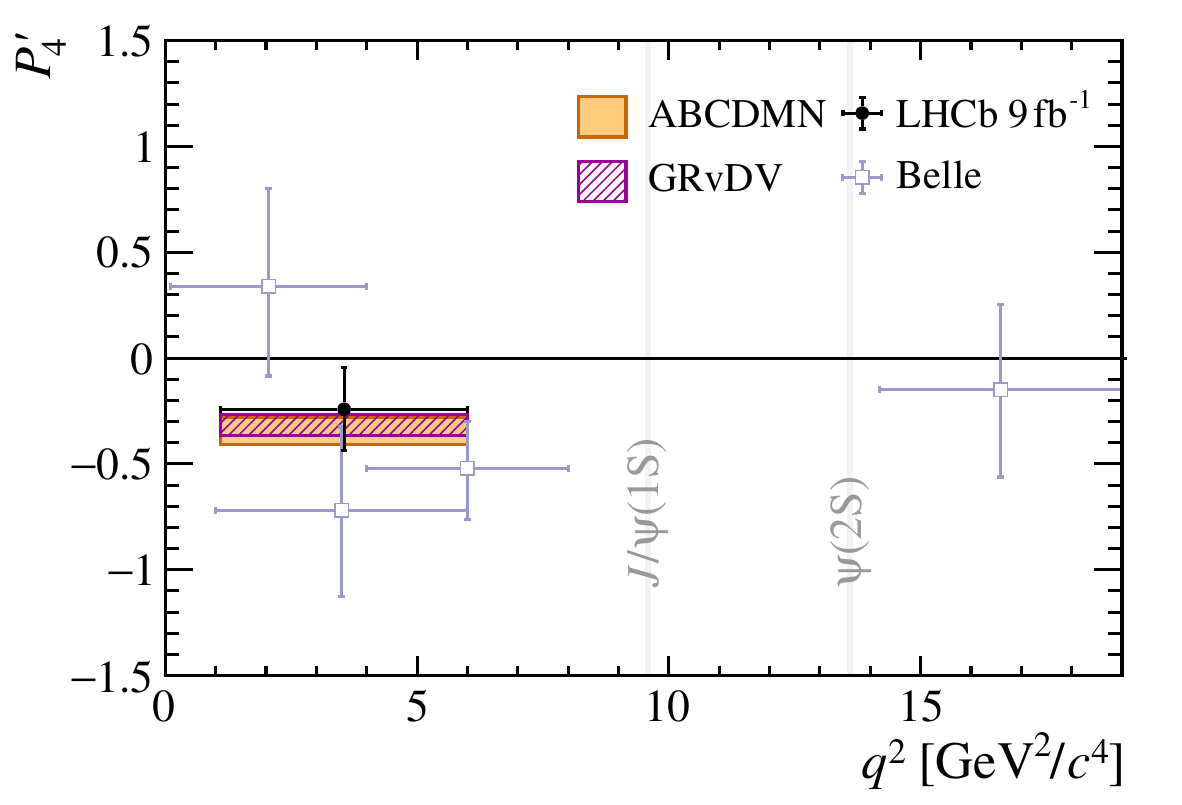}
    \includegraphics[width=.45\textwidth, trim={0 0 0 0},clip]{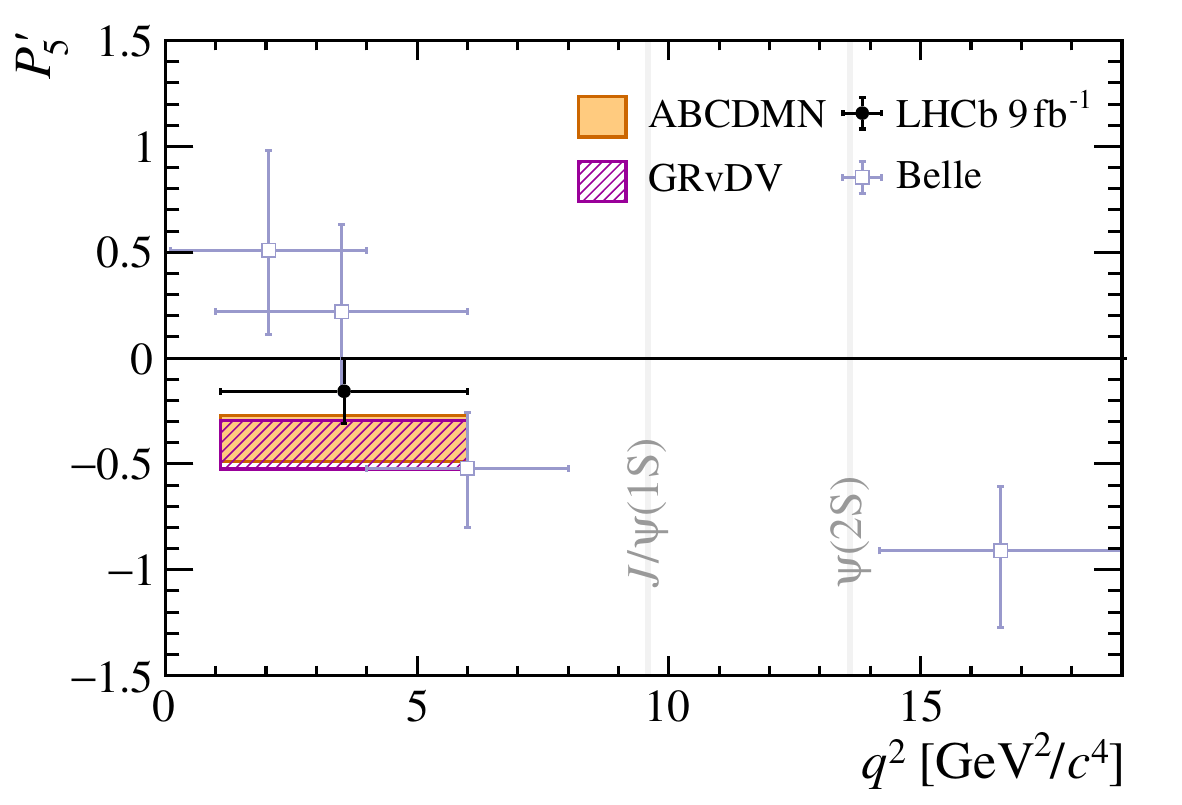} 
    \includegraphics[width=.45\textwidth, trim={0 0 0 0},clip]{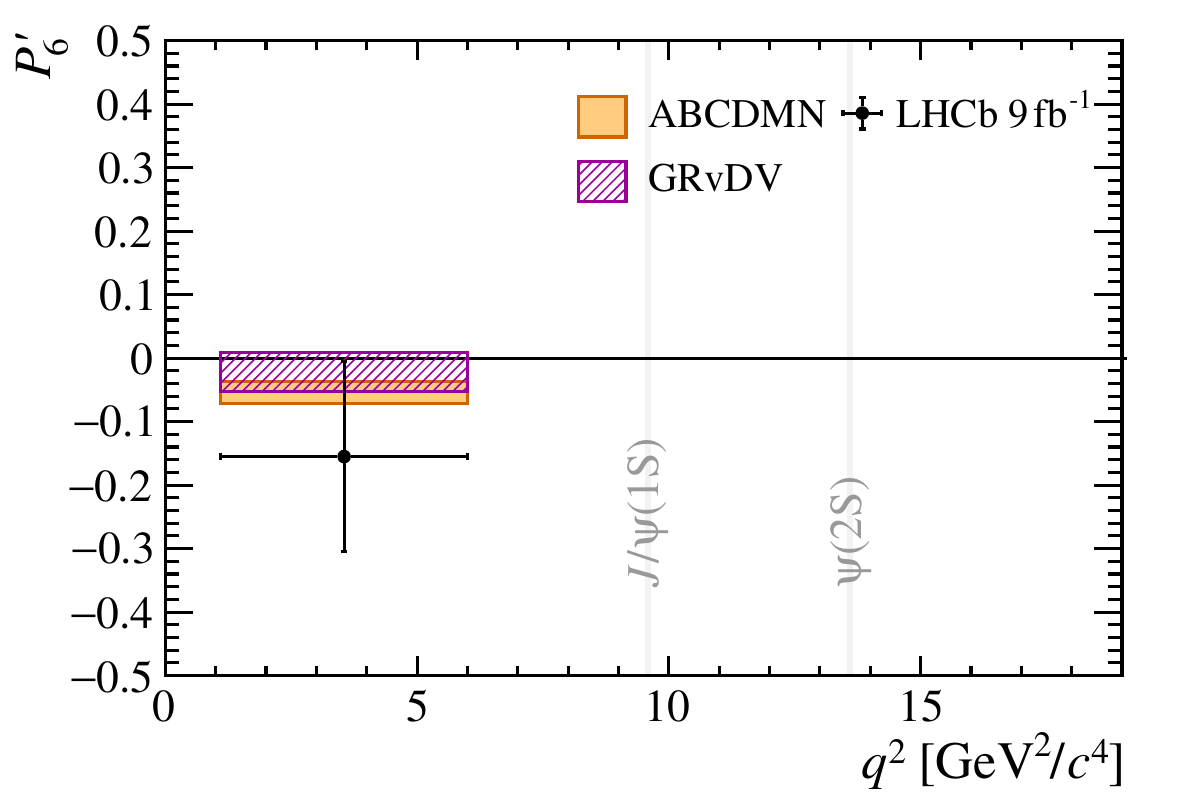} 
    \includegraphics[width=.45\textwidth, trim={0 0 0 0},clip]{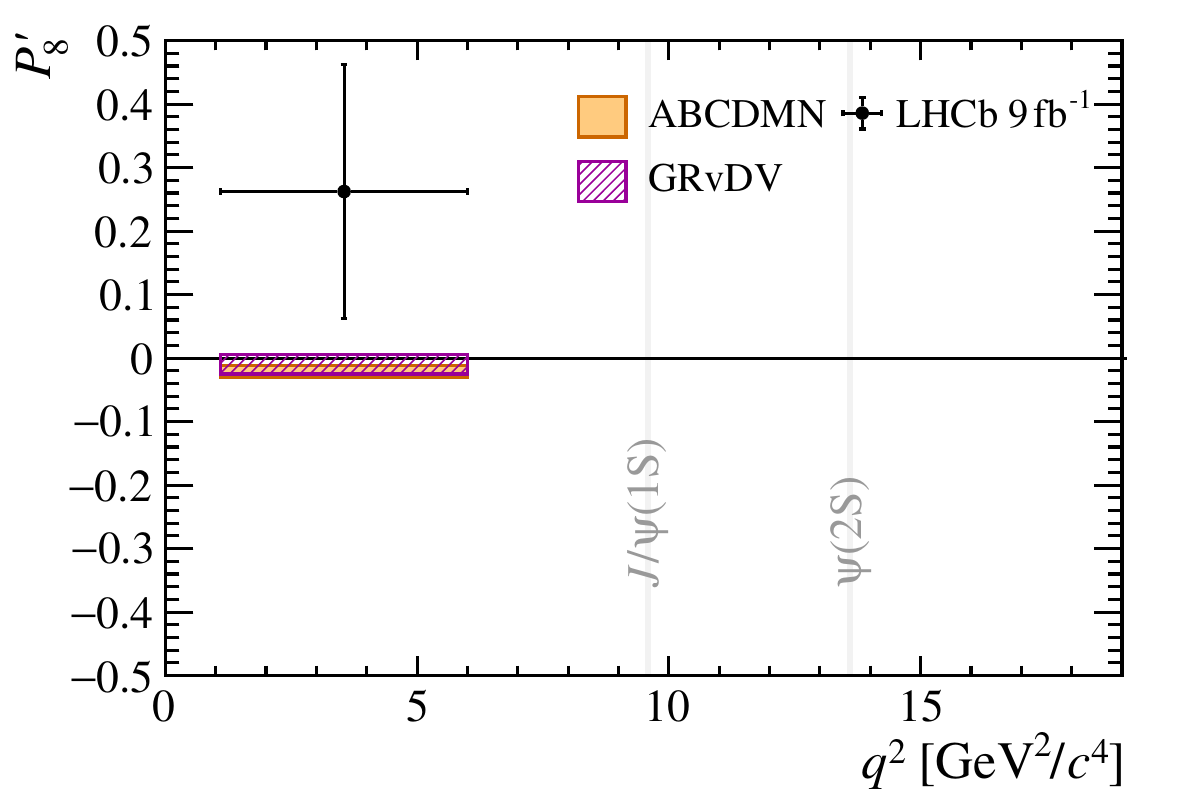} 
  \caption{Measured $\it{P}$-basis angular observables.
  The orange and hatched purple boxes correspond to SM predictions based on Ref.~\cite{Alguero:2023jeh} and Refs.~\cite{EOSAuthors:2021xpv,Gubernari:2022hxn}, respectively. 
  The values of $P_4^{\prime}$ and $P_5^{\prime}$ measured by Belle~\cite{Belle:2016fev} for the decays of $B^{+,0}\to K^{*+,*0} e^+e^-$ are shown in light blue.}
\label{fig:Pi_plots}
\end{figure}

\begin{figure}[!tb]
\centering
    \includegraphics[width=.45\textwidth, trim={0 0 0 0},clip]{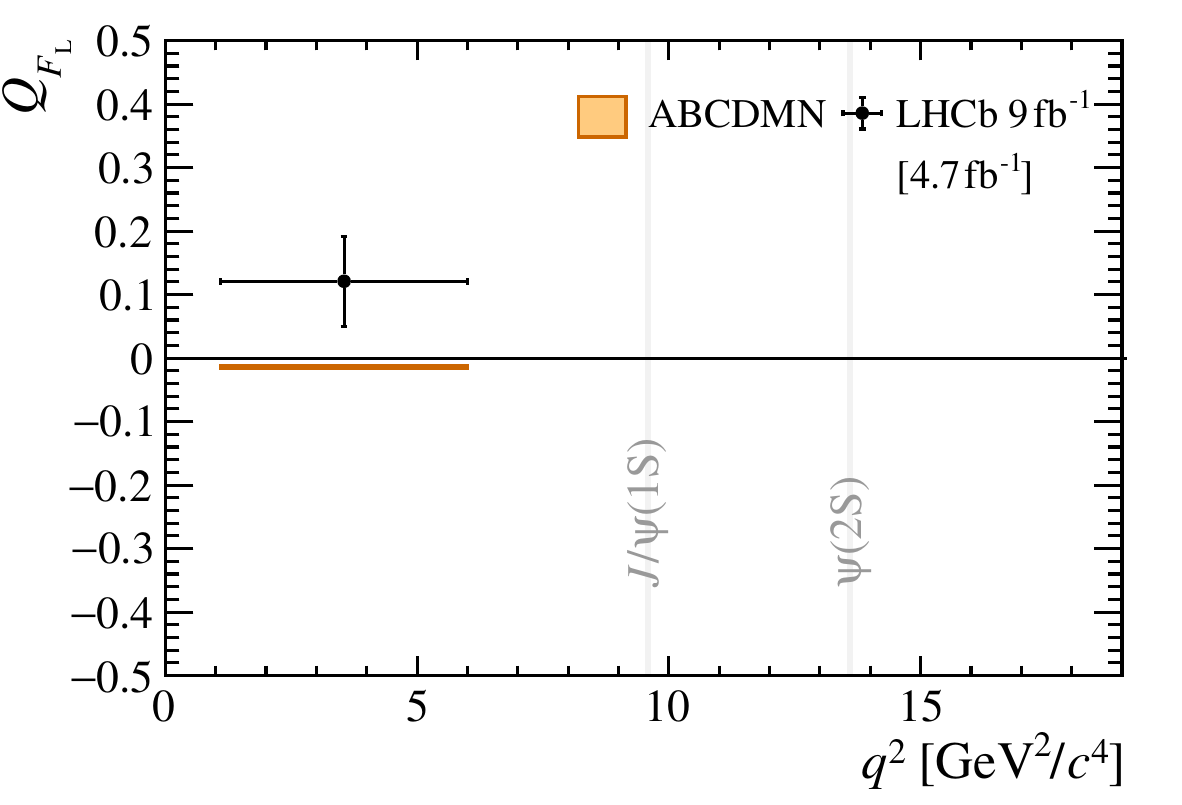} 
    \includegraphics[width=.45\textwidth, trim={0 0 0 0},clip]{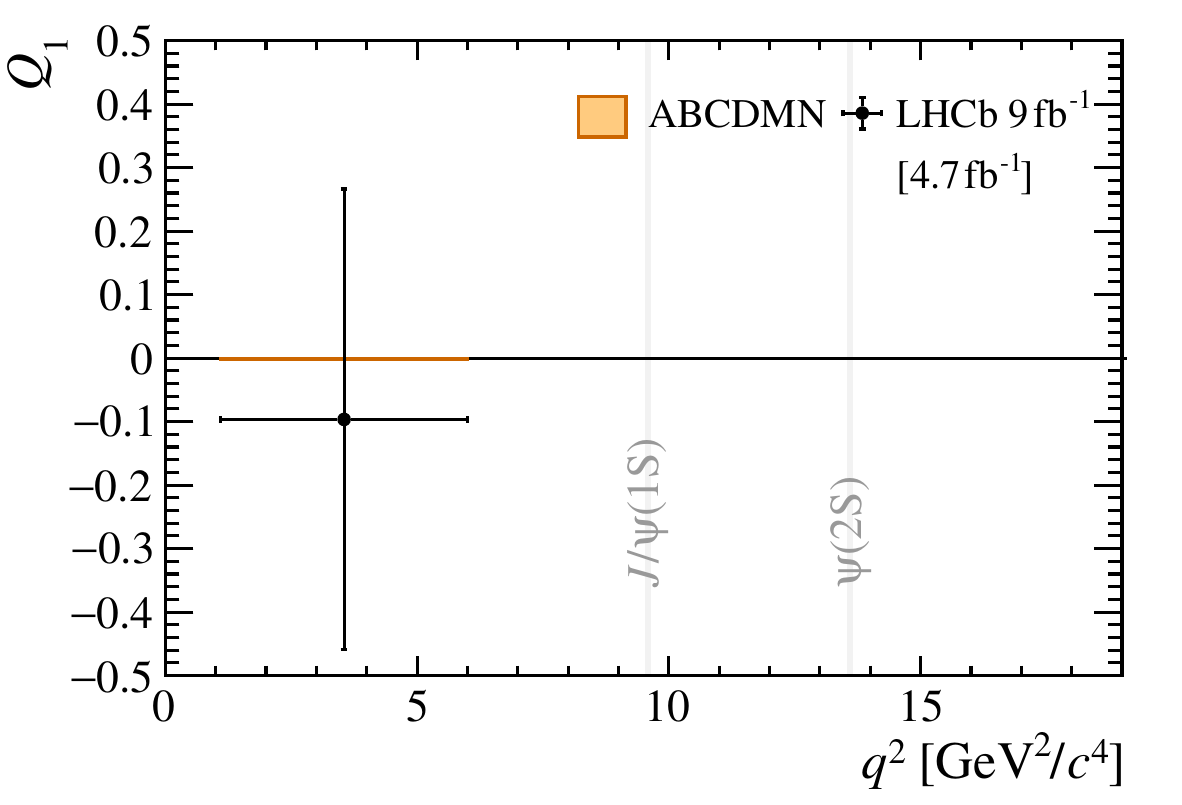} 
    \includegraphics[width=.45\textwidth, trim={0 0 0 0},clip]{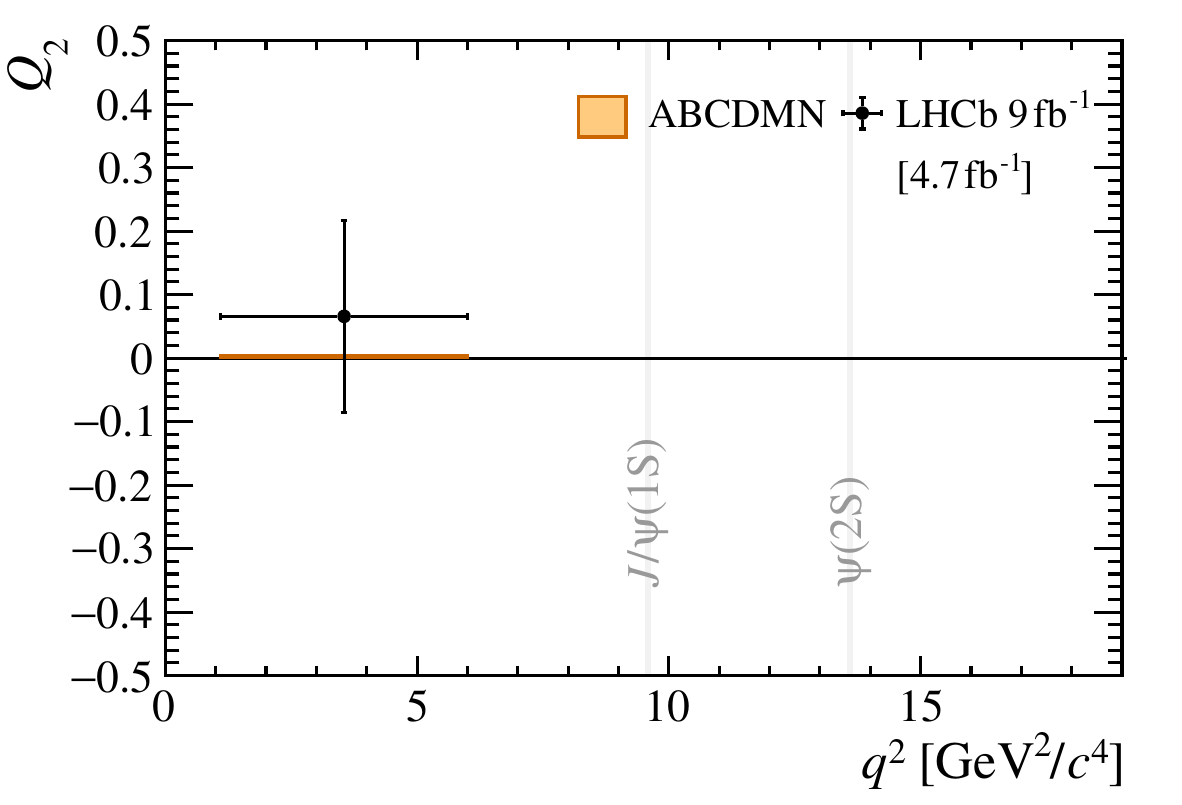} 
    \includegraphics[width=.45\textwidth, trim={0 0 0 0},clip]{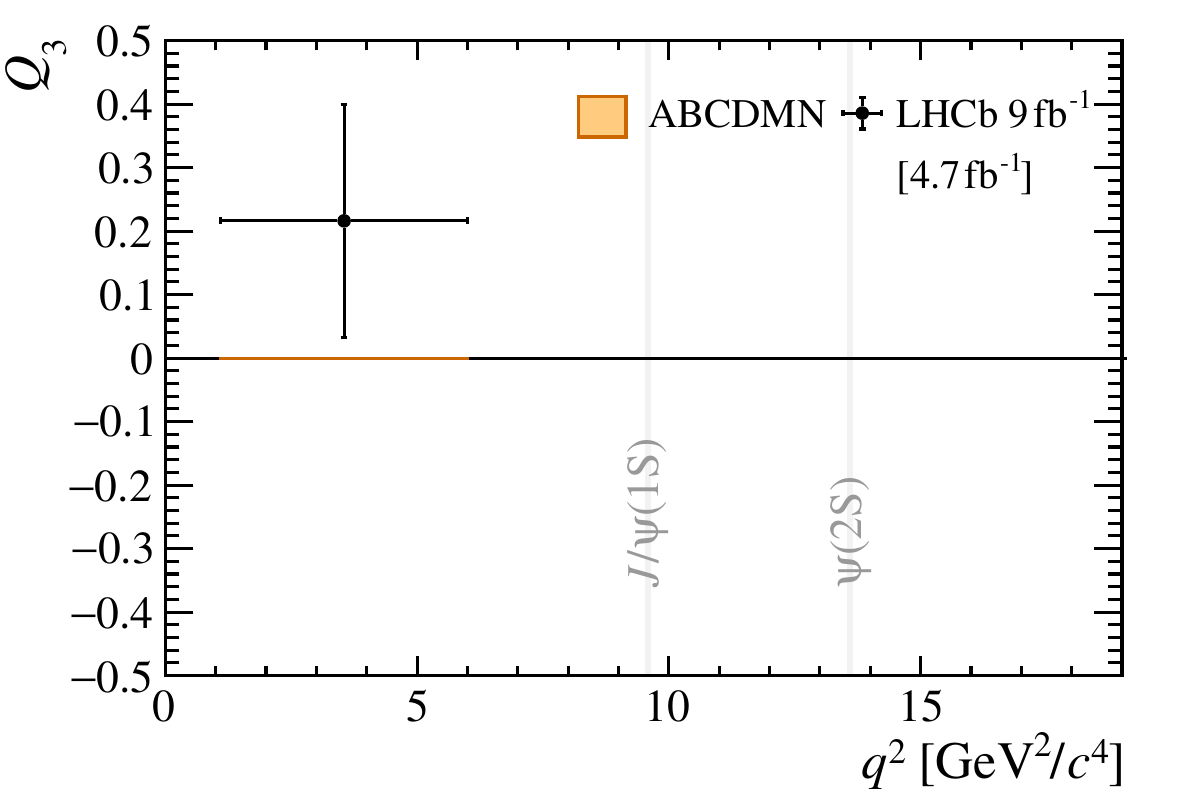} 
    \includegraphics[width=.45\textwidth, trim={0 0 0 0},clip]{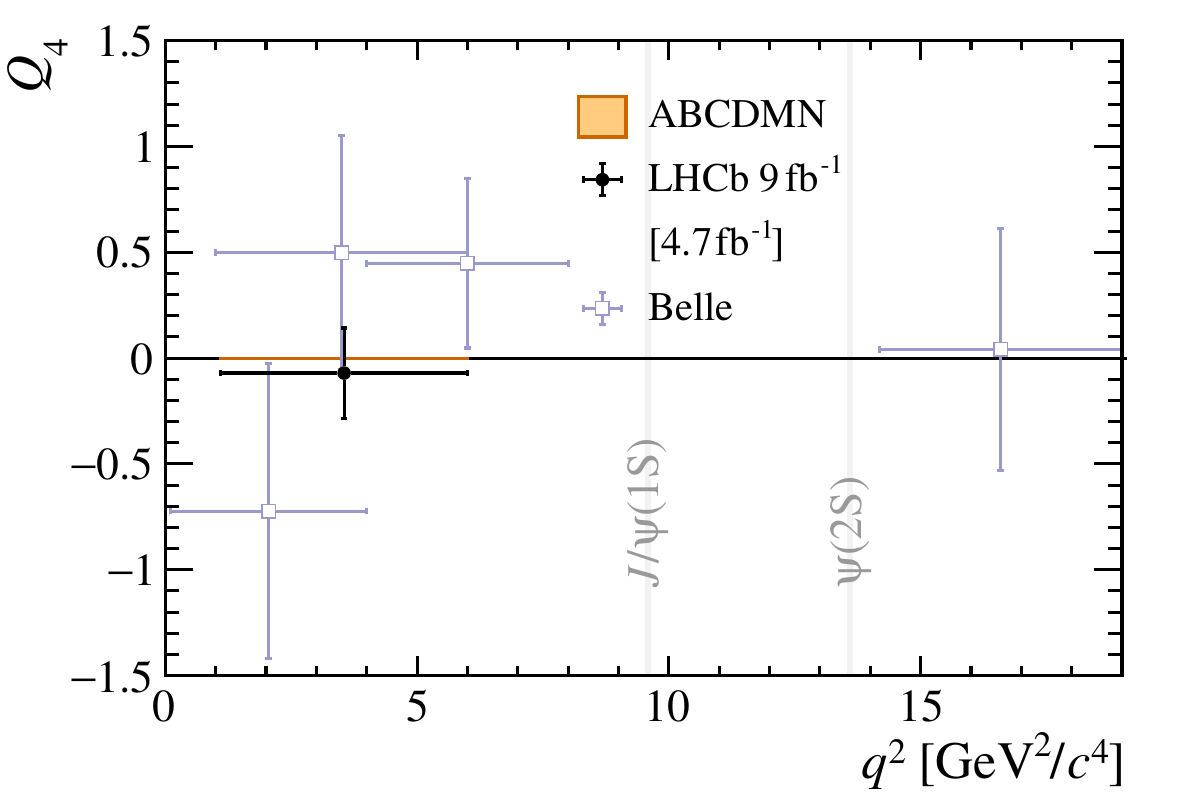} 
    \includegraphics[width=.45\textwidth, trim={0 0 0 0},clip]{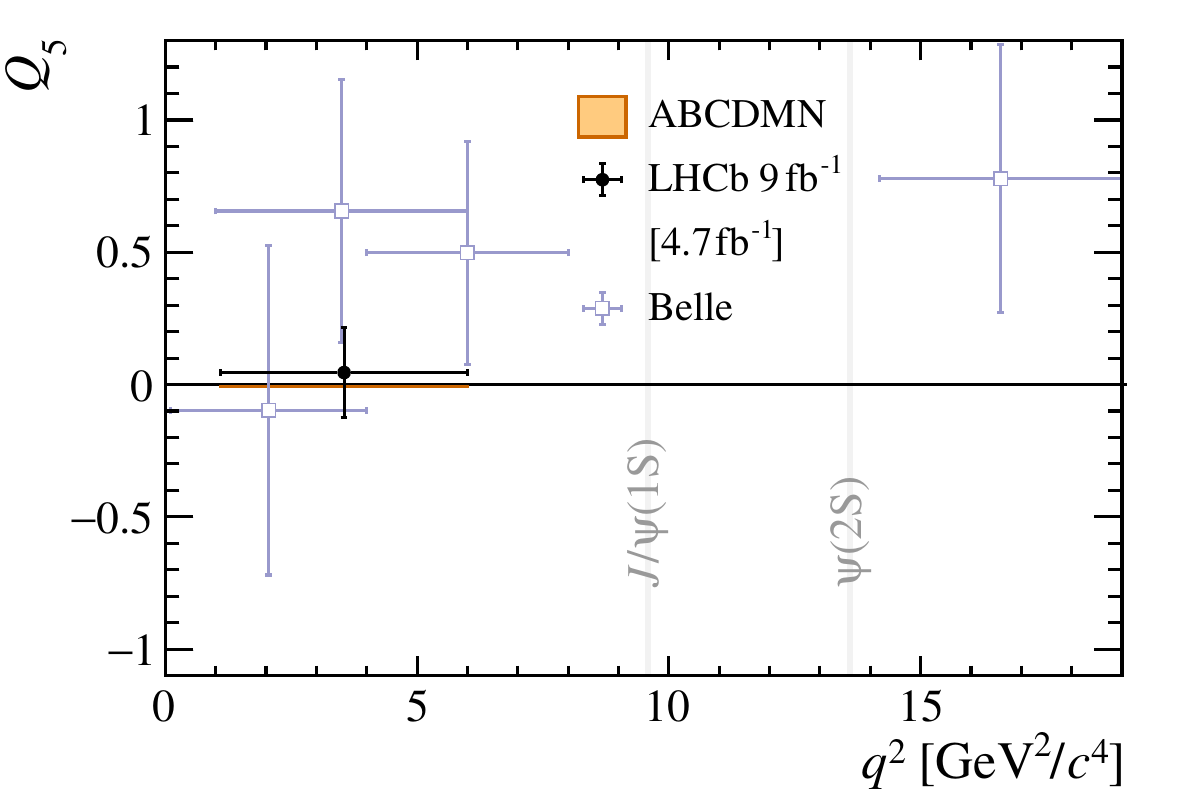} 
    \includegraphics[width=.45\textwidth, trim={0 0 0 0},clip]{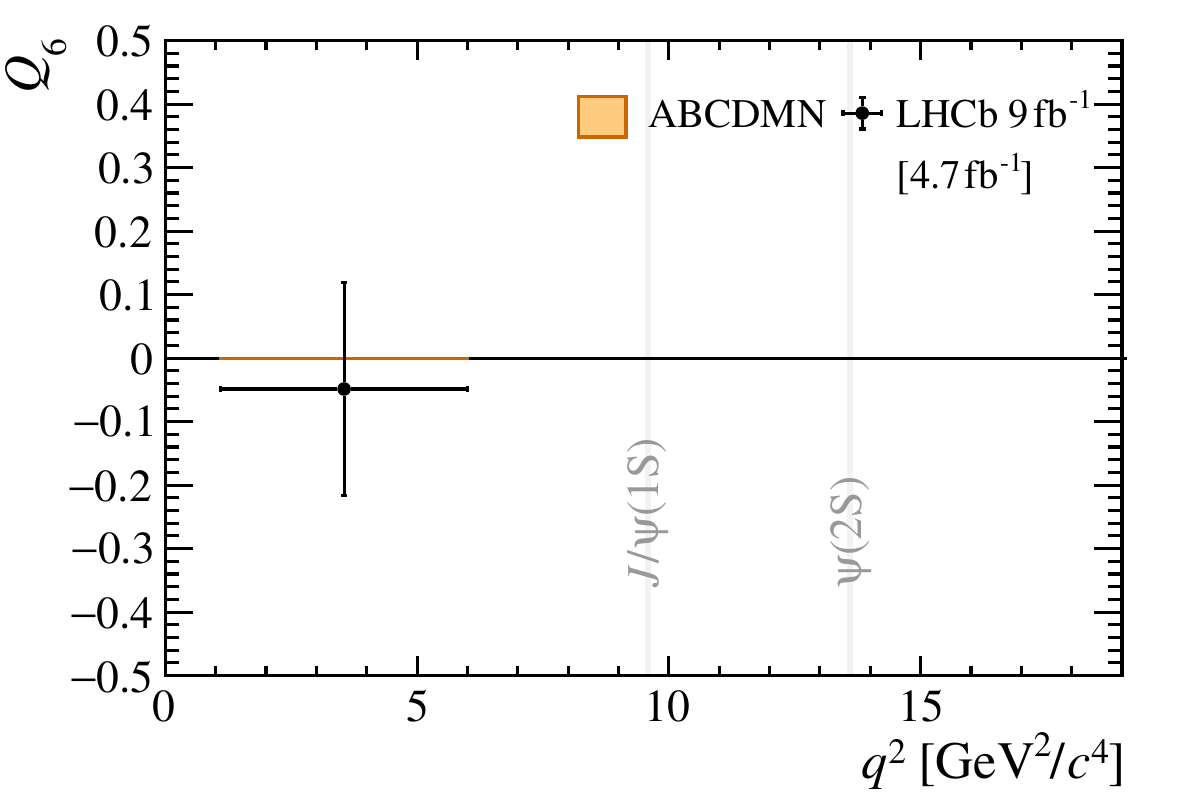} 
    \includegraphics[width=.45\textwidth, trim={0 0 0 0},clip]{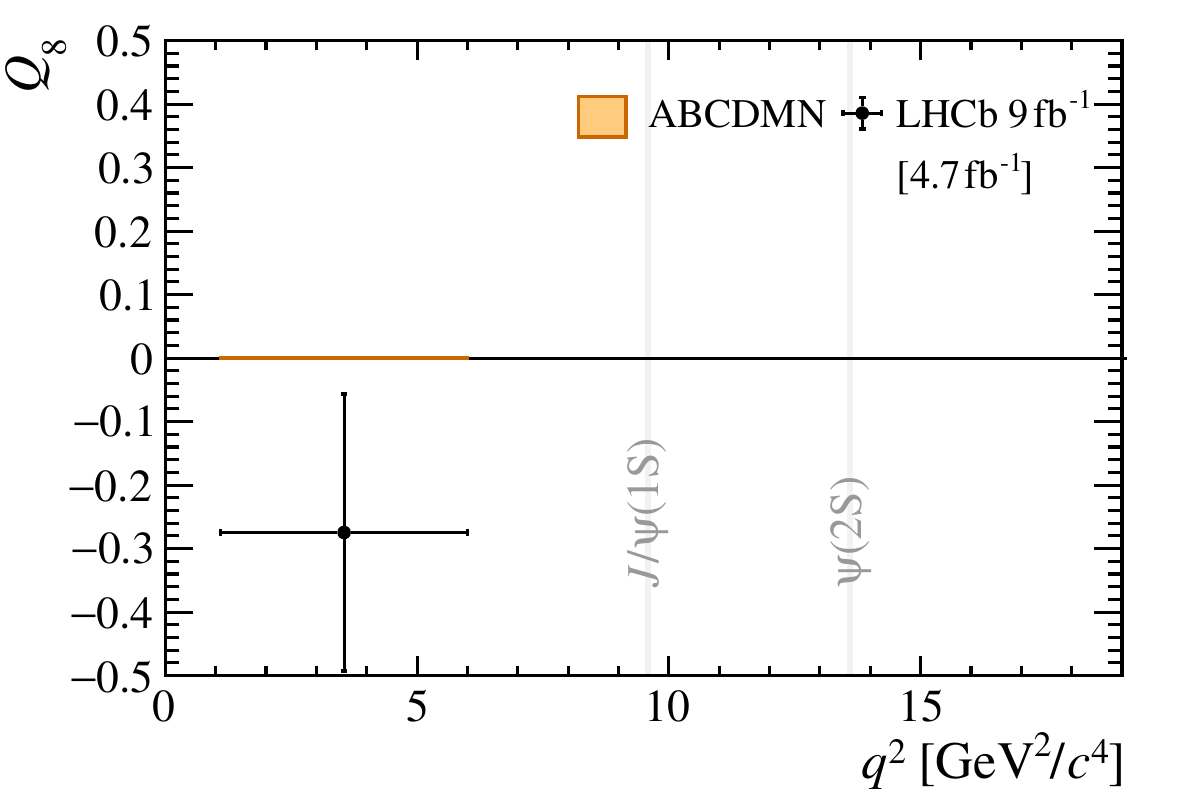} 
  \caption{Measured $Q_i$ LFU observables compared with the SM predictions based on Ref.~\cite{Alguero:2023jeh}.
  The values of $Q_4$ and $Q_5$ measured by Belle~\cite{Belle:2016fev} for the decays of $B^{+,0}\to K^{*+,*0} \ell^+\ell^-$, where $\ell=e,\,\mu$, are shown in light blue.}
\label{fig:Qi_plots}
\end{figure}

To determine the LFU observables $Q_{i}$, which are given by the differences between the $P$-basis angular observables of the muon and the electron modes~\cite{Capdevila:2016ivx}, the analysis strategy used in this measurement is applied to the $\Bz\to\Kstarz\mu^+\mu^-$ sample analysed in Ref.~\cite{LHCb-PAPER-2020-002}: information from the $\Km\pip$ system is not used, S-wave and interference terms are neglected, the acceptance functions are parametrised using trigonometric terms for $\phi$, and a weighted maximum-likelihood fit is performed to the $\Bz$ invariant mass and the three decay angles.
The result of this fit is shown in Appendix~\ref{sec:muon_mode_fits}.
After this alignment in the fit strategy, only the systematic uncertainties of the electron mode need to be considered, as those of the muon mode are negligibly small in comparison.
The $Q_i$ observables are summarised in Fig.~\ref{fig:summary_plots_Qi}, and the numerical values are given in Table~\ref{tab:observables_results_Qi}.
The statistical uncertainties of the $Q_i$ observables are obtained by summing the fit uncertainties of the electron and muon modes in quadrature. 
Most observables show good agreement with the LFU hypothesis. The largest difference of around $2\sigma$ is found for $Q_{\FL}$.
The individual angular and LFU observables are shown in Figs.~\ref{fig:Si_plots},~\ref{fig:Pi_plots} and~\ref{fig:Qi_plots}.

\begin{figure}[!tb]
  \begin{center}
    \includegraphics[width=.49\textwidth]{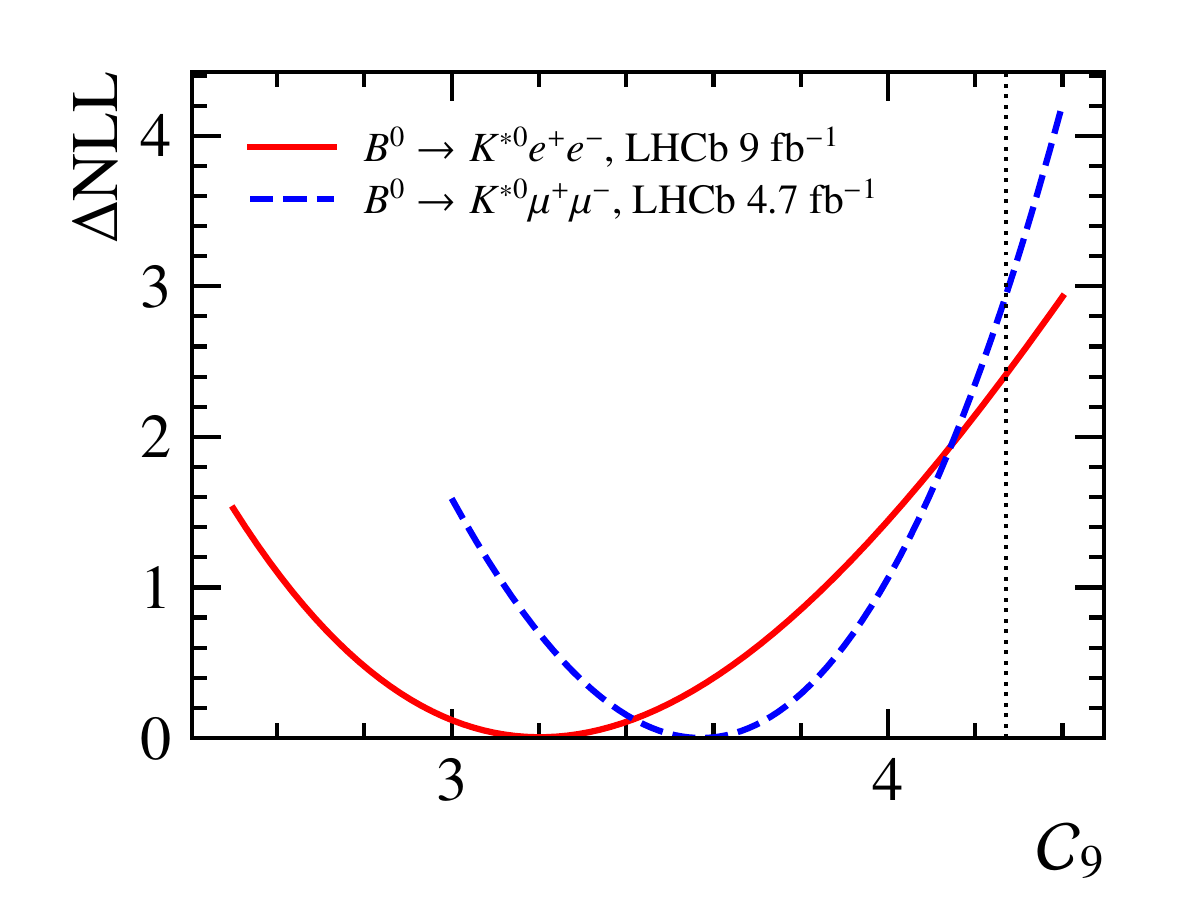}
    \includegraphics[width=.49\textwidth]{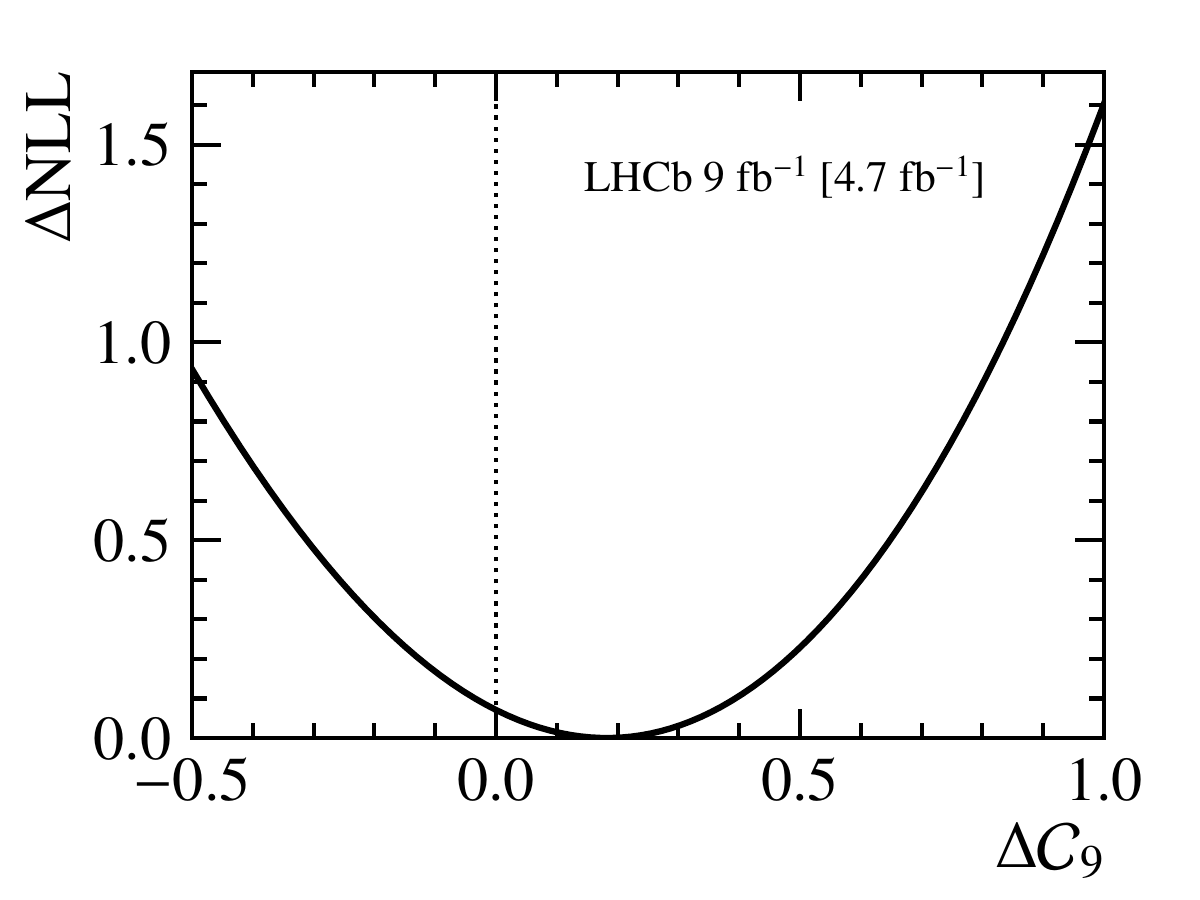}
  \end{center}
  \vspace{-6mm}
  \caption{Negative log-likelihood scan of (left) $C_9^e$ and $C_9^\mu$ and (right) $\Delta C_9 = C_9^{(\mu)} - C_9^{(e)}$.
  The dotted vertical line corresponds to the SM prediction~\cite{Bobeth:1999mk,Gorbahn:2004my}. 
  }
  \label{fig:C9}
\end{figure}

In order to quantify the agreement with the SM and evaluate the constraints imposed by this measurement on the LFU hypothesis, a global fit to the angular observables is performed, varying the ${C}_9$ coefficient, as motivated by Refs.~\cite{Greljo:2022jac,Alguero:2023jeh,Gubernari:2022hxn,Capdevila:2023yhq,Hurth:2023jwr}.
The signal decay amplitudes are parametrised based on Ref.~\cite{Kruger:2005ep} with the
local form-factor parameters taken from Ref.~\cite{Gubernari:2023puw}. Nonlocal hadronic contributions are treated as in Refs.~\cite{LHCb-PAPER-2023-032,LHCb-PAPER-2023-033}; they are modelled based on their analytic structure~\cite{Gubernari:2020eft} by means of a $z$-expansion truncated at second order, constrained to their theoretical predictions at $q^2 <0$~\cite{Gubernari:2022hxn} and fitted to the binned angular observables in the physical $q^2$ range.
All local and nonlocal hadronic contributions are known to be lepton-flavour universal and are therefore shared between the muon and electron modes.
Finally, the $C_9$ coefficient is varied independently for the $\Bz\to\Kstarz e^+e^-$ and $\Bz\to\Kstarz\mu^+\mu^-$ decay channels. 
The fit inputs consist of the angular observables in $\Bz\to\Kstarz e^+e^-$ decays and those measured in the five narrower bins up to $8\,\gevgevcccc$ for $\Bz\to\Kstarz\mu^+\mu^-$ decays~\cite{LHCb-PAPER-2020-002}.
Figure~\ref{fig:C9} (left) shows the result of the negative log-likelihood scan for $C_9^{(e)}$ and $C_9^{(\mu)}$, when both coefficients are varied independently.
A negative shift in the values of $C_9^{(e)}$ of the order of $-1$ with respect to the SM prediction~\cite{Bobeth:1999mk,Gorbahn:2004my} is required to describe the measured $\Bz\to\Kstarz e^+e^-$ angular observables, with a significance above $2\sigma$, which is similar to what is found for the muon channel.
The correlation between $C_9^{(e)}$ and $C_9^{(\mu)}$ is found to be around 40\%, indicating a residual correlation between the two results due to the common choice of the amplitude model.
Finally, in order to quantify the compatibility of the results with the LFU hypothesis, a negative log-likelihood scan of $\Delta C_9 = C_9^{(\mu)} - C_9^{(e)}$ is made, and shown in Fig.~\ref{fig:C9} (right).
In this case, since the focus is purely on the detection of possible LFU-breaking effects, the theoretical inputs of Ref.~\cite{Gubernari:2022hxn} are removed, and only the differences between the angular observables of the muon and electron channels measured in the region of $1.1$--$6.0\gevgevcccc$ (Table~\ref{tab:observables_results_Qi}) are used as inputs.
The resulting $\Delta C_9$ value is found to be compatible with zero within one standard deviation, which is fully consistent with the LFU hypothesis in $\bsll $ transitions.

\begin{table}[!tb]
\sisetup{separate-uncertainty}
\centering
\caption{Values for the $Q_i$ LFU observables given by the differences between the muon and electron $\it{P}$-basis angular observables. The first uncertainty is statistical and the second is systematic.
}
\begin{tabular}{l r }
  \multicolumn{2}{c}{}  \\ \hline
 \multicolumn{2}{c}{LFU observables}  \\ \hline
 $Q_{\FL}$ &$ 0.12  \pm 0.05 \pm 0.05 $\\ 
 $Q_{1}$ &$ -0.10  \pm 0.26 \pm 0.25 $\\
 $Q_{4}$ &$ -0.07  \pm 0.17 \pm 0.13 $\\
 $Q_{5}$ &$ 0.05  \pm 0.13 \pm 0.11 $\\
 $Q_{2}$ &$ 0.07  \pm 0.10 \pm 0.12 $\\
 $Q_{6}$ &$ -0.05  \pm 0.14 \pm 0.10 $\\
 $Q_{8}$ &$ -0.28  \pm 0.17 \pm 0.14 $\\
 $Q_{3}$ &$ 0.22  \pm 0.14 \pm 0.11 $\\ \hline
\end{tabular}
\label{tab:observables_results_Qi}
\end{table}

\clearpage

\section{Conclusions}
\label{sec:conclusions}
This paper presents an angular analysis of $\Bz\to\Kstarz e^+e^-$ decays performed in the central-$q^2$ region of $1.1$--$6.0\gevgevcccc$ using $\lhcb$ $pp$ data collected between 2011 and 2018, corresponding to an integrated luminosity of $9\invfb$. 
Angular observables are extracted using a weighted maximum-likelihood fit to the invariant-mass and angular distributions of $\Bz\to\Kstarz e^+e^-$ decays, where the weights correct for distortions of the signal distribution caused by acceptance and resolution effects.
The results presented here are the most precise to date. 
Overall, the set of angular observables show good agreement with the SM predictions. Discrepancies at the level of $2\sigma$ are observed for $\FL$ and $\AFB$, which are consistent with the hypothesis of a negative shift in the value of the Wilson coefficient $C_9$ reported by global analyses of other $\bsll $ transitions~\cite{Greljo:2022jac,Alguero:2023jeh,Gubernari:2022hxn,Capdevila:2023yhq,Hurth:2023jwr}. 
Finally, no strong sign of LFU violation is observed when the angular observables of the $\Bz\to\Kstarz e^+ e^-$ and $\Bz\to\Kstarz \mu^+ \mu^-$ decays are analysed together.

\section*{Acknowledgements}
\noindent We express our gratitude to our colleagues in the CERN
accelerator departments for the excellent performance of the LHC. We
thank the technical and administrative staff at the LHCb
institutes.
We acknowledge support from CERN and from the national agencies:
ARC (Australia);
CAPES, CNPq, FAPERJ and FINEP (Brazil); 
MOST and NSFC (China); 
CNRS/IN2P3 (France); 
BMBF, DFG and MPG (Germany); 
INFN (Italy); 
NWO (Netherlands); 
MNiSW and NCN (Poland); 
MCID/IFA (Romania); 
MICIU and AEI (Spain);
SNSF and SER (Switzerland); 
NASU (Ukraine); 
STFC (United Kingdom); 
DOE NP and NSF (USA).
We acknowledge the computing resources that are provided by ARDC (Australia), 
CBPF (Brazil),
CERN, 
IHEP and LZU (China),
IN2P3 (France), 
KIT and DESY (Germany), 
INFN (Italy), 
SURF (Netherlands),
Polish WLCG (Poland),
IFIN-HH (Romania), 
PIC (Spain), CSCS (Switzerland), 
and GridPP (United Kingdom).
We are indebted to the communities behind the multiple open-source
software packages on which we depend.
Individual groups or members have received support from
Key Research Program of Frontier Sciences of CAS, CAS PIFI, CAS CCEPP, 
Fundamental Research Funds for the Central Universities,  and Sci. \& Tech. Program of Guangzhou (China);
Minciencias (Colombia);
EPLANET, Marie Sk\l{}odowska-Curie Actions, ERC and NextGenerationEU (European Union);
A*MIDEX, ANR, IPhU and Labex P2IO, and R\'{e}gion Auvergne-Rh\^{o}ne-Alpes (France);
Alexander-von-Humboldt Foundation (Germany);
ICSC (Italy); 
Severo Ochoa and Mar\'ia de Maeztu Units of Excellence, GVA, XuntaGal, GENCAT, InTalent-Inditex and Prog. ~Atracci\'on Talento CM (Spain);
SRC (Sweden);
the Leverhulme Trust, the Royal Society and UKRI (United Kingdom).

\newpage

\section*{Appendices}

\appendix

\section{Large $\boldsymbol{q^2}$ region analysis}
\label{sec:large_q2}
The use of the constrained $q^2$ variable to define the measurement region suppresses background from the radiative tail of the $\Bz \to \Kstarz \jpsi (\to e^+ e^-)$ mode (Fig.~\ref{fig:jpsi_leak_comb}), and allows the analysis strategy discussed in Secs.~\ref{sec:effective_acceptance} to~\ref{sec:angular_fit} to be applied directly to the larger $q^2$ region of \mbox{1.1--7.0\gevgevcccc}, resulting in a gain in the signal yield of around 20\%. The results of the measurement in this extended region are presented and discussed in this self-contained Appendix.

Compared to the baseline measurement in the region of $1.1$--$6.0$\gevgevcccc, the only differences are the signal and background models, which are obtained from data or simulation samples within the extended region. 
The same effective acceptance functions discussed in Sec.~\ref{sec:effective_acceptance}, which are parametrised in the $q^2$ region of 0.5--8.0\gevgevcccc, are used. 
The results of the parameterisation validation are shown in Fig.~\ref{fig:lq2_rare_mode_acceptance_validation_display}.
The data fit projections are shown in Fig.~\ref{fig:data_fit_q2c_large_P}. All systematic uncertainties are quantified using the methods discussed in Sec.~\ref{sec:systematic_uncertainties}, and summarised in Tables~\ref{tab:rare_mode_syst_S_lq2c} and~\ref{tab:rare_mode_syst_P_lq2c}. The measured observable values are displayed in Fig.~\ref{fig:summary_plots_lq2}, and their corresponding numerical values are given in Table~\ref{tab:lq2_observables_results}. 
The correlations among the observables are given in Tables~\ref{tab:corr_lq2_S} and~\ref{tab:corr_lq2_P}, for the statistical uncertainties, and Tables~\ref{tab:corr_syst_lq2_S} and~\ref{tab:corr_syst_lq2_P}, for the systematic uncertainties, respectively.

To calculate the $Q_i$ observables, a muon-mode fit is performed using the strategy detailed in Sec.~\ref{sec:results}, but in the extended $q^2$ region. The result of this fit is shown in Fig.~\ref{fig:muon_lq2}. The LFU observables are displayed in Fig.~\ref{fig:summary_plots_Qi_lq2} and given in Table~\ref{tab:lq2_observables_results_Qi}. Figures showing all observables individually, including results from both $q^2$ regions, are given in Appendix~\ref{sec:additional_figures}.

\subsection{Validation of the effective acceptance functions}
\label{sec:lq2_app_effective_acceptance_function}

\begin{figure}[b!]
\centering
  \begin{tabular}{@{}cccc@{}}
    \includegraphics[width=.48\textwidth]{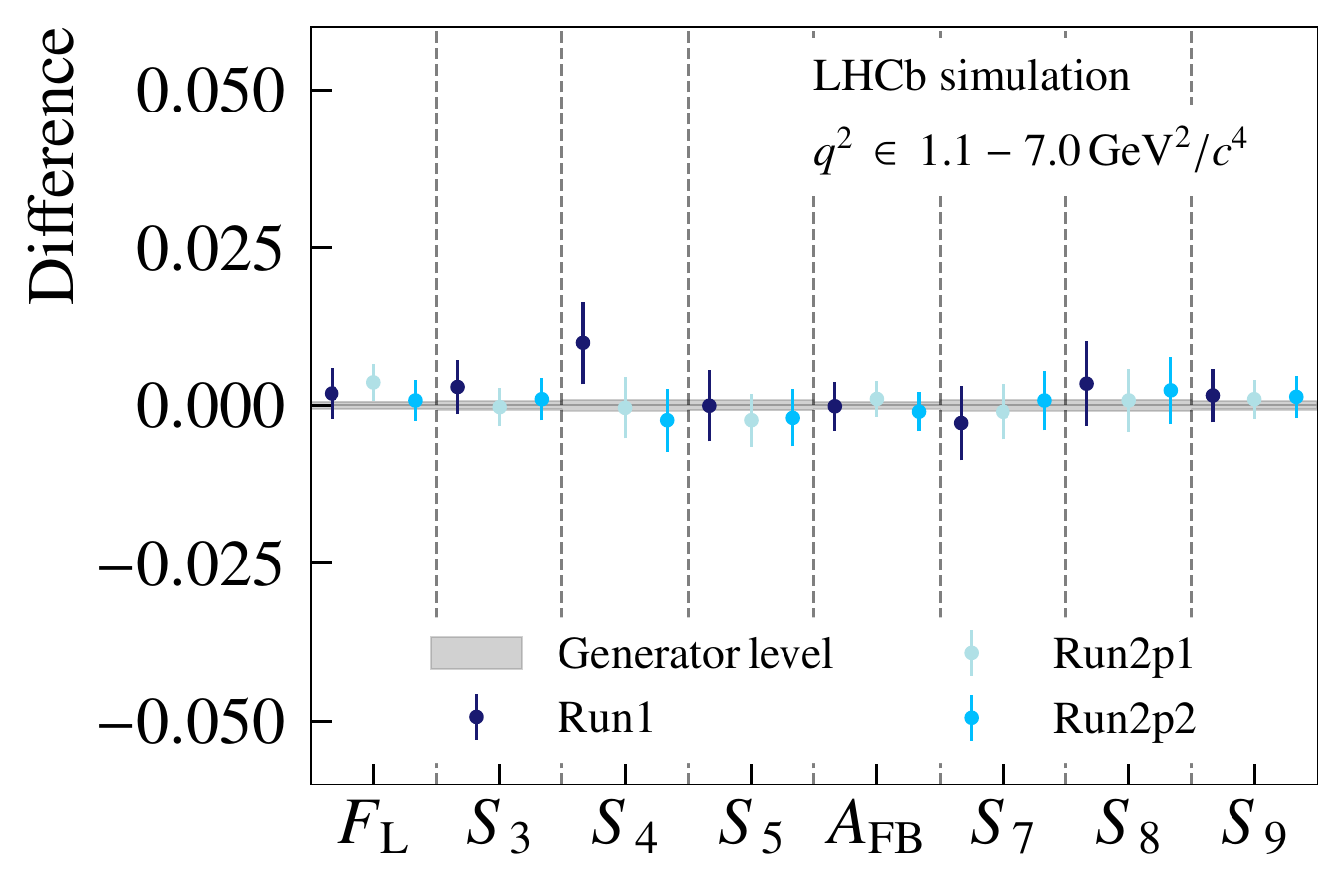} &
    \includegraphics[width=.48\textwidth]{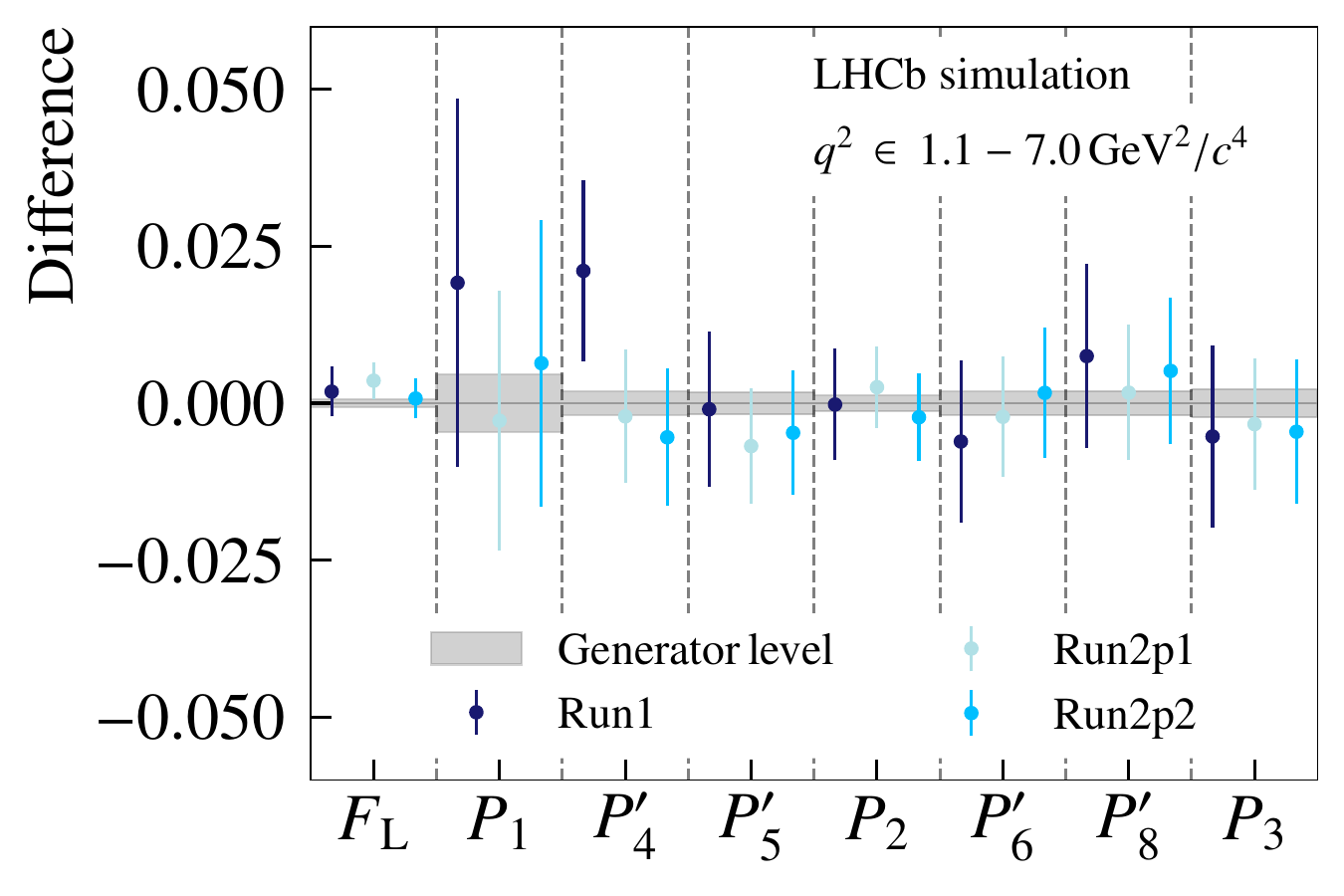} & 
   \end{tabular}
  \caption{Differences between the observable values found from fits to the signal simulation samples with acceptance correction weights, and the fit to the generator-level sample (centred at zero) in the large $q^2$ region.}
\label{fig:lq2_rare_mode_acceptance_validation_display}
\end{figure}

The results of the validation of the effective acceptance parametrisation for the large $q^2$ region are shown in Fig.~\ref{fig:lq2_rare_mode_acceptance_validation_display}.

 \clearpage

\subsection{Fit results in the large $\boldsymbol{q^2}$ region}

\begin{figure}[!b]
\centering
    \includegraphics[width=.45\textwidth, trim={0 0 0 0},clip]{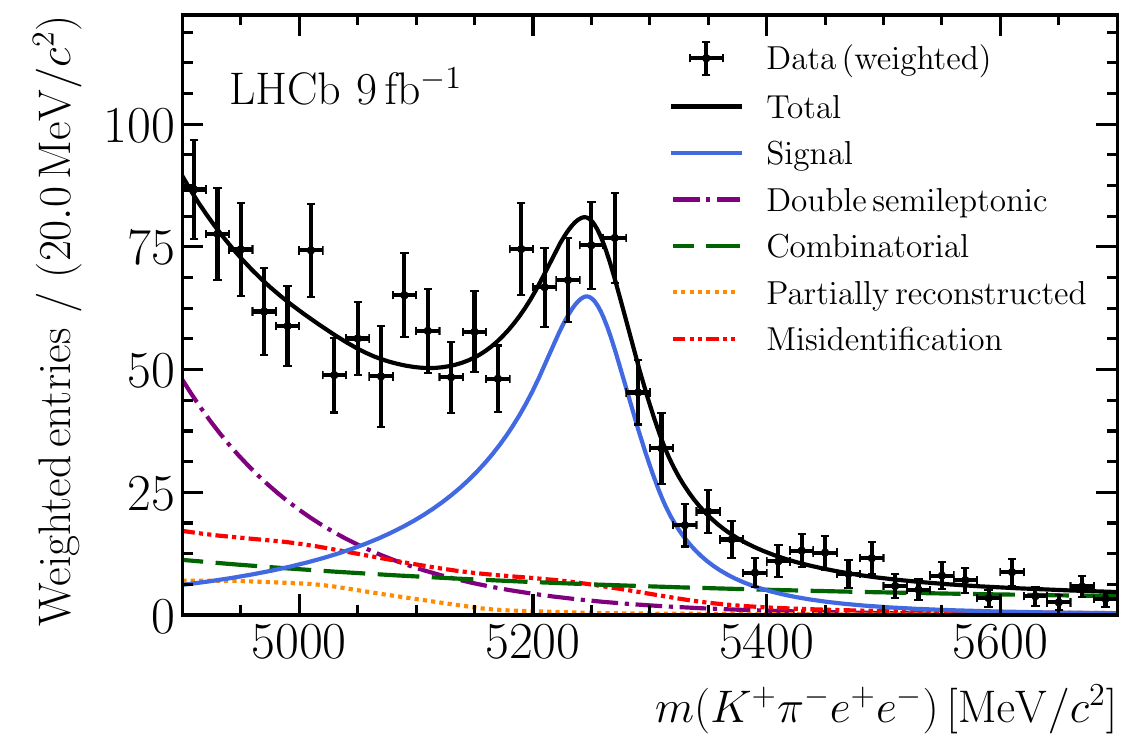} 
    \includegraphics[width=.45\textwidth, trim={0 0 0 0},clip]{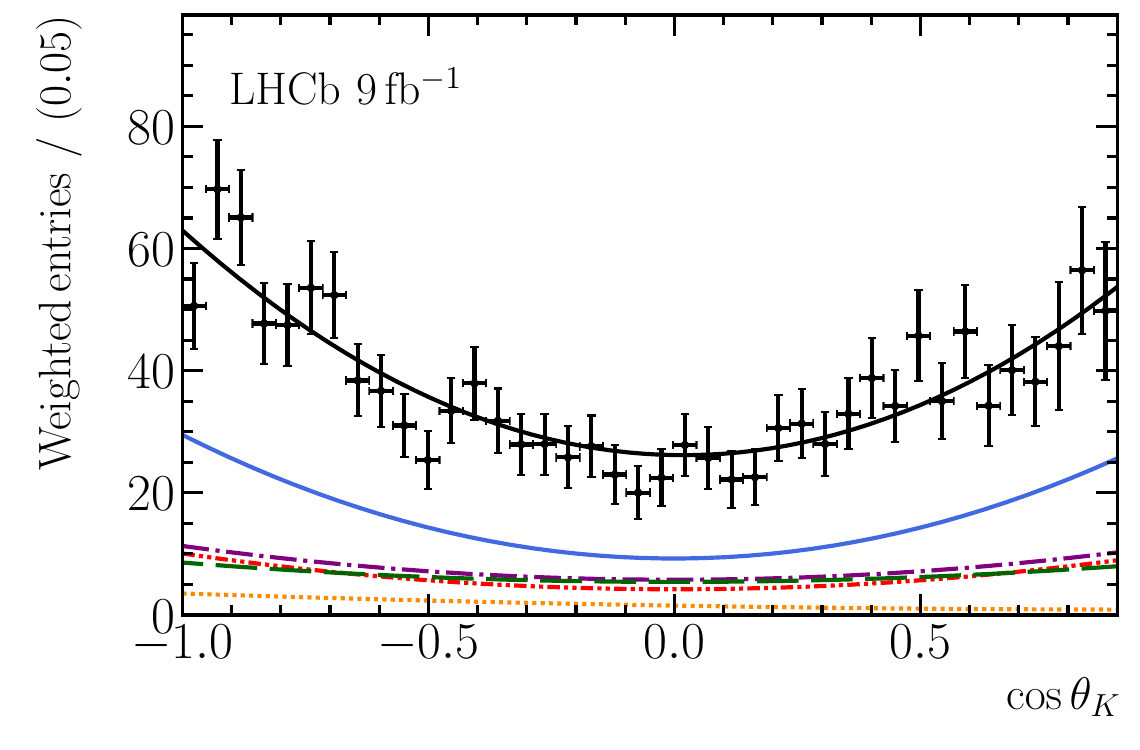} 
    \includegraphics[width=.45\textwidth, trim={0 0 0 0},clip]{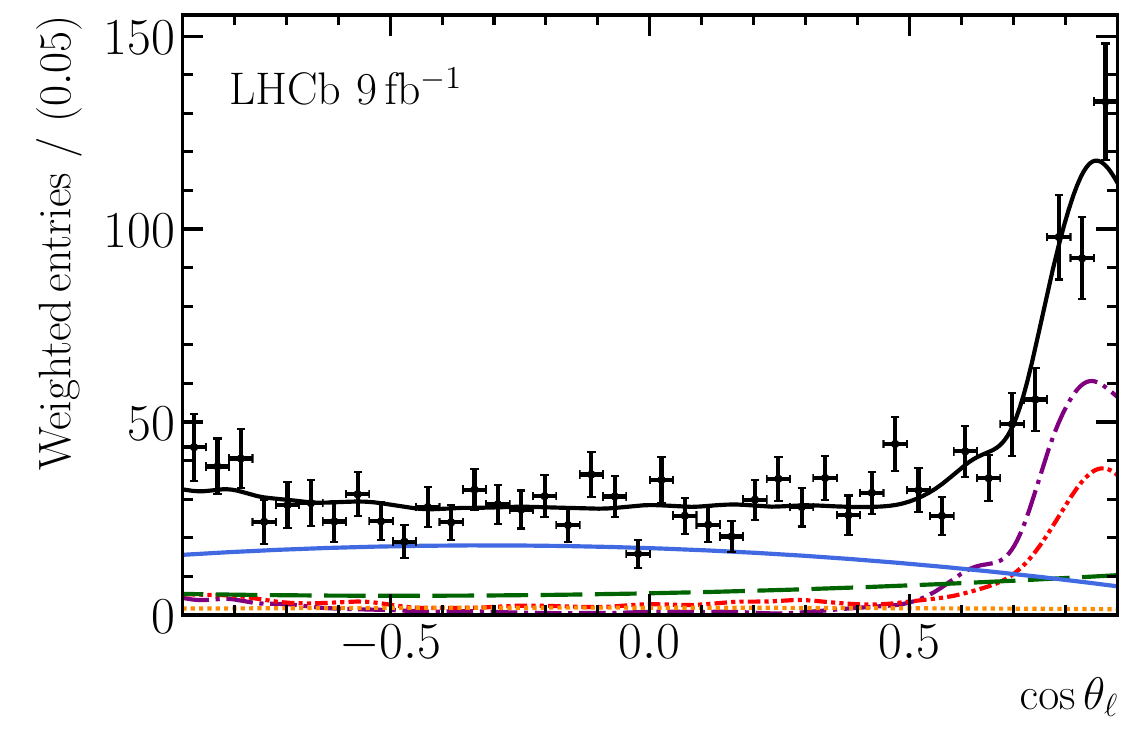} 
    \includegraphics[width=.45\textwidth, trim={0 0 0 0},clip]{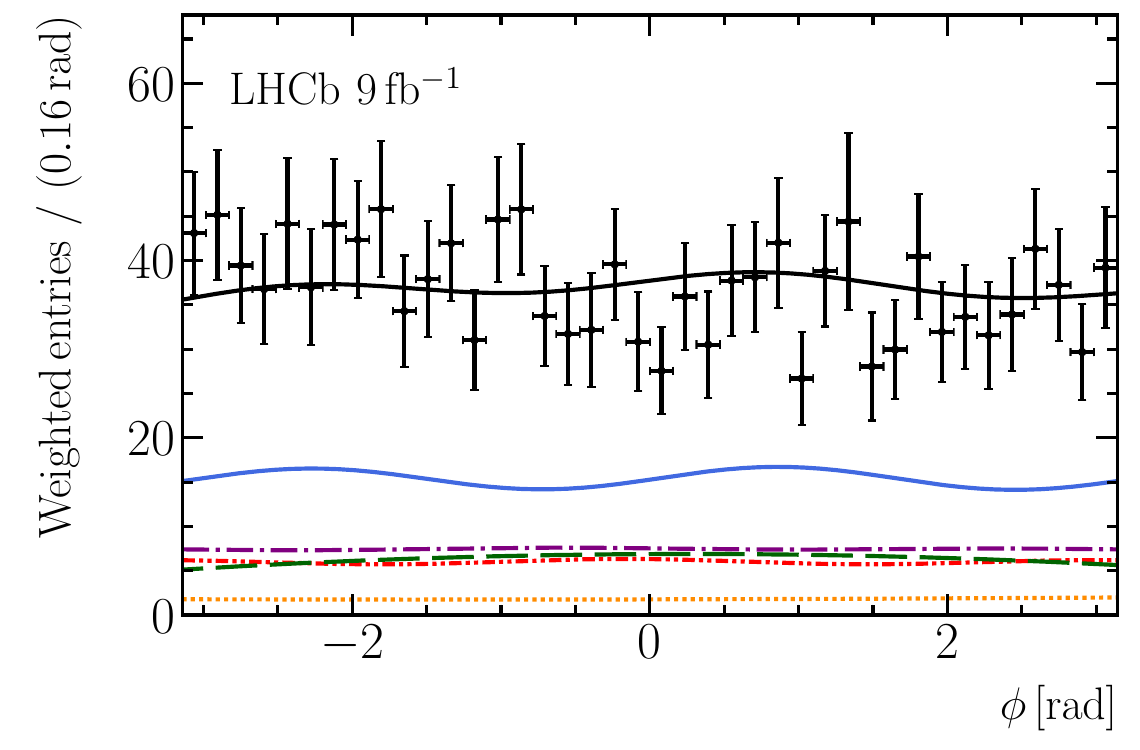} 
  \caption{
  Invariant-mass and angular distributions of selected candidates in the large $q^2$ region after the application of effective acceptance weights. 
  The signal distribution is shown with a solid blue line, and the background components are shown with dashed, dotted and dash-dotted lines. The solid black line corresponds to the full fit function.
  }
\label{fig:data_fit_q2c_large_P}
\end{figure}

The result of the weighted maximum-likelihood fit performed to the $\Bz$ invariant-mass and angular distributions of signal candidates selected in the large $q^{2}$ region of 1.1--7.0\gevgevcccc is shown in Fig.~\ref{fig:data_fit_q2c_large_P}.

\clearpage

\subsection{Systematic uncertainties}

Systematic uncertainties associated with the observables of the large $q^2$ region are summarised in Tables~\ref{tab:rare_mode_syst_S_lq2c} and~\ref{tab:rare_mode_syst_P_lq2c}.

\begin{table}[!b]
\caption{Summary of the systematic uncertainties on the $\it{S}$-basis angular observables of the large $q^2$ region. All values are given as fractions of the statistical uncertainties.}
\label{tab:rare_mode_syst_S_lq2c}
\vspace{0.2cm}
\sisetup{separate-uncertainty}
\centering
\begin{tabular}{l|cccccccc}
\multicolumn{1}{l}{Source} & \multicolumn{1}{c}{$\FL$} & \multicolumn{1}{c}{$S_{3}$} & \multicolumn{1}{c}{$S_{4}$} & \multicolumn{1}{c}{$S_{5}$} & \multicolumn{1}{c}{$\AFB$} & \multicolumn{1}{c}{$S_{7}$} & \multicolumn{1}{c}{$S_{8}$} & \multicolumn{1}{c}{$S_{9}$} \\\hline

\multicolumn{1}{l}{Comb and DSL backgrounds} & \multicolumn{1}{c}{0.67} & \multicolumn{1}{c}{0.41} & \multicolumn{1}{c}{0.32} & \multicolumn{1}{c}{0.42} & \multicolumn{1}{c}{0.49} & \multicolumn{1}{c}{0.16} & \multicolumn{1}{c}{0.45} & \multicolumn{1}{c}{0.31} \\

\multicolumn{1}{l}{Part. reco. background} & \multicolumn{1}{c}{0.20} & \multicolumn{1}{c}{0.05} & \multicolumn{1}{c}{0.09} & \multicolumn{1}{c}{0.12} & \multicolumn{1}{c}{0.08} & \multicolumn{1}{c}{0.04} & \multicolumn{1}{c}{0.04} & \multicolumn{1}{c}{0.05} \\

\multicolumn{1}{l}{Misid. had. background} & \multicolumn{1}{c}{0.39} & \multicolumn{1}{c}{0.26} & \multicolumn{1}{c}{0.11} & \multicolumn{1}{c}{0.19} & \multicolumn{1}{c}{0.16} & \multicolumn{1}{c}{0.36} & \multicolumn{1}{c}{0.13} & \multicolumn{1}{c}{0.14} \\

\multicolumn{1}{l}{Effective acceptance} & \multicolumn{1}{c}{0.43} & \multicolumn{1}{c}{0.26} & \multicolumn{1}{c}{0.48} & \multicolumn{1}{c}{0.45} & \multicolumn{1}{c}{0.38} & \multicolumn{1}{c}{0.55} & \multicolumn{1}{c}{0.44} & \multicolumn{1}{c}{0.23} \\

\multicolumn{1}{l}{Signal mass modelling} & \multicolumn{1}{c}{0.26} & \multicolumn{1}{c}{0.06} & \multicolumn{1}{c}{0.09} & \multicolumn{1}{c}{0.12} & \multicolumn{1}{c}{0.14} & \multicolumn{1}{c}{0.05} & \multicolumn{1}{c}{0.05} & \multicolumn{1}{c}{0.06} \\

\multicolumn{1}{l}{$J/\psi$ backgrounds} & \multicolumn{1}{c}{0.39} & \multicolumn{1}{c}{0.06} & \multicolumn{1}{c}{0.07} & \multicolumn{1}{c}{0.09} & \multicolumn{1}{c}{0.32} & \multicolumn{1}{c}{0.05} & \multicolumn{1}{c}{0.05} & \multicolumn{1}{c}{0.09} \\

\multicolumn{1}{l}{S-wave component} & \multicolumn{1}{c}{0.38} & \multicolumn{1}{c}{0.05} & \multicolumn{1}{c}{0.16} & \multicolumn{1}{c}{0.17} & \multicolumn{1}{c}{0.07} & \multicolumn{1}{c}{0.17} & \multicolumn{1}{c}{0.02} & \multicolumn{1}{c}{0.07} \\

\multicolumn{1}{l}{$B^+$ veto} & \multicolumn{1}{c}{0.51} & \multicolumn{1}{c}{0.12} & \multicolumn{1}{c}{0.15} & \multicolumn{1}{c}{0.18} & \multicolumn{1}{c}{0.27} & \multicolumn{1}{c}{0.16} & \multicolumn{1}{c}{0.15} & \multicolumn{1}{c}{0.11} \\

\multicolumn{1}{l}{Fit bias} & \multicolumn{1}{c}{0.01} & \multicolumn{1}{c}{0.01} & \multicolumn{1}{c}{0.03} & \multicolumn{1}{c}{0.06} & \multicolumn{1}{c}{0.03} & \multicolumn{1}{c}{0.02} & \multicolumn{1}{c}{0.00} & \multicolumn{1}{c}{0.01} \\\hline

\multicolumn{1}{l}{Total} & \multicolumn{1}{c}{1.20} & \multicolumn{1}{c}{0.57} & \multicolumn{1}{c}{0.65} & \multicolumn{1}{c}{0.72} & \multicolumn{1}{c}{0.78} & \multicolumn{1}{c}{0.73} & \multicolumn{1}{c}{0.67} & \multicolumn{1}{c}{0.44} \\\hline
\end{tabular}
\end{table}
\begin{table}[!tb]
\caption{Summary of the systematic uncertainties on the $\it{P}$-basis angular observables of the large $q^2$ region. All values are given as fractions of the statistical uncertainties.}
\label{tab:rare_mode_syst_P_lq2c}
\vspace{0.2cm}
\sisetup{separate-uncertainty}
\centering
\begin{tabular}{l|cccccccc}
\multicolumn{1}{l}{Source} & \multicolumn{1}{c}{$\FL$} & \multicolumn{1}{c}{$P_{1}$} & \multicolumn{1}{c}{$P_{4}^{\prime}$} & \multicolumn{1}{c}{$P_{5}^{\prime}$} & \multicolumn{1}{c}{$P_{2}$} & \multicolumn{1}{c}{$P_{6}^{\prime}$} & \multicolumn{1}{c}{$P_{8}^{\prime}$} & \multicolumn{1}{c}{$P_{3}$} \\\hline

\multicolumn{1}{l}{Comb. and DSL backgrounds} & \multicolumn{1}{c}{0.67} & \multicolumn{1}{c}{0.72} & \multicolumn{1}{c}{0.48} & \multicolumn{1}{c}{0.62} & \multicolumn{1}{c}{0.71} & \multicolumn{1}{c}{0.20} & \multicolumn{1}{c}{0.52} & \multicolumn{1}{c}{0.57} \\

\multicolumn{1}{l}{Part. reco. background} & \multicolumn{1}{c}{0.20} & \multicolumn{1}{c}{0.12} & \multicolumn{1}{c}{0.14} & \multicolumn{1}{c}{0.21} & \multicolumn{1}{c}{0.11} & \multicolumn{1}{c}{0.06} & \multicolumn{1}{c}{0.05} & \multicolumn{1}{c}{0.12} \\

\multicolumn{1}{l}{Misid. had. background} & \multicolumn{1}{c}{0.39} & \multicolumn{1}{c}{0.48} & \multicolumn{1}{c}{0.17} & \multicolumn{1}{c}{0.25} & \multicolumn{1}{c}{0.29} & \multicolumn{1}{c}{0.42} & \multicolumn{1}{c}{0.16} & \multicolumn{1}{c}{0.28} \\

\multicolumn{1}{l}{Effective acceptance} & \multicolumn{1}{c}{0.43} & \multicolumn{1}{c}{0.43} & \multicolumn{1}{c}{0.54} & \multicolumn{1}{c}{0.51} & \multicolumn{1}{c}{0.52} & \multicolumn{1}{c}{0.62} & \multicolumn{1}{c}{0.49} & \multicolumn{1}{c}{0.35} \\

\multicolumn{1}{l}{Signal mass modelling} & \multicolumn{1}{c}{0.27} & \multicolumn{1}{c}{0.13} & \multicolumn{1}{c}{0.16} & \multicolumn{1}{c}{0.19} & \multicolumn{1}{c}{0.24} & \multicolumn{1}{c}{0.06} & \multicolumn{1}{c}{0.06} & \multicolumn{1}{c}{0.12} \\

\multicolumn{1}{l}{$J/\psi$ backgrounds} & \multicolumn{1}{c}{0.39} & \multicolumn{1}{c}{0.15} & \multicolumn{1}{c}{0.09} & \multicolumn{1}{c}{0.15} & \multicolumn{1}{c}{0.64} & \multicolumn{1}{c}{0.06} & \multicolumn{1}{c}{0.05} & \multicolumn{1}{c}{0.18} \\

\multicolumn{1}{l}{S-wave component} & \multicolumn{1}{c}{0.38} & \multicolumn{1}{c}{0.10} & \multicolumn{1}{c}{0.23} & \multicolumn{1}{c}{0.22} & \multicolumn{1}{c}{0.20} & \multicolumn{1}{c}{0.22} & \multicolumn{1}{c}{0.02} & \multicolumn{1}{c}{0.14} \\

\multicolumn{1}{l}{$B^+$ veto} & \multicolumn{1}{c}{0.52} & \multicolumn{1}{c}{0.30} & \multicolumn{1}{c}{0.24} & \multicolumn{1}{c}{0.33} & \multicolumn{1}{c}{0.36} & \multicolumn{1}{c}{0.19} & \multicolumn{1}{c}{0.19} & \multicolumn{1}{c}{0.27} \\

\multicolumn{1}{l}{Fit bias} & \multicolumn{1}{c}{0.01} & \multicolumn{1}{c}{0.00} & \multicolumn{1}{c}{0.03} & \multicolumn{1}{c}{0.07} & \multicolumn{1}{c}{0.07} & \multicolumn{1}{c}{0.03} & \multicolumn{1}{c}{0.01} & \multicolumn{1}{c}{0.01} \\\hline

\multicolumn{1}{l}{Total} & \multicolumn{1}{c}{1.21} & \multicolumn{1}{c}{1.04} & \multicolumn{1}{c}{0.85} & \multicolumn{1}{c}{0.98} & \multicolumn{1}{c}{1.23} & \multicolumn{1}{c}{0.84} & \multicolumn{1}{c}{0.77} & \multicolumn{1}{c}{0.82} \\\hline

\end{tabular}
\end{table}

\clearpage

\subsection{Fit to the $\boldsymbol{\Bz\to \Kstarz \mu^+\mu^-}$ decay}
\begin{figure}[!b]
\centering
    \includegraphics[width=.45\textwidth, trim={0 0 0 0},clip]{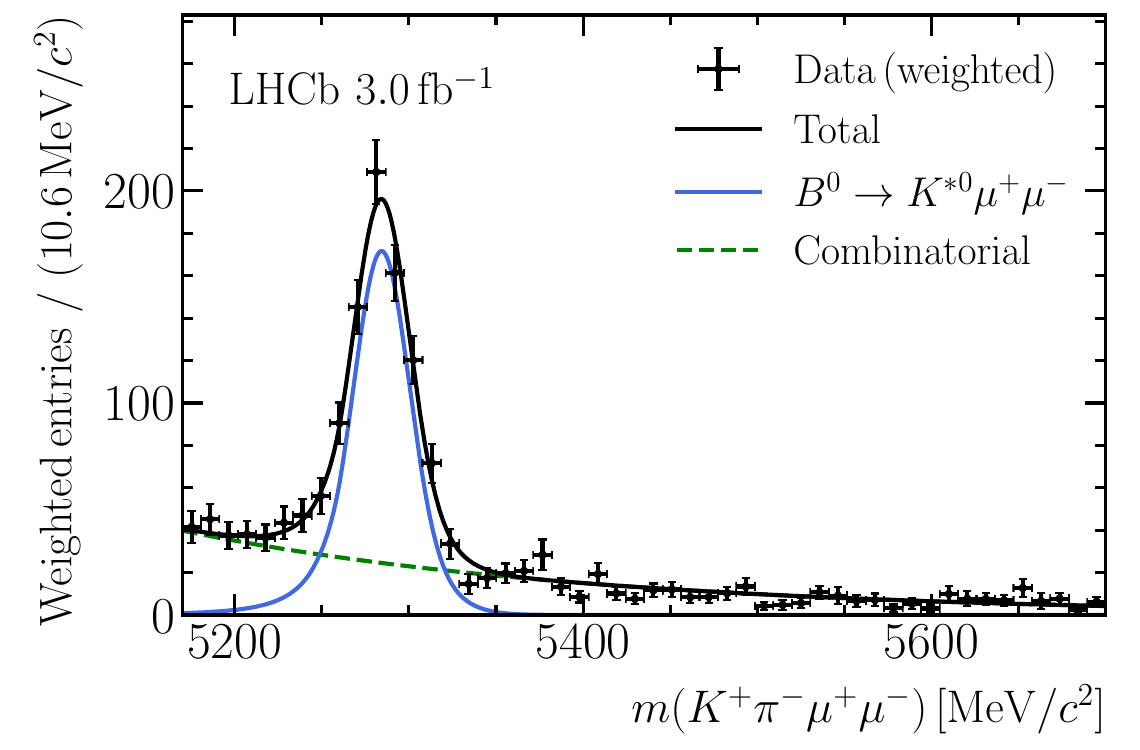} 
    \includegraphics[width=.45\textwidth, trim={0 0 0 0},clip]{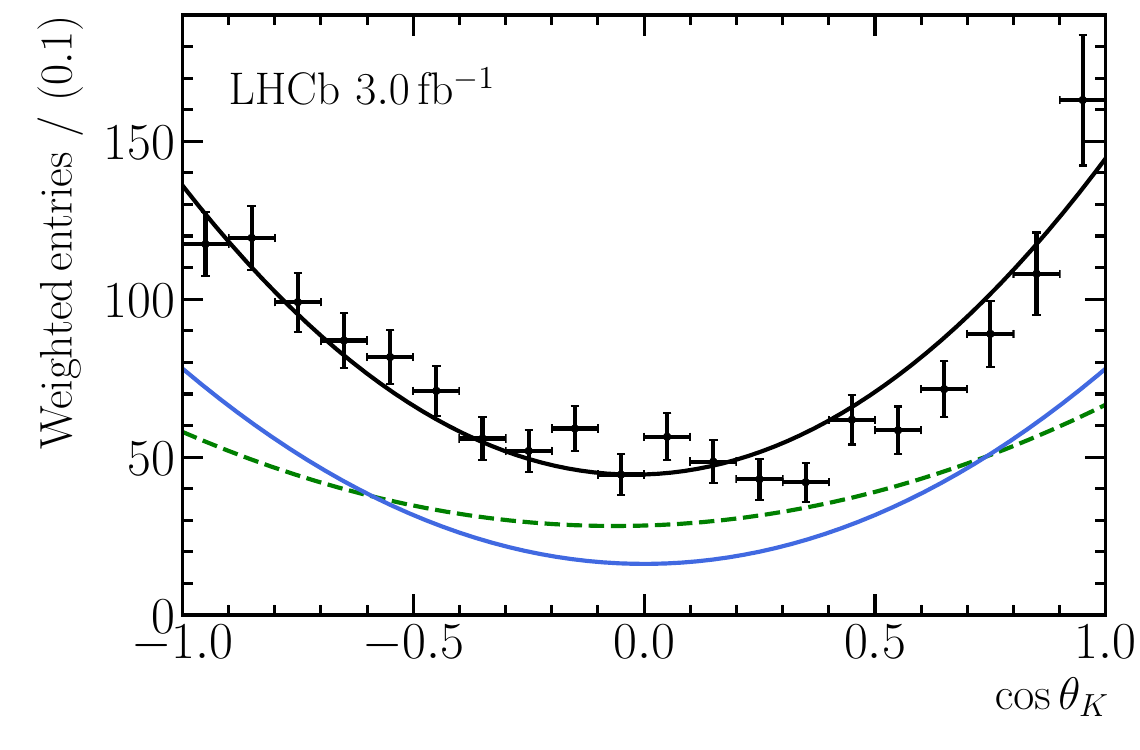} 
    \includegraphics[width=.45\textwidth, trim={0 0 0 0},clip]{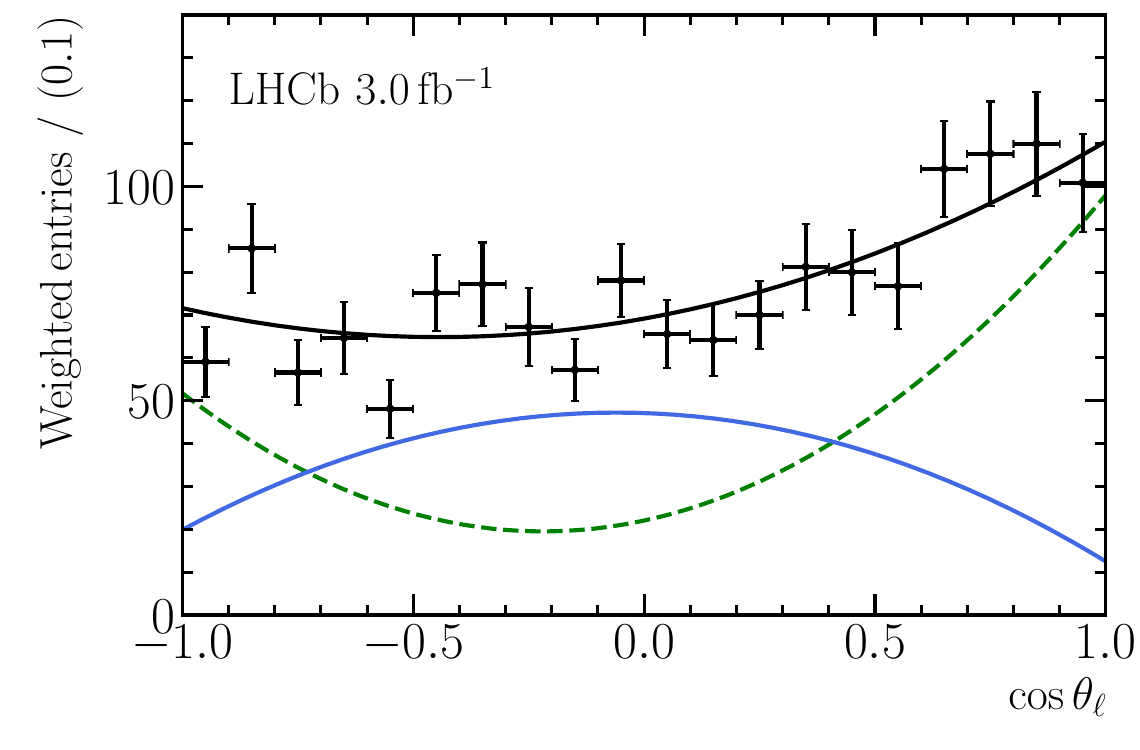} 
    \includegraphics[width=.45\textwidth, trim={0 0 0 0},clip]{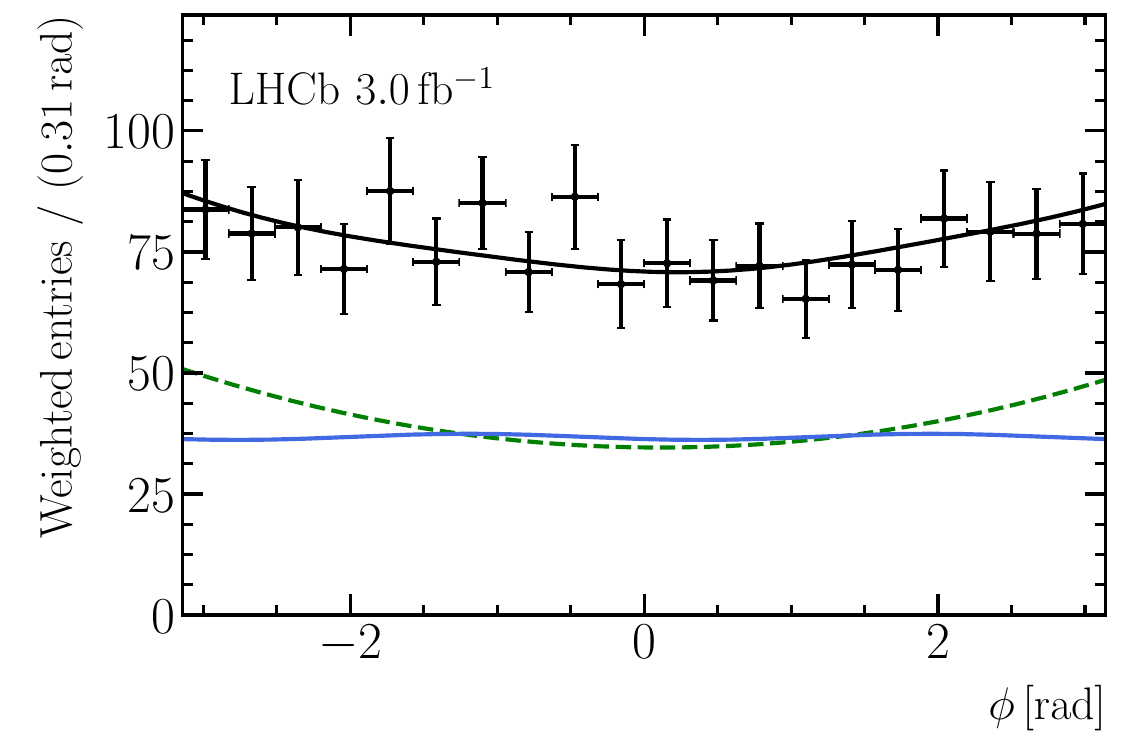} 

    \includegraphics[width=.45\textwidth, trim={0 0 0 0},clip]{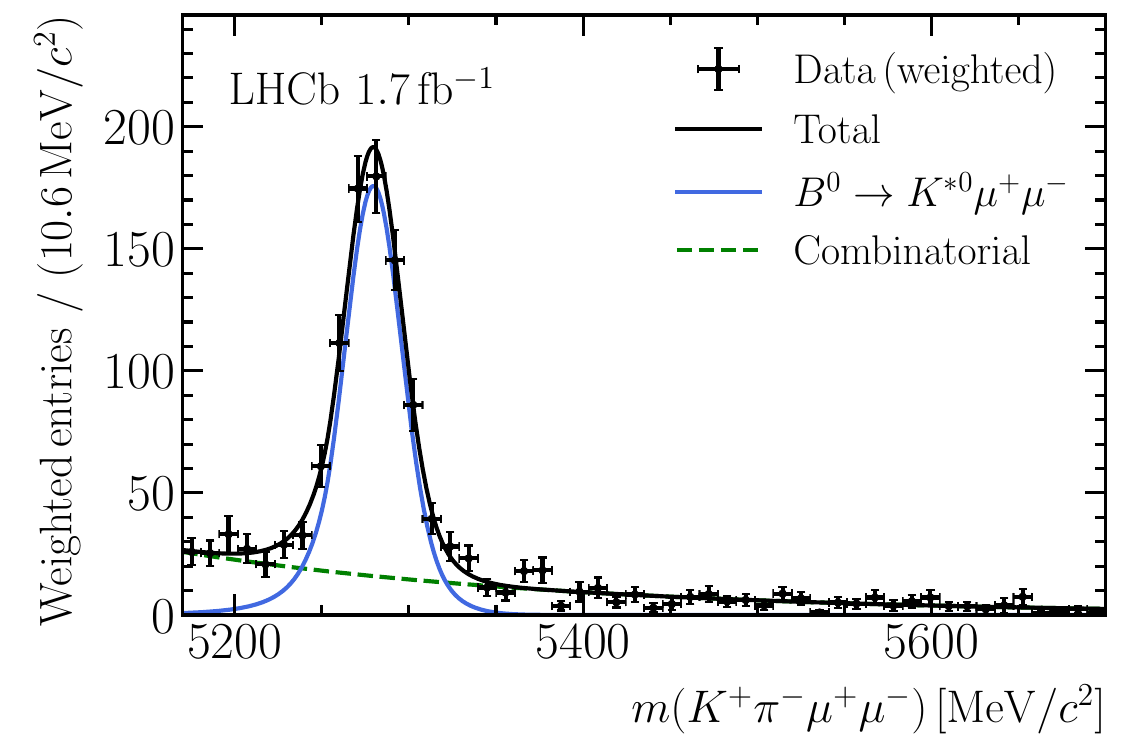} 
    \includegraphics[width=.45\textwidth, trim={0 0 0 0},clip]{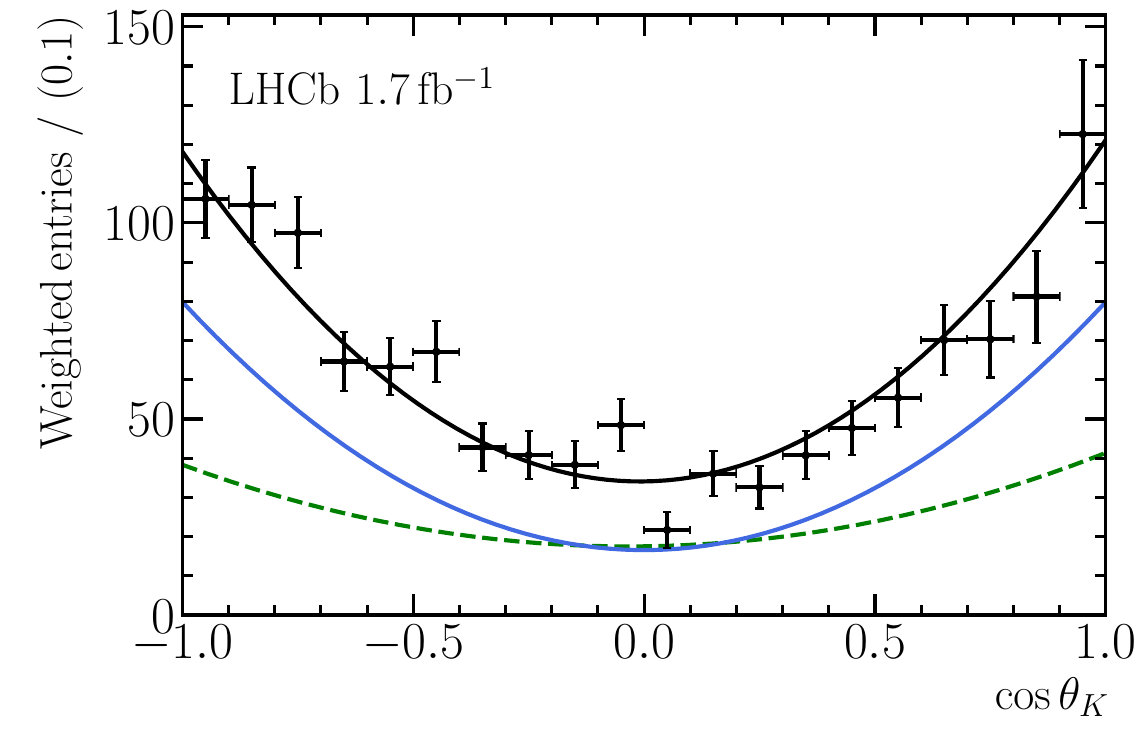} 
    \includegraphics[width=.45\textwidth, trim={0 0 0 0},clip]{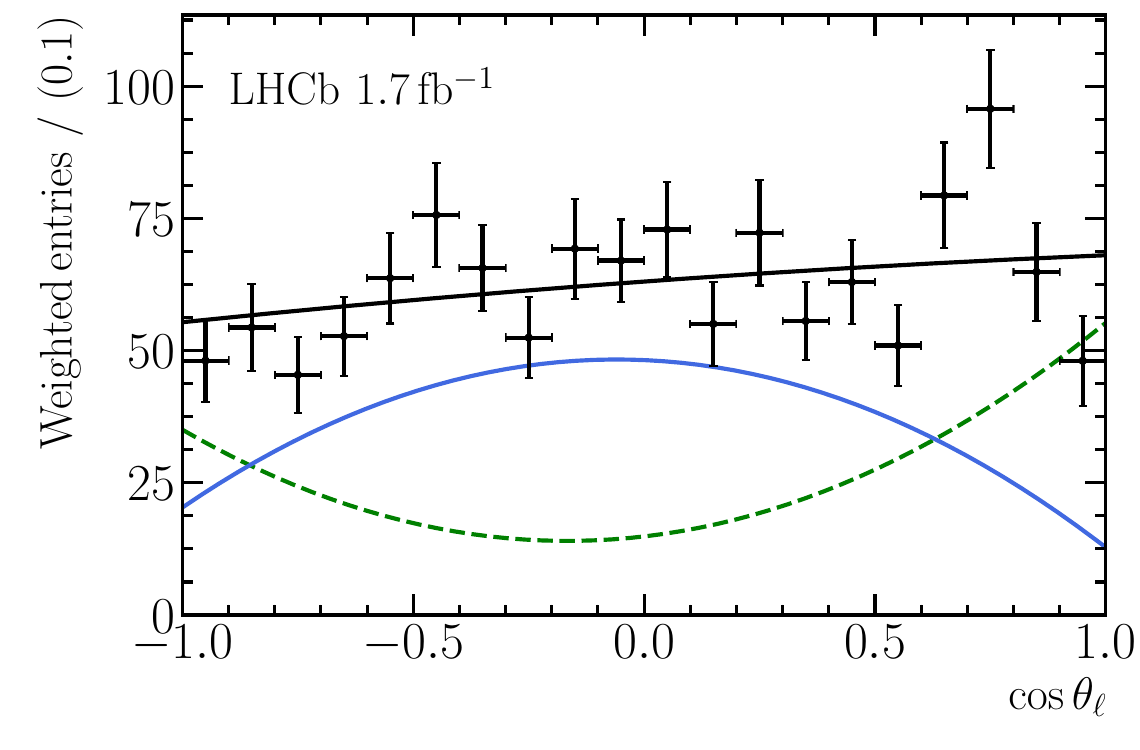} 
    \includegraphics[width=.45\textwidth, trim={0 0 0 0},clip]{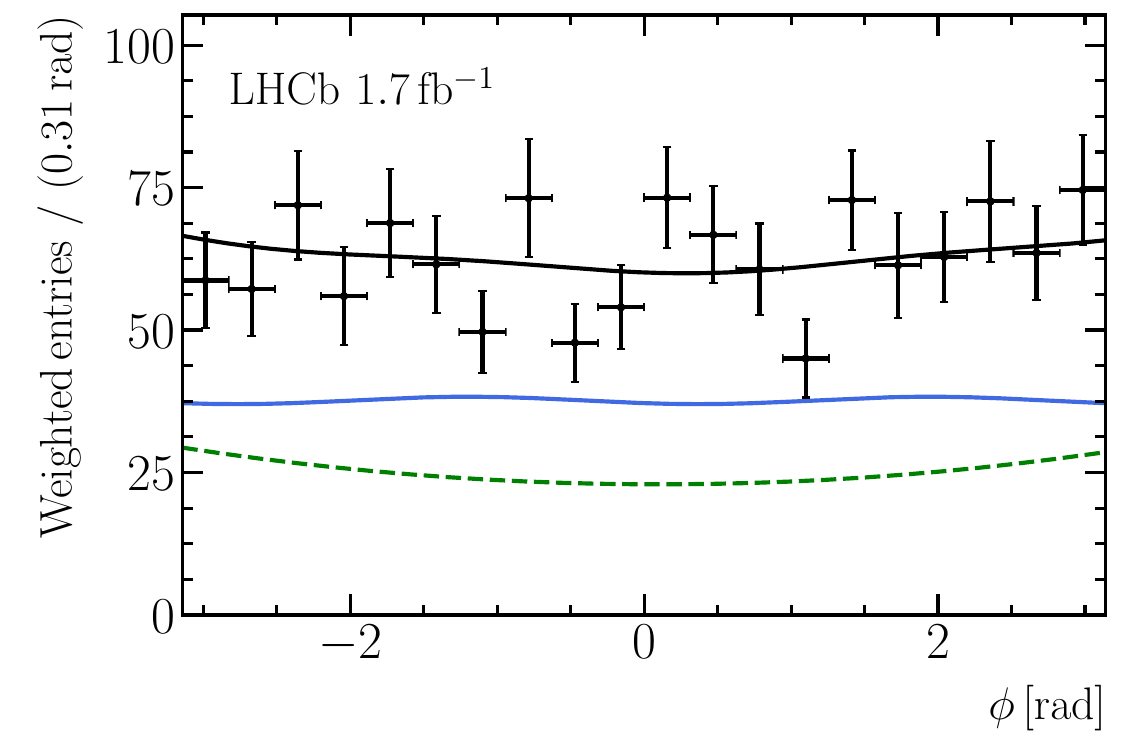}  
  \caption{
  Weighted invariant-mass and angular distributions of $\Bz\to \Kstarz \mu^+\mu^-$ candidates in the (top two rows) Run1 and (bottom two rows) 2016 data samples within the large $q^2$ region. The signal distribution is shown with a solid blue line, and the combinatorial background is shown with a dashed green line. The solid black line corresponds to the full fit function.
  }
\label{fig:muon_lq2}
\end{figure}

The results of the weighted maximum-likelihood fit performed to the $\Bz\to \Kstarz \mu^+\mu^-$ candidates in the unconstrained $q^2$ region of 1.1--7.0\gevgevcccc are shown in Fig.~\ref{fig:muon_lq2}. 

\clearpage

\subsection{Results for the large $\boldsymbol{q^2}$ region}

\begin{figure}[!b]
\centering
    \includegraphics[width=.45\textwidth, trim={0 0 0 0},clip]{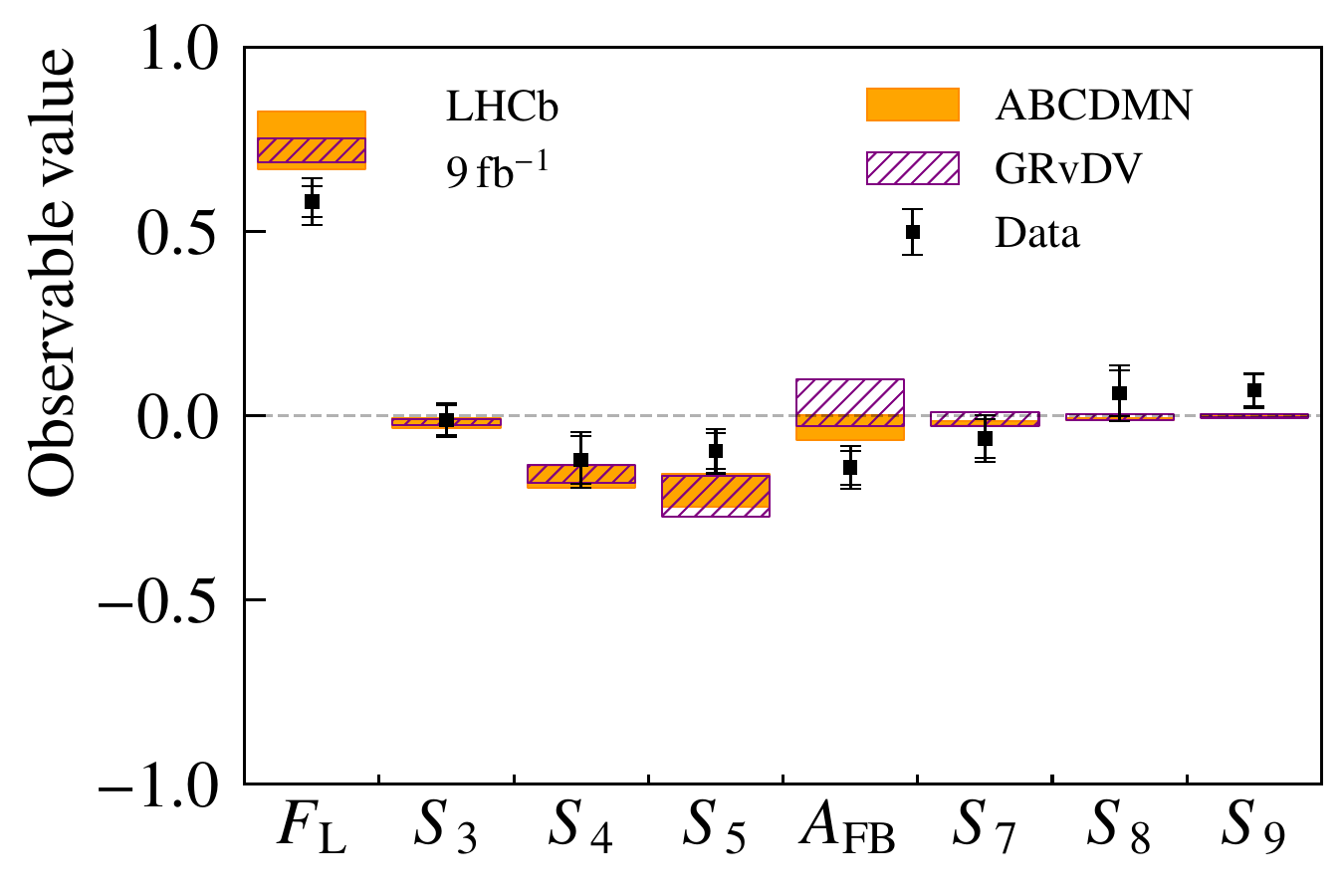} 
    \includegraphics[width=.45\textwidth, trim={0 0 0 0},clip]{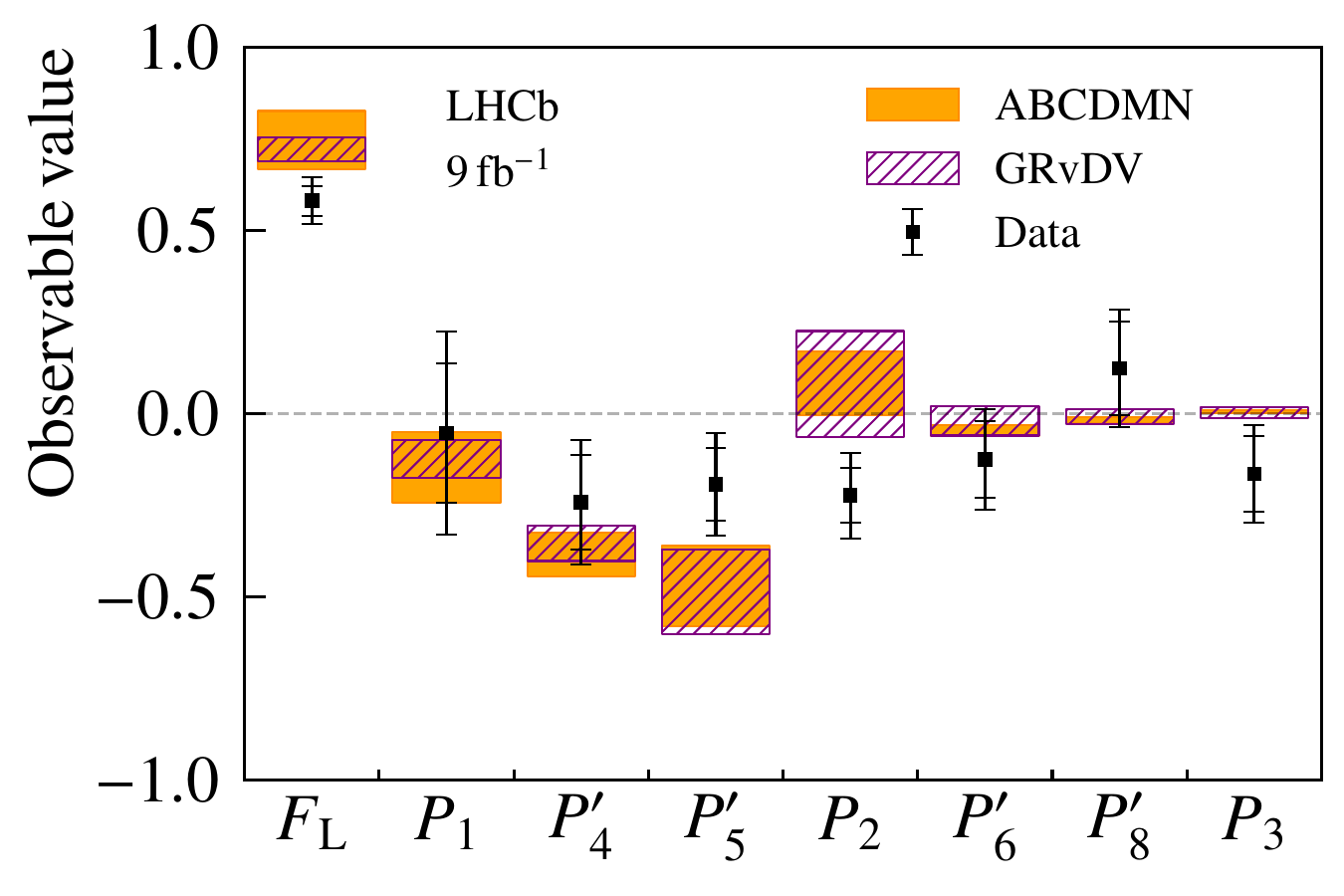} 
  \caption{The (left) $\it{S}$- and (right) $\it{P}$-basis angular observables of the large $q^2$ region. The overlapping error bars show statistical and total uncertainties. The orange and hatched purple boxes correspond to SM predictions based on Ref.~\cite{Alguero:2023jeh} and Refs.~\cite{EOSAuthors:2021xpv,Gubernari:2022hxn}, respectively. 
  }
\label{fig:summary_plots_lq2}
\end{figure}

The angular observables measured in the extended $q^2$ region of 1.1--7.0\gevgevcccc are summarised in Fig.~\ref{fig:summary_plots_lq2}, and their numerical values are given in Table~\ref{tab:lq2_observables_results}. In general, these values are in good agreement with both sets of SM predictions, with differences of around $2\sigma$ or less. 
Observables that differ with respect to one (or both) predictions at a level of $1.5\sigma$ or more include $\FL$, $S_5$, $P_5^{\prime}$, $\AFB$, $P_2$ and $S_9$. 

\begin{figure}[!tb]
\centering
    \includegraphics[width=.45\textwidth, trim={0 0 0 0},clip]{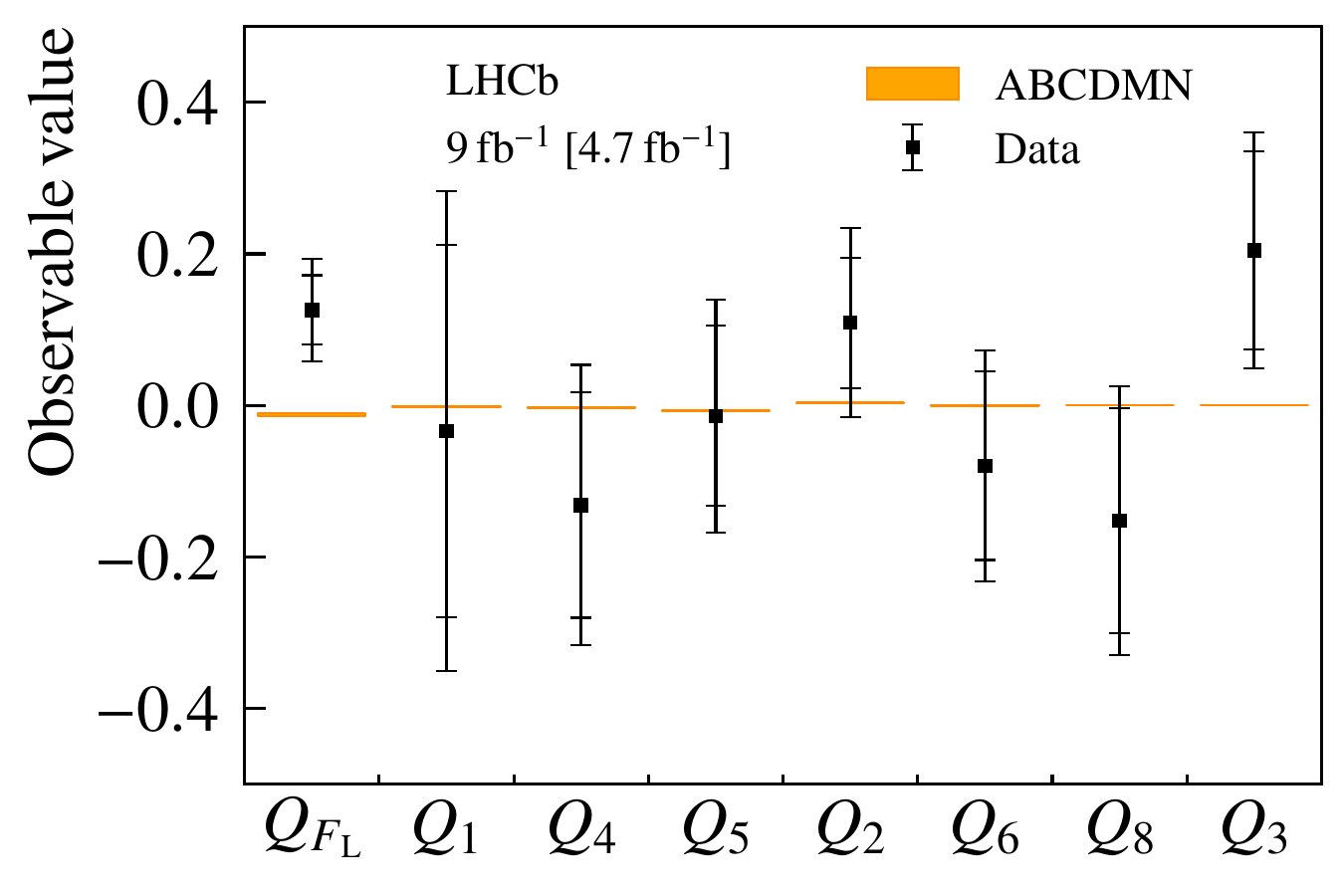} 
  \caption{LFU observables $Q_{i}$ calculated using the $\it{P}$-basis angular observables of the muon and electron modes in the large $q^2$ region.
  The overlapping error bars show statistical and total uncertainties. The SM predictions (orange boxes) are based on Ref.~\cite{Alguero:2023jeh}. 
  }
\label{fig:summary_plots_Qi_lq2}
\end{figure}

The $Q_i$ observables are summarised in Fig.~\ref{fig:summary_plots_Qi_lq2}, and Table~\ref{tab:lq2_observables_results_Qi}. In this case, with the exception of $Q_{\FL}$, which shows a tension of around $2\sigma$, all other values are compatible with the SM prediction at less than $1.5\sigma$.

\begin{table}[!b]
\sisetup{separate-uncertainty}
\centering
\caption{Values for the (left) $\it{S}$- and (right) $\it{P}$-basis angular observables of the large $q^2$ region. The first uncertainty is statistical and the second is systematic.}
\begin{tabular}{l r l r}
\multicolumn{4}{c}{}  \\ \hline
 \multicolumn{4}{c}{Angular observables}  \\ \hline
 $\FL$ &$ 0.58 \pm 0.04  \pm 0.05 $& $\FL$ &$ 0.58 \pm 0.04  \pm 0.05 $\\ 
 $S_{3}$ &$ -0.01 \pm 0.04  \pm 0.02 $ & $P_{1}$ &$ -0.05 \pm 0.19  \pm 0.20 $\\
 $S_{4}$ &$ -0.12 \pm 0.06  \pm 0.04 $ & $P_{4}^{\prime}$ &$ -0.24 \pm 0.13  \pm 0.11 $\\
 $S_{5}$ &$ -0.10 \pm 0.05  \pm 0.04 $ & $P_{5}^{\prime}$ &$ -0.19 \pm 0.10  \pm 0.10 $\\
 $\AFB$ &$ -0.14 \pm 0.05  \pm 0.04 $ & $P_{2}$ &$ -0.22 \pm 0.07  \pm 0.09 $\\
 $S_{7}$ &$ -0.06 \pm 0.05  \pm 0.04 $ & $P_{6}^{\prime}$ &$ -0.13 \pm 0.11  \pm 0.09 $\\
 $S_{8}$ &$ 0.06 \pm 0.06  \pm 0.04 $ & $P_{8}^{\prime}$ &$ 0.12 \pm 0.13  \pm 0.10 $\\
 $S_{9}$ &$ 0.07 \pm 0.04  \pm 0.02 $ & $P_{3}$ &$ -0.17 \pm 0.10  \pm 0.09 $\\ \hline
\end{tabular}
\label{tab:lq2_observables_results}
\end{table}

\begin{table}[!tb]
\sisetup{separate-uncertainty}
\centering
\caption{Values for the $Q_i$ LFU observables given by the differences between the muon and electron $\it{P}$-basis angular observables in the large $q^2$ region. The first uncertainty is statistical and the second is systematic.
}
\begin{tabular}{l r }
\multicolumn{2}{c}{}  \\ \hline
 \multicolumn{2}{c}{LFU observables}  \\ \hline
$Q_{\FL}$ &$ 0.13  \pm 0.05 \pm 0.05 $\\ 
 $Q_{1}$ &$ -0.03  \pm 0.25 \pm 0.20 $\\
 $Q_{4}$ &$ -0.13  \pm 0.15 \pm 0.11 $\\
 $Q_{5}$ &$ -0.01  \pm 0.12 \pm 0.10 $\\
 $Q_{2}$ &$ 0.11  \pm 0.09 \pm 0.09 $\\
 $Q_{6}$ &$ -0.08  \pm 0.12 \pm 0.09 $\\
 $Q_{8}$ &$ -0.15  \pm 0.15 \pm 0.10 $\\
 $Q_{3}$ &$ 0.21  \pm 0.13 \pm 0.09 $\\ \hline
\end{tabular}
\label{tab:lq2_observables_results_Qi}
\end{table}

\clearpage

\subsection{Correlation among angular observables}

Correlations among observables of the large $q^2$ region are given in Tables~\ref{tab:corr_lq2_S} and~\ref{tab:corr_lq2_P}. Correlations of the systematic uncertainties among observables are given in Tables~\ref{tab:corr_syst_lq2_S} and~\ref{tab:corr_syst_lq2_P}.

\begin{table}[!b]
\sisetup{separate-uncertainty}
\centering
\caption{Correlation matrix for the $\it{S}$-basis angular observables of the large $q^2$ region.}
\begin{tabular}{l S[table-format = 2.2] S[table-format = 2.2] S[table-format = 2.2] S[table-format = 2.2] S[table-format = 2.2] S[table-format = 2.2] S[table-format = 2.2] S[table-format = 2.2]}
\multicolumn{1}{l}{} & \multicolumn{1}{c}{$\FL$} & \multicolumn{1}{c}{$S_{3}$} & \multicolumn{1}{c}{$S_{4}$} & \multicolumn{1}{c}{$S_{5}$} & \multicolumn{1}{c}{$\AFB$} & \multicolumn{1}{c}{$S_{7}$ } & \multicolumn{1}{c}{$S_{8}$ } & \multicolumn{1}{c}{$S_{9}$ } \\ \hline 
 $\FL$          & 1.00 & 0.02 & -0.05 & -0.01 & 0.09 & -0.05 & -0.03 & -0.05\\ 
 $S_{3}$          &        & 1.00 & -0.05 & -0.03 & 0.04 & 0.05 & -0.05 & 0.02 \\
 $S_{4}$          &        &        & 1.00 & -0.10 & -0.14 & -0.05 & 0.06 & 0.04 \\
 $S_{5}$          &        &        &        & 1.00 & -0.07 & 0.06 & -0.02 & -0.04 \\
 $\AFB$ &        &        &        &        & 1.00 & 0.03 & -0.04 & -0.01 \\
 $S_{7}$ &        &        &        &        &        & 1.00 & -0.06 & -0.13 \\
 $S_{8}$ &        &        &        &        &        &        & 1.00 & -0.04 \\
 $S_{9}$ &        &        &        &        &        &        &        & 1.00 \\
\end{tabular}
\label{tab:corr_lq2_S}
\end{table}

\begin{table}[!tb]
\sisetup{separate-uncertainty}
\centering
\caption{Correlation matrix for the $\it{P}$-basis angular observables of the large $q^2$ region.}
\begin{tabular}{l S[table-format = 2.2] S[table-format = 2.2] S[table-format = 2.2] S[table-format = 2.2] S[table-format = 2.2] S[table-format = 2.2] S[table-format = 2.2] S[table-format = 2.2]}
\multicolumn{1}{l}{} & \multicolumn{1}{c}{$\FL$} & \multicolumn{1}{c}{$P_{1}$} & \multicolumn{1}{c}{$P_{2}$} & \multicolumn{1}{c}{$P_{3}$} & \multicolumn{1}{c}{$P_{4}^{\prime}$} & \multicolumn{1}{c}{$P_{5}^{\prime}$ } & \multicolumn{1}{c}{$P_{6}^{\prime}$ } & \multicolumn{1}{c}{$P_{8}^{\prime}$ } \\ \hline 
 $\FL$          & 1.00 & 0.00 & -0.18 & -0.11 & -0.07 & -0.04 & -0.06 & -0.02\\ 
 $P_{1}$          &        & 1.00 & 0.04 & -0.02 & -0.05 & -0.03 & 0.05 & -0.05 \\
 $P_{2}$          &        &        & 1.00 & 0.03 & -0.13 & -0.06 & 0.05 & -0.03 \\
 $P_{3}$          &        &        &        & 1.00 & -0.03 & 0.04 & 0.14 & 0.04 \\
 $P_{4}^{\prime}$ &        &        &        &        & 1.00 & -0.10 & -0.05 & 0.06 \\
 $P_{5}^{\prime}$ &        &        &        &        &        & 1.00 & 0.06 & -0.02 \\
 $P_{6}^{\prime}$ &        &        &        &        &        &        & 1.00 & -0.06 \\
 $P_{8}^{\prime}$ &        &        &        &        &        &        &        & 1.00 \\
\end{tabular}
\label{tab:corr_lq2_P}
\end{table}

\begin{table}[!tb]
\sisetup{separate-uncertainty}
\centering
\caption{Correlation matrix of the systematic uncertainties for the $\it{S}$-basis angular observables of the large $q^2$ region.}
\begin{tabular}{l S[table-format = 2.2] S[table-format = 2.2] S[table-format = 2.2] S[table-format = 2.2] S[table-format = 2.2] S[table-format = 2.2] S[table-format = 2.2] S[table-format = 2.2]}
\multicolumn{1}{l}{} & \multicolumn{1}{c}{$\FL$} & \multicolumn{1}{c}{$S_{3}$} & \multicolumn{1}{c}{$S_{4}$} & \multicolumn{1}{c}{$S_{5}$} & \multicolumn{1}{c}{$\AFB$} & \multicolumn{1}{c}{$S_{7}$ } & \multicolumn{1}{c}{$S_{8}$ } & \multicolumn{1}{c}{$S_{9}$ } \\ \hline 
 $\FL$          & 1.00 & 0.01 & -0.10 & -0.15 & -0.18 & -0.02 & 0.02 & -0.04\\ 
 $S_{3}$          &        & 1.00 & 0.00 & -0.05 & -0.00 & 0.00 & 0.02 & 0.01 \\
 $S_{4}$          &        &        & 1.00 & 0.34 & 0.00 & -0.04 & 0.00 & 0.00 \\
 $S_{5}$          &        &        &        & 1.00 & 0.06 & 0.00 & -0.04 & -0.00 \\
 $\AFB$ &        &        &        &        & 1.00 & -0.02 & -0.01 & 0.01 \\
 $S_{7}$ &        &        &        &        &        & 1.00 & 0.11 & -0.05 \\
 $S_{8}$ &        &        &        &        &        &        & 1.00 & -0.01 \\
 $S_{9}$ &        &        &        &        &        &        &        & 1.00 \\
\end{tabular}
\label{tab:corr_syst_lq2_S}
\end{table}

\begin{table}[!tb]
\sisetup{separate-uncertainty}
\centering
\caption{Correlation matrix of the systematic uncertainties for the $\it{P}$-basis angular observables of the large $q^2$ region.}
\begin{tabular}{l S[table-format = 2.2] S[table-format = 2.2] S[table-format = 2.2] S[table-format = 2.2] S[table-format = 2.2] S[table-format = 2.2] S[table-format = 2.2] S[table-format = 2.2]}
\multicolumn{1}{l}{} & \multicolumn{1}{c}{$\FL$} & \multicolumn{1}{c}{$P_{1}$} & \multicolumn{1}{c}{$P_{2}$} & \multicolumn{1}{c}{$P_{3}$} & \multicolumn{1}{c}{$P_{4}^{\prime}$} & \multicolumn{1}{c}{$P_{5}^{\prime}$ } & \multicolumn{1}{c}{$P_{6}^{\prime}$ } & \multicolumn{1}{c}{$P_{8}^{\prime}$ } \\ \hline 
 $\FL$          & 1.00 & -0.04 & -0.05 & 0.02 & -0.20 & -0.29 & -0.02 & 0.01\\ 
 $P_{1}$          &        & 1.00 & 0.00 & -0.02 & 0.02 & -0.01 & 0.00 & 0.02 \\
 $P_{2}$          &        &        & 1.00 & 0.01 & -0.01 & 0.03 & -0.01 & 0.00 \\
 $P_{3}$          &        &        &        & 1.00 & -0.00 & 0.01 & 0.05 & 0.01 \\
 $P_{4}^{\prime}$ &        &        &        &        & 1.00 & 0.38 & -0.03 & 0.00 \\
 $P_{5}^{\prime}$ &        &        &        &        &        & 1.00 & 0.01 & -0.03 \\
 $P_{6}^{\prime}$ &        &        &        &        &        &        & 1.00 & 0.11 \\
 $P_{8}^{\prime}$ &        &        &        &        &        &        &        & 1.00 \\
\end{tabular}
\label{tab:corr_syst_lq2_P}
\end{table}

\clearpage

\subsection{Additional figures of the two $\boldsymbol{q^2}$ regions}
\label{sec:additional_figures}
\begin{figure}[!tb]
\centering
    \includegraphics[width=.45\textwidth, trim={0 0 0 0},clip]{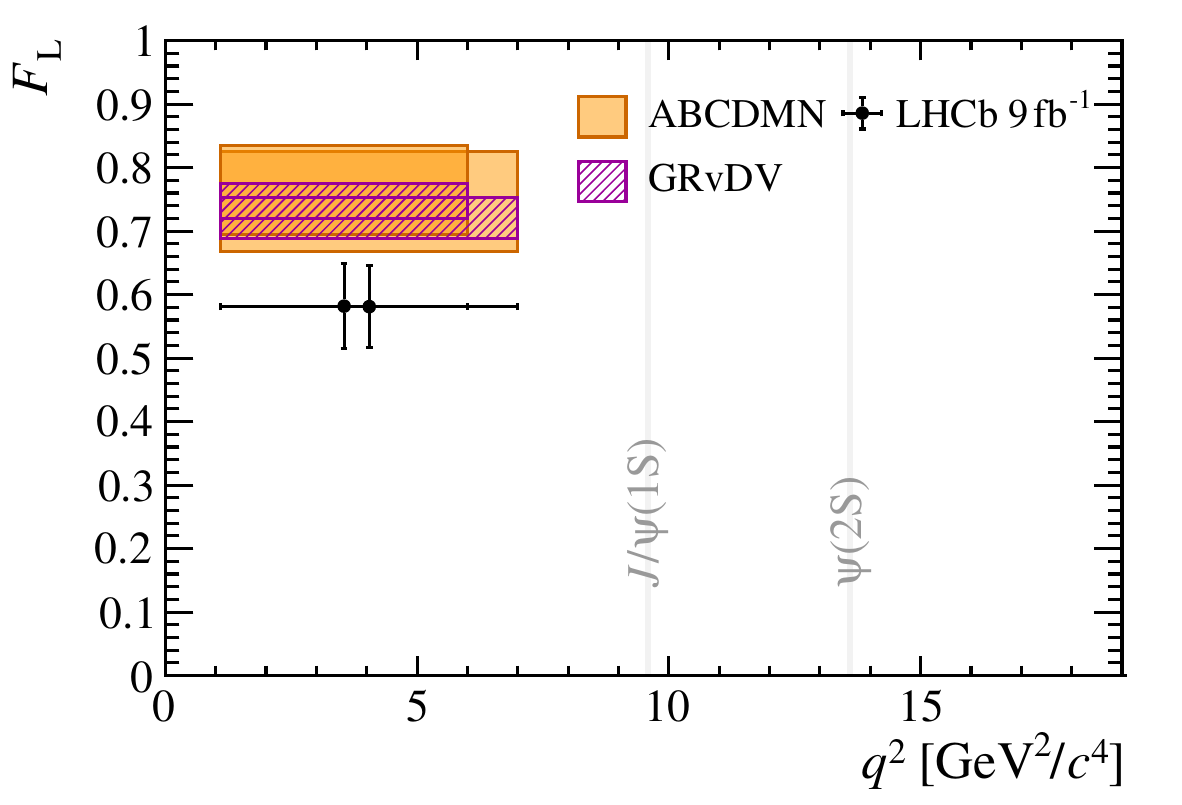} 
    \includegraphics[width=.45\textwidth, trim={0 0 0 0},clip]{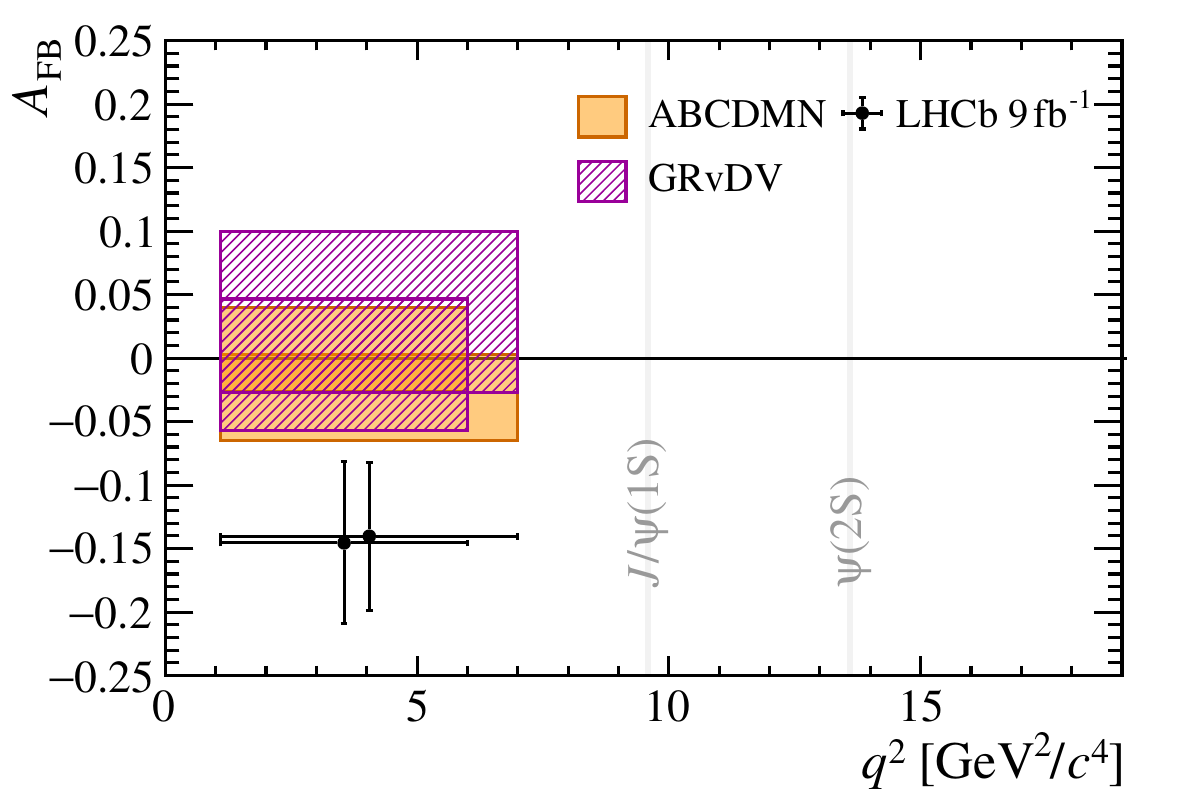} 
    \includegraphics[width=.45\textwidth, trim={0 0 0 0},clip]{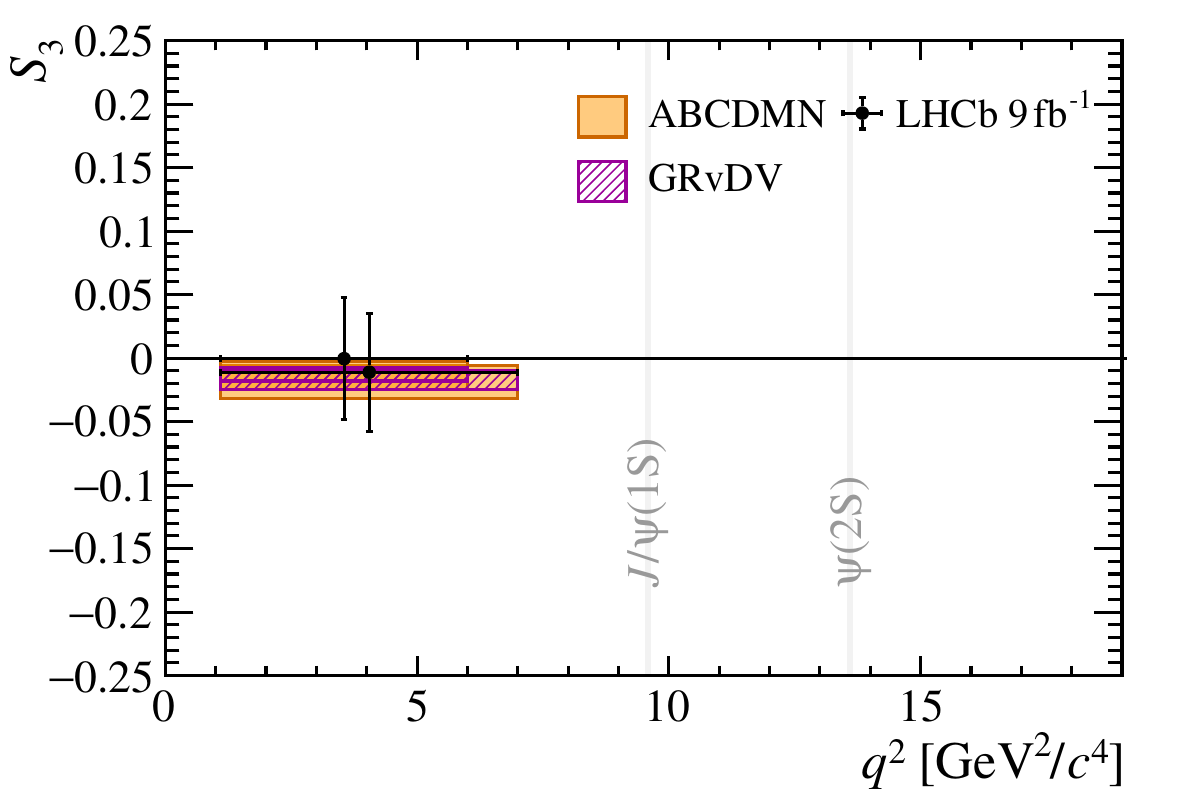} 
    \includegraphics[width=.45\textwidth, trim={0 0 0 0},clip]{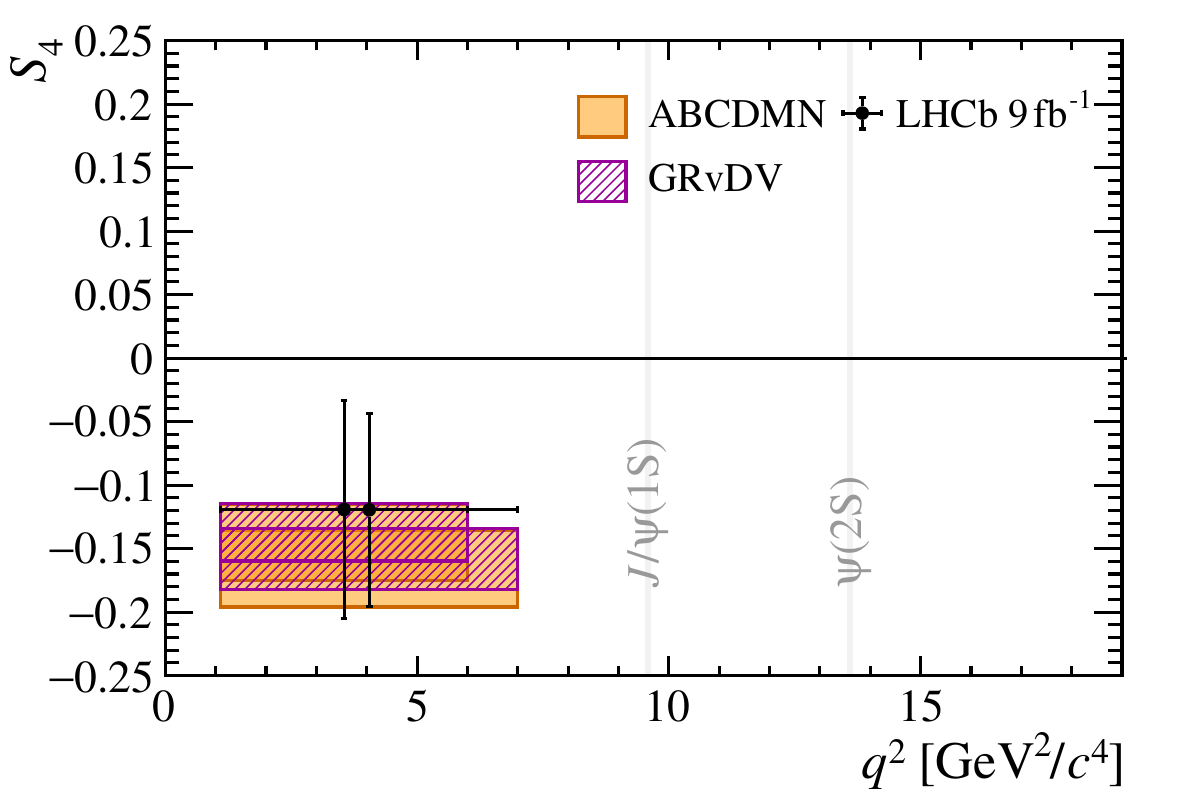} 
    \includegraphics[width=.45\textwidth, trim={0 0 0 0},clip]{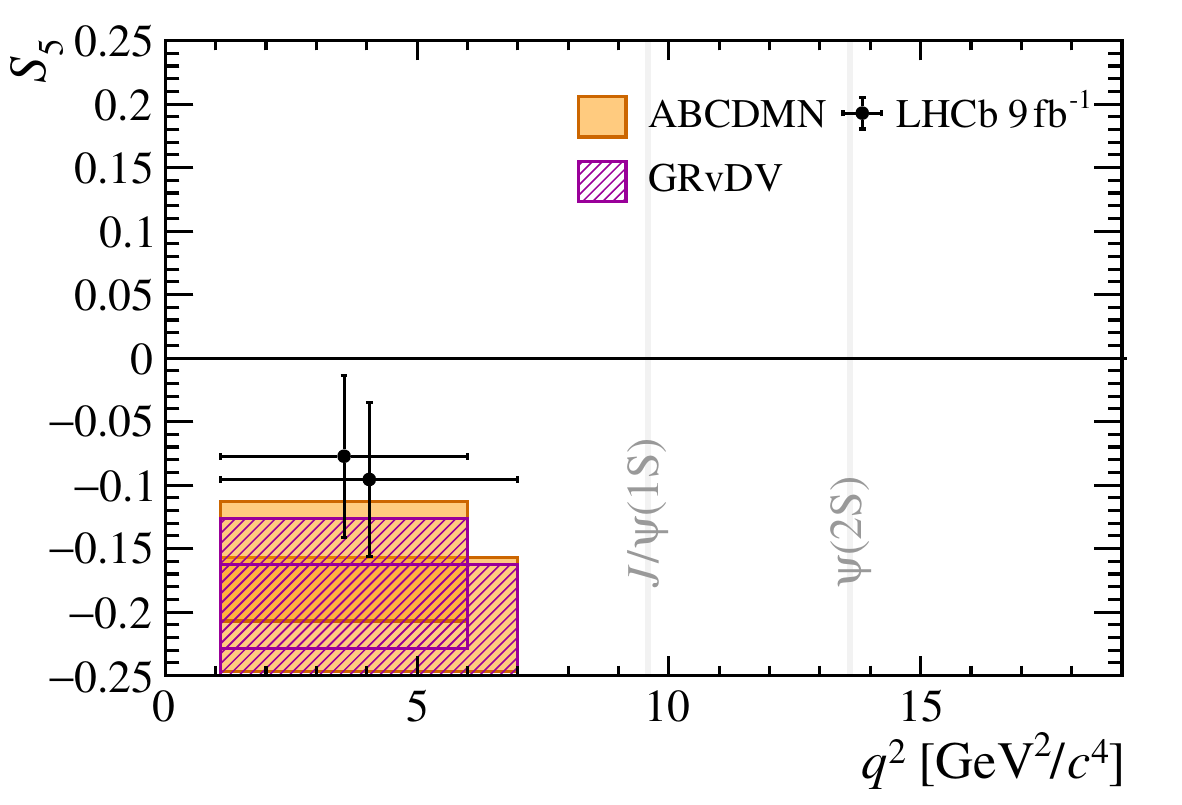} 
    \includegraphics[width=.45\textwidth, trim={0 0 0 0},clip]{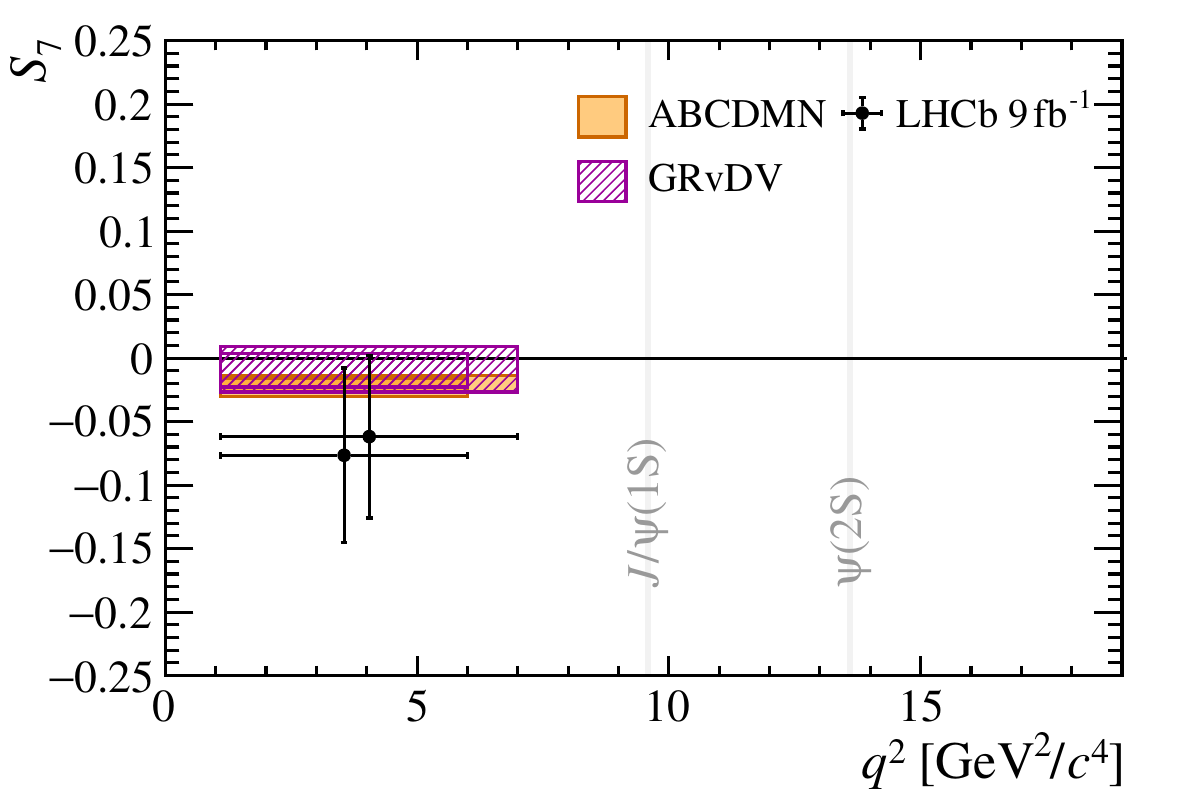} 
    \includegraphics[width=.45\textwidth, trim={0 0 0 0},clip]{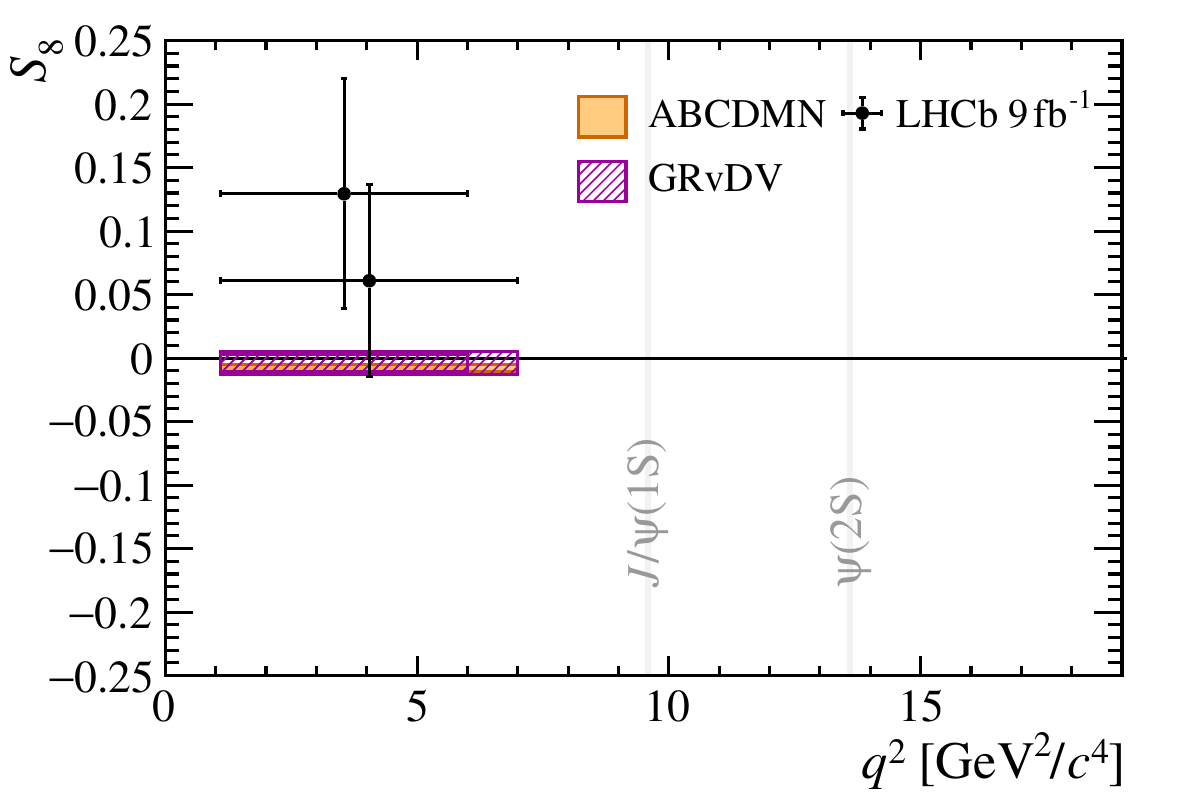} 
    \includegraphics[width=.45\textwidth, trim={0 0 0 0},clip]{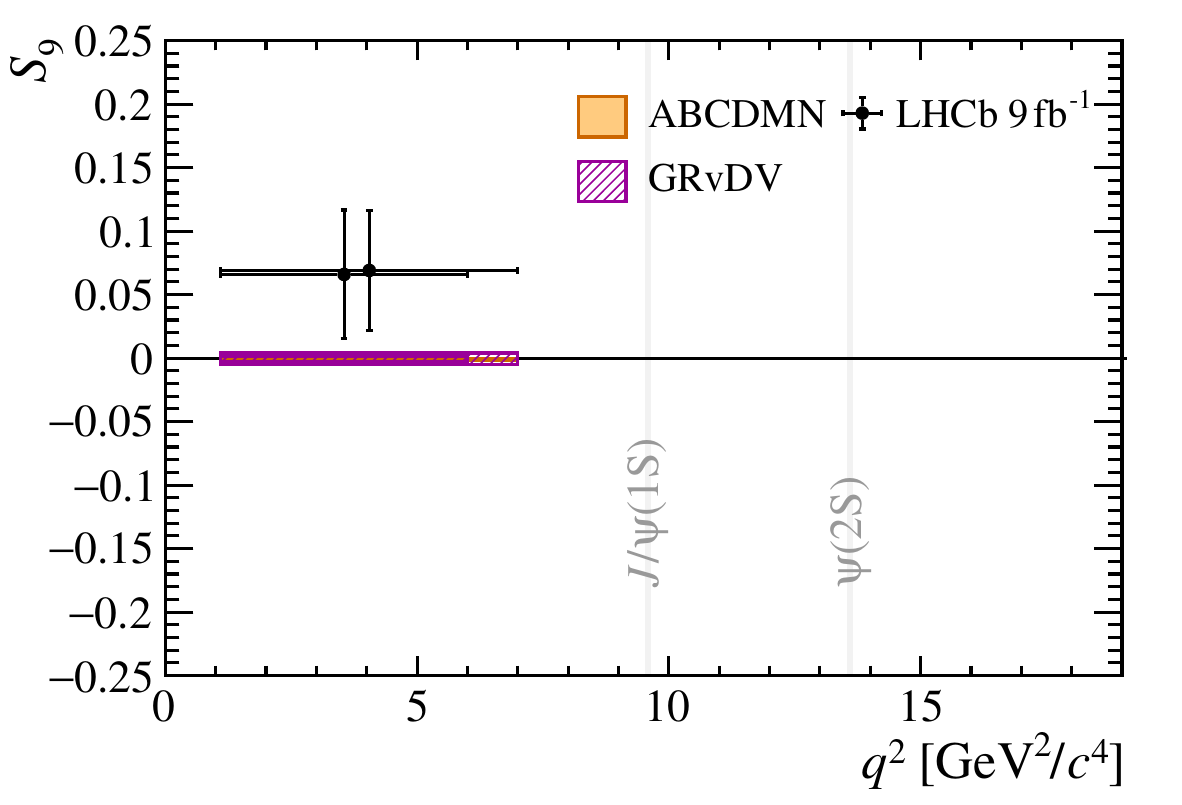} 
  \caption{Measured $\it{S}$-basis angular observables of the baseline and extended $q^2$ regions.
  The orange and hatched purple boxes correspond to SM predictions based on Ref.~\cite{Alguero:2023jeh} and Refs.~\cite{EOSAuthors:2021xpv,Gubernari:2022hxn}, respectively.
  }
\label{fig:Si_plots_app}
\end{figure}

\begin{figure}[!tb]
\centering
    \includegraphics[width=.45\textwidth, trim={0 0 0 0},clip]{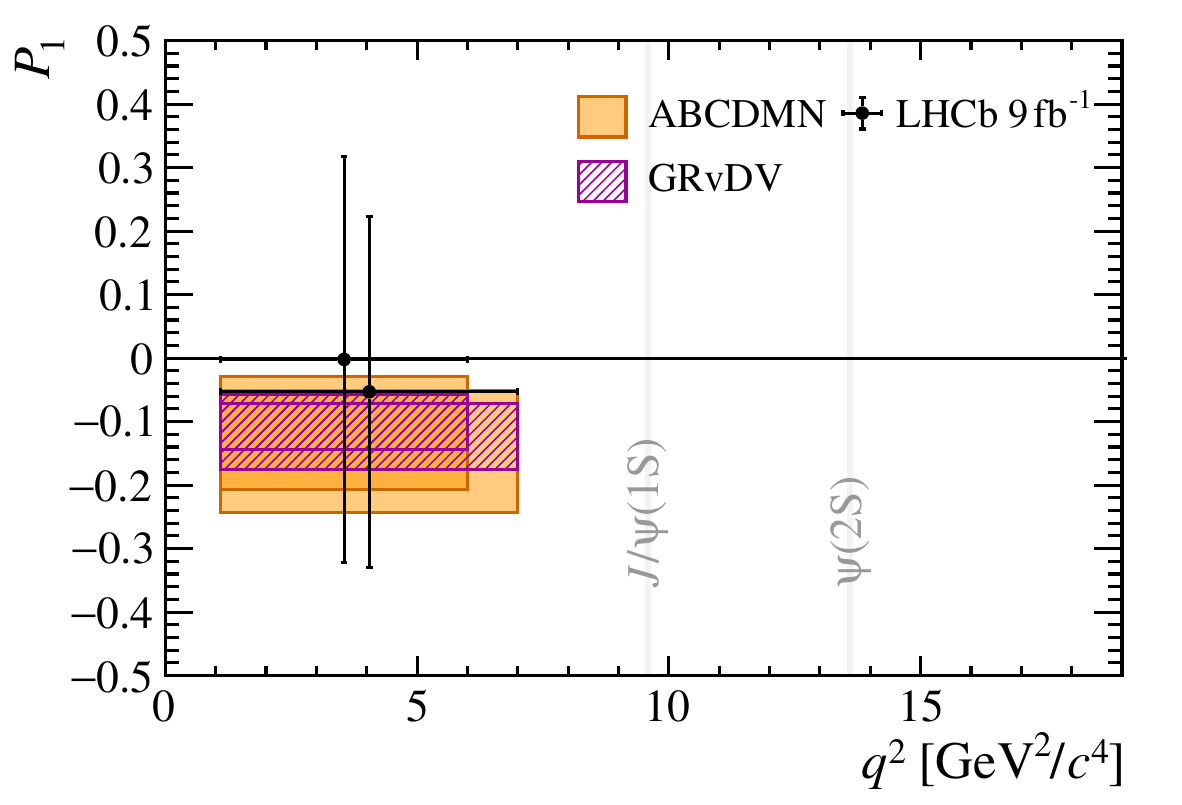} 
    \includegraphics[width=.45\textwidth, trim={0 0 0 0},clip]{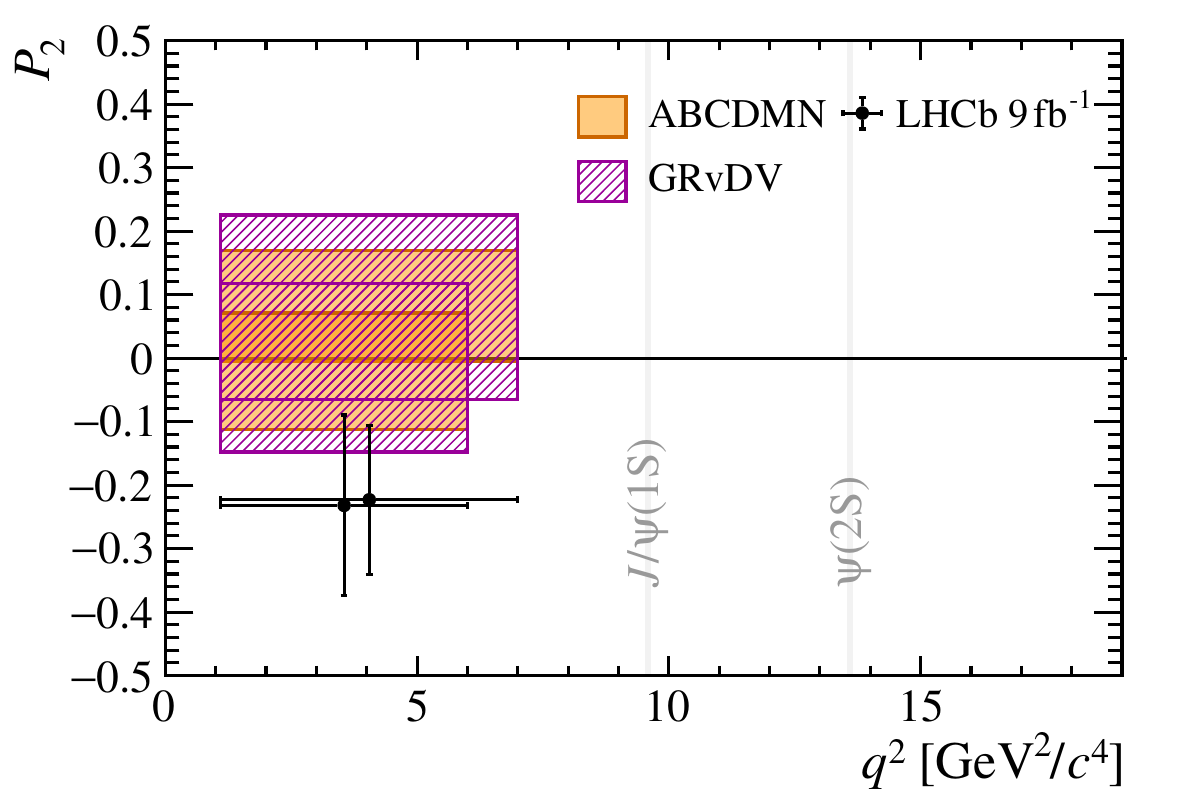} 
    \includegraphics[width=.45\textwidth, trim={0 0 0 0},clip]{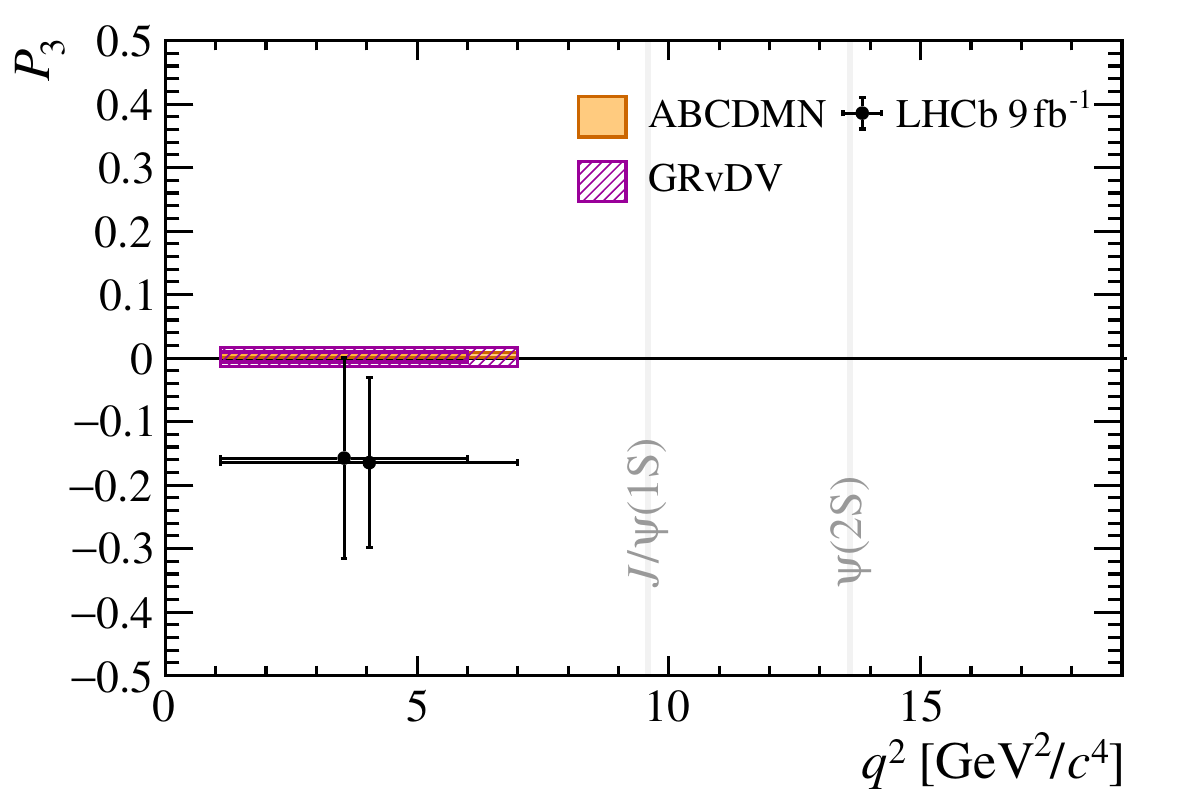} 
    \includegraphics[width=.45\textwidth, trim={0 0 0 0},clip]{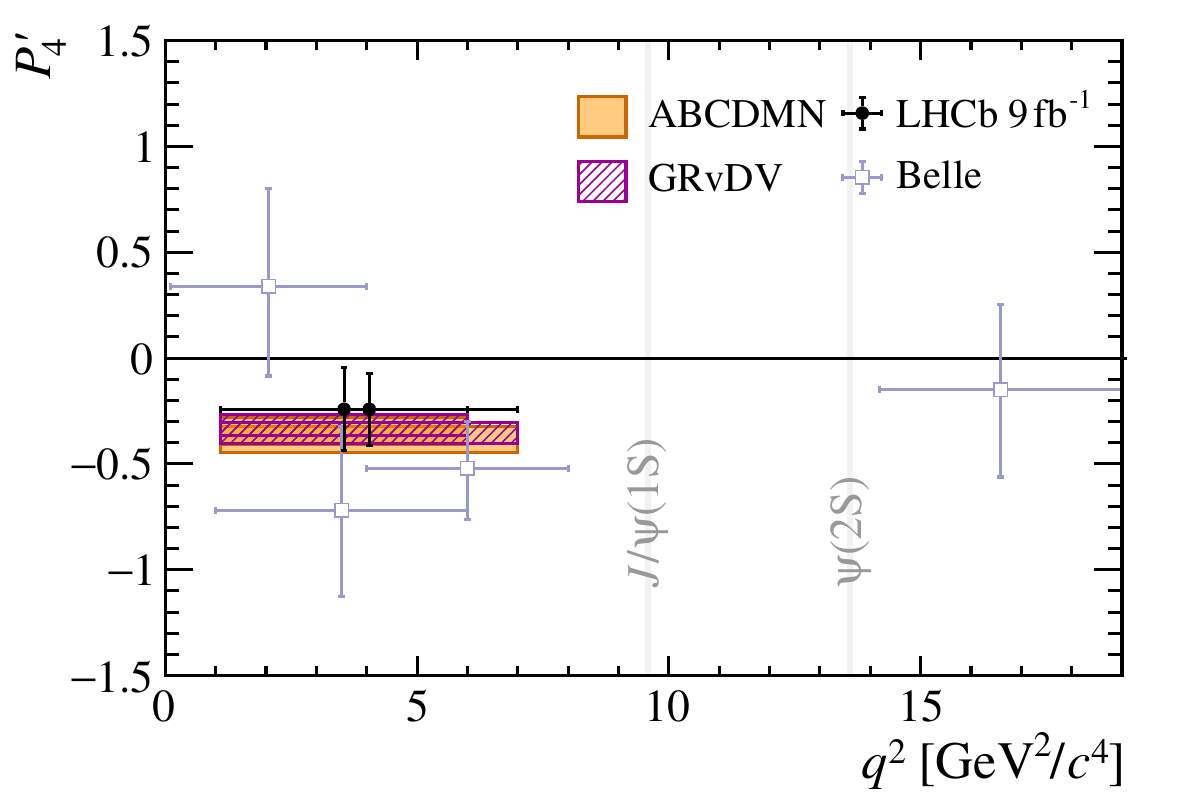} 
    \includegraphics[width=.45\textwidth, trim={0 0 0 0},clip]{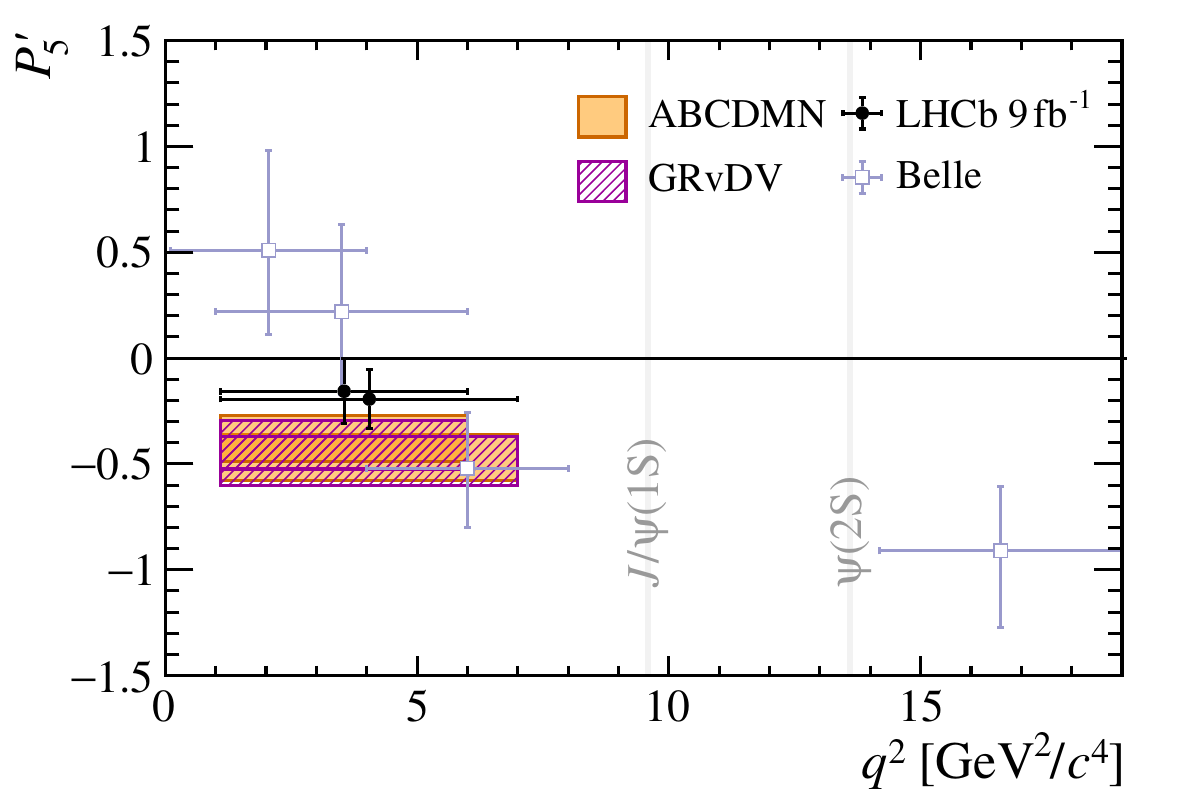} 
    \includegraphics[width=.45\textwidth, trim={0 0 0 0},clip]{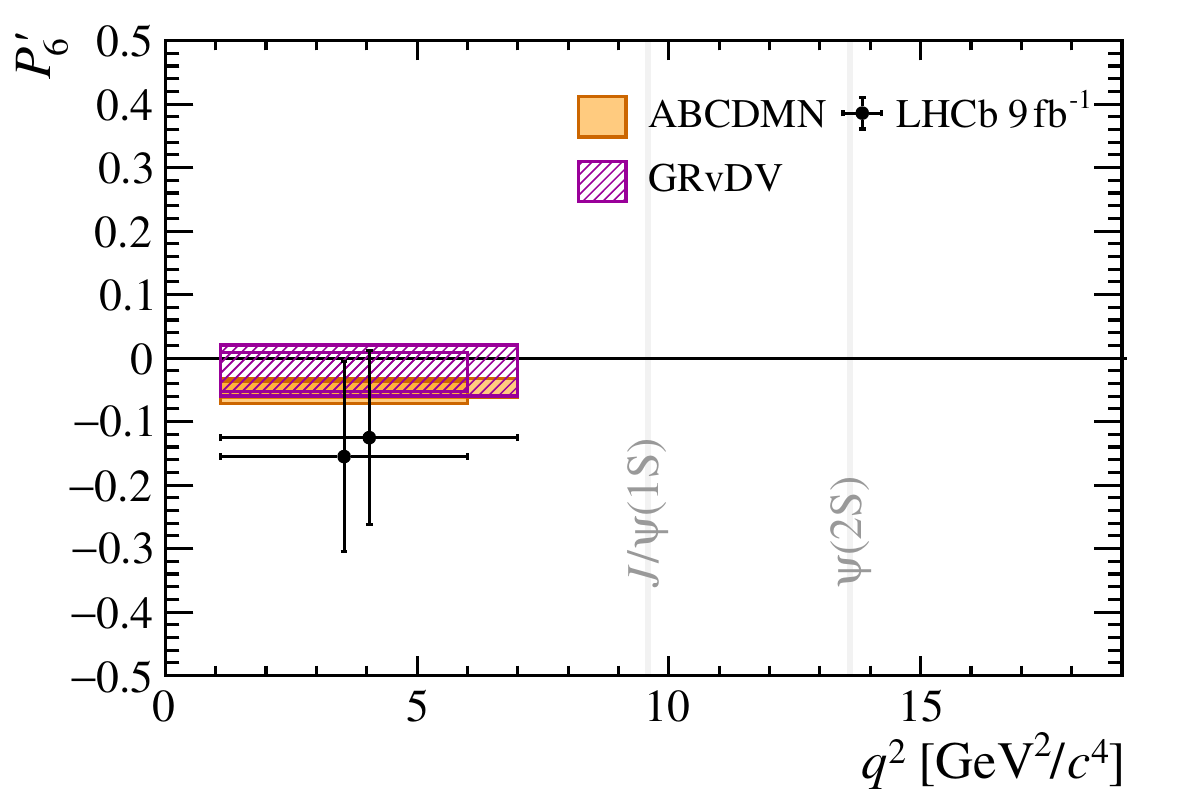} 
    \includegraphics[width=.45\textwidth, trim={0 0 0 0},clip]{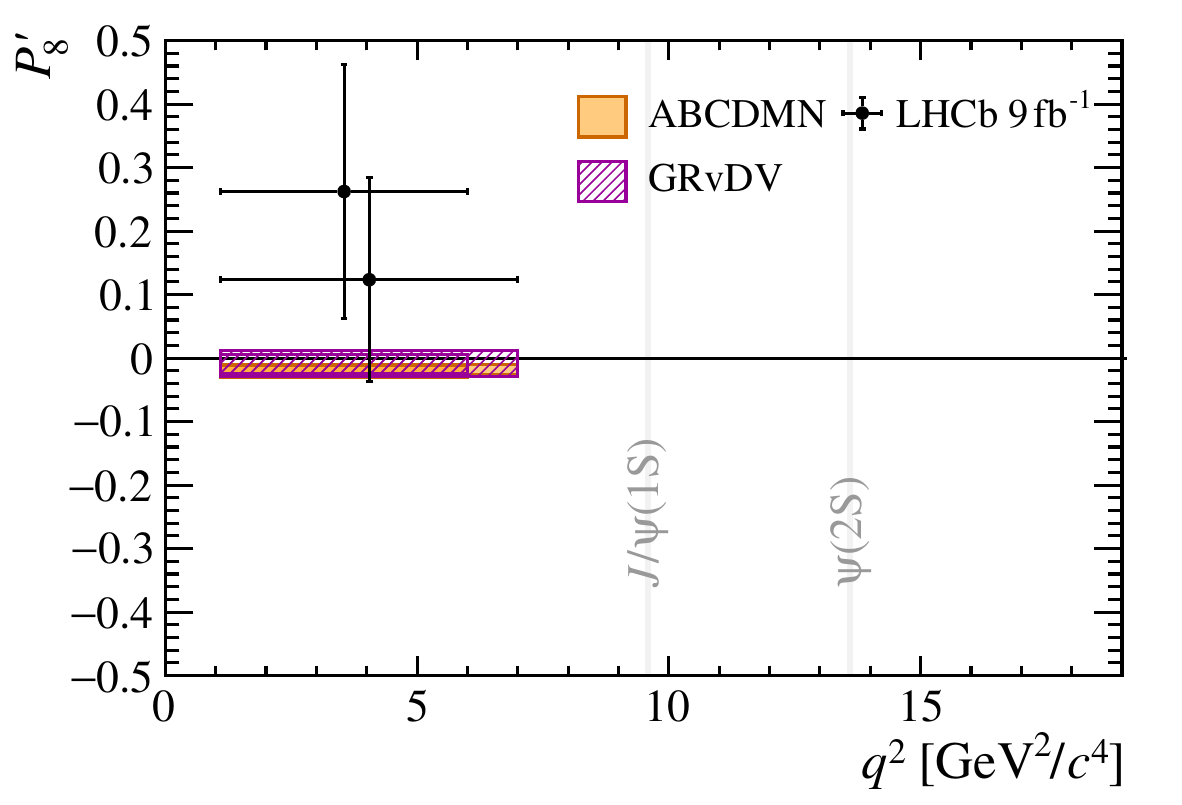} 
  \caption{Measured $\it{P}$-basis angular observables of the baseline and extended $q^2$ regions.
  The orange and hatched purple boxes correspond to SM predictions based on Ref.~\cite{Alguero:2023jeh} and Refs.~\cite{EOSAuthors:2021xpv,Gubernari:2022hxn}, respectively.
  The values of $P_4^{\prime}$ and $P_5^{\prime}$ measured by Belle~\cite{Belle:2016fev} for the decays of $B^{+,0}\to K^{*+,*0} e^+e^-$ are shown in light blue.}
\label{fig:Pi_plots_app}
\end{figure}

\begin{figure}[!tb]
\centering
    \includegraphics[width=.45\textwidth, trim={0 0 0 0},clip]{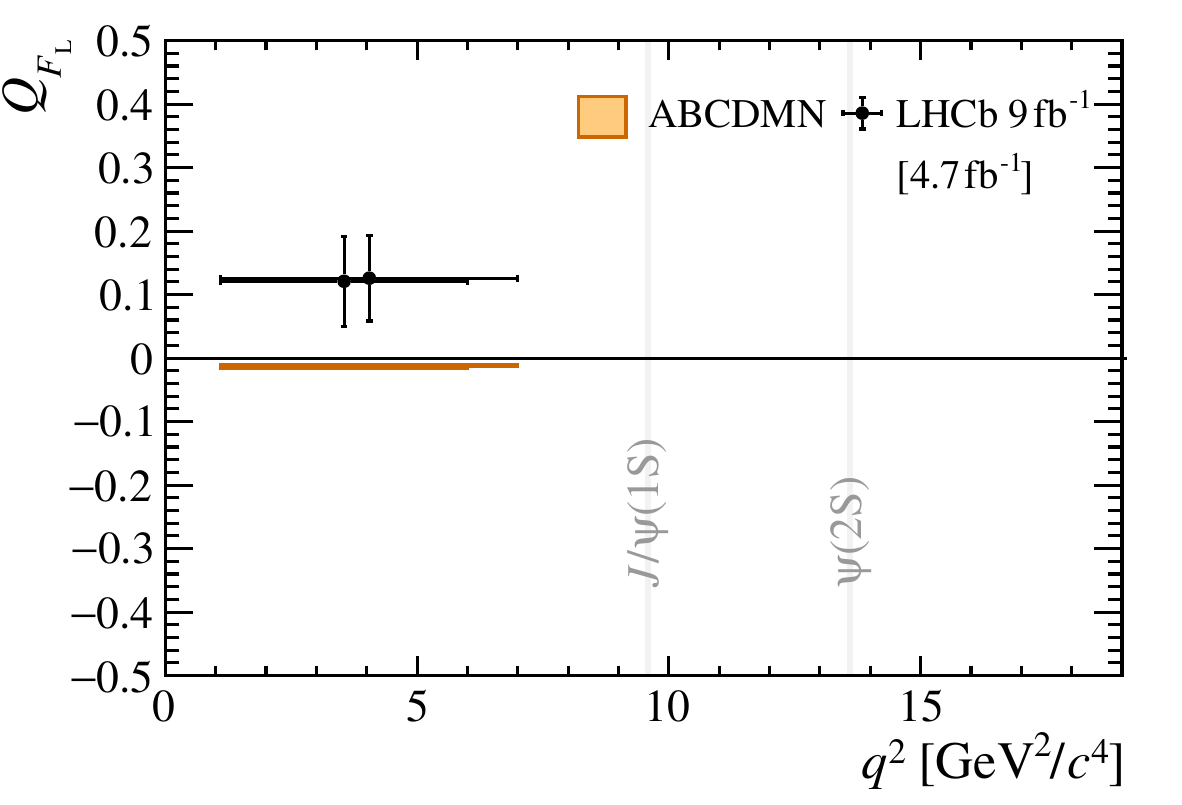} 
    \includegraphics[width=.45\textwidth, trim={0 0 0 0},clip]{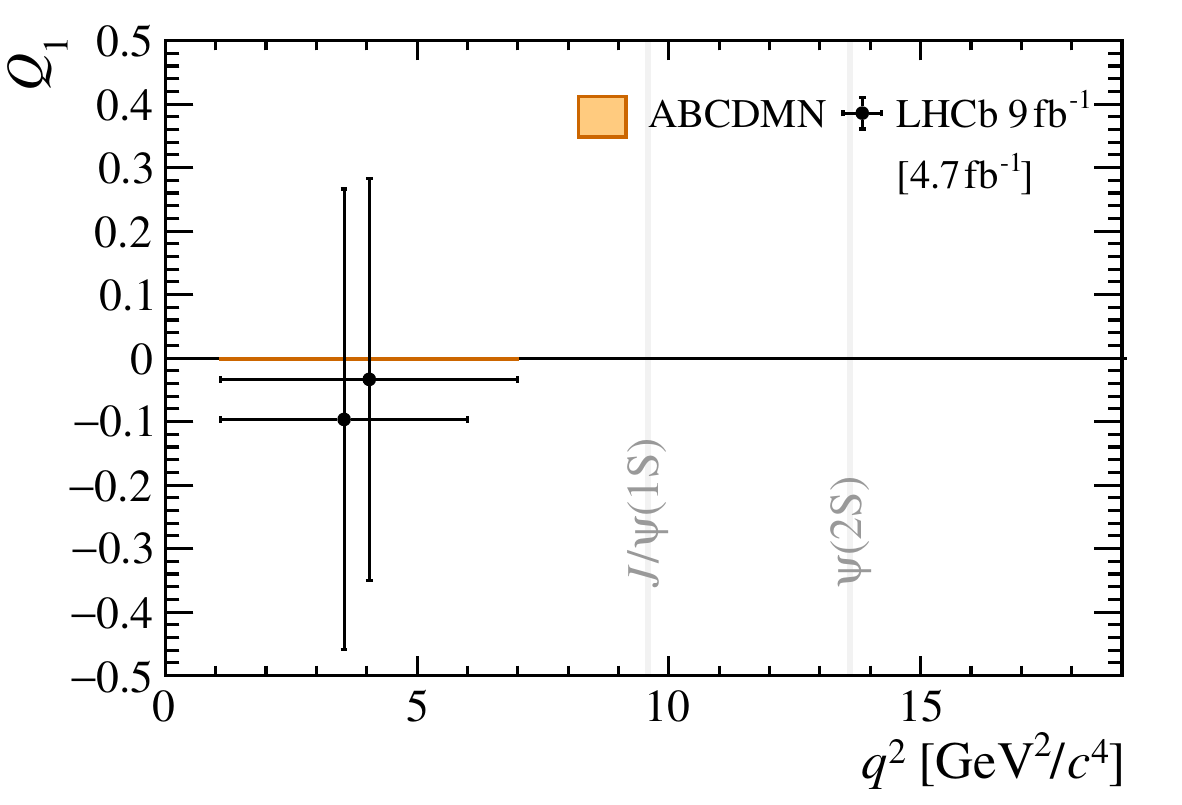} 
    \includegraphics[width=.45\textwidth, trim={0 0 0 0},clip]{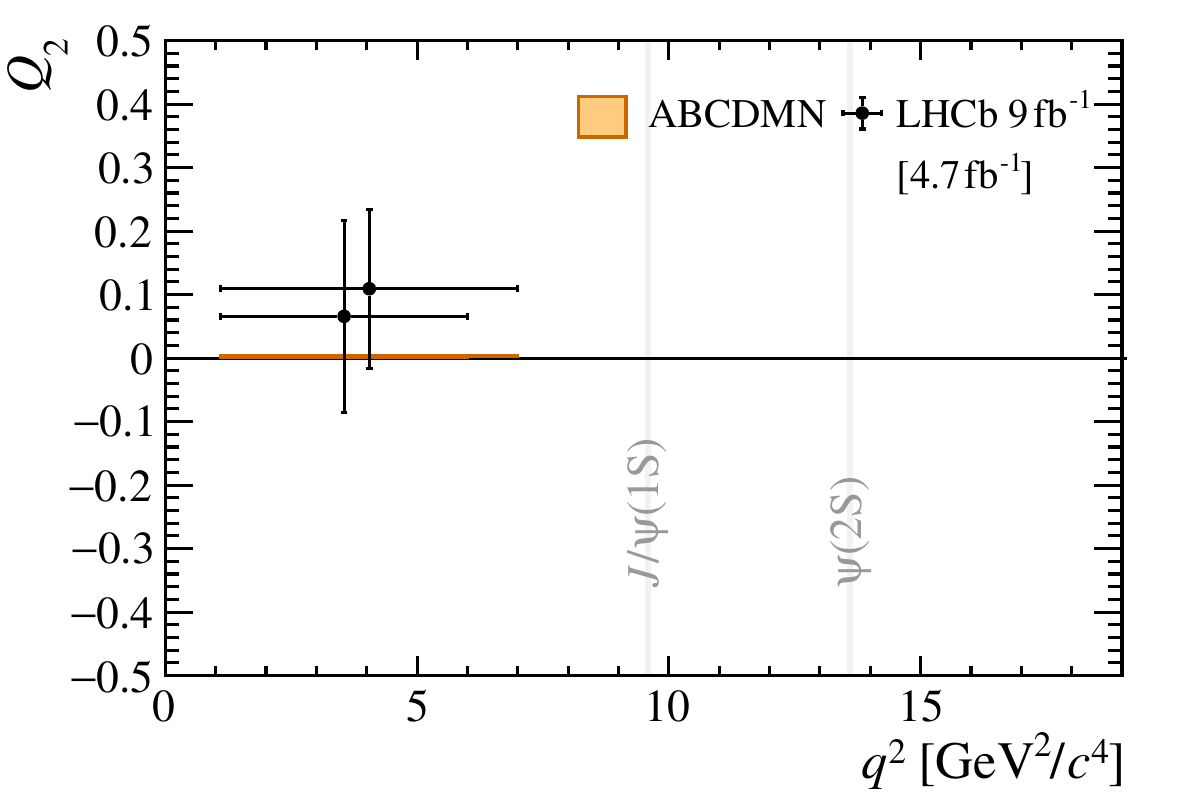} 
    \includegraphics[width=.45\textwidth, trim={0 0 0 0},clip]{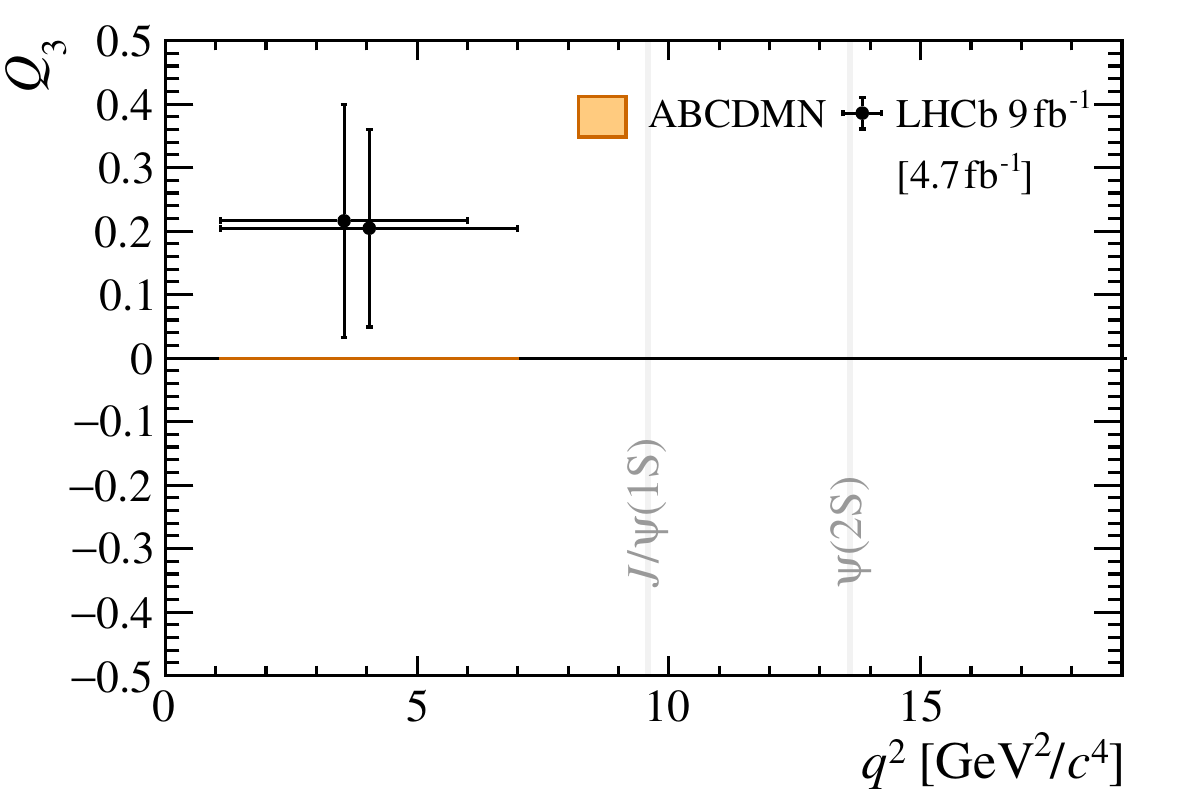} 
    \includegraphics[width=.45\textwidth, trim={0 0 0 0},clip]{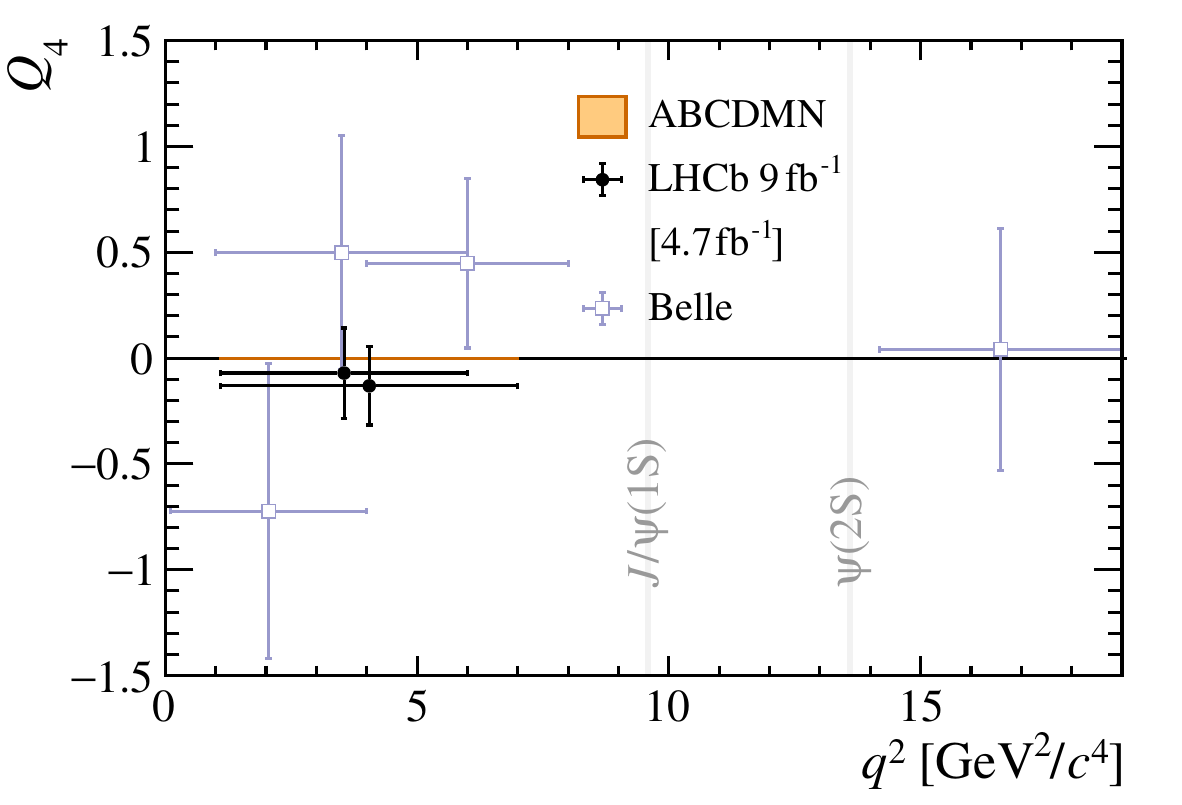} 
    \includegraphics[width=.45\textwidth, trim={0 0 0 0},clip]{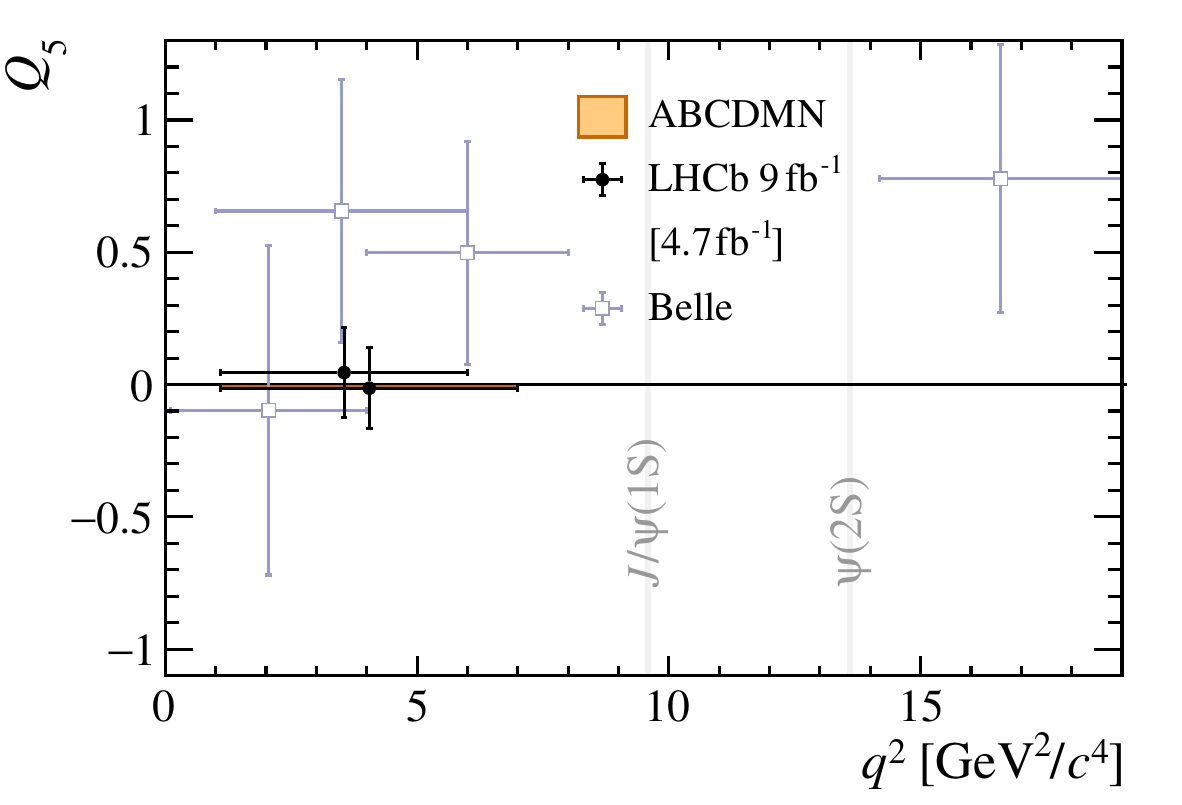} 
    \includegraphics[width=.45\textwidth, trim={0 0 0 0},clip]{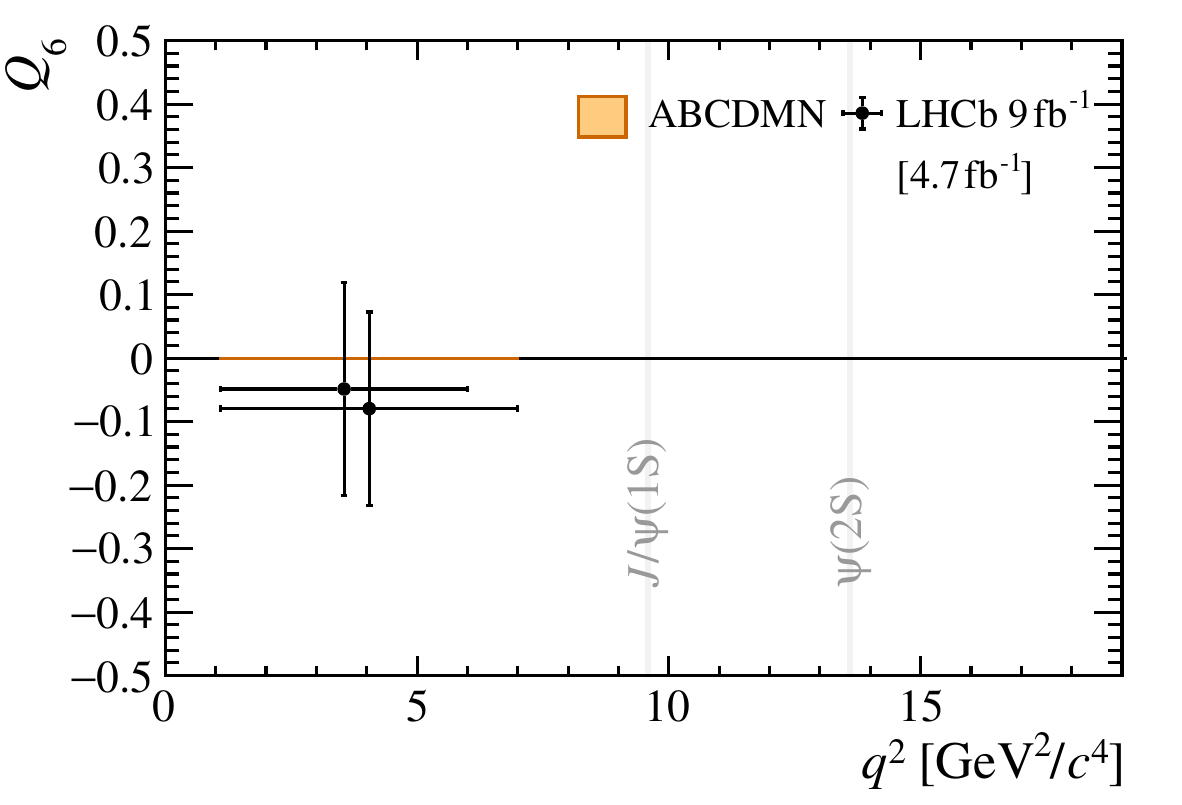} 
    \includegraphics[width=.45\textwidth, trim={0 0 0 0},clip]{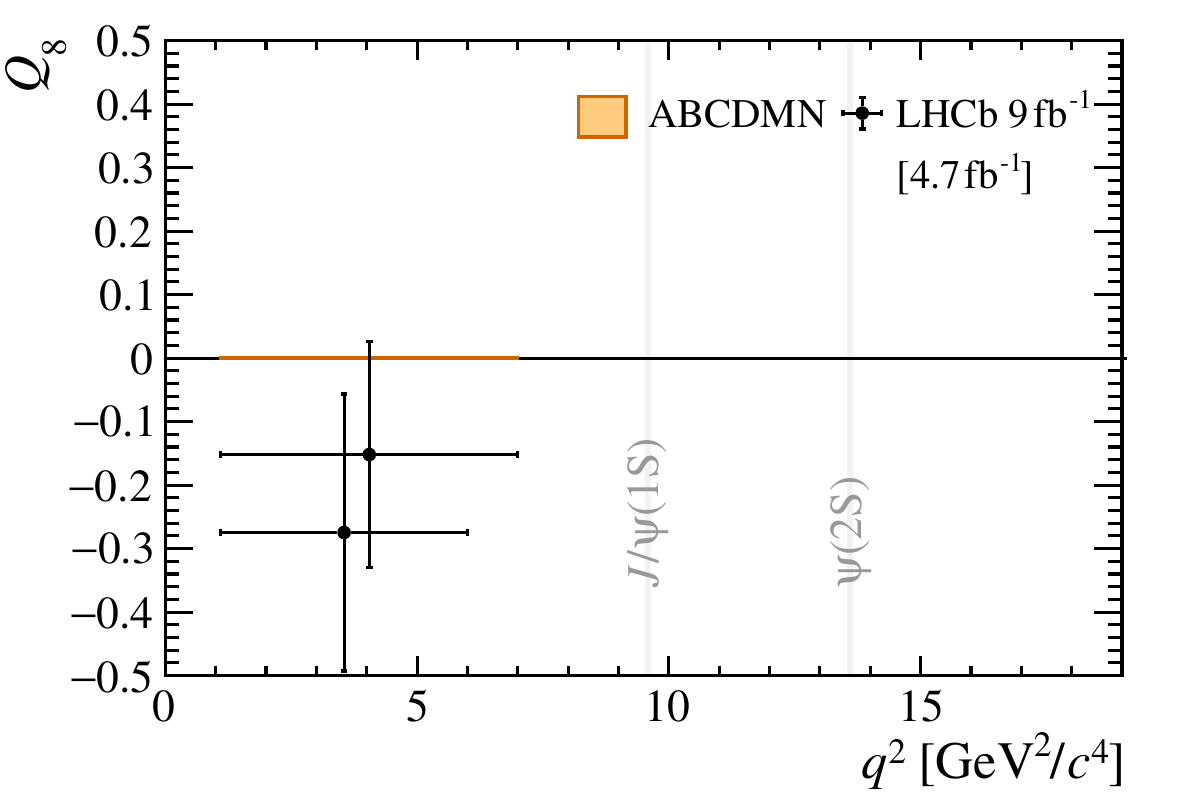} 
  \caption{Measured $Q_i$ LFU observables of the baseline and extended $q^2$ regions compared with the SM predictions based on Ref.~\cite{Alguero:2023jeh}.
  The values of $Q_4$ and $Q_5$ measured by Belle~\cite{Belle:2016fev} for the decays of $B^{+,0}\to K^{*+,*0} \ell^+\ell^-$, where $\ell=e,\,\mu$, are shown in light blue.}
\label{fig:Qi_plots_app}
\end{figure}

The results of the $S$- and $P$-basis angular observables measured in the baseline $q^2$ region of 1.1--6.0\gevgevcccc and the extended region of 1.1--7.0\gevgevcccc are summarised in Figs.~\ref{fig:Si_plots_app} and~\ref{fig:Pi_plots_app}, respectively, where the SM predictions from Refs.~\cite{Alguero:2023jeh, EOSAuthors:2021xpv,Gubernari:2022hxn} are also shown. The $Q_i$ LFU observables are shown in Fig.~\ref{fig:Qi_plots_app}. In this case, only the SM predictions from Ref.~\cite{Alguero:2023jeh} are shown, as Refs.~\cite{EOSAuthors:2021xpv,Gubernari:2022hxn} do not distinguish between the two lepton flavours and their predictions are therefore lepton-flavour universal by definition.

\clearpage

\section{Comparison between constrained and unconstrained $\boldsymbol{q^2}$ distributions}
\label{sec:q2_choice}
\begin{figure}[!b]
\centering
    \includegraphics[width=.7\textwidth, trim={0 0 0 0},clip]{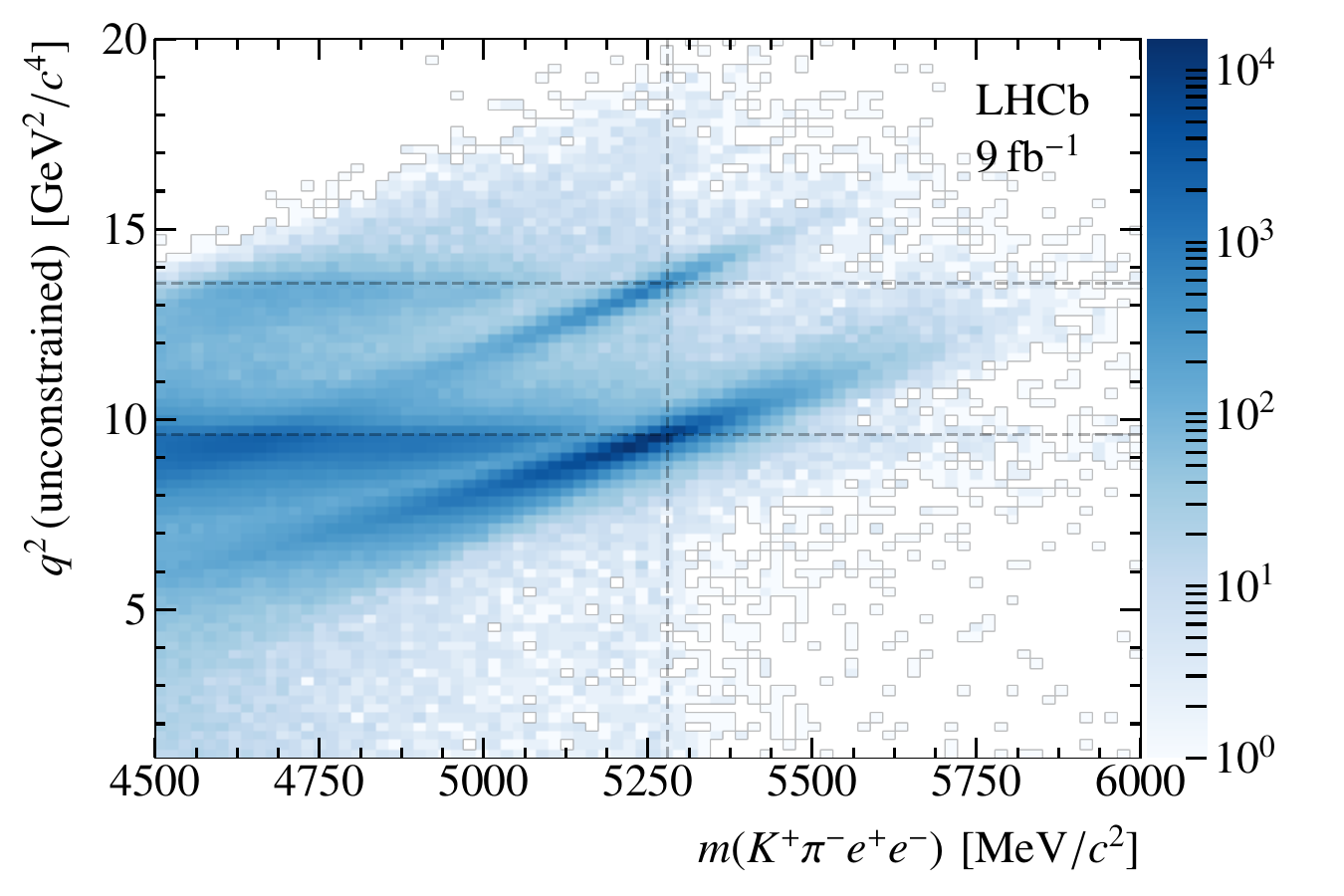} 
  \caption{
  Distribution of the unconstrained $q^2$ and \Bz invariant mass of signal candidates in data.
 Signal decays lie in a vertical band close to the known $\Bz$ mass (vertical dashed line)~\cite{PDG2024}. 
The decays $\Bz\to\Kstarz J/\psi(\to e^+e^-)$ and $\Bz\to\Kstarz\psi(2S)(\to e^+e^-)$ have an invariant mass close to that of the $\Bz$ meson, and $q^2$ values close to the square of the known $J/\psi$ and $\psi(2S)$ masses, respectively (horizontal dashed lines). 
They are visible as diagonal bands. 
The two horizontal bands contain combinatorial background comprised of genuine $J/\psi$ or $\psi(2S)$ mesons combined with random kaon and pion tracks.}
\label{fig:q2_vs_mass}
\end{figure}

\begin{figure}[!tb]
\centering
    \includegraphics[width=.45\textwidth, trim={0 0 0 0},clip]{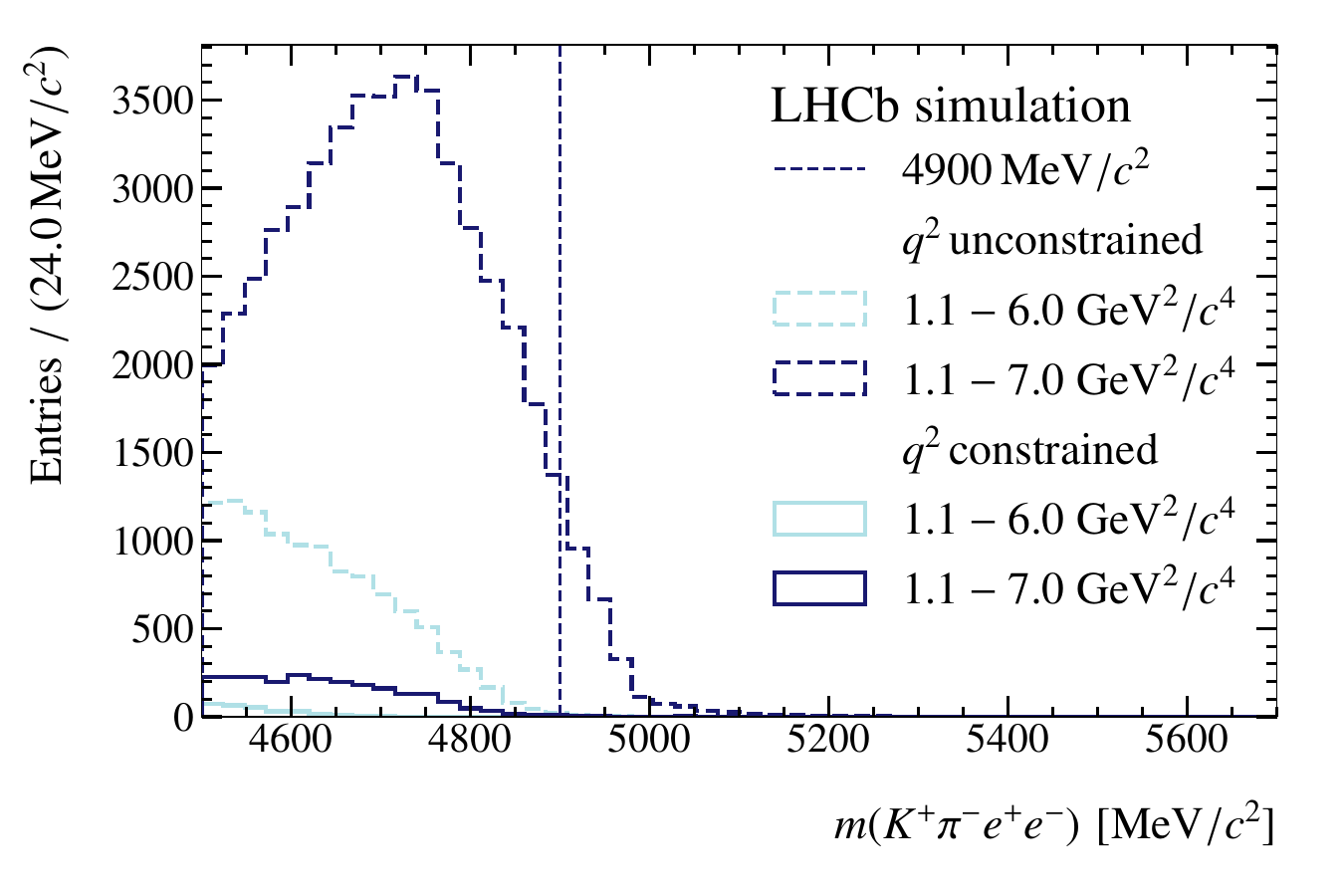} 
    \includegraphics[width=.45\textwidth, trim={0 0 0 0},clip]{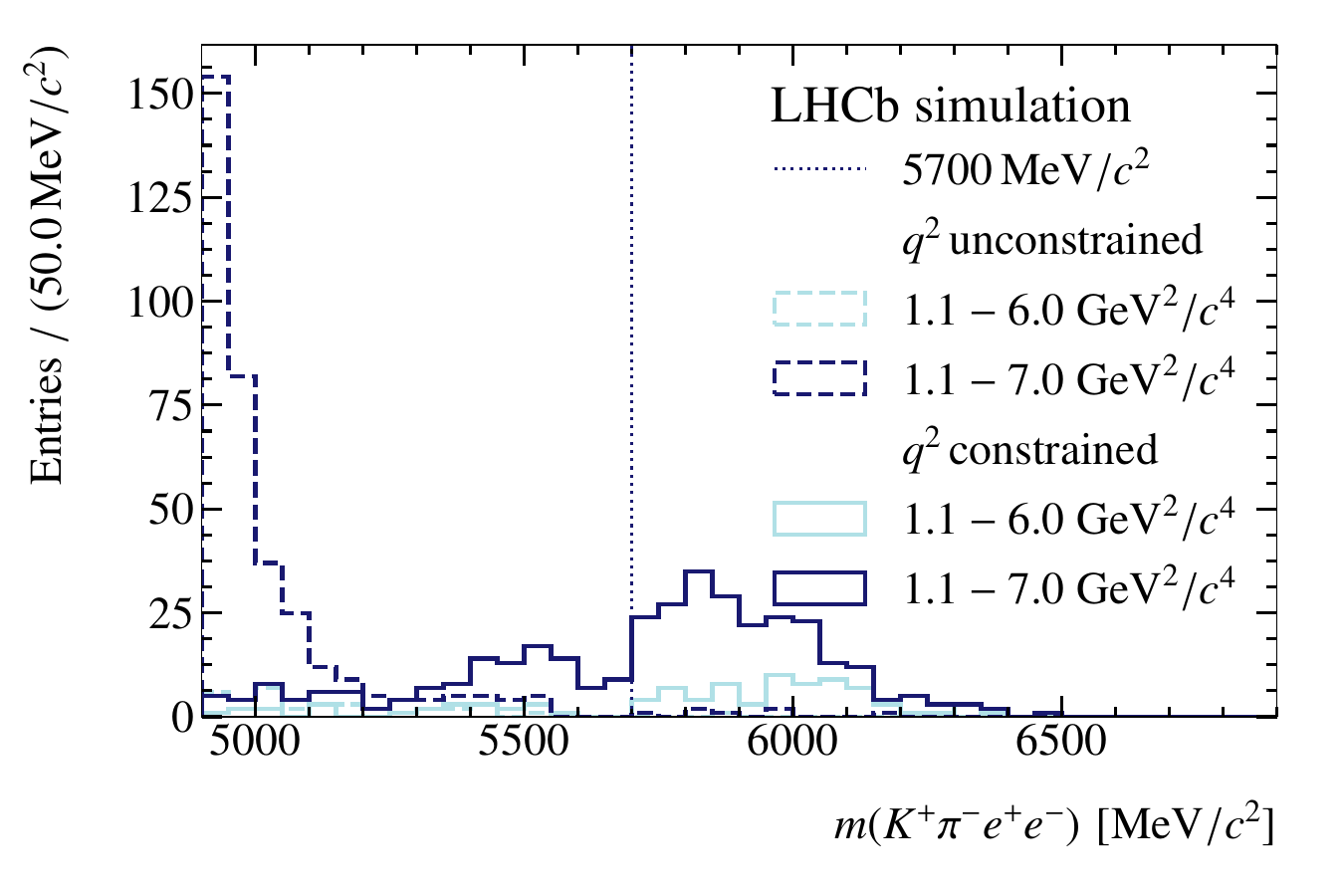} 
  \caption{Distributions of the \Bd invariant mass in the constrained and unconstrained $q^2$ regions for (left) simulated control-mode candidates and (right) simulated control-mode candidates where $\jpsi$ mesons are combined with random hadron tracks. The upper and lower mass limits of the signal region are shown by the dashed and dotted vertical lines, respectively.}
\label{fig:jpsi_leak_comb}
\end{figure}

The distribution of the $\Bz$ invariant mass versus the $q^2$ variable calculated without constraints for signal candidates that satisfy all selection requirements is shown in Fig.~\ref{fig:q2_vs_mass}. 
This figure can be compared to Fig.~\ref{fig:q2c_vs_mass}, which shows that the constraints applied in the calculation of $q^2$, namely that the \Bz candidate is required to originate from its associated PV and to have an invariant mass consistent with the known mass of the \Bz meson, shift the radiative tails of the charmonium decays away from the signal region. Therefore using the constrained $q^2$ variable reduces the leakage of control-mode candidates into the signal region. This is illustrated in Fig.~\ref{fig:jpsi_leak_comb} (left) using simulated control-mode candidates. However, the constraints introduce a source of background above the known $\Bz$ mass~\cite{PDG2024}, which is composed of $J/\psi$ mesons combined with random hadron tracks. The distribution of this background is illustrated 
in Fig.~\ref{fig:jpsi_leak_comb} (right).

\clearpage

\section{Effective acceptance functions}
\label{sec:effective_acceptance_functions}
\begin{figure}[!b]
\centering
    \includegraphics[width=.45\textwidth, trim={0 0 0 0},clip]{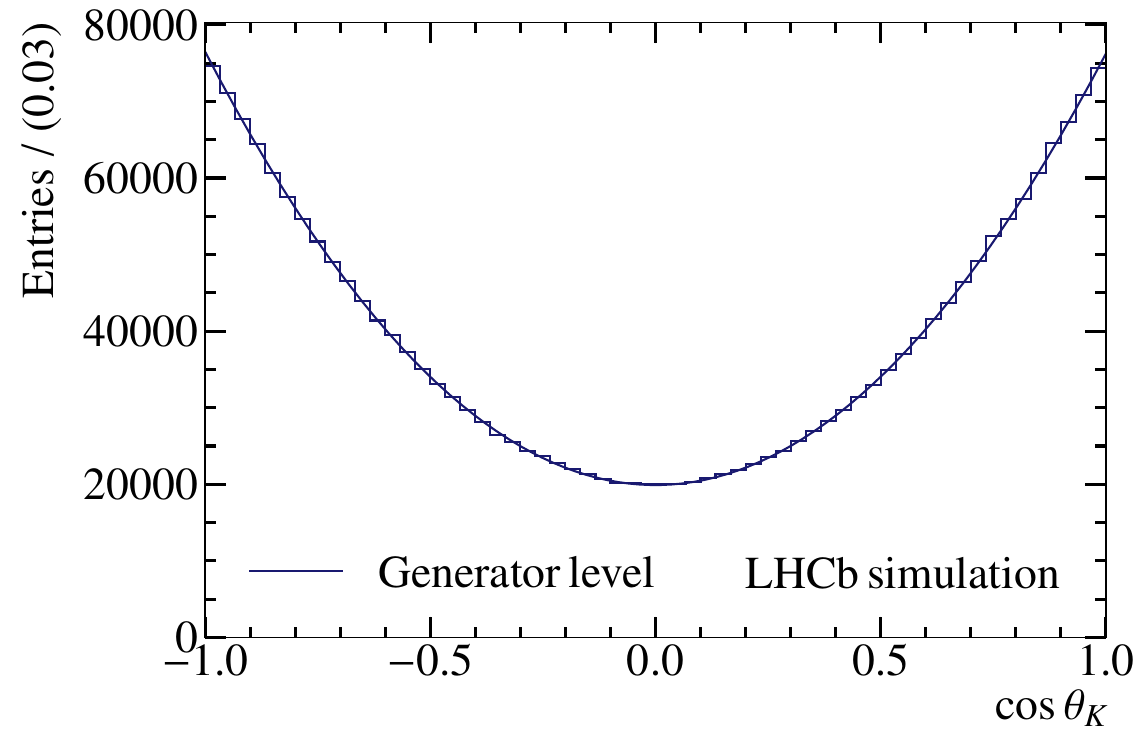} 
    \includegraphics[width=.45\textwidth, trim={0 0 0 0},clip]{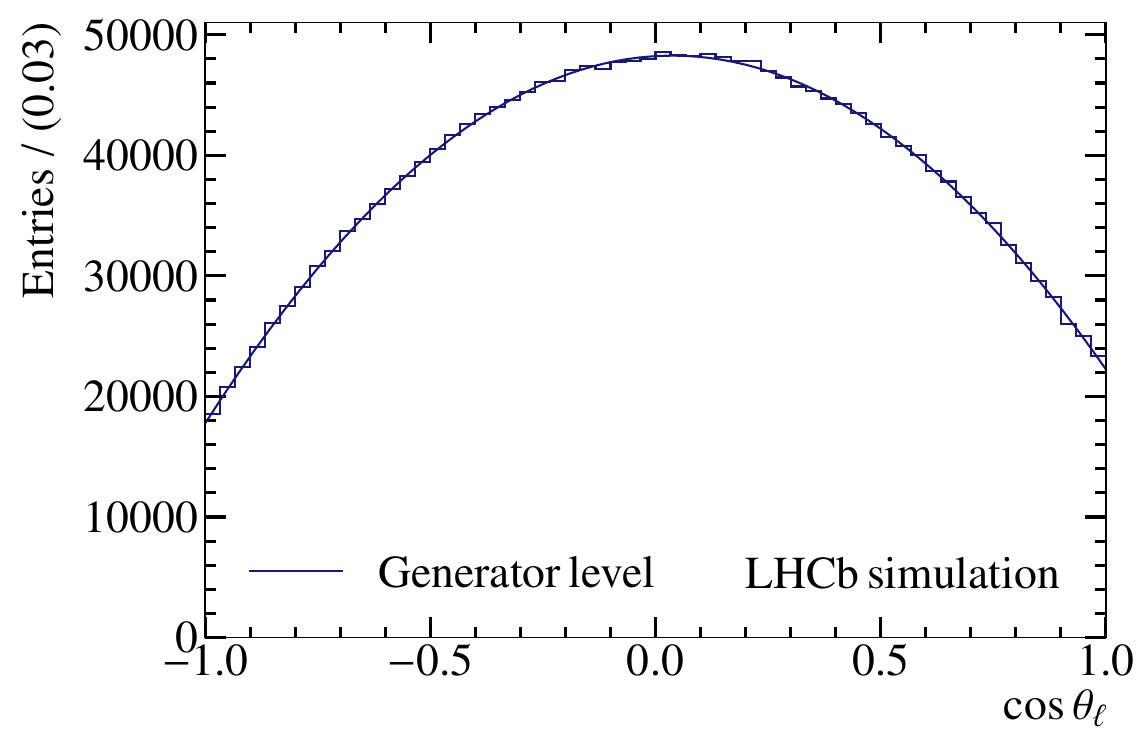} 
    \includegraphics[width=.45\textwidth, trim={0 0 0 0},clip]{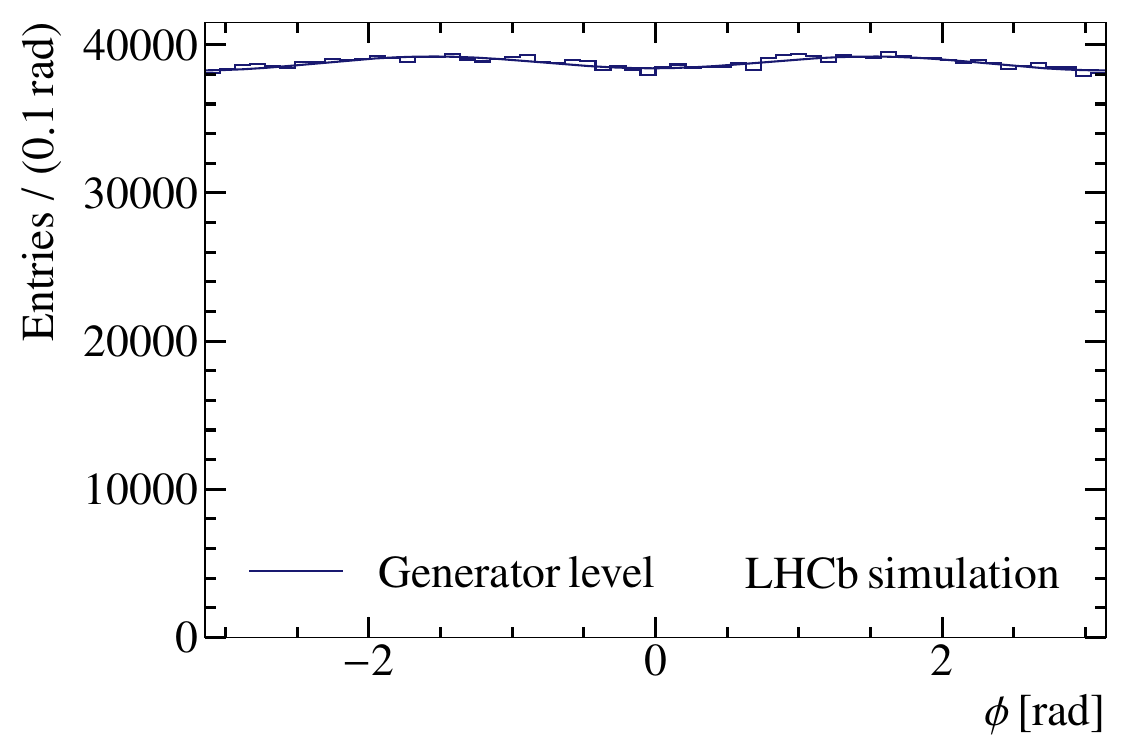} 
    \includegraphics[width=.45\textwidth, trim={0 0 0 0},clip]{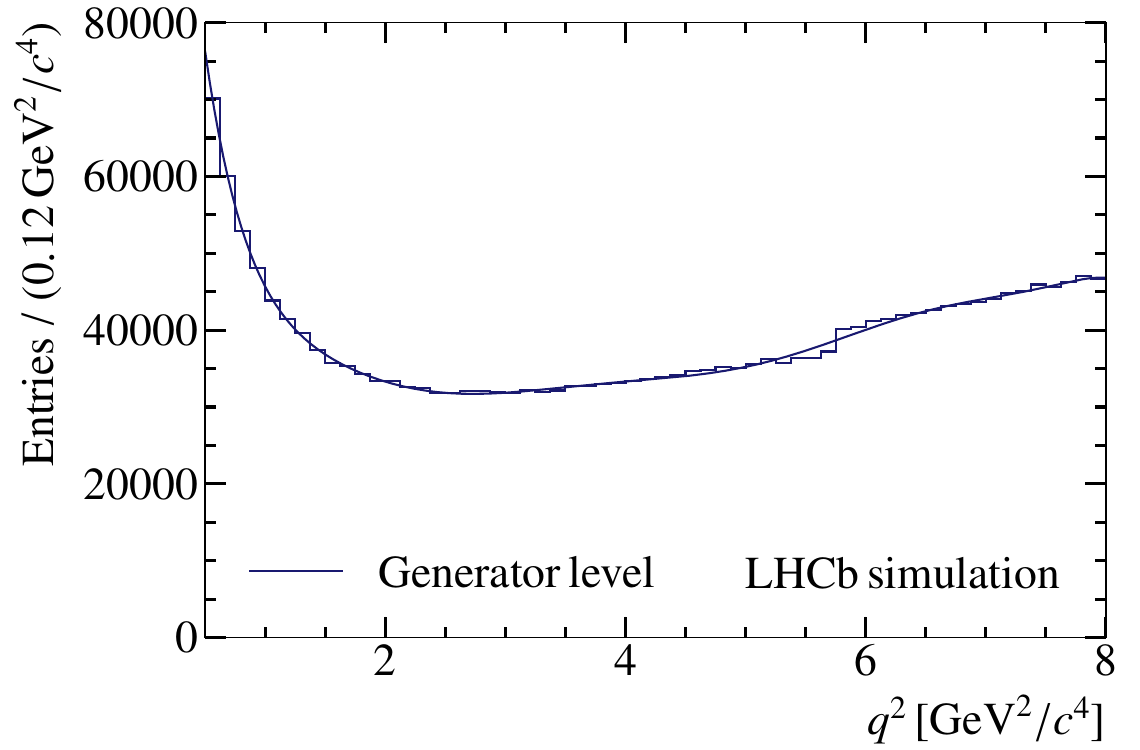} 
  \caption{Distribution of the generator-level sample and the projections of the four-dimensional function used to describe it. This function is used to produce inverse weights for the parametrisation of the effective acceptance functions.}
\label{fig:gen_lvl_projections}
\end{figure}

\begin{figure}[!tb]
\centering
    \includegraphics[width=.45\textwidth, trim={0 0 0 0},clip]{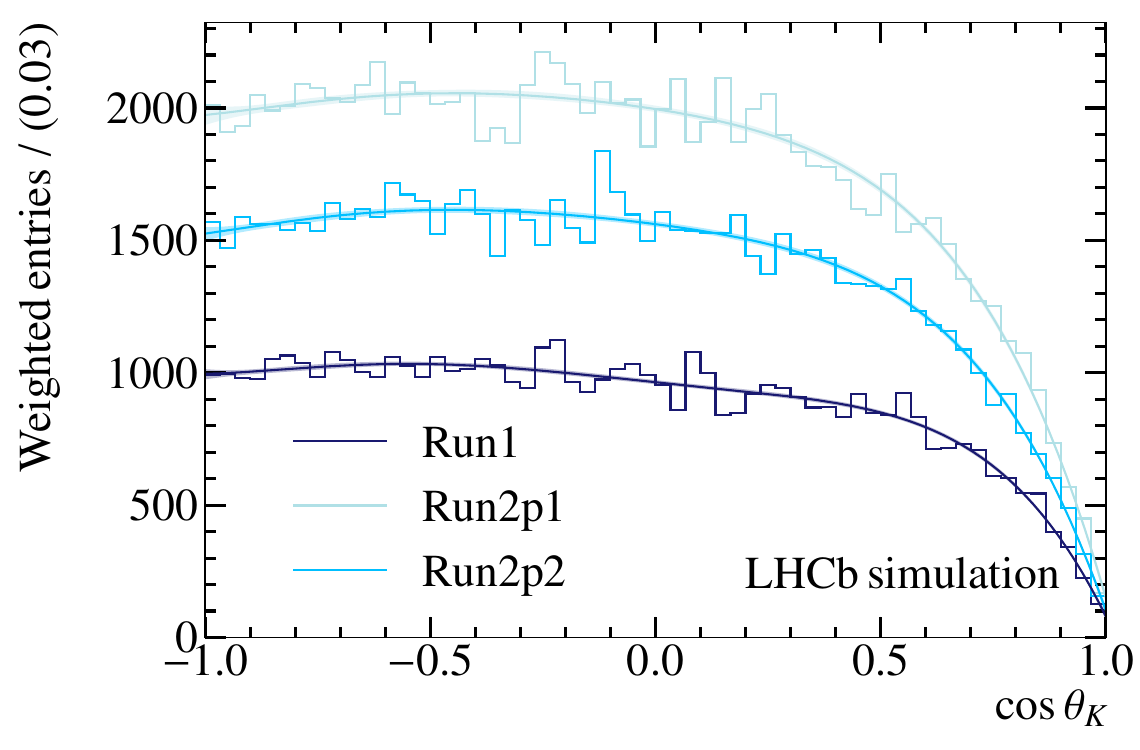} 
    \includegraphics[width=.45\textwidth, trim={0 0 0 0},clip]{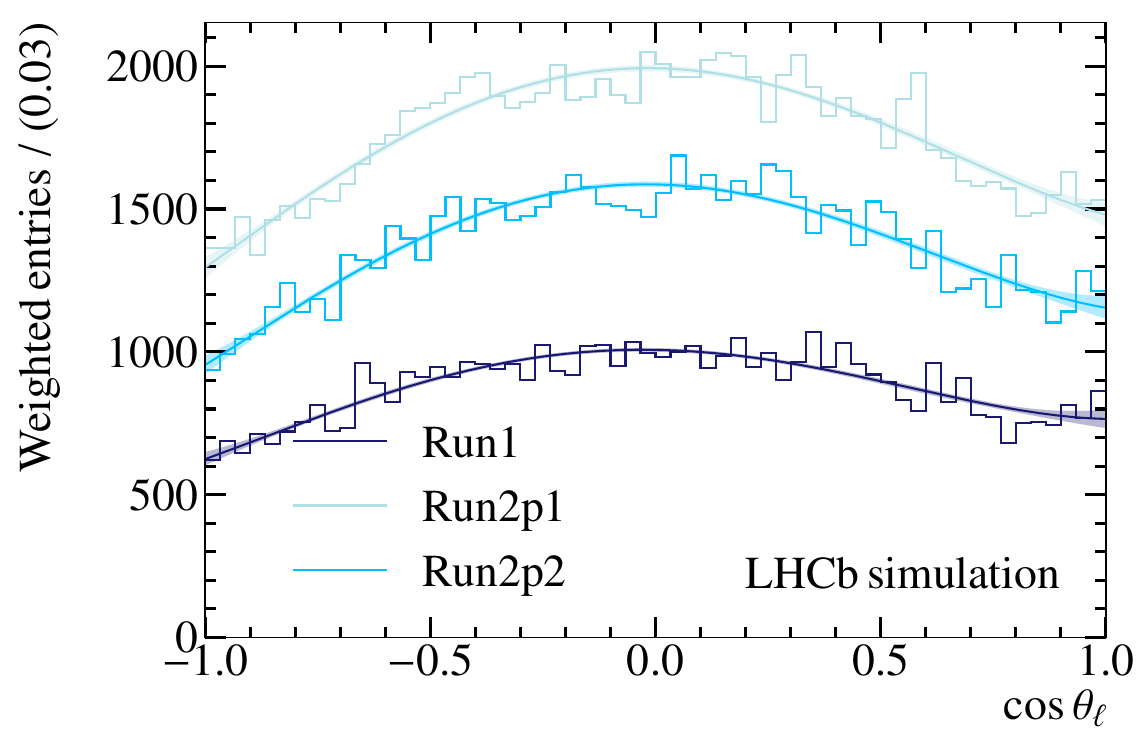} 
    \includegraphics[width=.45\textwidth, trim={0 0 0 0},clip]{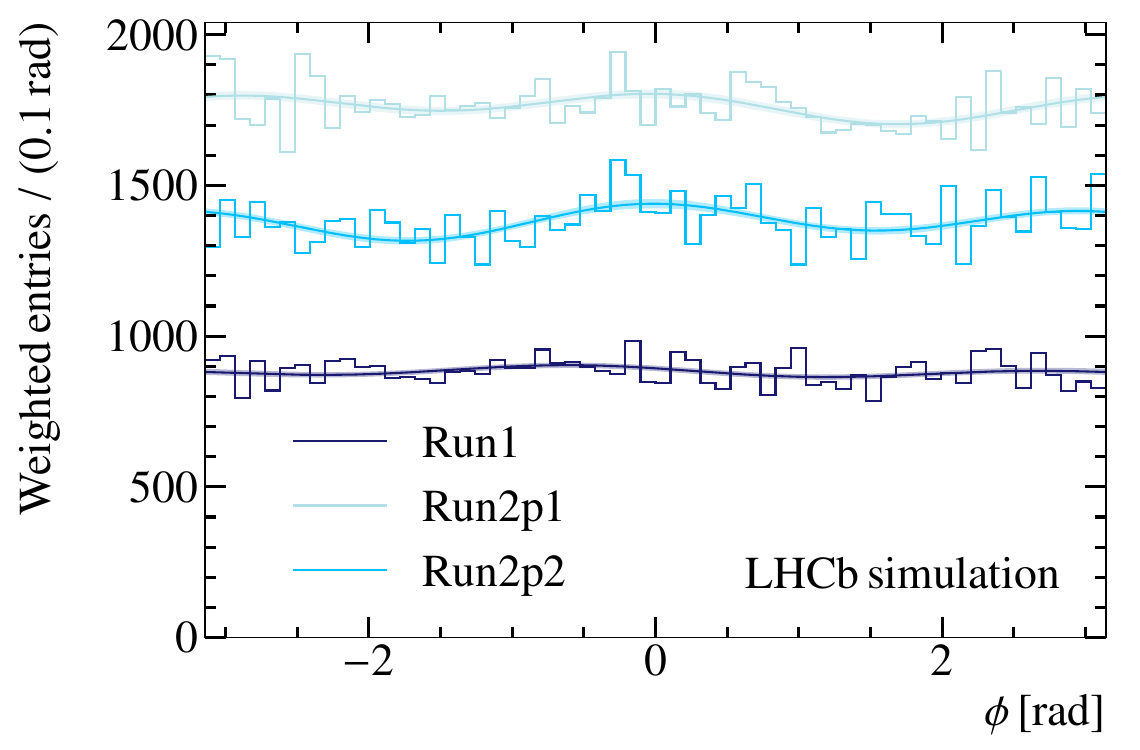} 
    \includegraphics[width=.45\textwidth, trim={0 0 0 0},clip]{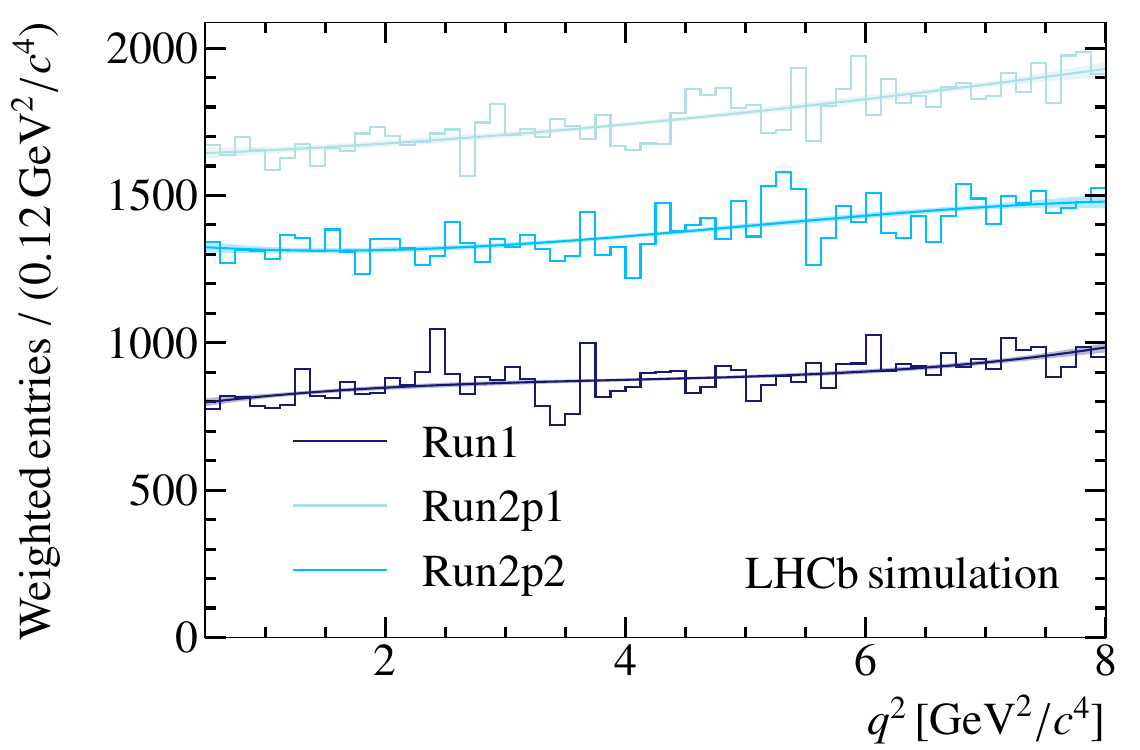} 
  \caption{Distributions of simulated signal decays for Run1, Run2p1 and Run2p2 periods weighted by the inverse of the output of the function that describes the generator-level sample, and projections of the 
  four-dimensional effective acceptance functions. The bands represent $1\sigma$ uncertainty and are obtained by bootstrapping the samples used for parametrisation.
  The normalisation of the distributions is arbitrary and is related to the size of the simulation samples.
  }
\label{fig:acceptance_projections}
\end{figure}

\begin{figure}[!tb]
\centering
  \begin{tabular}{@{}cccc@{}}
    \includegraphics[width=.48\textwidth]{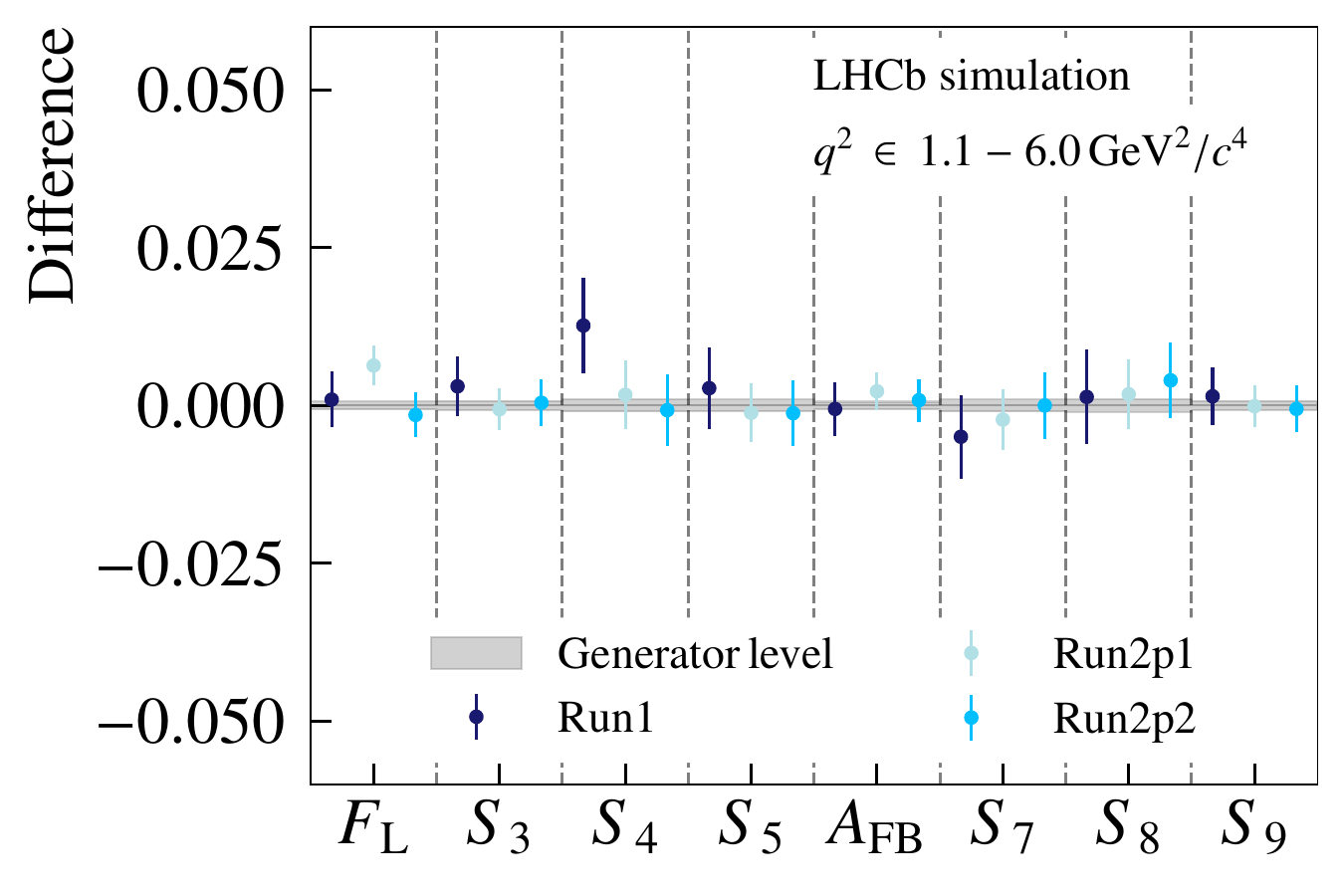} &
    \includegraphics[width=.48\textwidth]{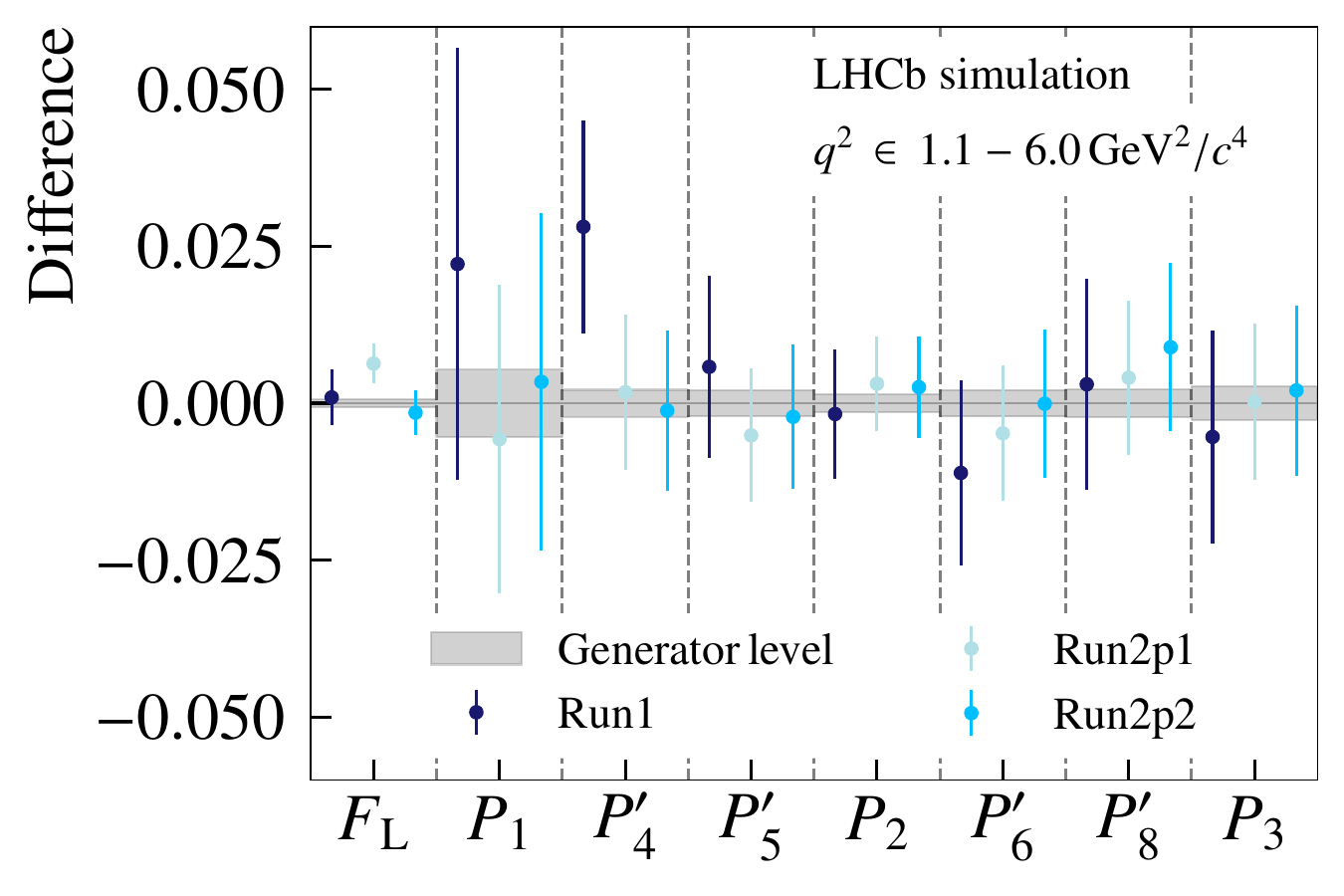} & \\
   \end{tabular}
  \caption{Differences between the observable values found from fits to the simulated signal decays with effective acceptance weights, and the fit to the generator-level sample (centred at zero).}
\label{fig:rare_mode_acceptance_validation_display}
\end{figure}

The projections of the four-dimensional function that parametrises the generator-level sample are shown in Fig.~\ref{fig:gen_lvl_projections}.
The projections of the effective acceptance functions for the three run periods are shown in Fig.~\ref{fig:acceptance_projections}.
The results of the effective acceptance parametrisation validation are shown in Fig.~\ref{fig:rare_mode_acceptance_validation_display}.

\clearpage

\section{Model components}
\label{sec:component_models}
\begin{figure}[!b]
\centering
    \includegraphics[width=.45\textwidth, trim={0 0 0 0},clip]{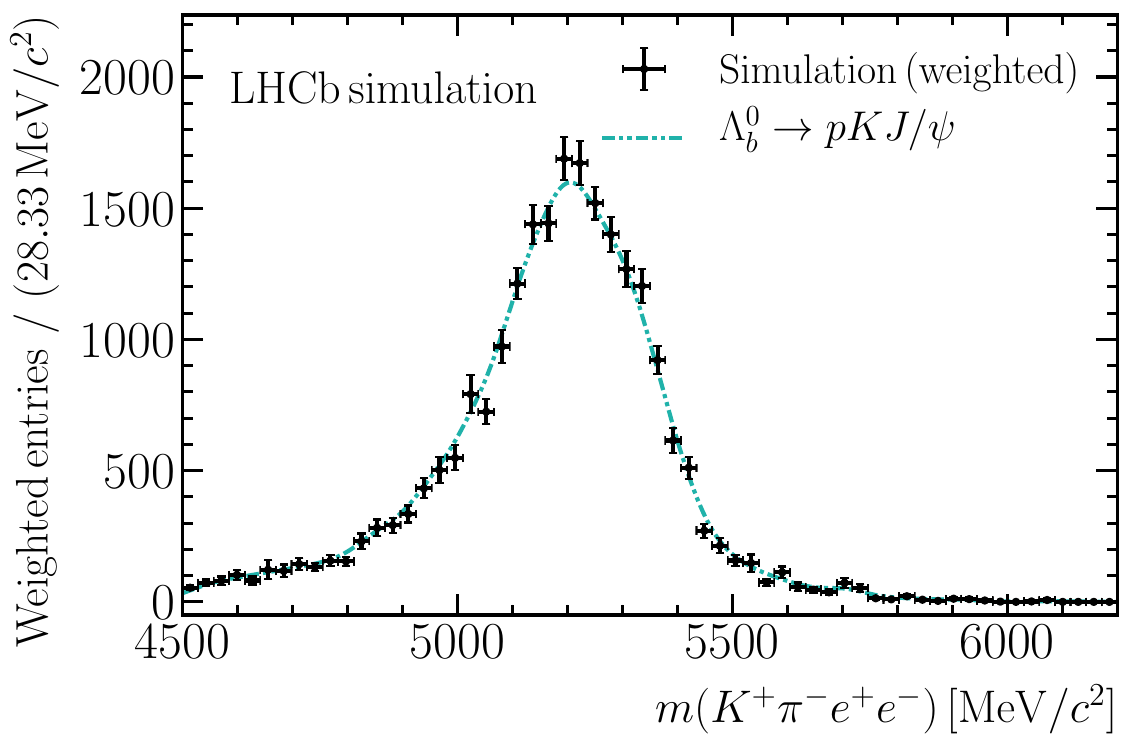} 
 \caption{Distribution of simulated phase-space \mbox{$\Lambda_b^0\rightarrow p\Km J/\psi(\rightarrow e^+e^-)$} decays reconstructed as $B^0\rightarrow \Kstarz J/\psi(\rightarrow e^+e^-)$ and the model used to describe this component in the control-mode fit. Data-driven correction weights are included to reproduce the resonant structures in the $p\Km$ spectrum.
 }
\label{fig:data_fit_control_Lb_mass}
\end{figure}

\begin{figure}[!b]
\centering
    \includegraphics[width=.45\textwidth, trim={0 0 0 0},clip]{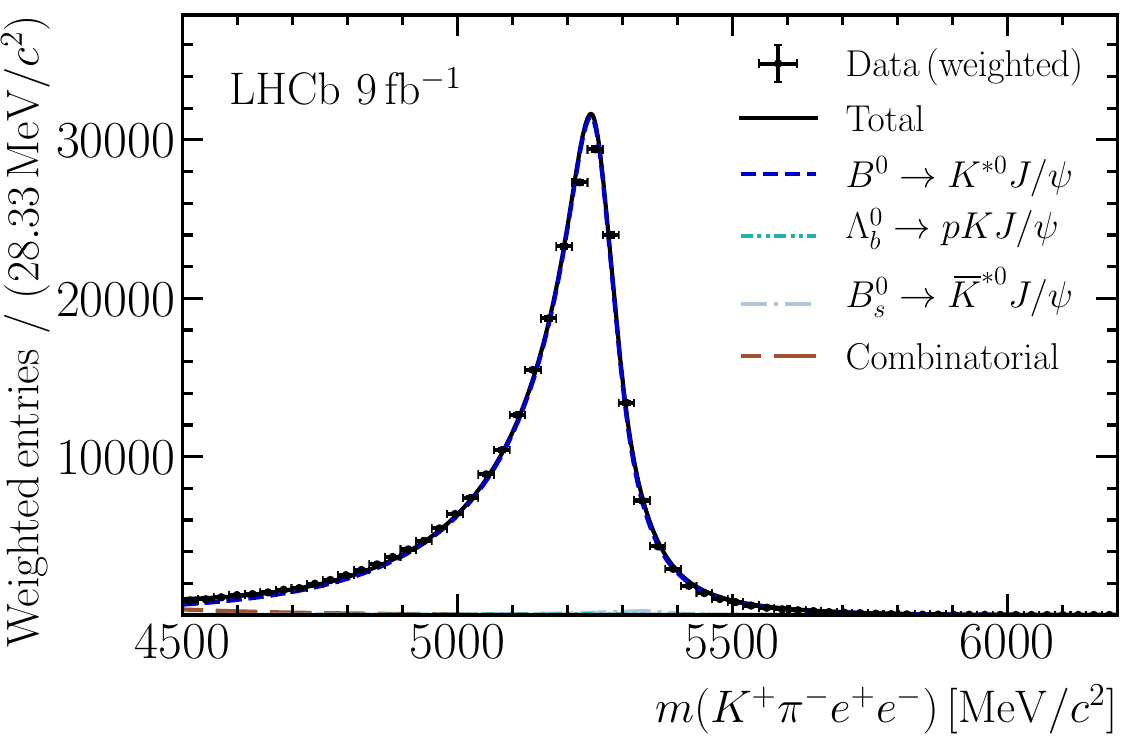} 
 \caption{Distribution of the $\Bz$ invariant mass of the control-mode candidates and the result of the fits.
 }
\label{fig:data_fit_control_mass}
\end{figure}

\begin{figure}[!tb]
\centering
    \includegraphics[width=.45\textwidth, trim={0 0 0 0},clip]{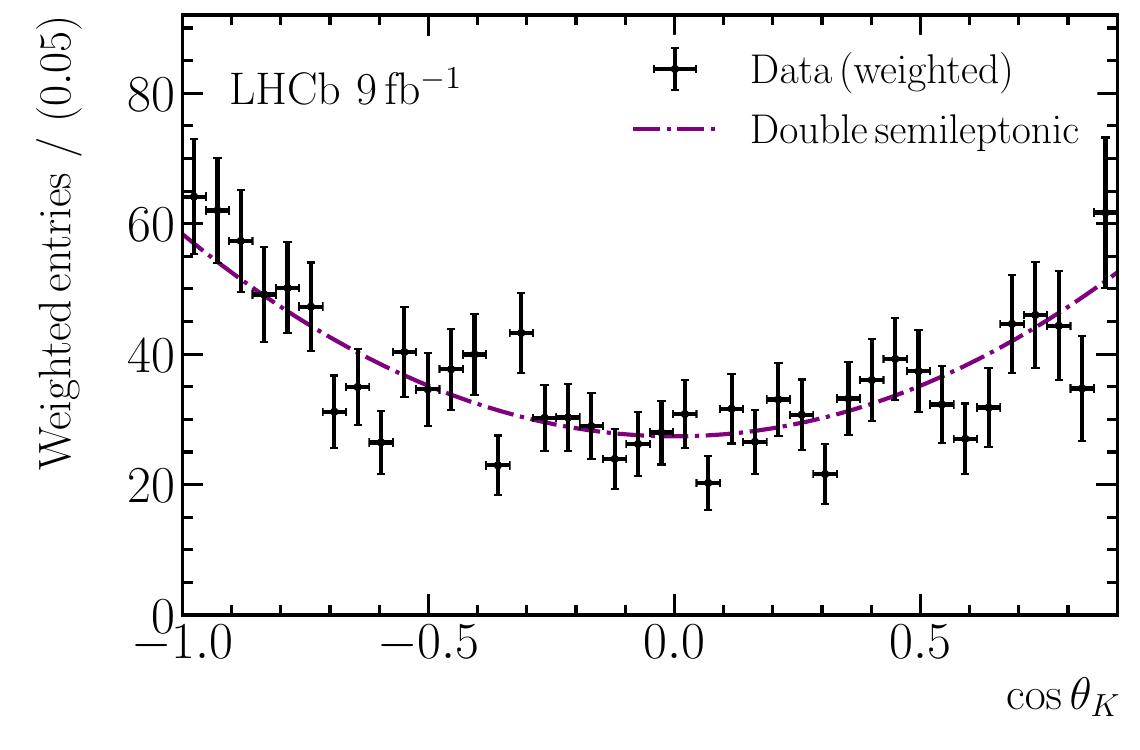} 
    \includegraphics[width=.45\textwidth, trim={0 0 0 0},clip]{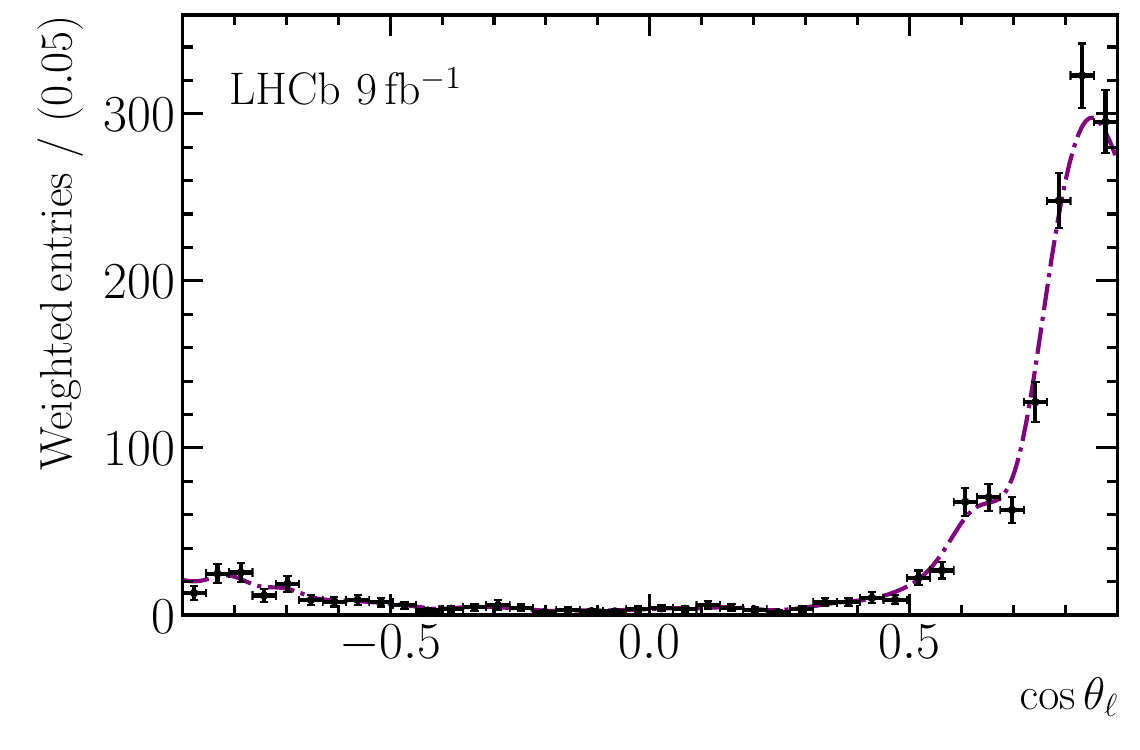} 
    \includegraphics[width=.45\textwidth, trim={0 0 0 0},clip]{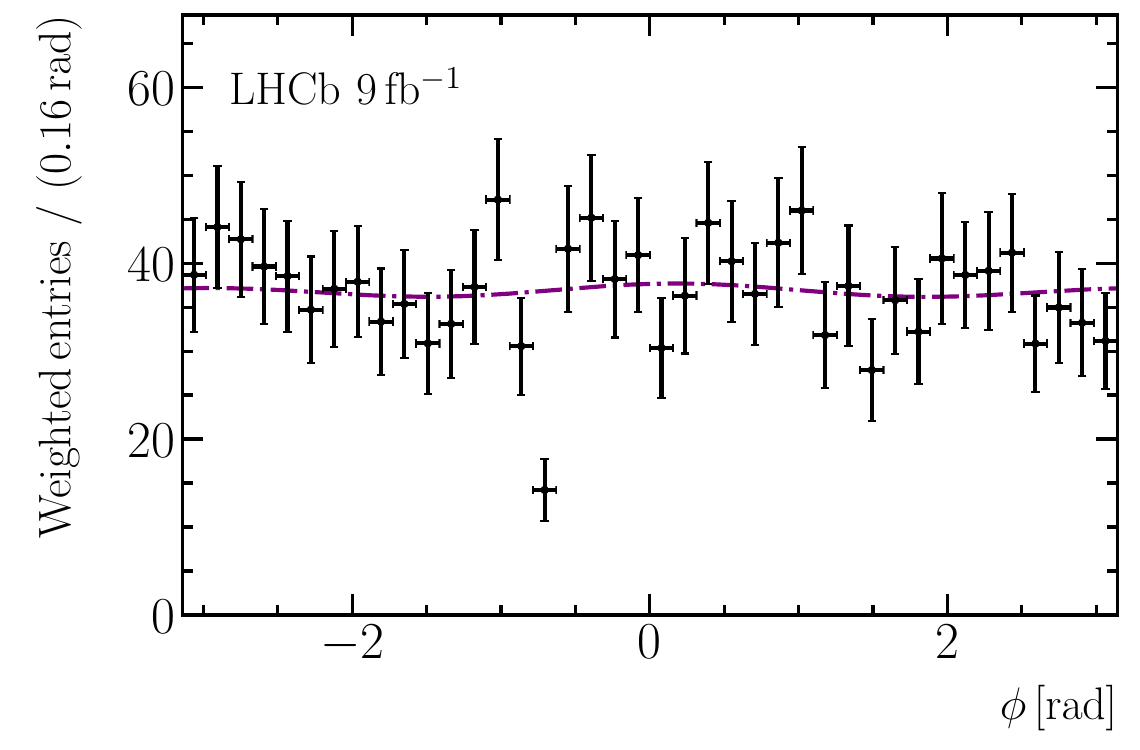} 
  \caption{Distribution of $\Kp\pim e^{+}\mu^{-}$ candidates selected by a stringent BDT requirement and the model used to describe the angular distribution of the DSL background.}
\label{fig:DSL_step1_sq2}
\end{figure}

\begin{figure}[!tb]
\centering
    \includegraphics[width=.45\textwidth, trim={0 0 0 0},clip]{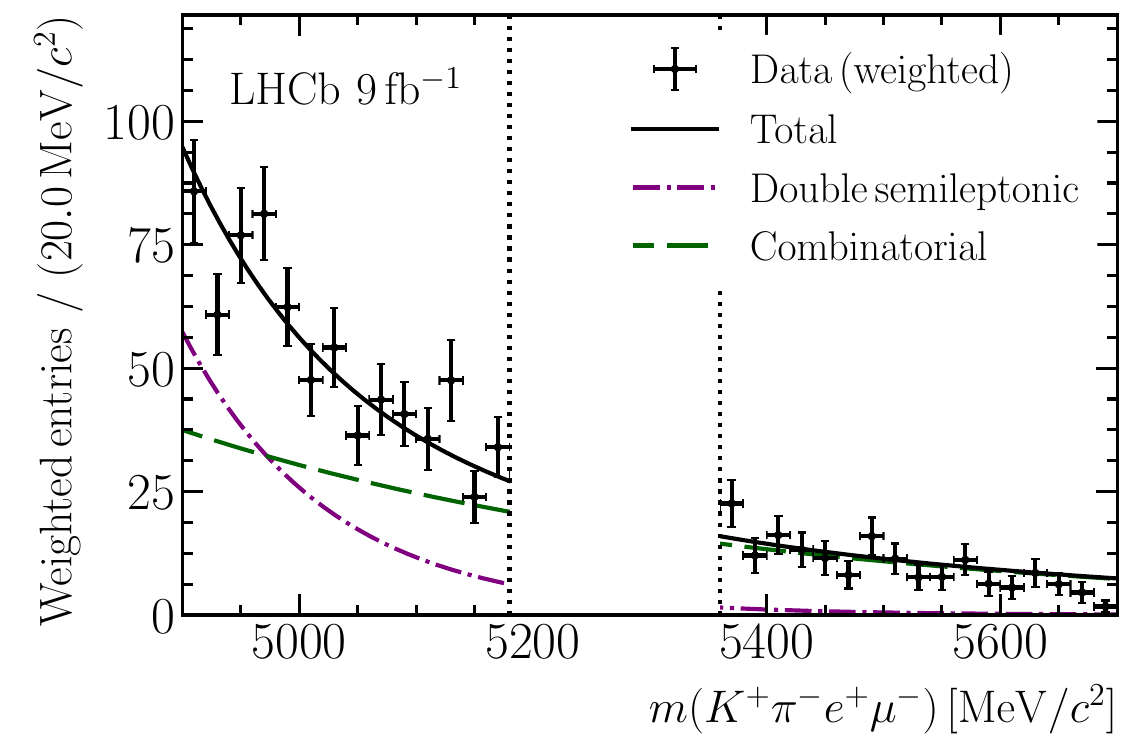} 
    \includegraphics[width=.45\textwidth, trim={0 0 0 0},clip]{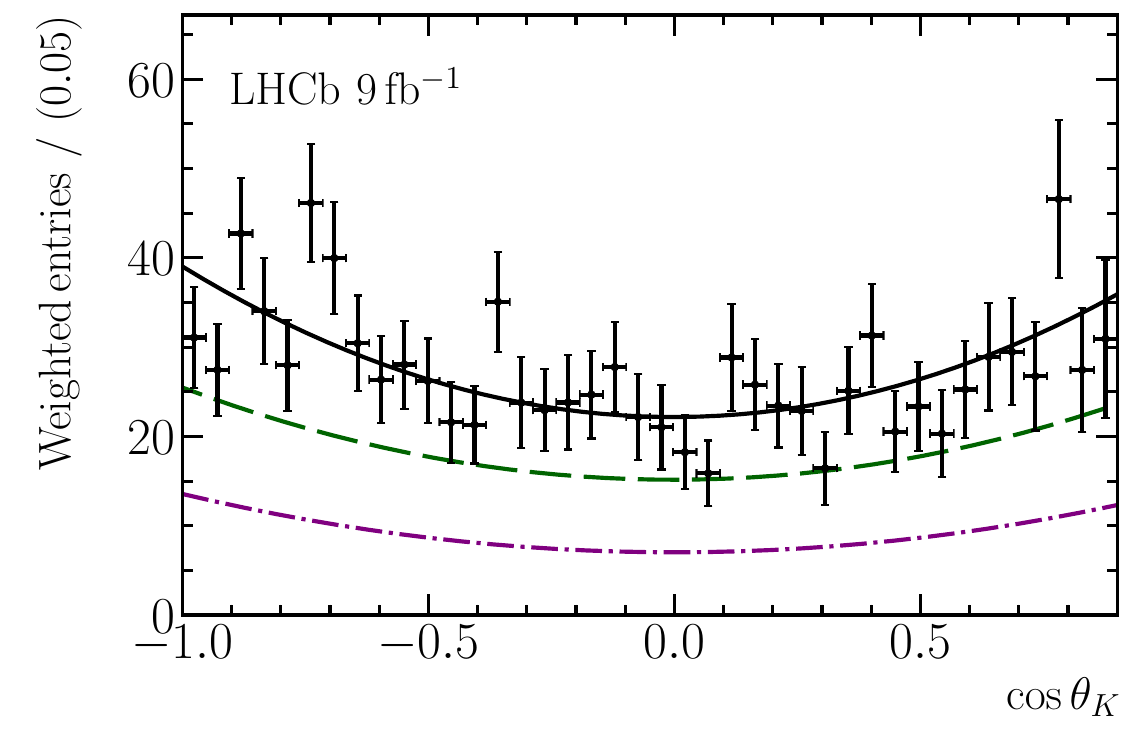} 
    \includegraphics[width=.45\textwidth, trim={0 0 0 0},clip]{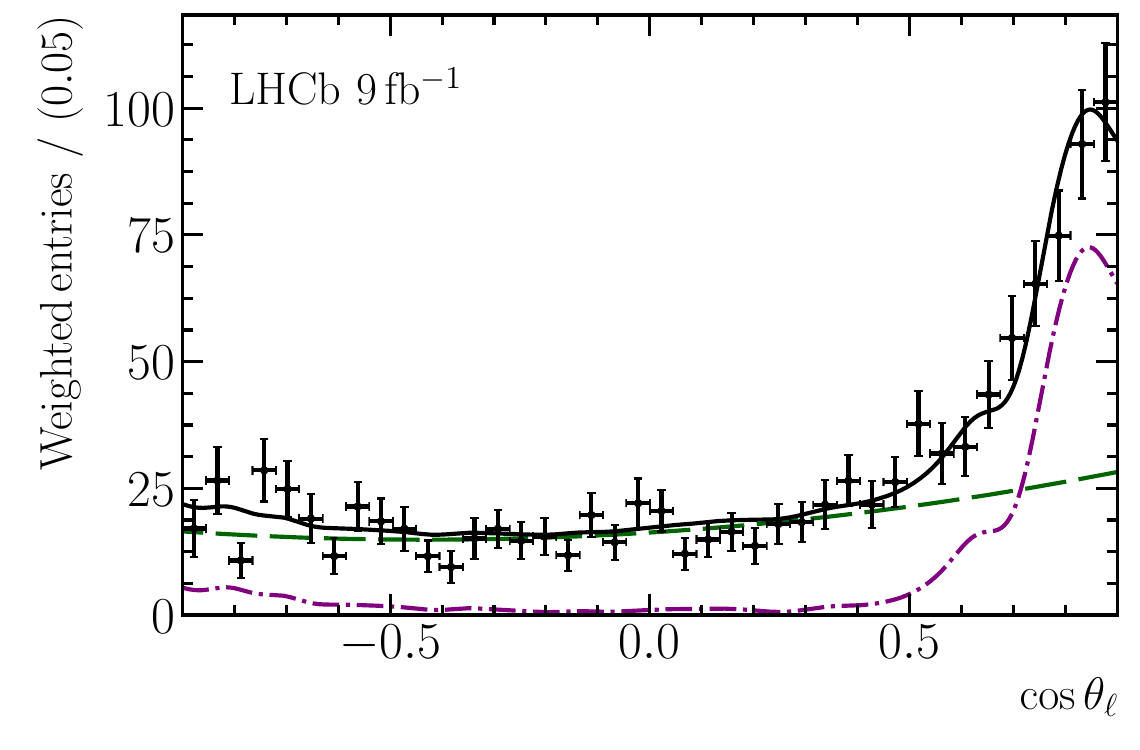} 
    \includegraphics[width=.45\textwidth, trim={0 0 0 0},clip]{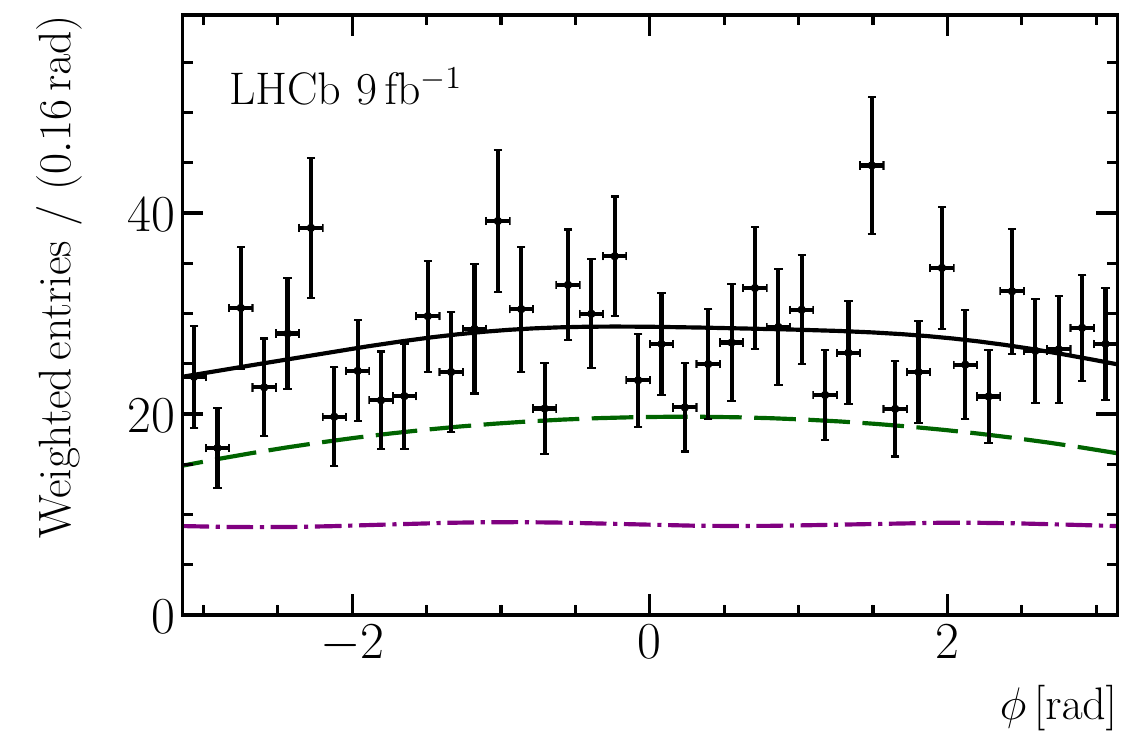} 
  \caption{Distribution of $\Kp\pim e^{+}\mu^{-}$ candidates selected by the baseline BDT requirement and the models used to describe the invariant-mass and angular distributions of the DSL and combinatorial backgrounds.
  }
\label{fig:DSL_step2_sq2}
\end{figure}

\begin{figure}[!tb]
\centering
    \includegraphics[width=.45\textwidth, trim={0 0 0 0},clip]{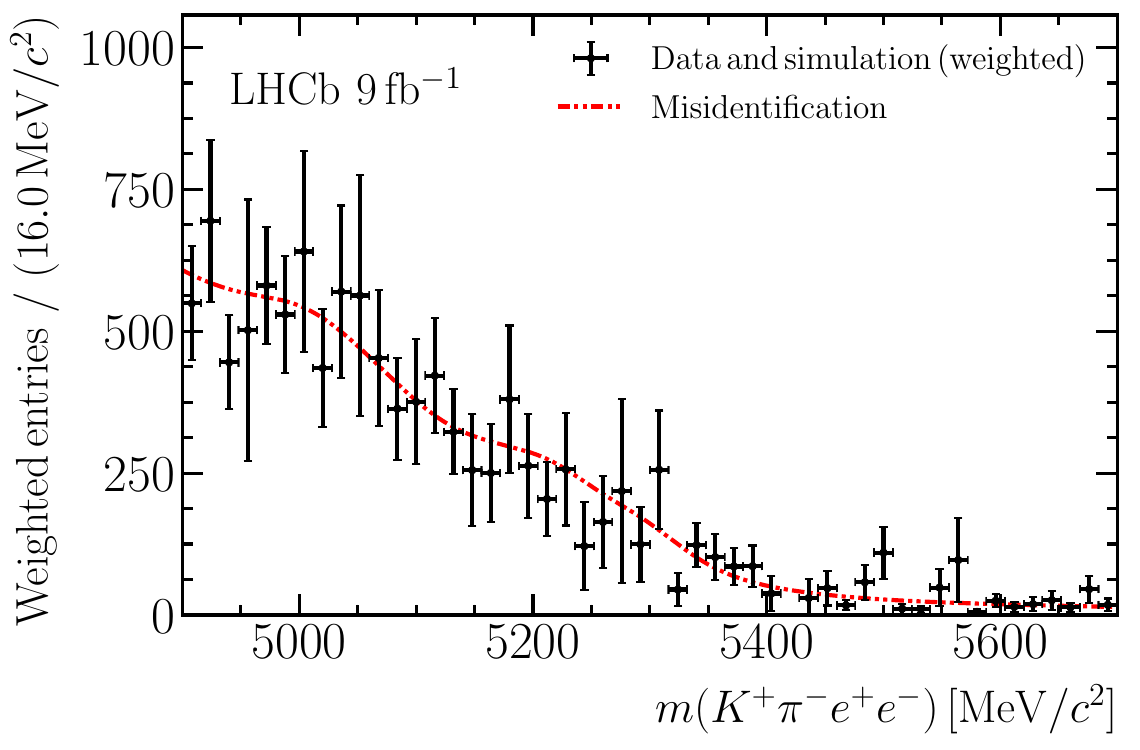} 
    \includegraphics[width=.45\textwidth, trim={0 0 0 0},clip]{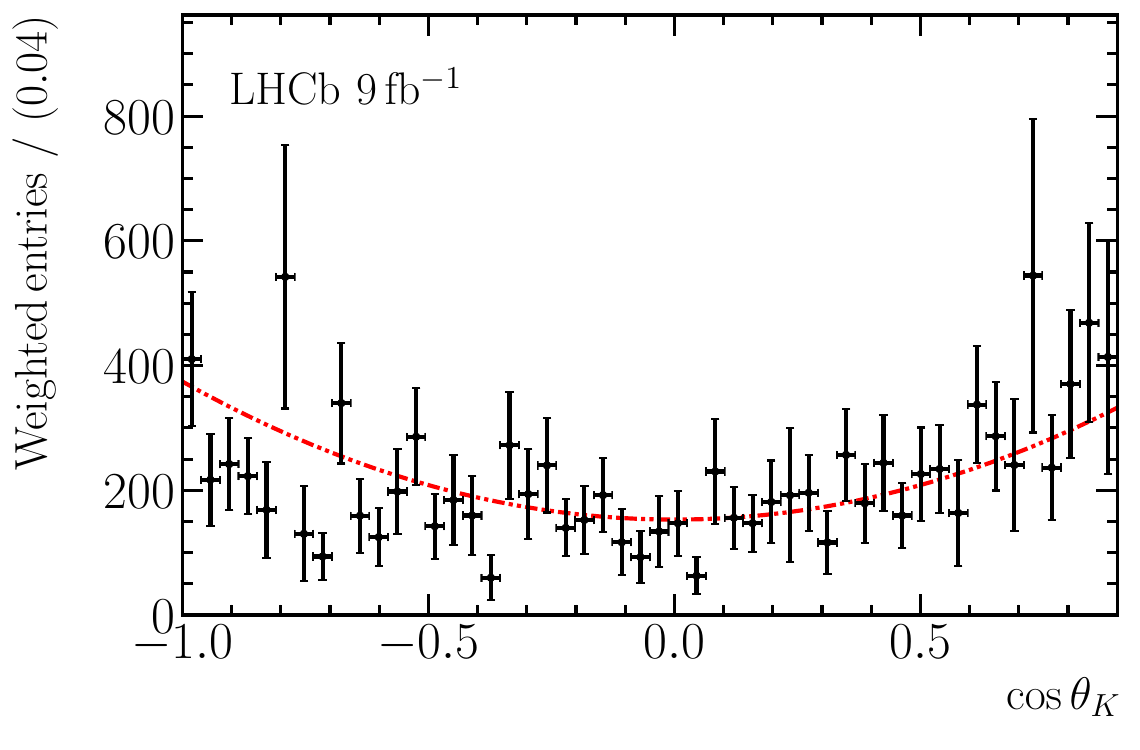} 
    \includegraphics[width=.45\textwidth, trim={0 0 0 0},clip]{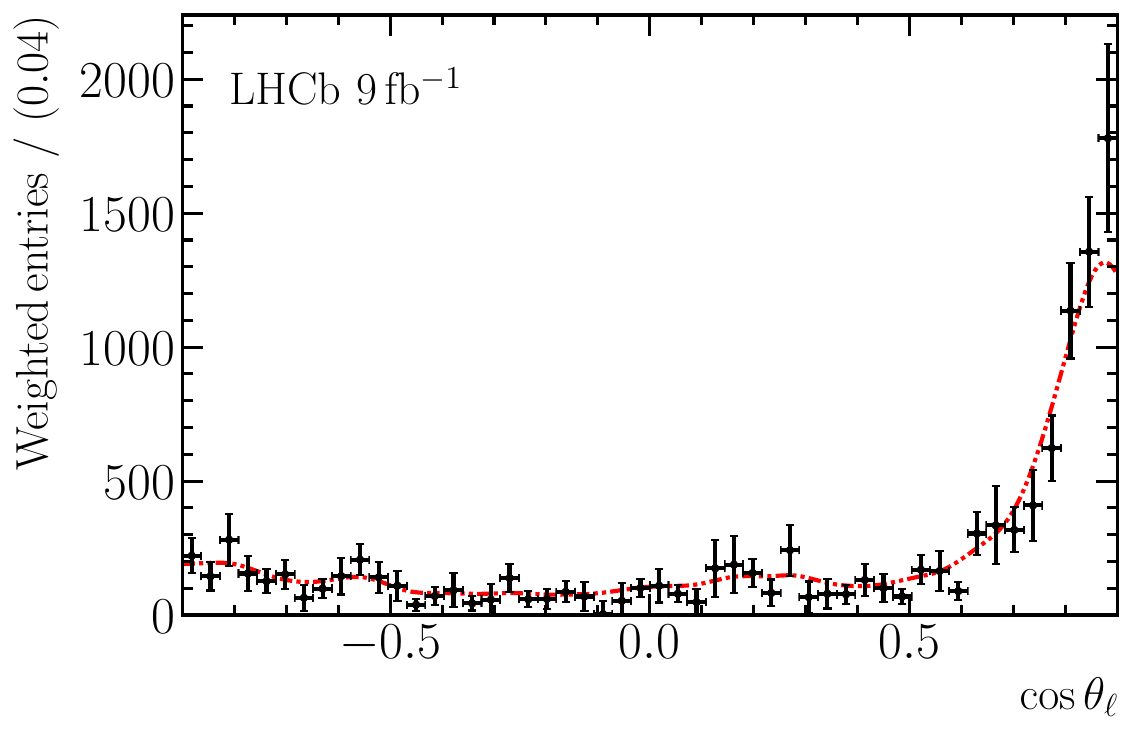} 
    \includegraphics[width=.45\textwidth, trim={0 0 0 0},clip]{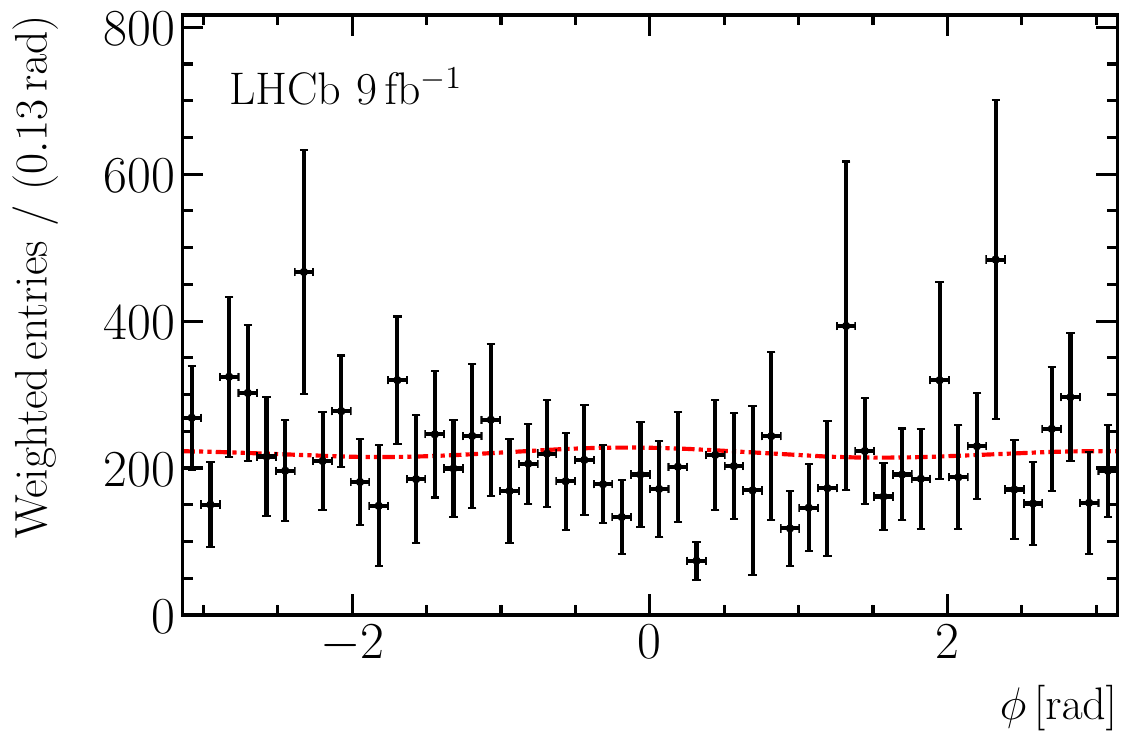} 
  \caption{Distribution of data and simulation candidates and the model that describes the invariant-mass and angular distributions of the misidentified hadronic backgrounds.
  Weights obtained from transfer functions are included, and 
  simulation is used to subtract contributions from residual signal and control-mode decays.  
  }
\label{fig:Had_sq2}
\end{figure}

\begin{figure}[!tb]
\centering
    \includegraphics[width=.45\textwidth, trim={0 0 0 0},clip]{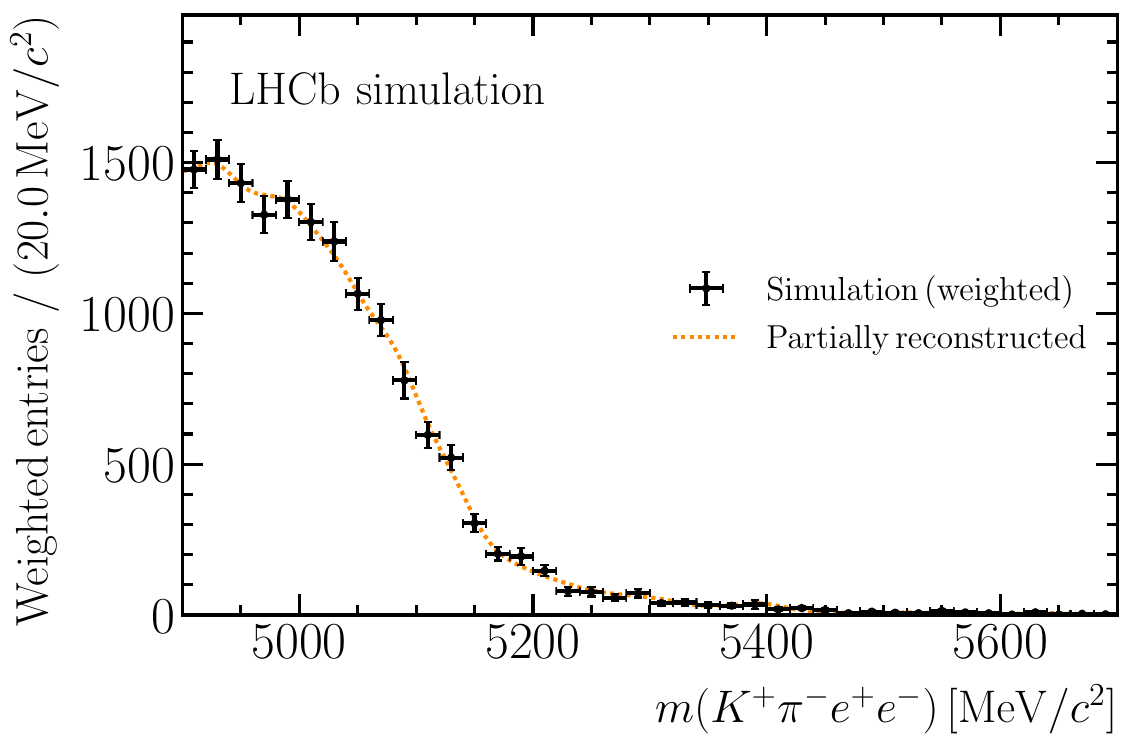} 
    \includegraphics[width=.45\textwidth, trim={0 0 0 0},clip]{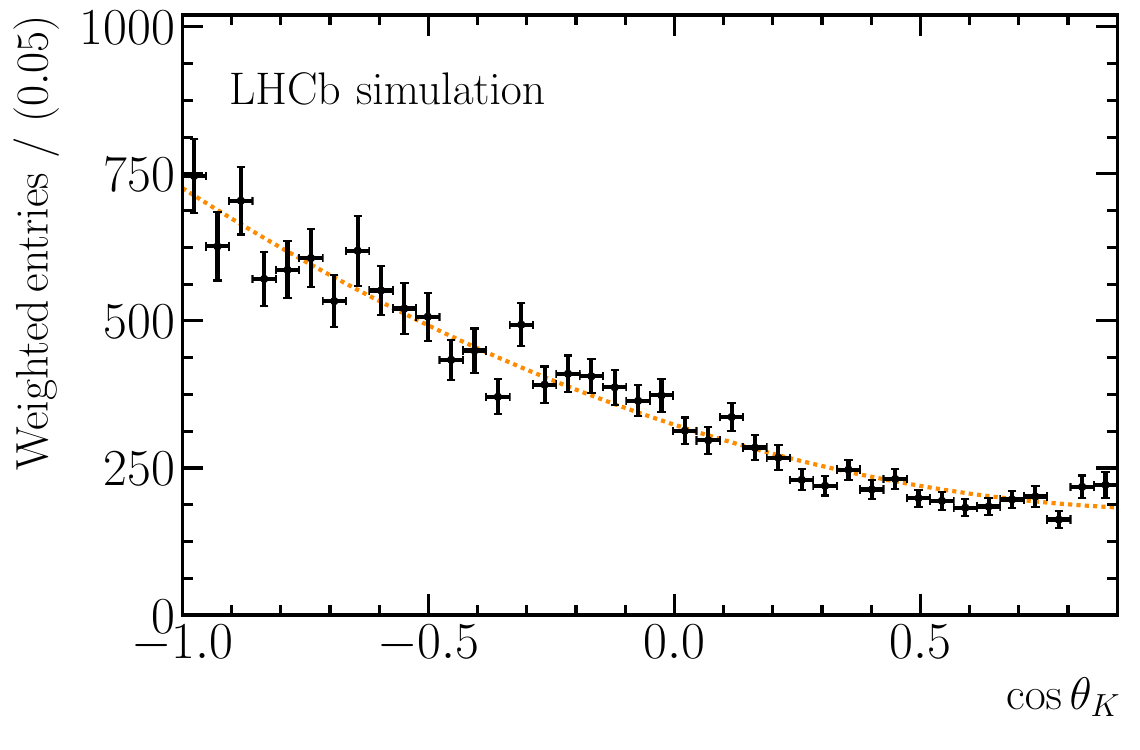} 
    \includegraphics[width=.45\textwidth, trim={0 0 0 0},clip]{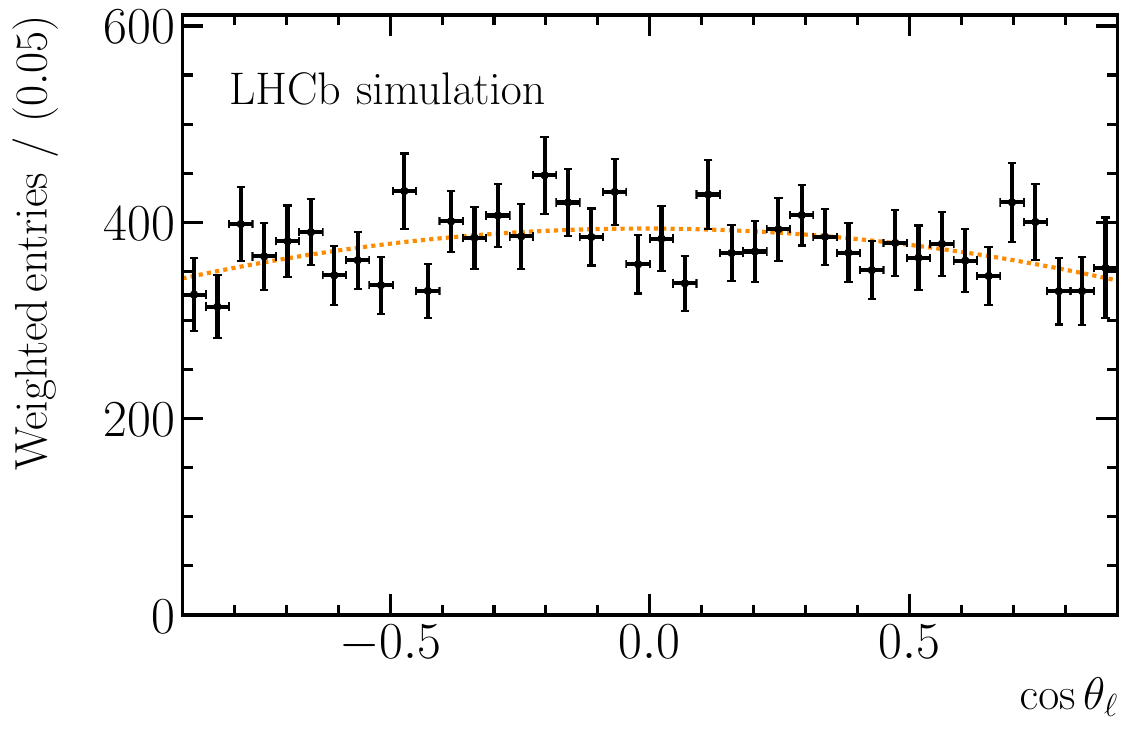} 
    \includegraphics[width=.45\textwidth, trim={0 0 0 0},clip]{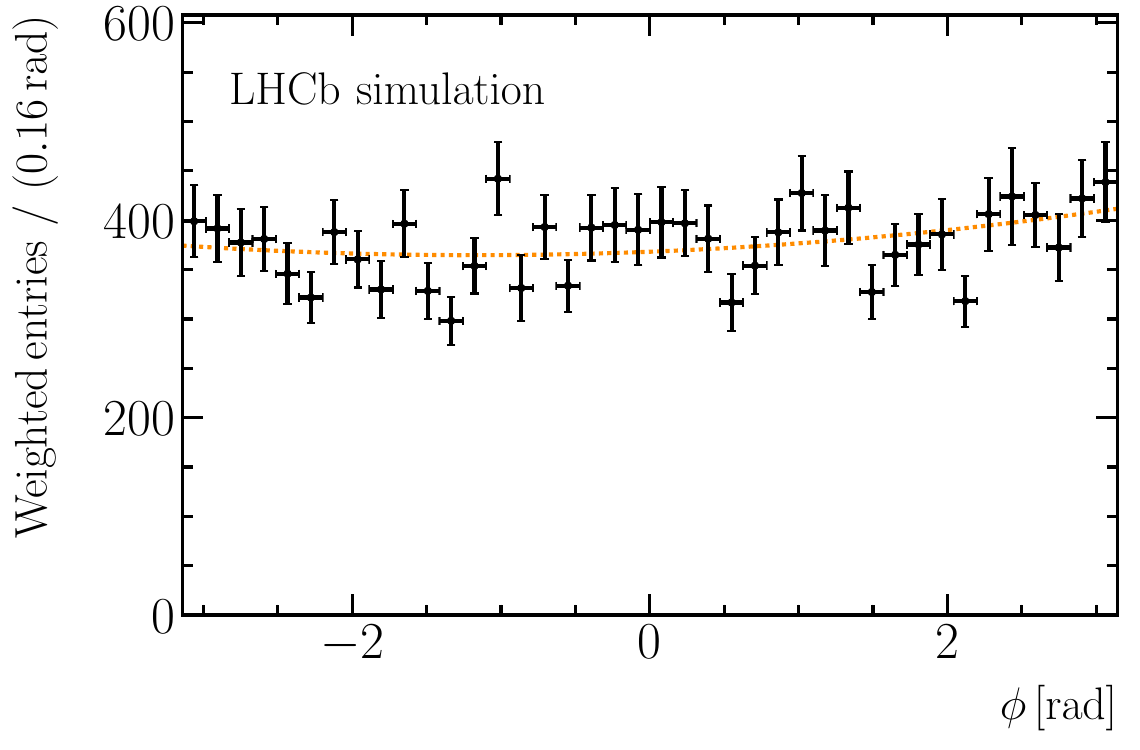} 
  \caption{Distribution of simulated phase-space $B^{+}\rightarrow K^{+}\pi^{+}\pi^{-}e^{+}e^{-}$ decays reconstructed as \mbox{$B^0\rightarrow \Kstarz e^+e^-$} and the model used to describe the partially reconstructed background. Data-driven correction weights are included to reproduce the resonant structures in the $\Kp\pip\pim$ spectrum.
  }
\label{fig:PR_sq2}
\end{figure}

Models used to describe background components and the result of the control-mode fits are illustrated in the following. In all cases, the samples of the three run periods are combined, their corresponding models are added accordingly, and only the resulting function is shown. Effective acceptance weights are always included, and additional correction weights are used in specific cases.
The model used to describe the mass distribution of the misidentified $\Lambda_b^0\rightarrow p\Km J/\psi(\rightarrow e^+e^-)$ decays is shown in Fig.~\ref{fig:data_fit_control_Lb_mass}. The result of the mass fits to the control mode used to determine the shift and scaling parameters of the signal mass model is shown in Fig.~\ref{fig:data_fit_control_mass}. The DSL angular model is shown in Fig.~\ref{fig:DSL_step1_sq2} and the full combinatorial and DSL models are shown in Fig.~\ref{fig:DSL_step2_sq2}. The models used for the misidentified hadronic and partially reconstructed backgrounds are shown in Figs.~\ref{fig:Had_sq2} and~\ref{fig:PR_sq2}, respectively.

\clearpage

\section{Correlation among angular observables}
\label{sec:app_correlation_matrices}
Correlation matrices for the statistical uncertainties of the angular observables are given in Tables~\ref{tab:corr_sq2_S} and~\ref{tab:corr_sq2_P}.
Correlations of the systematic uncertainties among observables are shown in Tables~\ref{tab:corr_syst_sq2_S} and~\ref{tab:corr_syst_sq2_P}.
Statistical correlations are small in almost all cases, with the largest value of 20\% found for $\FL$ and $P_2$. Systematic correlations are generally small, although they can be substantial for some observables. Pairs of observables which have correlations exceeding 20\% are $\FL$ and $\AFB$, $S_4$ and $S_5$, $\FL$ and $P_4^{\prime}$, $\FL$ and $P_5^{\prime}$, and $P_4^{\prime}$ and $P_5^{\prime}$.

\begin{table}[!b]
\sisetup{separate-uncertainty}
\centering
\caption{Correlation matrix for the $\it{S}$-basis angular observables.}
\begin{tabular}{l S[table-format = 2.2] S[table-format = 2.2] S[table-format = 2.2] S[table-format = 2.2] S[table-format = 2.2] S[table-format = 2.2] S[table-format = 2.2] S[table-format = 2.2]}
\multicolumn{1}{l}{} & \multicolumn{1}{c}{$\FL$} & \multicolumn{1}{c}{$S_{3}$} & \multicolumn{1}{c}{$S_{4}$} & \multicolumn{1}{c}{$S_{5}$} & \multicolumn{1}{c}{$\AFB$} & \multicolumn{1}{c}{$S_{7}$ } & \multicolumn{1}{c}{$S_{8}$ } & \multicolumn{1}{c}{$S_{9}$ } \\ \hline 
 $\FL$          & 1.00 & 0.01 & -0.07 & -0.00 & 0.06 & -0.01 & -0.04 & -0.06\\ 
 $S_{3}$          &        & 1.00 & -0.07 & -0.02 & 0.05 & 0.10 & -0.08 & -0.01 \\
 $S_{4}$          &        &        & 1.00 & -0.10 & -0.10 & -0.07 & 0.09 & 0.09 \\
 $S_{5}$          &        &        &        & 1.00 & -0.05 & 0.06 & -0.04 & -0.03 \\
 $\AFB$ &        &        &        &        & 1.00 & 0.11 & -0.07 & -0.06 \\
 $S_{7}$ &        &        &        &        &        & 1.00 & -0.07 & -0.14 \\
 $S_{8}$ &        &        &        &        &        &        & 1.00 & -0.01 \\
 $S_{9}$ &        &        &        &        &        &        &        & 1.00 \\
\end{tabular}
\label{tab:corr_sq2_S}
\end{table}

\begin{table}[!tb]
\sisetup{separate-uncertainty}
\centering
\caption{Correlation matrix for the $\it{P}$-basis angular observables.}
\begin{tabular}{l S[table-format = 2.2] S[table-format = 2.2] S[table-format = 2.2] S[table-format = 2.2] S[table-format = 2.2] S[table-format = 2.2] S[table-format = 2.2] S[table-format = 2.2]}
\multicolumn{1}{l}{} & \multicolumn{1}{c}{$\FL$} & \multicolumn{1}{c}{$P_{1}$} & \multicolumn{1}{c}{$P_{2}$} & \multicolumn{1}{c}{$P_{3}$} & \multicolumn{1}{c}{$P_{4}^{\prime}$} & \multicolumn{1}{c}{$P_{5}^{\prime}$ } & \multicolumn{1}{c}{$P_{6}^{\prime}$ } & \multicolumn{1}{c}{$P_{8}^{\prime}$ } \\ \hline 
 $\FL$          & 1.00 & 0.02 & -0.20 & -0.08 & -0.09 & -0.02 & -0.02 & -0.01\\ 
 $P_{1}$          &        & 1.00 & 0.04 & 0.01 & -0.07 & -0.02 & 0.10 & -0.08 \\
 $P_{2}$          &        &        & 1.00 & 0.06 & -0.07 & -0.05 & 0.11 & -0.06 \\
 $P_{3}$          &        &        &        & 1.00 & -0.08 & 0.03 & 0.14 & 0.02 \\
 $P_{4}^{\prime}$ &        &        &        &        & 1.00 & -0.10 & -0.07 & 0.09 \\
 $P_{5}^{\prime}$ &        &        &        &        &        & 1.00 & 0.06 & -0.03 \\
 $P_{6}^{\prime}$ &        &        &        &        &        &        & 1.00 & -0.07 \\
 $P_{8}^{\prime}$ &        &        &        &        &        &        &        & 1.00 \\
\end{tabular}
\label{tab:corr_sq2_P}
\end{table}

\begin{table}[!tb]
\sisetup{separate-uncertainty}
\centering
\caption{Correlation matrix of the systematic uncertainties for the $\it{S}$-basis angular observables.}
\begin{tabular}{l S[table-format = 2.2] S[table-format = 2.2] S[table-format = 2.2] S[table-format = 2.2] S[table-format = 2.2] S[table-format = 2.2] S[table-format = 2.2] S[table-format = 2.2]}
\multicolumn{1}{l}{} & \multicolumn{1}{c}{$\FL$} & \multicolumn{1}{c}{$S_{3}$} & \multicolumn{1}{c}{$S_{4}$} & \multicolumn{1}{c}{$S_{5}$} & \multicolumn{1}{c}{$\AFB$} & \multicolumn{1}{c}{$S_{7}$ } & \multicolumn{1}{c}{$S_{8}$ } & \multicolumn{1}{c}{$S_{9}$ } \\ \hline 
 $\FL$          & 1.00 & 0.01 & -0.11 & -0.16 & -0.23 & -0.01 & 0.02 & -0.05\\ 
 $S_{3}$          &        & 1.00 & -0.02 & -0.08 & -0.01 & 0.01 & 0.01 & 0.02 \\
 $S_{4}$          &        &        & 1.00 & 0.35 & 0.01 & -0.03 & 0.01 & -0.01 \\
 $S_{5}$          &        &        &        & 1.00 & 0.08 & -0.00 & -0.03 & -0.01 \\
 $\AFB$ &        &        &        &        & 1.00 & -0.02 & -0.01 & 0.02 \\
 $S_{7}$ &        &        &        &        &        & 1.00 & 0.10 & -0.05 \\
 $S_{8}$ &        &        &        &        &        &        & 1.00 & -0.01 \\
 $S_{9}$ &        &        &        &        &        &        &        & 1.00 \\
\end{tabular}
\label{tab:corr_syst_sq2_S}
\end{table}

\begin{table}[!tb]
\sisetup{separate-uncertainty}
\centering
\caption{Correlation matrix of the systematic uncertainties for the $\it{P}$-basis angular observables.}
\begin{tabular}{l S[table-format = 2.2] S[table-format = 2.2] S[table-format = 2.2] S[table-format = 2.2] S[table-format = 2.2] S[table-format = 2.2] S[table-format = 2.2] S[table-format = 2.2]}
\multicolumn{1}{l}{} & \multicolumn{1}{c}{$\FL$} & \multicolumn{1}{c}{$P_{1}$} & \multicolumn{1}{c}{$P_{2}$} & \multicolumn{1}{c}{$P_{3}$} & \multicolumn{1}{c}{$P_{4}^{\prime}$} & \multicolumn{1}{c}{$P_{5}^{\prime}$ } & \multicolumn{1}{c}{$P_{6}^{\prime}$ } & \multicolumn{1}{c}{$P_{8}^{\prime}$ } \\ \hline 
 $F_{L}$          & 1.00 & -0.03 & -0.15 & 0.03 & -0.23 & -0.33 & -0.02 & 0.01\\ 
 $P_{1}$          &        & 1.00 & 0.00 & -0.02 & -0.02 & -0.05 & -0.01 & 0.01 \\
 $P_{2}$          &        &        & 1.00 & 0.01 & 0.06 & 0.12 & 0.01 & -0.00 \\
 $P_{3}$          &        &        &        & 1.00 & 0.00 & 0.01 & 0.04 & 0.01 \\
 $P_{4}^{\prime}$ &        &        &        &        & 1.00 & 0.42 & -0.01 & 0.01 \\
 $P_{5}^{\prime}$ &        &        &        &        &        & 1.00 & 0.01 & -0.02 \\
 $P_{6}^{\prime}$ &        &        &        &        &        &        & 1.00 & 0.10 \\
 $P_{8}^{\prime}$ &        &        &        &        &        &        &        & 1.00 \\
\end{tabular}
\label{tab:corr_syst_sq2_P}
\end{table}

\clearpage

\section{Fit to the $\boldsymbol{\Bz\to \Kstarz \mu^+\mu^-}$ decay}
\label{sec:muon_mode_fits}
\begin{figure}[!b]
\centering
    \includegraphics[width=.45\textwidth, trim={0 0 0 0},clip]{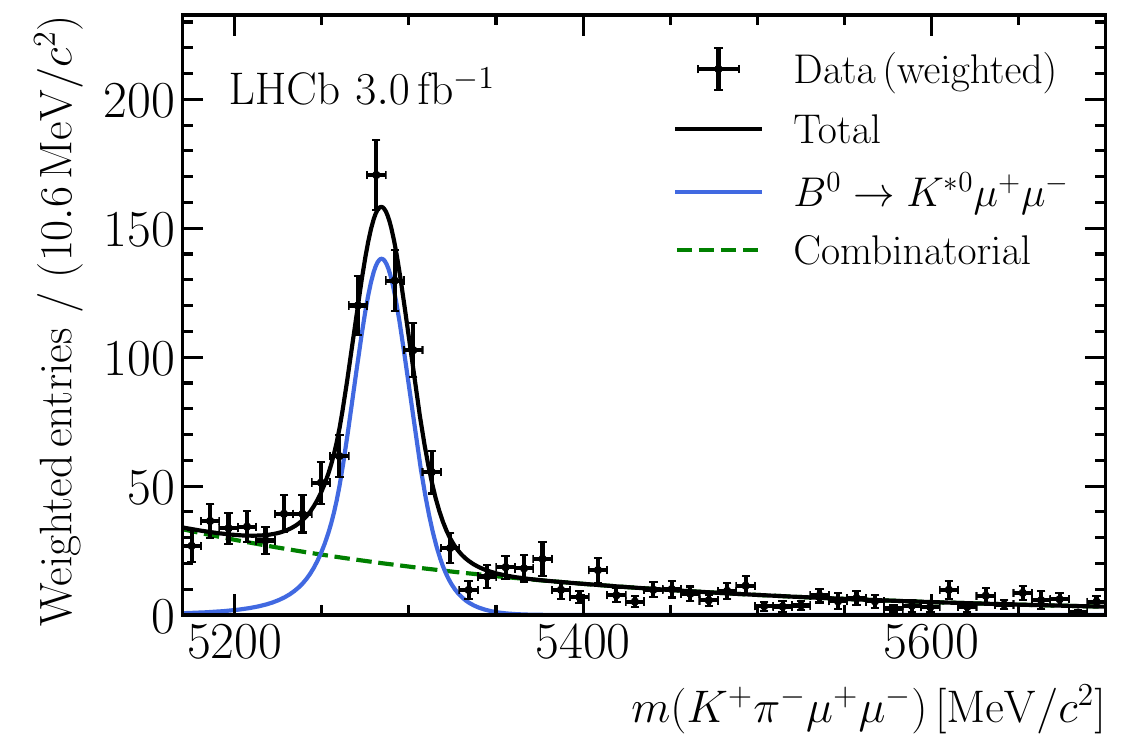} 
    \includegraphics[width=.45\textwidth, trim={0 0 0 0},clip]{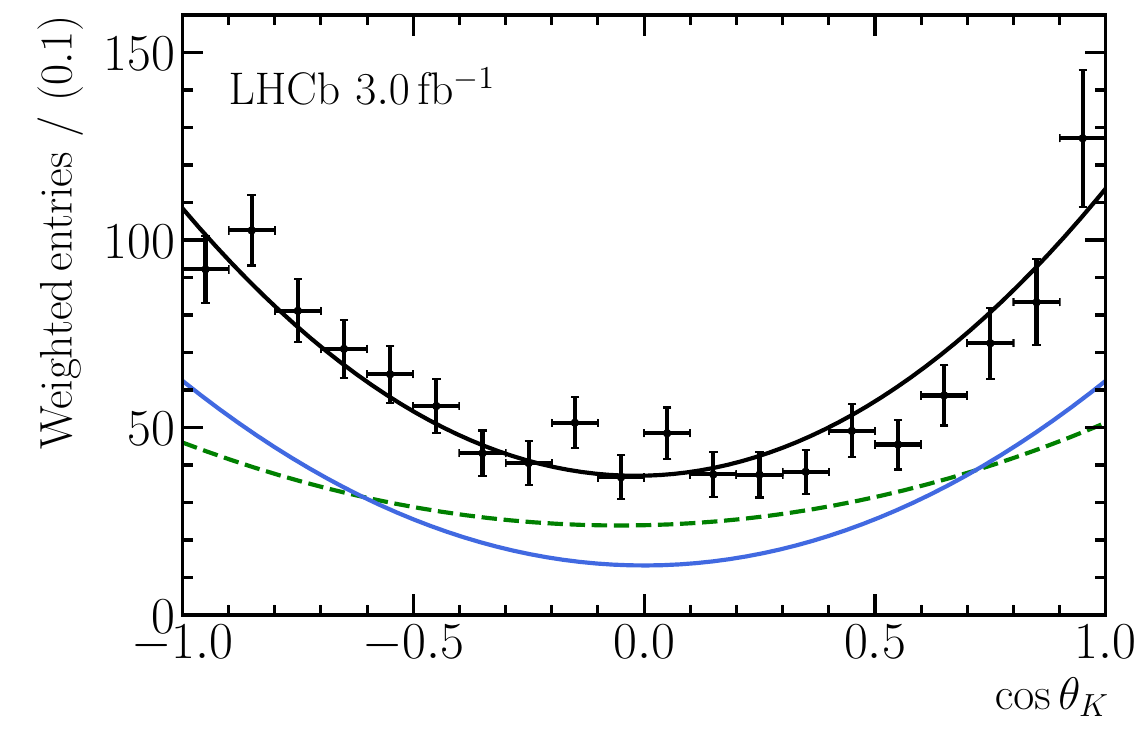} 
    \includegraphics[width=.45\textwidth, trim={0 0 0 0},clip]{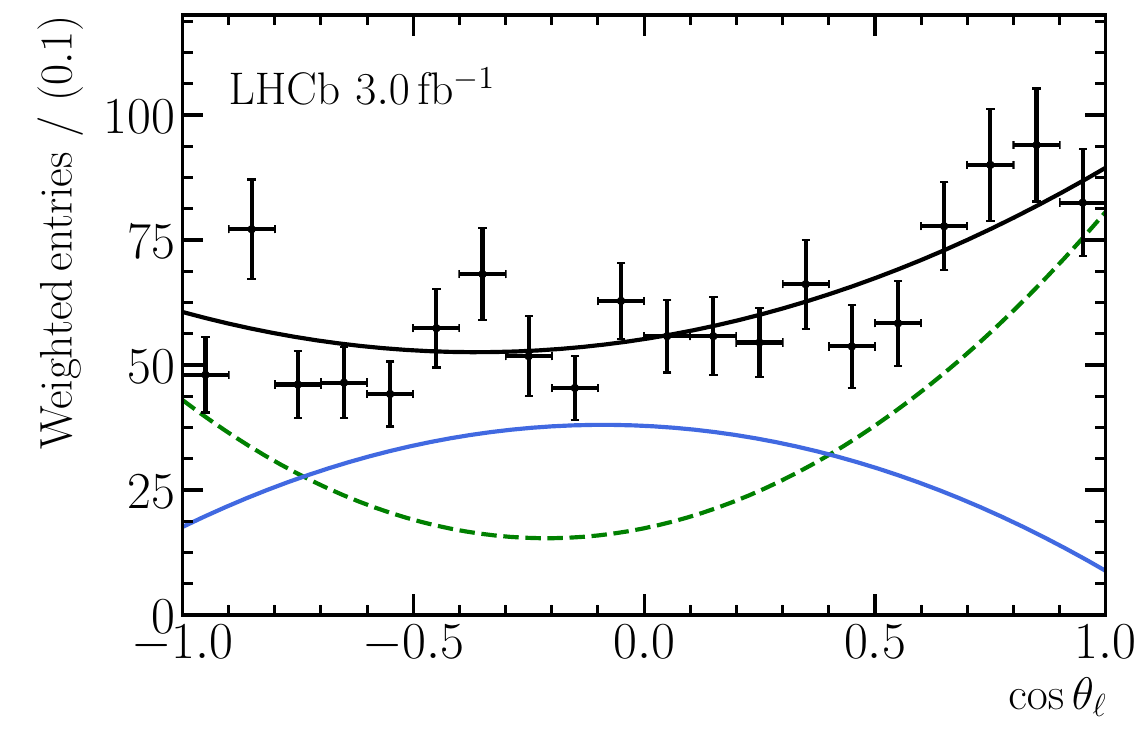} 
    \includegraphics[width=.45\textwidth, trim={0 0 0 0},clip]{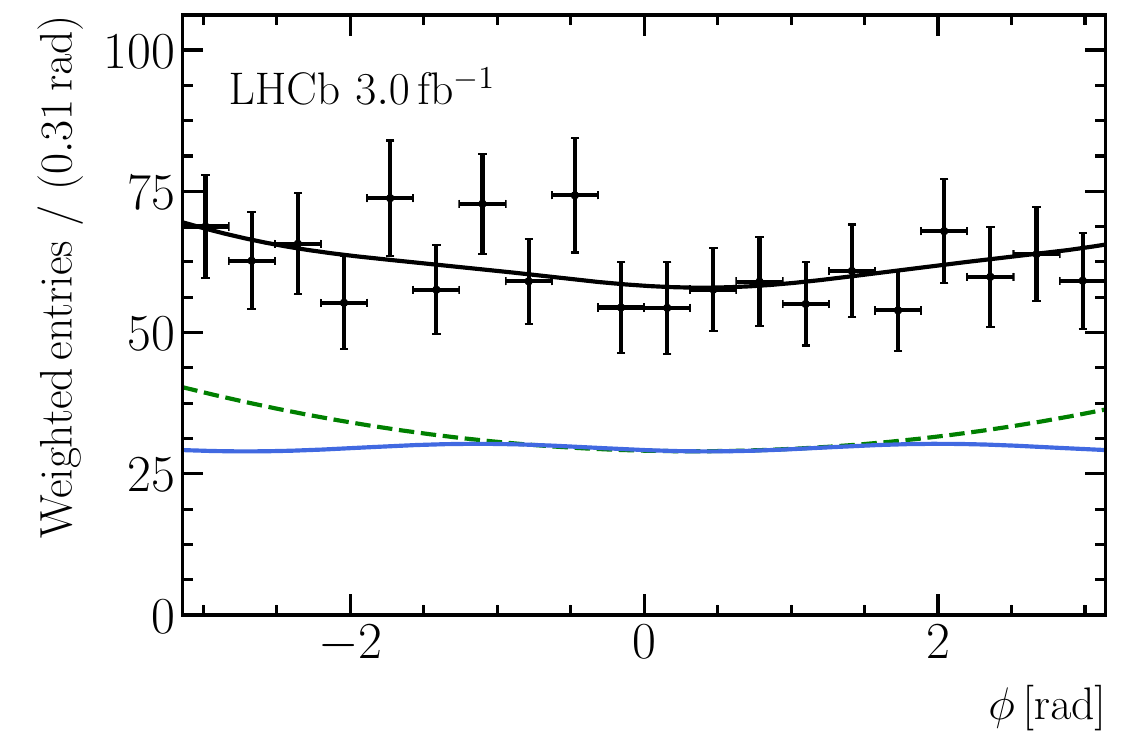} 

    \includegraphics[width=.45\textwidth, trim={0 0 0 0},clip]{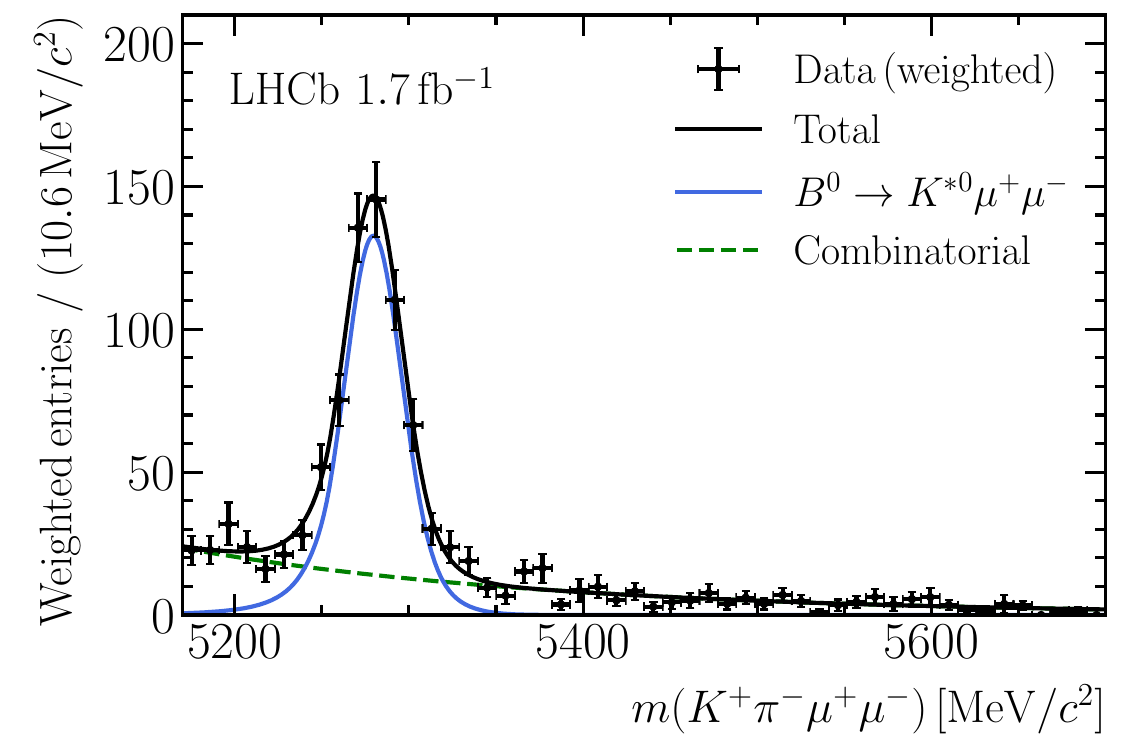} 
    \includegraphics[width=.45\textwidth, trim={0 0 0 0},clip]{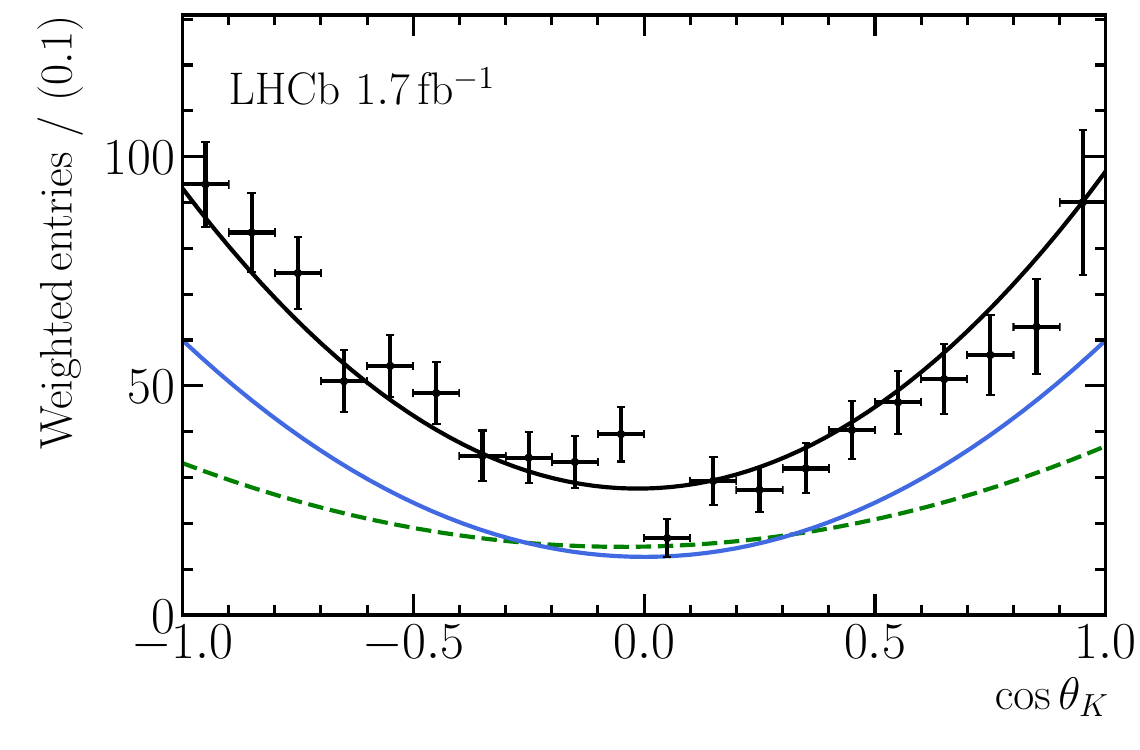} 
    \includegraphics[width=.45\textwidth, trim={0 0 0 0},clip]{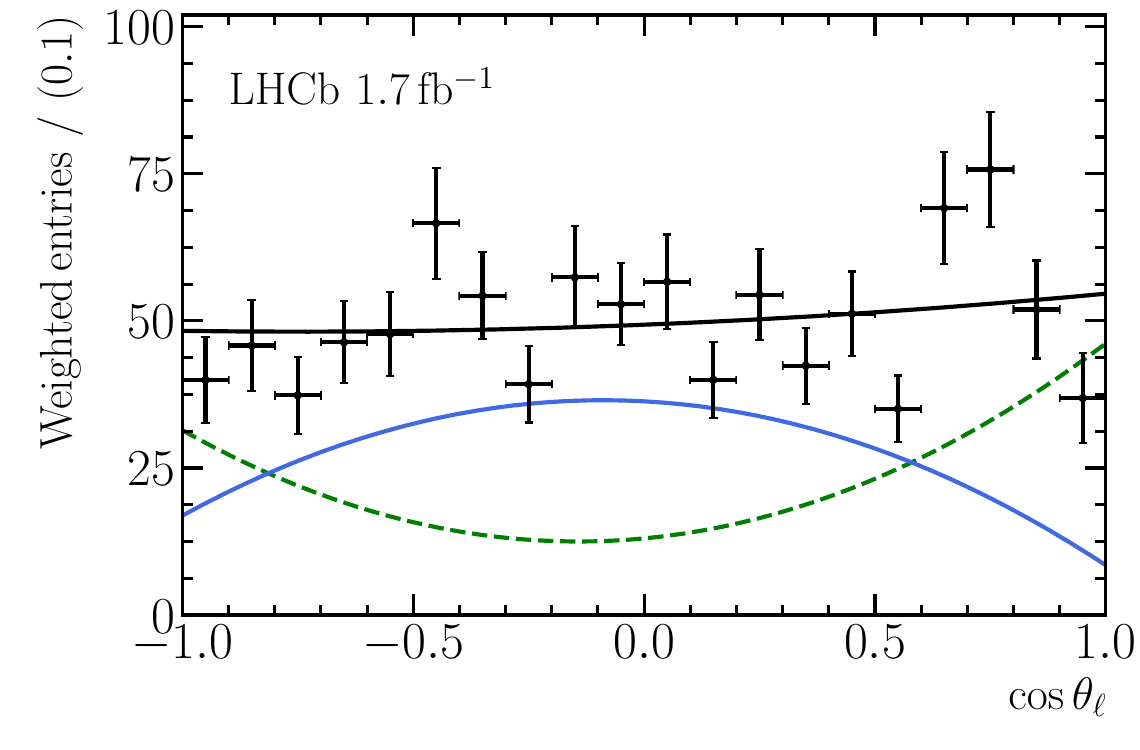} 
    \includegraphics[width=.45\textwidth, trim={0 0 0 0},clip]{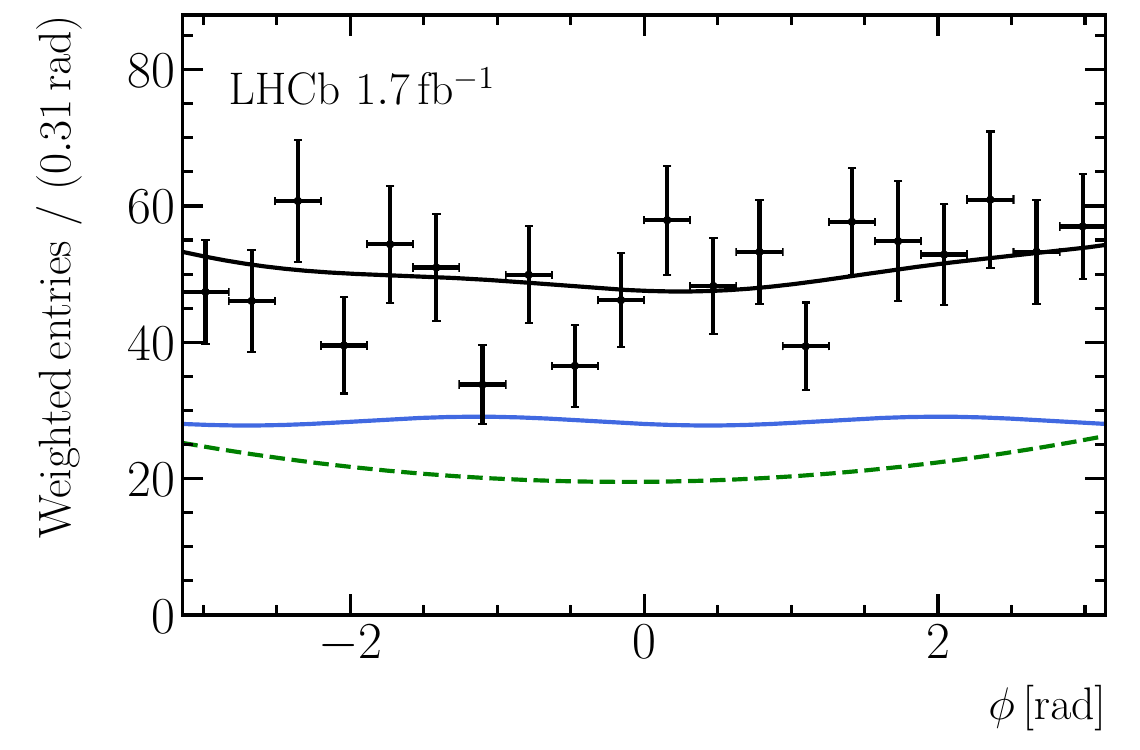}  
  \caption{Weighted invariant-mass and angular distributions of $\Bz\to \Kstarz \mu^+\mu^-$ candidates in the (top two rows) Run1 and (bottom two rows) 2016 data samples. The signal distribution is shown with a solid blue line, and the combinatorial background is shown with a dashed green line. The solid black line corresponds to the full fit function.
  }
\label{fig:muon_sq2}
\end{figure}

The result of the fit to the $\Bz\to \Kstarz \mu^+\mu^-$ candidates is shown in Fig.~\ref{fig:muon_sq2}. The data samples used are the same as those analysed in Ref.~\cite{LHCb-PAPER-2020-002}.
The fit strategy is aligned to that of the electron mode. In particular, information from the $\Kp\pim$ system is not used, the S-wave and interference terms are neglected, and the effective acceptance functions are parametrised using trignometric terms for $\phi$. The $q^2$ variable calculated without constraints is used to define the measurement region.

\clearpage

\addcontentsline{toc}{section}{References}
\bibliographystyle{LHCb}
\bibliography{main,standard,LHCb-PAPER,LHCb-CONF,LHCb-DP,LHCb-TDR}

\newpage
\centerline
{\large\bf LHCb collaboration}
\begin
{flushleft}
\small
R.~Aaij$^{37}$\lhcborcid{0000-0003-0533-1952},
A.S.W.~Abdelmotteleb$^{56}$\lhcborcid{0000-0001-7905-0542},
C.~Abellan~Beteta$^{50}$\lhcborcid{0009-0009-0869-6798},
F.~Abudin{\'e}n$^{56}$\lhcborcid{0000-0002-6737-3528},
T.~Ackernley$^{60}$\lhcborcid{0000-0002-5951-3498},
A. A. ~Adefisoye$^{68}$\lhcborcid{0000-0003-2448-1550},
B.~Adeva$^{46}$\lhcborcid{0000-0001-9756-3712},
M.~Adinolfi$^{54}$\lhcborcid{0000-0002-1326-1264},
P.~Adlarson$^{81}$\lhcborcid{0000-0001-6280-3851},
C.~Agapopoulou$^{14}$\lhcborcid{0000-0002-2368-0147}
% ...

{\footnotesize \it

% ...
$^{14}$Universit{\'e} Paris-Saclay, CNRS/IN2P3, IJCLab, Orsay, France\\
% ...
$^{37}$Nikhef National Institute for Subatomic Physics, Amsterdam, Netherlands\\
% ...
$^{46}$Instituto Galego de F{\'\i}sica de Altas Enerx{\'\i}as (IGFAE), Universidade de Santiago de Compostela, Santiago de Compostela, Spain\\
% ...
$^{50}$Physik-Institut, Universit{\"a}t Z{\"u}rich, Z{\"u}rich, Switzerland\\
% ...
$^{54}$H.H. Wills Physics Laboratory, University of Bristol, Bristol, United Kingdom\\
% ...
$^{56}$Department of Physics, University of Warwick, Coventry, United Kingdom\\
% ...
$^{59}$School of Physics and Astronomy, University of Glasgow, Glasgow, United Kingdom\\
$^{60}$Oliver Lodge Laboratory, University of Liverpool, Liverpool, United Kingdom\\
% ...
$^{68}$Syracuse University, Syracuse, NY, United States\\
% ...
$^{81}$Department of Physics and Astronomy, Uppsala University, Uppsala, Sweden, associated to $^{59}$\\
% ...
}
\end{flushleft}

\end{document}